\documentclass{elsart}
\usepackage{graphics}
\input{epsf}
\usepackage{times,latexsym,euscript,amstext,amssymb,amsbsy}

\hoffset   =- .5 cm
\def\beqn{\begin{eqnarray}}
\def\eeqn{\end{eqnarray}}
\def\barr{\begin{array}}
\def\earr{\end{array}}
\def\btab{\begin{tabular}}
\def\etab{\end{tabular}}
\def\bite{\begin{itemize}}
\def\eite{\end{itemize}}
\def\bcen{\begin{center}}
\def\ecen{\end{center}}

\def\rmqp{{ \hbox{\rm q}'} }
\def\rmq{{\hbox{\rm q}}}
\def\qt0{\tilde{q}_0}

\def\calm{{\mathcal M}}
\def\dcalm{\Delta{\mathcal M}}

\def\kdagger{K\hspace{-0.3cm}/}

\def\qqs{q\!\cdot\!q'}

\def\bd{\mathbf{d}}
\def\bP{\mathbf{P}}
\def\br{\mathbf{r}}
\def\bEo{\mathbf{E}_{\omega}}
\def\be{\begin{equation}}
\def\ee{\end{equation}}
\def\hate{\hat{\mathbf{e}}}
\def\bq{\mathbf{q}}
\def\bvare{\boldsymbol{\varepsilon}}
\def\bsig{\boldsymbol{\sigma}}
\def\bmu{\boldsymbol{\mu}}
\def\sc{\scriptsize}

\begin{document}
\begin{frontmatter}

\title{Dispersion relations in real and virtual Compton scattering}

\author{D. Drechsel $^1$, B. Pasquini $^{2,3}$, M. Vanderhaeghen $^1$}
\address{$^1$ Institut f\"ur Kernphysik, Johannes Gutenberg-Universit\"at,
D-55099 Mainz, Germany}
\address{$^2$ ECT* - European Centre for Theoretical Studies in Nuclear
Physics and Related Areas, I-38050 Villazzano (Trento), Italy;
and INFN, Trento}
\address{$^3$ Dipartimento di Fisica, Universit\`a degli Studi di Trento,
I-38050 Povo (Trento)}

\begin{abstract}
A unified presentation is given on the use of dispersion relations
in the real and virtual Compton scattering processes off the
nucleon. The way in which dispersion relations for Compton
scattering amplitudes establish connections between low energy
nucleon structure quantities, such as polarizabilities or
anomalous magnetic moments, and the nucleon excitation spectrum is
reviewed. 
We discuss various sum rules for forward real and virtual Compton
scattering, such as the Gerasimov-Drell-Hearn sum rule and its
generalizations, the Burkhardt-Cottingham sum rule, as well as sum
rules for forward nucleon polarizabilities, and review their
experimental status. 
Subsequently, we address the general case of real Compton scattering (RCS).
Various types of dispersion relations for RCS are presented 
as tools for extracting nucleon polarizabilities from the RCS data. 
The information on nucleon polarizabilities gained in this way 
is reviewed and the nucleon structure information encoded 
in these quantities is discussed.
The dispersion relation formalism is then extended to virtual
Compton scattering (VCS). The information on generalized nucleon
polarizabilities extracted from recent VCS experiments is
described, along with its interpretation in nucleon structure
models. As a summary, the physics content of the existing data is
discussed and some perspectives for future theoretical and
experimental activities in this field are presented.
\end{abstract}


\end{frontmatter}

\vspace{1.5cm}
{\it to appear in Physics Reports}

\newpage

\tableofcontents
\section{Introduction}

The internal structure of the strongly interacting particles has
been an increasingly active area of experimental and theoretical
research over the past 5 decades. Precision experiments at high
energy have clearly established Quantum Chromodynamics (QCD) as
the underlying gauge theory describing the interaction between
quarks and gluons, the elementary constituents of hadronic matter.
However, the "running" coupling constant of QCD grows at low
energies, and these constituents are confined to colorless
"hadrons", the mesons and baryons, which are the particles
eventually observed by the detection devices. Therefore, we have
to live with a dichotomy: The small value of the coupling constant
at high energies allows for an interpretation of the experiments
in terms of perturbative QCD, while the large value at low
energies calls for a description in terms of the hadronic degrees
of freedom, in particular in the approach developed as Chiral
Perturbation Theory.
\newline
\indent Between these two regions, at excitation energies between
a few hundred~MeV and 1-2~GeV, lies the interesting region of
nucleon resonance structures which is beyond the scope of either
perturbation scheme. There is some hope that this regime will
eventually be described by numerical solutions of QCD through
lattice gauge calculations. At present, however, our understanding
of resonance physics is still mostly based on phenomenology. In
the absence of a descriptive theory it is essential to extract new
and precise hadronic structure information, and in this quest
electromagnetic probes have played a decisive role. In particular,
high precision Compton scattering experiments have become possible
with the advent of modern electron accelerators with high current
and duty factor, and of laser backscattering facilities, and in
combination with high precision and large acceptance detectors.
This intriguing new window offers, among other options, the
possibility for precise and detailed investigations of the nucleon
polarizability as induced by the applied electromagnetic multipole
fields.
\newline
\indent The polarizability of a composite system is an elementary
structure constant, just as are its size and shape. In a
macroscopic medium, the electric and magnetic dipole
polarizabilities are related to the dielectric constant and the
magnetic permeability, and these in turn determine the index of
refraction. These quantities can be studied by considering an
incident electromagnetic wave inducing dipole oscillations in the
constituent atoms or molecules of a target medium. These
oscillations then emit dipole radiation leading, by way of
interference with the incoming wave, to  the complex amplitude of
the transmitted wave. A general feature of these processes is the
dispersion relation of Kronig and Kramers~\cite{Kro26}, which
connects the real refraction index as function of the frequency
with a weighted integral of the extinction coefficient over all
frequencies.
\newline
\indent Dispersion theory in general relies on a few basic
principles of physics: relativistic covariance, causality and
unitarity. As a first step a complete set of amplitudes has to be
constructed, in accordance with relativity and without kinematical
singularities. Next, causality requires certain analytic
properties of the amplitudes, which allow for a continuation of
the scattering amplitudes into the complex plane and lead to
dispersion relations connecting the real and imaginary parts of
these amplitudes. Finally, the imaginary parts can be replaced by
absorption cross sections by the use of unitarity, and as a result
we can, for example, complete the Compton amplitudes from
experimental information on photoabsorption and photo-induced
reactions.
\newline
\indent In the following Sec.~2 we first discuss the classical
theory of dispersion and absorption in a medium, and briefly
compare the polarizability of macroscopic matter and microscopic
systems, atoms and nucleons. This is followed by a review of
forward  Compton scattering and its connection to total absorption
cross sections. Combining dispersion relations and low energy
theorems, we obtain sum rules for certain combinations of the
polarizabilities and other ground state properties, e.g., the
Gerasimov-Drell-Hearn sum rule for real
photons~\cite{Ger65,Dre65}, and the much debated
Burkhardt-Cottingham sum rule for virtual Compton
scattering~\cite{BC70} as obtained from radiative electron
scattering.
\newline
\indent 
We then address the general case of real Compton
scattering in Sec.~3. Besides the electric and magnetic (dipole)
polarizabilities of a scalar system, the spin of the nucleon leads
to four additional spin or vector polarizabilities, and higher
multipole polarizabilities will appear with increasing photon
energy. We show how these polarizabilities can be obtained from
photon scattering and photoexcitation processes through a combined
analysis based on dispersion theory. The results of such an
analysis are then compared in detail with the experimental data
and predictions from theory. In the Sec.~4 we discuss the more
general case of virtual Compton scattering, which can be achieved
by radiative electron-proton scattering. Such experiments have
become possible only very recently. The non-zero four-momentum
transfer squared of the virtual photon allows us to study
generalized polarizabilities as function of four-momentum transfer
squared and therefore, in some sense, to explore the spatial
distribution of the polarization effects. In the last Section, we
summarize the pertinent features of our present knowledge on the
nucleon polarizability and conclude by outlining some remaining
challenges for future work.
\newline
\indent
This review is largely based on dispersion theory whose
development is related to Heisenberg's idea that the interaction
of particles can be described by their behavior at large
distances, i.e., in terms of the S matrix~\cite{Hei43}. The
practical consequences of this program were worked out by
Mandelstam and others~\cite{Man62}. An excellent primer for the
beginner is the textbook of Nussenzveig~\cite{Nus72}. In order to
feel comfortable on Mandelstam planes and higher Riemann sheets,
the review of Hoehler~\cite{Hoe83} is an absolute must for the
practitioner. Concerning the structure aspect of our review, we
refer the reader to a general treatise of the electromagnetic
response of hadronic systems by Boffi {\it et al.}~\cite{Bof96},
and to the recent book of Thomas and Weise~\cite{Tho01}, which is
focused on the structure aspects of the nucleon.

\section{Forward dispersion relations and sum rules for real and
virtual Compton scattering}

\subsection{Classical theory of dispersion and absorption in a medium}

The classical theory of Lorentz describes the dispersion in a
medium in terms of electrons bound by a harmonic force. In the
presence of a monochromatic external field, $\bEo$, the equations
of motion take the form
\beqn
\label{DDeq2.1.1}
\left (\frac{\partial^2}{\partial t^2}+2\gamma_j
\frac{\partial}{\partial t}+\omega_j^2 \right ) {\br}(t) =
-\frac{e}{m}\,{\bEo}\,e^{-i\omega t}\ ,
\eeqn
with $-e$ the charge\footnote{In Sec. 2.1 we shall use Gaussian
units as in most of the literature on theoretical electrodynamics,
i.e., the fine structure constant takes the form
$\alpha_{em}=e^2/c\hbar\approx1/137$ and the classical electron
radius is $r_{cl}=e^2/mc^2$. In all later sections the
Heaviside-Lorentz units will be used in order to concur with the
standard notation of particle physics.} and $m$ the mass of the
electron, and $\gamma_j>0$ and $\omega_j>0$ the damping constant
and oscillator frequency, respectively, of a specific bound state
$j$. The stationary solution for the displacement is then given by
\beqn
\label{DDeq2.1.2}
{\br}_j(t) = - \frac{e\,{\bEo}\,
e^{-i\omega t}}
{m(\omega_j^2-2i\gamma_j\,\omega-\omega^2)}\ ,
\eeqn
and the polarization $\bP$ is obtained by summing the
individual dipole moments ${\bd}_j=-e\,{\br}_j$ over all electrons
and oscillator frequencies in the medium,
\beqn
\label{DDeq2.1.3}
{\bP}(t) = \sum_j N_j \frac{e^2\,{\bEo}\,e^{-i\omega t}}
{m(\omega_j^2-2i\gamma_j\,\omega-\omega^2)} =
{\bP}_{\omega}\,e^{-i\omega t}\ ,
\eeqn
where $N_j $ is the number of electrons per unit volume, in the
state $j$. The dielectric susceptibility $\chi$ is defined by
\beqn
\label{DDeq2.1.4}
{\bP}_{\omega} = \chi(\omega){\bEo}\ ,
\eeqn
with
\beqn
\label{DDeq2.1.5}
\chi(\omega) = \frac{e^2}{m}\sum_j
\frac{N_j}{\omega_j^2-2i\,\gamma_j\,\omega-\omega^2}\ .
\eeqn
We observe at this point that $\chi(\omega)$
\begin{enumerate}
\item[(I)]
is square integrable in the upper half-plane $(I_+)$ for any line
parallel to the real $\omega$~axis, and
\item[(II)]
has singularities only in the lower-half plane $(I_-)$ in the form
of pairs of poles at
\beqn
\label{DDeq2.1.6}
\omega_{\pm} = \pm\sqrt{\omega_j^2-\gamma_j^2}
-i\gamma_j\ .
\eeqn
\end{enumerate}
According to Titchmarsh's theorem these observations have the
following consequences:
\newline
\indent
The Fourier transform
\beqn
\label{DDeq2.1.7}
\chi(t) = \frac{1}{2\pi}\int_{-\infty}^{\infty}\chi(\omega)\,
e^{-i\omega t}\,d\omega
\eeqn
is causal, i.e., the dielectric susceptibility and the polarization
of the medium build up only after the electric field is applied,
and the real and imaginary parts of $\chi$ are Hilbert transforms,
\begin{eqnarray}
{\mbox{Re}}\,\chi(\omega) & = & \frac{1}{\pi} {\mathcal P}
\int_{-\infty}^{\infty}\frac{{\mbox{Im}}\chi(\omega')}
{\omega'-\omega}\, d\omega'\ , \nonumber \\
{\mbox{Im}}\,\chi(\omega) & = & - \frac{1}{\pi} {\mathcal P}
\int_{-\infty}^{\infty}\frac{{\mbox{Re}}\,\chi(\omega')}
{\omega'-\omega}\, d\omega' \ ,
\label{DDeq2.1.8}
\end{eqnarray}
where ${\mathcal P}$ denotes the principal value integral.
\newline
\indent
Applying the convolution theorem for Fourier transforms to
Eq.~(\ref{DDeq2.1.4}), we obtain
\beqn
\label{DDeq2.1.9}
{\bP}(t) = \int_{-\infty}^{\infty} \chi(t-t')\,{\mathbf{E}}(t')\,dt'\ ,
\eeqn
with general time profiles ${\bP}(t)$ and ${\mathbf{E}}(t)$ of medium
polarization and external field, respectively, constructed
according to Eq.~(\ref{DDeq2.1.7}).
\newline
\indent The proof of causality follows from integrating the
dielectric susceptibility over a contour $C_+$ along the real
$\omega$ axis, for $-R\le\omega\le R$, and closed by a large half
circle with radius $R$ in the upper part of the complex
$\omega$-plane. Since no singularities appear within this contour,
\beqn
\label{DDeq2.1.10}
\oint_{C+}\chi(\omega)\,e^{-i\omega\tau}d\omega = 0\ .
\eeqn
We make contact with the Fourier transform of
Eq.~(\ref{DDeq2.1.7}) by blowing up the contour
$(R\rightarrow\infty)$ and studying the convergence along the half
circle. According to our observation (I) the function $\chi$
itself is square integrable in $I_+$, and therefore the
convergence depends on the behavior of the exponential function
exp$(-i\omega\tau)$, which depends on the sign of $\tau$. In the
case of $\tau<0$ the convergence is improved by the exponential,
and the contribution of the half-circle vanishes in the limit
$R\rightarrow\infty$. Combining Eqs.~(\ref{DDeq2.1.7}) and
(\ref{DDeq2.1.10}), we then obtain
\beqn
\label{DDeq2.1.11}
\chi(\tau) = 0 \ \ \ \ \ {\mbox{for}}\ \tau<0\ ,
\eeqn
which enforces causality, as becomes obvious by inspecting
Eq.~(\ref{DDeq2.1.9}): The electric field ${\mathbf{E}}(t')$ will affect
the polarizability ${\mathbf{P}}(t)$ only at some later time,
$\tau=t-t'>0$. For such time, $\tau>0$, the contour integral
$C_+$ is of course useless for our purpose, because the
exponential overrides the convergence of $\chi$ in $I_+$.
Therefore, the contour has to be closed in the lower half-plane,
which picks up the contributions from the
singularities in $I_-$. We note in passing that Eq.~(\ref{DDeq2.1.11})
describes the nonrelativistic causality condition, which has to be
sharpened by the postulate of relativity that no signal can move
faster than the velocity of light. Furthermore, causality is found to
be a direct consequence of  analyticity of the Green function
$\chi(\omega)$, which in the Lorentz model results from the choice
of $\gamma_j$. For $\gamma_j<0$, the poles of $\chi$ would have moved to the upper
half-plane of $\omega$, and the result would be an acausal
response, $\chi(\tau)>0$ for $\tau<0$ and $\chi(\tau)=0$ for
$\tau>0$.
\newline
\indent
Next let us study the symmetry properties of $\chi$ under the
(``crossing'') transformation $\omega\rightarrow-\omega$. The real
($\chi_R$) and imaginary ($\chi_I$) parts of this function can be
read off Eq.~(\ref{DDeq2.1.5}),
\beqn
\label{DDeq2.1.12}
\chi_R(\omega) = -\frac{e^2}{m}\,\sum_j N_j
\frac{\omega^2-\omega^2_j}{(\omega^2-\omega^2_j)^2+4
\gamma_j^2\,\omega^2}\ ,
\eeqn
\beqn
\label{DDeq2.1.13}
\chi_I(\omega) = \frac{e^2}{m}\,\sum_j
N_j \frac{2\gamma_j\omega}{(\omega^2-\omega^2_j)^2+4
\gamma_j^2\,\omega^2}\ , \eeqn
and the crossing relations for real $\omega$ values are
\beqn
\label{DDeq2.1.14}
\chi_R(-\omega) = \chi_R(\omega)\ ,\ \ \
\chi_I(-\omega) = -\chi_I(\omega)\ .
\eeqn
This makes it possible to cast
Eq.~(\ref{DDeq2.1.8}) into the form
\beqn
\chi_R(\omega)  =  \frac{2}{\pi} {\mathcal{P}}
\int_{0}^{\infty} \frac{\omega'\chi_I(\omega')}{\omega'^2-\omega^2}
\,d\omega' \ ,\ \chi_I(\omega) = -
\frac{2}{\pi}\,\omega\,{\mathcal{P}}
\int_{0}^{\infty} \frac{\chi_R(\omega')}{\omega'^2-\omega^2}
\,d\omega'\ .
\label{DDeq2.1.15}
\eeqn
The crossing relations Eq.~(\ref{DDeq2.1.14}) can be combined and
extended to complex values of $\omega$ by
\beqn
\label{DDeq2.1.16}
\chi(-\omega^{\ast}) = \chi^{\ast}(\omega) \ .
\eeqn
In particular, $\chi$ is real on the imaginary axis and takes on
complex conjugate values at points situated mirror-symmetrically
to this axis. The dielectric susceptibility can be expressed by
the dielectric constant $\varepsilon$,
\beqn
\label{DDeq2.1.17}
\chi(\omega) = \frac{\varepsilon(\omega)-1}{4\pi}\ ,
\eeqn
which in turn is related to the refraction index $n$ and the
phase velocity $v_P$ in the medium,
\beqn
\label{DDeq2.1.18}
\upsilon_P(\omega) = \frac{\omega}{k(\omega)} =
\frac{c}{n(\omega)} = \frac{c}
{\sqrt{\varepsilon(\omega)\,\mu(\omega)}}\ ,
\eeqn
where $k$ is the wave number, and $\mu$ the
magnetic permeability  of the medium.
\newline
\noindent
In the case of $\mu=1$, it is obvious that also
$(\varepsilon-1)$ and hence $(n^2-1)$ obey the dispersion relations of
Eq.~(\ref{DDeq2.1.15}). In a gas of low density, the refraction index
is close to 1, and we can approximate $(n^2-1)$ by $2(n-1)$. The
result is the Lorentz dispersion formula for the oscillator
model, to be obtained from Eqs.~(\ref{DDeq2.1.5}), (\ref{DDeq2.1.17}) and
(\ref{DDeq2.1.18}),
\beqn
\label{DDeq2.1.20}
n(\omega) = 1+2\pi\frac{e^2}{m}\sum_j \frac{N_j}
{\omega_j^2-2i\gamma_j\,\omega-\omega^2}\ .
\eeqn
Let us now discuss the connection between absorption and
dispersion on the microscopic level. Suppose that a monochromatic
plane wave hits a homogeneous and isotropic medium at $x=0$ and
leaves the slab of matter at $x=\Delta x$. The incoming wave is
denoted by
\beqn \label{DDeq2.1.21}
{\mathbf{E}}_{\mbox{\scriptsize{in}}}(x,t) =  e^{i(kx-\omega t)}\,
E_0\,{\hat{\mathbf{e}}}_0\ , \eeqn
with the linear dispersion $\omega=ck$ and the polarization vector
$\hat{\mathbf{e}}_0$.
\newline
\indent
Having passed the slab of matter with the dispersion of
Eq.~(\ref{DDeq2.1.20}), the wave function is
\begin{eqnarray}
{\mathbf{E}}_{\mbox{\scriptsize{out}}} (\Delta x,t) & = &
e^{i\frac{\omega}{c}n(\omega)\Delta x} \,e^{-i\omega
t}\,E_0\,{\hat{\mathbf{e}}}_0 \nonumber \\
& = & e^{i\frac{\omega}{c}(n_R-1)\Delta x}
\,e^{-\frac{\omega}{c}n_I\Delta x}
{\mathbf{E}}_{\mbox{\scriptsize{in}}} (\Delta x,t) \ .
\label{DDeq2.1.22}
\end{eqnarray}
The imaginary part of $n$ is associated with absorption, which
defines an extinction coefficient $\kappa$, such that
the intensity drops like
$|{\mathbf{E}}_{\mbox{\scriptsize{out}}}|^2= e^{-\kappa\Delta
x}|{\mathbf{E}}_{\mbox{\scriptsize{in}}}|^2$. On the other hand
the extinction coefficient is related to the product of the
total absorption cross section $\sigma_T$ for an
individual constituent (e.g., a $^1$H atom) and the number of
constituents per volume $N$, and therefore
\beqn
\label{DDeq2.1.23}
\kappa(\omega) = 2\omega n_I/c =
N\sigma_T(\omega)\ .
\eeqn
Further on the elementary level, the incident light wave excites
dipole oscillations of the constituents with electric dipole
moments
\beqn
\label{DDeq2.1.24}
{\bd}(t) = \alpha{\mathbf{E}}_{\mbox{\scriptsize{in}}}(0,t)\ ,
\eeqn
with $\alpha=\alpha(\omega)$ the electric dipole
polarizability of a
constituent. We note that here and in the following the dipole
approximation has been used such that we can neglect retardation
effects and evaluate the incoming wave at $x=0$. Within the slab
of matter, the dipole moments radiate, thus giving rise to an
induced electric field ${\mathbf{E}}_s$.
\newline
\indent
The field due to the individual dipole at ${\mathbf{r}}',$ measured at a point
${\mathbf{r}}=x\,{\hat{\mathbf{e}}}_x$ in beam direction, is
\beqn
\label{DDeq2.1.25}
{\mathbf{e}}_s=\alpha k^2\,E_0\frac{e^{i(k\rho-\omega t)}}{\rho}\,
(\hat{\boldsymbol{\varrho}}\times{\mathbf{\hat{e}}}_0)
\times\hat{\boldsymbol{\varrho}}\ ,
\eeqn
with $\hat{\boldsymbol{\varrho}}=(\mathbf{r}'-\mathbf{r})/
|\mathbf{r}'-\mathbf{r}|$ and $\rho =|\mathbf{r}'-\mathbf{r}|$.
\newline
\indent
In particular, forward scattering is obtained in the limit
$kx \gg 1$. Since the incoming field is polarized
perpendicularly to this axis, we find
\beqn
\label{DDeq2.1.26}
{\mathbf{e}}_s\,(\theta=0)=\alpha k^2\,\frac{e^{i(k\rho-\omega t)}}{\rho}\,
E_0{\mathbf{\hat{e}}}_0\ ,
\eeqn
and by definition the forward scattering amplitude
\beqn
\label{DDeq2.1.27}
f(k,\theta=0) = \alpha k^2\ .
\eeqn
The total field due to the dipole oscillations, ${\mathbf{E}}_s$,
is obtained by
integrating Eq.~(\ref{DDeq2.1.25}) over the volume of the slab and
multiplying with $N$, the number of particles per volume. The
result for small $\Delta x$ is
\beqn
\label{DDeq2.1.28}
{\mathbf{E}}_{\mbox{\scriptsize{out}}}=
{\mathbf{E}}_{\mbox{\scriptsize{in}}}+{\mathbf{E}}_s\approx
(1+2\pi ik\,\Delta x\,N\alpha)
{\mathbf{E}}_{\mbox{\scriptsize{in}}}\ .
\eeqn
A comparison of Eqs.~(\ref{DDeq2.1.27}) and (\ref{DDeq2.1.28}) with the
macroscopic form, Eq. (\ref{DDeq2.1.22}), expanded for small $\Delta x$,
yields the connection between the refractive index and the forward
scattering amplitude,
\beqn
\label{DDeq2.1.29}
n(\omega)-1=2\pi\,N\,\alpha(\omega) =
\frac{2\pi N}{k^2}\,f(k,\theta=0)\ .
\eeqn
From Eqs.~(\ref{DDeq2.1.23}) and (\ref{DDeq2.1.29}) we obtain
the optical theorem,
\beqn
\label{DDeq2.1.29a}
{\mbox{Im}}\,f(\omega) = \frac{\omega}{4\pi}\sigma_T (\omega)\ ,
\eeqn
and since $f/k^2$ is proportional to $(n-1)$ and $\chi$,
there follows a dispersion relation for ${\mbox{Re}}\,f$ analogous to
Eq.~(\ref{DDeq2.1.15}),
\beqn
\label{DDeq2.1.30}
{\mbox{Re}}\,f(\omega)  =
\frac{2\omega^2}{\pi}\, {\mathcal{P}} \int_0^{\infty}
\frac{{\mbox{Im}}\,f(\omega')}{\omega'(\omega'^2-\omega^2)}
\,d\omega'
=
\frac{\omega^2}{2\pi^2}\, {\mathcal{P}} \int_0^{\infty}
\frac{\sigma_T (\omega')}{\omega'^2-\omega^2} \,d\omega'\ ,
\eeqn
where we have
set $c=\hbar=1$ here and in the following. Historically,
Eq.~(\ref{DDeq2.1.30}) expressed in terms of $n(\omega)-1$, was
first derived by Kronig and Kramers~\cite{Kro26}.
We also note that without the
crossing symmetry, Eq.~(\ref{DDeq2.1.14}), the dispersion integral
would also need information about the cross section at negative
energies, which of course is not available.
\newline
\indent
In order to prepare
for the specific content of this review, several comments are in
order:
\begin{enumerate}
\item[(I)]
The derivation of the Kramers-Kronig dispersion relation started
from a neutral system, an atom like the hydrogen atom. Since the
total charge is zero, the electromagnetic field can only excite
the internal degrees of freedom, while the center of mass remains
fixed. As a consequence the scattering amplitude
$f(\omega)=\mathcal{O}(\omega^2)$, which leads to a differential
cross section
\[
\frac{d\sigma}{d\Omega} =
|f(\omega)|^2=\mathcal{O}(\omega^4)\ .
\]
The result is Rayleigh scattering which among other things
explains the blue sky. However, for charged systems like ions,
electrons or protons,
also the center of mass will be
accelerated by the electromagnetic field, and the scattering
amplitude takes the general form
\beqn
\label{DDeq2.1.31}
{\mbox{Re}}\,f(\omega,0) = -\frac{Q_{\rm{tot}}^2}{M_{\rm{tot}}}
+\mathcal{O}(\omega^2)\ .
\eeqn
The additional ``Thomson'' term due to $c.m.$ motion results in a
finite scattering amplitude for $\omega=0$ and depends only on the
total charge $Q_{\rm{tot}}$ and the total mass $M_{\rm{tot}}$.
\item[(II)]
We have defined the electric dipole polarizability $\alpha$
as a complex function
$\alpha(\omega)$ whose real and imaginary parts can be calculated
directly from the total absorption cross section
$\sigma_T(\omega)$. In the Lorentz model this cross section starts
as $\omega^2$ for small $\omega$.
In reality, however, the total absorption cross section has a
threshold energy $\omega_0$. The absorption spectrum of, say,
hydrogen is given by a series of discrete levels
$(\omega_{1s\rightarrow1p}=10.2$~eV, etc.) followed by a continuum
for $\omega\ge 13.6$~eV. As a result $\sigma_T(\omega)$ vanishes
in a range $0\le\omega<\omega_0$, and therefore
$\alpha(\omega)={\mbox{Re}}\,\alpha(\omega)$ can be expanded in a
Taylor series in the vicinity of the origin,
\beqn
\label{DDeq2.1.32}
\alpha(\omega) = \frac{1}{2\pi^2}\, 
\int_{\omega_0}^{\infty} \frac{\sigma_T(\omega'^2)}{\omega'^2}\,
d\omega' + \frac{\omega^2}{2\pi^2}\, 
\int_{\omega_0}^{\infty} \frac{\sigma_T(\omega'^2)}{\omega'^4}\,
d\omega' + \ldots
\eeqn
In the following chapters we shall use the term ``polarizability''
or more exactly ``static polarizability''
only for the first term of the expansion. Moreover, in the
dipole expansion used in Eq.~(\ref{DDeq2.1.24}), this first term
is solely determined by electric dipole $(E1)$ radiation,
\beqn
\label{DDeq2.1.33}
\alpha \equiv \alpha(\omega=0) =
\frac{1}{2\pi^2}\,
\int_{\omega_0}^{\infty} \frac{\sigma_T(\omega'^2)}{\omega'^2}\,
d\omega'\ge 0\ .
\eeqn
The terms ${\mathcal{O}}(\omega^2)$ in Eq.~(\ref{DDeq2.1.32})
are then the first order retardation effects for $E1$ radiation,
and the full function $\alpha (\omega)$ will be called the
``dynamical polarizability'' of the system.

\item[(III)]
Finally, the Lorentz model discards magnetic effects because of
the small velocities involved in atomic systems. In a general
derivation, the first term on the $rhs$ of Eq.~(\ref{DDeq2.1.32})
equals the sum of the electric ($\alpha$) and magnetic ($\beta$)
dipole polarizabilities, while the second term describes the
retardation of these dipole polarizabilities and the static
quadrupole polarizabilities.

\end{enumerate}

Let us finally discuss the polarizability for some specific cases.
The Hamiltonian for an electron bound by a harmonic restoring
force, as in the Lorentz model of Eq.~(\ref{DDeq2.1.1}), takes the
form
\beqn
\label{DDeq2.1.34}
H=\frac{{\mathbf{p}}^2}{2m}+\frac{m\omega_0^2}{2}\,{\br}^2+e{\br}\cdot
{\mathbf{E}}\ ,
\eeqn
where the electric field ${\mathbf{E}}$ is assumed to be static
and uniform. Substituting ${\br}={\br}'+\Delta{\br}$ and
${\mathbf{p}}={\mathbf{p}}'$, where
$\Delta{\br}$ is the displacement due to the electric field, we
may rewrite this equation as
\beqn
\label{DDeq2.1.35}
H=\frac{{\mathbf{p}}'^2}{2m}+\frac{m\omega_0^2}{2}\,{\br}'^2+\Delta E\ .
\eeqn
The displacement $\Delta{\br}$ leads to an induced dipole moment
${\bd}$ and an energy shift $\Delta E$,
\beqn
\label{DDeq2.1.36}
{\bd} = -e\Delta{\br} = \frac{e^2}{m\omega_0^2}\,{\mathbf{E}}\ ,
\ \ \ \Delta E=-\frac{e^2}{2m\omega_0^2}\,{\mathbf{E}}^2\ .
\eeqn
The induced dipole moment ${\bd}$ and the energy shift $\Delta E$
are both proportional to the polarizability,
$\alpha=e^2/m\omega_0^2$,
which can also be read off Eqs.~(\ref{DDeq2.1.2}) and
(\ref{DDeq2.1.24}) in the limit $\omega\rightarrow0$. In fact, the
relation
\beqn
\label{DDeq2.1.37}
\alpha=\frac{\delta{\bd}}{\delta{\mathbf{E}}} =
-\frac{\delta^2\Delta E}{(\delta{\mathbf{E}})^2}
\eeqn
is quite general and even survives in quantum mechanics. As a
result we can calculate the energy of such a system by second
order perturbation theory. The perturbation to first order (linear
Stark effect) vanishes for a system with good parity, and if the
system is also spherically symmetric, the second order (quadratic
Stark effect) yields,
\beqn
\label{DDeq2.1.38}
\Delta E = -\sum_{n>0}\,\frac{|<n|e \, z|0>|^2}
{\epsilon_n-\epsilon_0}\,{\mathbf{E}}^2\ ,
\eeqn
where $\epsilon_n$ are the energies of the eigenstate $|n>$.
Equations~(\ref{DDeq2.1.37}) and (\ref{DDeq2.1.38})
immediately yield the static electric dipole
polarizability,
\beqn
\label{DDeq2.1.39}
\alpha = 2\sum_{n>0}\,\frac{|<n|e \, z|0>|^2}
{\epsilon_n-\epsilon_0}\ .
\eeqn
As an example for a classical extended object we quote the
electric ($\alpha$) and magnetic ($\beta$) dipole
polarizabilities of small dielectric
or permeable spheres of radius $a$~\cite{Jac75},
\beqn
\label{DDeq2.1.40}
\alpha=\frac{\epsilon-1}{\epsilon+2}\,a^3\ ,\ \ \
\beta=\frac{\mu-1}{\mu+2}\,a^3\ .
\eeqn
The same quantities for a perfectly conducting sphere are obtained
in the limits $\epsilon\rightarrow\infty$ and $\mu\rightarrow0$,
respectively,
\beqn
\label{DDeq2.1.41}
\alpha=a^3\ ,\ \ \
\beta=-\frac{1}{2}\,a^3\ .
\eeqn
The electric polarizability of the conducting sphere is
essentially the volume of the sphere, up to a factor $4\pi/3$. Due
to the different boundary conditions, the magnetic polarizability
is negative, which corresponds to diamagnetism $(\mu<1)$.
In this case the currents and with them the magnetizations are
induced against the direction of the applied field according to
Lenz's law. A permeable sphere can be diamagnetic or paramagnetic
$(\mu>1)$, in the latter case the magnetic moments are already
preformed and become aligned in the presence of the external field.
While the magnetic polarizabilities of atoms and molecules are usually
very small because of $|\mu-1|\lesssim10^{-2}$, electric
polarizabilities may be quite large compared to the volume. For
example, the static dielectric constant of water
$\varepsilon=81$ leads to a nearly perfect conductor; in the
visible range this constant is down to $\varepsilon=1.8$ with the
consequence that the index of refraction is $n=1.34$.

A quantum
mechanical example is the hydrogen atom in nonrelativistic
description. Its ground state has good parity and spherical
symmetry and therefore Eq.~(\ref{DDeq2.1.38} ) applies. In
this case it is even
possible to perform the sum over the excited states and to obtain
the closed expression~\cite{Mer70}
\beqn
\label{DDeq2.1.42}
\alpha\,(^1H) = \frac{9}{2}\,a_B^3\ ,
\eeqn
where $a_B$ is the Bohr radius. The $rms$ radius of $^1H$ is
$<r^2>=3a^2_B$, the radius of an equivalent hard sphere is given
by $R^2=5a^2_B$, and as a result the hydrogen atom is a pretty good
conductor, $\alpha$/volume $\approx1/10$.
\newline
\indent
In the following
sections we report on the polarizabilities of the nucleon. As
compared to hydrogen and other atoms, we shall find that the
nucleon is a dielectric medium with $\varepsilon\approx1.001$,
i.e., a very good insulator. Furthermore, magnetic effects are
$a\ priori$ of
the same order as the electric ones, because the charged
constituents, the quarks move with velocities close to the velocity of
light. However the diamagnetic effects of the pion
cloud and the paramagnetic effects of the quark core of the
nucleon tend to cancel, with the result of a relatively small
net value of
$\beta$. We shall see that ``virtual'' light allows one to
gain information about the spatial distribution of the
polarization densities, which will be particularly interesting to
resolve the interfering effects of para- and diamagnetism.
Furthermore, the nucleon has a spin and therefore appears as an
anisotropic object in the case of polarized nucleons. This leads
to additional spin polarizabilities whose closest parallel in
classical physics is the Faraday effect.


\subsection{Real Compton scattering (RCS) : nucleon polarizabilities
and the GDH sum rule}

In this section we discuss the forward scattering of a real photon
by a nucleon. The incident photon is characterized by the Lorentz
vectors of momentum, $q=(q_0,\,\mathbf{q})$ and polarization,
$\varepsilon_{\lambda}=(0,\,\boldsymbol{\varepsilon}_{\lambda})$,
with $q\cdot q=0$ (real photon) and $\varepsilon_{\lambda}\cdot
q=0$ (transverse polarization). If the photon moves in the
direction of the z-axis, ${\mathbf{q}}=q_0\,{\hate}_z$, the two
polarization vectors may be taken as
\beqn
\label{DDeq2.2.1} \bvare_{\pm} = \mp
\frac{1}{\sqrt{2}}\,(\hate_x\pm i\hate_y)\ ,
\eeqn
corresponding to circularly polarized light with helicities
$\lambda=+1$ (right-handed) and $\lambda=-1$ (left-handed). The
kinematics of the outgoing photon is then described by the
corresponding primed quantities. \newline \indent For the purpose
of dispersion relations we choose the $lab$ frame, and introduce
the notation $q_0^{\mbox{\sc{lab}}}=\nu$ for the photon energy in
that system. The total {\it c.m.} energy $W$ is expressed in terms of
$\nu$ as : $W^2 = M^2 + 2 M \nu$, where $M$ is the nucleon mass.
The forward Compton amplitude then takes the form
\beqn
\label{DDeq2.2.2}
T(\nu,\,\theta=0) = \bvare'^{\ast}\cdot\bvare\,f(\nu)+
i\,\bsig\cdot(\bvare'^{\ast}\times\bvare)\,g(\nu)\ .
\eeqn
This is the most general expression that is

\begin{enumerate}
\item[(I)] constructed from the independent vectors
$\bvare',\ \bvare,\ \bq'=\bq$ (forward scattering!), and $\bsig$
(the proton spin operator),
\item[(II)] linear in $\bvare'$ and $\bvare$,
\item[(III)] obeying the transverse gauge,
$\bvare'\cdot\bq'=\bvare\cdot\bq=0$, and
\item[(IV)] invariant under rotational and parity transformations.
\end{enumerate}

\begin{figure}
\vspace{-0.5cm}
\epsfxsize=15cm
\epsfysize=5cm
\centerline{\epsffile{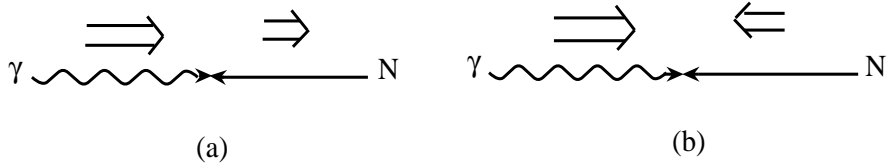}}
\vspace{-1.5cm}
\caption{Spin and helicity of a double polarization experiment.
The
arrows $\Longrightarrow$ denote the spin projections on the photon
momentum, the arrows $\longrightarrow$ the momenta of the
particles.
The spin projection and helicity of the photon is assumed to be
$\lambda=1$. The spin projection and helicity of the target
nucleon $N$ are denoted by $S_z$ and $h$, respectively,
and the eigenvalues
of the excited system $N^{\ast}$ by the corresponding
primed quantities. \newline
a) Helicity 3/2: Transition $N(S_z=1/2,\ h=-1/2)\rightarrow
N^{\ast}(S_z=h=3/2)$, which changes the helicity by 2 units.
\newline
b) Helicity 1/2: Transition $N(S_z=-1/2,\
h=+1/2)\rightarrow N^{\ast}(S_z=h=+1/2)$, which conserves the helicity.
\label{DDfig2.2.1}}
\end{figure}

Furthermore, the Compton amplitude has to be invariant under
photon crossing, corresponding to the fact that each graph with
emission of the final-state photon after the absorption of the
incident photon has to be accompanied by a graph with the opposite
time order, i.e. absorption following emission (``crossed
diagram''). This symmetry requires that the amplitude $T$ of
Eq.~(\ref{DDeq2.2.2}) be invariant under the transformation
$\varepsilon'\leftrightarrow\varepsilon$ and
$\nu\leftrightarrow-\nu$, with the result that $f$ is an even and
$g$ an odd function,
\beqn
\label{DDeq2.2.3}
f(\nu) = f(-\nu)\ ,\ \ \ g(\nu) =-g(-\nu)\ .
\eeqn
These two amplitudes can be determined by scattering circularly
polarized photons (e.g., helicity $\lambda=1$) off nucleons
polarized along or opposite to the photon momentum $\bq$. The
former situation (Fig.~\ref{DDfig2.2.1}a) leads to an intermediate
state with helicity $3/2$. Since this requires a total spin
$S\ge3/2$, the transition can only take place on a correlated 3-quark system.
The transition of Fig.~\ref{DDfig2.2.1}b, on the other hand, is
helicity conserving and possible for an individual quark, and
therefore should dominate in the realm of deep inelastic
scattering. Denoting the Compton scattering amplitudes for the two
experiments indicated in Fig.~\ref{DDfig2.2.1} by $T_{3/2}$ and
$T_{1/2}$, we find $f(\nu)=(T_{1/2}+T_{3/2})/2$ and
$g(\nu)=(T_{1/2}-T_{3/2})/2$. In a similar way we define the total
absorption cross section as the spin average over the two helicity
cross sections,
\beqn \label{DDeq2.2.4}
\sigma_T=\frac{1}{2}\,(\sigma_{3/2}+\sigma_{1/2})\ ,
\eeqn
and the transverse-transverse interference term by the helicity
difference,
\beqn \label{DDeq2.2.5} \sigma'_{TT} =
\frac{1}{2}\,(\sigma_{3/2}-\sigma_{1/2})\ .
\eeqn
The optical theorem expresses the unitarity of the scattering
matrix
by relating the absorption cross sections to the imaginary part of
the respective forward scattering amplitude,
\begin{eqnarray}
\mbox{Im}\ f(\nu) & = & \frac{\nu}{8\pi}
(\sigma_{1/2}(\nu)+\sigma_{3/2}(\nu)) =
\frac{\nu}{4\pi}\,\sigma_T (\nu)\ , \nonumber \\
\mbox{Im}\ g(\nu) & = & \frac{\nu}{8\pi}
(\sigma_{1/2}(\nu)-\sigma_{3/2}(\nu)) =
-\frac{\nu}{4\pi}\,\sigma'_{TT} (\nu)\ .
\label{DDeq2.2.6}
\end{eqnarray}

\begin{figure}[h]
\vspace{-1.3cm}
\epsfysize=14cm
\centerline{\epsffile{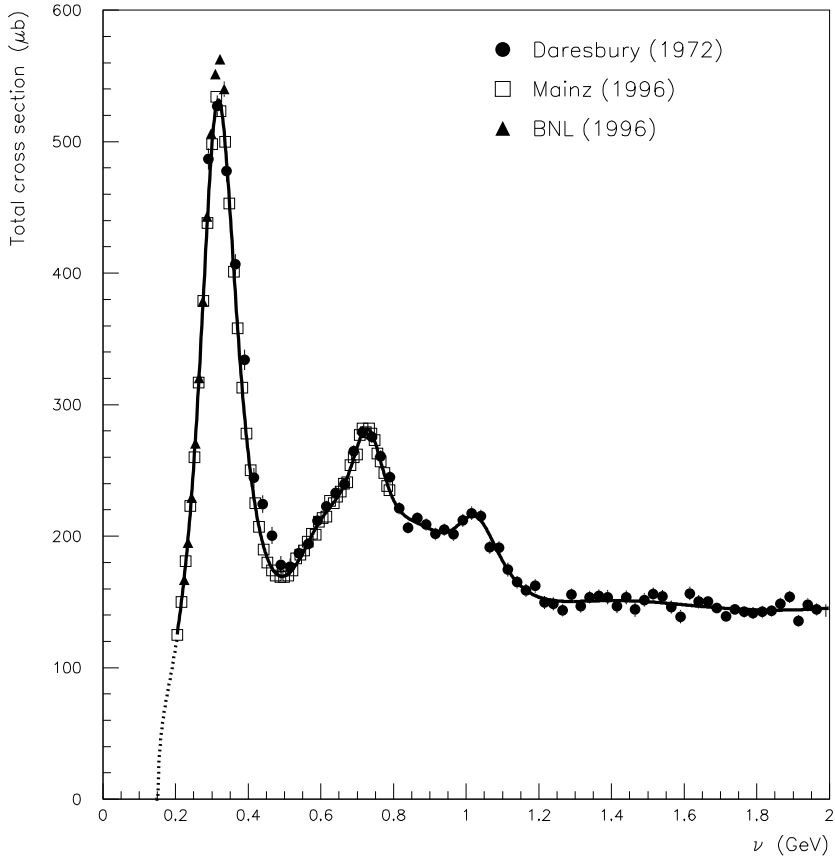}}
\vspace{-.7cm}
\caption{The total absorption cross section $\sigma_T(\nu)$ for
the proton. The fit to the data is described in Ref.~\cite{Bab98}, 
where also the references to the data can be found \label{DDfig2.2.2}.}
\end{figure}

\begin{figure}[h]
\vspace{-.3cm}
\epsfysize=14cm
\centerline{\epsffile{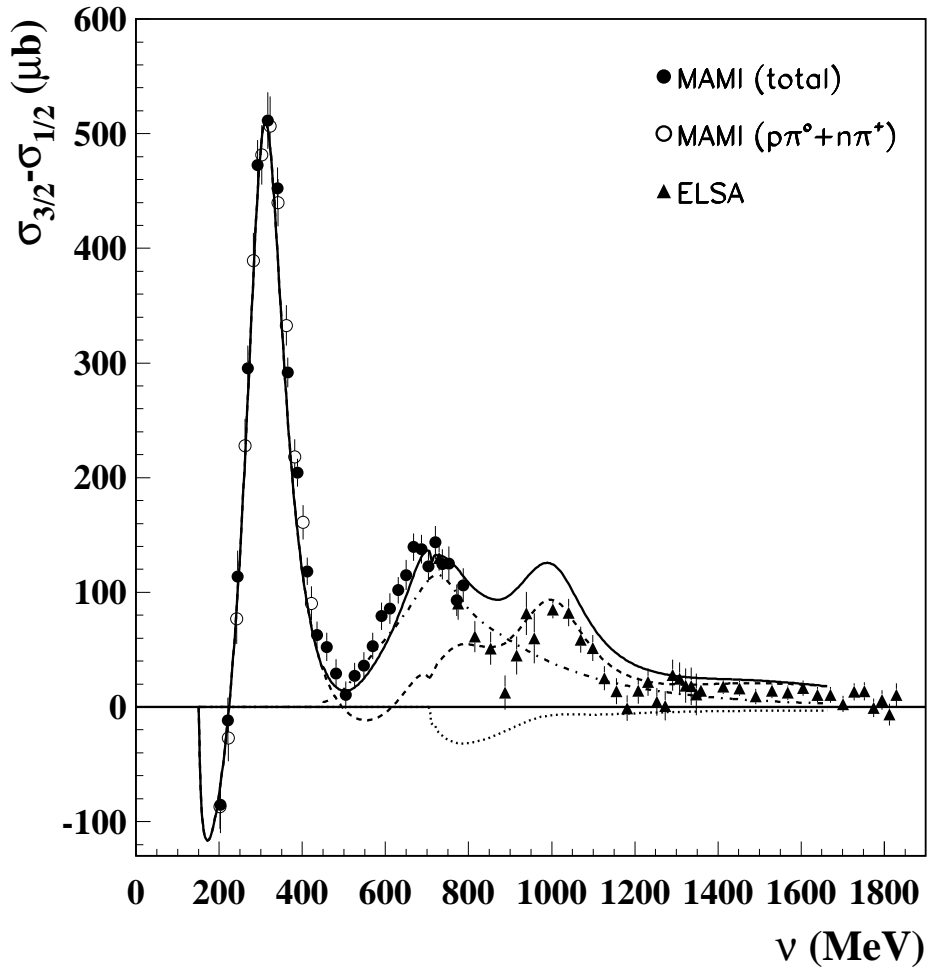}}
\vspace{-.25cm}
\caption{The helicity difference
$\sigma_{3/2}(\nu)-\sigma_{1/2}(\nu)$ for the proton.
The calculations include the contribution of $\pi N$ intermediate
states (dashed curve) \cite{Dre99}, $\eta N$ intermediate state
(dotted curve) \cite{Dre01}, and the $\pi\pi N$ intermediate
states
(dashed-dotted curve) \cite{Hol01,Hol02}. The total sum of these
contributions is shown by the full curves. The MAMI data are
from Ref.~\cite{Ahr00,Ahr01} and the (preliminary) ELSA data from 
Ref.~\cite{Hel02}.}
\label{DDfig2.2.3}
\end{figure}

\begin{figure}
\epsfysize=5.5cm
\centerline{\epsffile{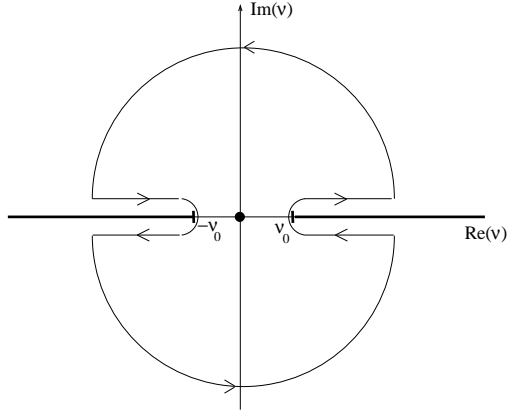}}
\vspace{0.cm} 
\caption{The contour $C$ for the dispersion integral Eq.~(\ref{DDeq2.2.7}). 
The physical point lies at $\nu + i \varepsilon$ and 
approaches the Re $\nu$ axis in the limit $\varepsilon\rightarrow+0$.
Singularities lie on the real axis, poles in the $s$- and $u$-channel
contributions with intermediate nucleon states at $\nu=0$, and
cuts for $|\nu|\ge\nu_0$ due to production of a pion or heavier
systems ($2\pi,\ K,$ etc.). In addition, there occur resonances in
the lower half-plane on the second Riemann sheet.} 
\label{DDfig2.2.4}
\end{figure}

Due to the smallness of the fine structure constant
$\alpha_{em}$ we may neglect all purely
electromagnetic processes in this context, such as photon
scattering to finite angles or electron-positron pair production 
in the Coulomb field of the proton.
Instead, we shall consider only the coupling of the photon to the
hadronic channels, which start at the threshold for pion
production, i.e., at a photon $lab$ energy $\nu_0
=m_{\pi}(1+m_{\pi}/2M)\approx150$~MeV. We shall return to this point
later in the context of the GDH integral. 
\newline
\indent The total photoabsorption cross section $\sigma_T$ is
shown in Fig.~\ref{DDfig2.2.2}. It clearly exhibits 3 resonance
structures on top of a strong background. These structures
correspond, in order, to concentrations of magnetic dipole
strength $(M1)$ in the region of the $\Delta (1232)$ resonance,
electric dipole strength $(E1)$ near the resonances
$N^{\ast}(1520)$ and $N^{\ast}(1535)$, and electric quadrupole $(E2)$ 
strength near the $N^{\ast}(1675)$. Since the absorption cross
sections are the input for the dispersion integrals, we have to
discuss the convergence for large $\nu$. For energies above the
resonance region $(\nu\gtrsim 1.66$~GeV which is equivalent to
total {\it c.m.} energy 
$W \gtrsim 2 $~GeV), $\sigma_T$ is very slowly decreasing and reaches a
minimum of about 115~$\mu$b around $W = 10 $~GeV. At the highest
energies, $W \simeq 200$~GeV (corresponding with $\nu \simeq 2
\cdot 10^4$~GeV), experiments at DESY ~\cite{DESY} have measured
an increase with energy of the form $\sigma_T \sim W^{0.2}$, in
accordance with Regge parametrizations through a soft pomeron
exchange mechanism \cite{Cud00}. Therefore, it can not be expected
that the unweighted integral over $\sigma_T$ converges.
\newline
\indent 
Recently, also the helicity difference has been measured.
The first measurement was carried out at MAMI (Mainz) for photon
energies in the range 200~MeV$ < \nu < 800~$MeV~\cite{Ahr00,Ahr01}. As
shown in Fig.~\ref{DDfig2.2.3}, this difference fluctuates much
more strongly than the total cross section $\sigma_T$. The
threshold region is dominated by S-wave pion production, i.e.,
intermediate states with spin $1/2$ and, therefore, mostly
contributes to the cross section $\sigma_{1/2}$. In the region of
the $\Delta (1232)$ with spin $J=3/2$, both helicity cross
sections contribute, but since the transition is essentially $M1$,
we find $\sigma_{3/2}/\sigma_{1/2}\approx3$, and $\sigma'_{TT}$
becomes large and positive. Figure~\ref{DDfig2.2.3} also shows
that $\sigma_{3/2}$ dominates the proton photoabsorption cross section
in the second and third resonance regions. It was in fact one of
the early successes of the quark model to predict this fact by a
cancellation of the convection and spin currents in the case of
$\sigma_{1/2}$ \cite{Cop69,Kon80}.
\newline
\indent 
The GDH collaboration has now extended the measurement
into the energy range up to 3~GeV at ELSA (Bonn)~\cite{Hel02}.
These preliminary data show a small positive value of
$\sigma'_{TT}$ up to $\nu\approx2$~GeV, with some indication of a
cross-over to negative values, as has been predicted from an
extrapolation of DIS data~\cite{Bia99}. This is consistent with
the fact that the helicity-conserving cross section $\sigma_{1/2}$
should dominate in DIS, because an individual quark cannot
contribute to $\sigma_{3/2}$ due to its spin. However, the
extrapolation from DIS to real photons should be taken with a
grain of salt.
\newline
\indent
Having studied the behavior of the absorption cross sections, we
are now in a position to set up dispersion relations. A generic
form starts from a Cauchy integral with contour $C$ shown in
Fig.~\ref{DDfig2.2.4},
\beqn
\label{DDeq2.2.7}
f(\nu+i\varepsilon) = \frac{1}{2\pi i}\oint_C\,\frac{f(\nu')}
{\nu'-\nu-i\varepsilon}\,d\nu'\ ,
\eeqn
where $\nu\ge0$ and $\varepsilon>0$, i.e., in the limit
$\varepsilon\rightarrow0$ the singularity approaches a physical point
at $\nu'=\nu>0$. The contour is closed in the upper half-plane by
a large circle of radius $R$ that eventually goes to infinity.
Since we want to neglect this contribution eventually, the cross
sections have to converge for $\nu\rightarrow\infty$ sufficiently
well. As we have seen before, this requirement is certainly not
fulfilled by $\sigma_T(\nu)$, and for this reason we have to
subtract the dispersion relation for $f$. If we subtract at
$\nu=0$, i.e., consider $f(\nu)-f(0)$, we also remove the 
nucleon pole terms at $\nu=0$. The remaining contribution comes from the
cuts along the real axis, which may be expressed in terms of the
discontinuity of ${\mbox{Im}}\ f$ across the cut for a contour as
shown in Fig.~\ref{DDfig2.2.4} or simply by an integral over
${\mbox{Im}}\ f$ as we approach the axis from above. By use of the
crossing relation and the optical theorem, the subtracted
dispersion integral can then be expressed in terms of the cross
section,
\beqn
\label{DDeq2.2.8}
{\mbox{Re}}\ f(\nu) = f(0)+\frac{\nu^2}{2\pi^2}\,{\mathcal{P}}\,
\int_{\nu_0}^{\infty}\,\frac{\sigma_T(\nu')}
{\nu'^2-\nu^2}\,d\nu'\ .
\eeqn
Though the dispersion integral is clearly dominated by hadronic
reactions, the subtraction is also necessary for a charged lepton,
because the integral over $\sigma_{T}$ also diverges
(logarithmically) for a purely electromagnetic process. We note
that in a hypothetical world where this integral would converge,
the charge could be predicted from the absorption cross section.
\newline
\indent
For the odd function $g(\nu)$ we may expect the existence of an
unsubtracted dispersion relation,
\beqn
\label{DDeq2.2.9}
{\mbox{Re}}\ g(\nu) = \frac{\nu}{4\pi^2}\,{\mathcal{P}}\,
\int_{\nu_0}^{\infty}\,\frac{\sigma_{1/2}(\nu')-\sigma_{3/2}(\nu')}
{\nu'^2-\nu^2}\,\nu'd\nu'\ .
\eeqn
\newline
\indent
If the integrals exist, the relations Eq.~(\ref{DDeq2.2.8}) and
(\ref{DDeq2.2.9}) can be expanded into a Taylor series at the
origin, which should converge up to the lowest threshold,
$\nu=\nu_0$~:
\begin{eqnarray}
{\mbox{Re}}\ f(\nu) \,&=&\, f(0)
 + \sum_{n=1}
\left (\frac{1}{2\pi^2}\,\int_{\nu_0}^{\infty}\,
\frac{\sigma_T(\nu')}{\nu'^{2n}}\,d\nu'\right )\,\nu^{2n} \, , 
\label{DDeq2.2.10} \\
{\mbox{Re}}\ g(\nu) \,&=&\,
 \sum_{n=1}
\left (\frac{1}{4\pi^2}\,\int\,
\frac{\sigma_{1/2}(\nu')-\sigma_{3/2}(\nu')}{(\nu')^{2n-1}}
\,d\nu' \right )
\,\nu^{2n-1}\ .
\label{DDeq2.2.11}
\end{eqnarray}
The expansion coefficients in brackets parametrize the
electromagnetic response of the medium, e.g., the nucleon. These
Taylor series may be compared to the predictions of the low energy
theorem (LET) of Low~\cite{Low54}, and Gell-Mann and
Goldberger~\cite{Gel54} who showed that the leading and
next-to-leading terms of the expansions are fixed by the global
properties of the system. These properties are the mass $M$, the
charge $e\,e_N$, and the anomalous magnetic moment
$(e/2M)\kappa_N$ for a particle with spin $1/2$ like the nucleon
(i.e., $e_p=1,\ e_n=0,\ \kappa_p=1.79,\ \kappa_n=-1.91)$. The
predictions of the LET start from the observation that the leading
term for $\nu\rightarrow0$ is described by the Born terms, because
these have a pole structure in that limit. If constructed from a
Lorentz, gauge invariant and crossing symmetrical theory, the
leading and next-to-leading order terms are completely determined
by the Born terms,
\begin{eqnarray}
f(\nu) & = & -\frac{e^2\,e_N^2}{4\pi M} + (\alpha+\beta)
\,\nu^2+ {\mathcal{O}}(\nu^4) \ , \label{DDeq2.2.12} \\
g(\nu) & = & -\frac{e^2\kappa^2_N}{8\pi M^2}\,\nu +
\gamma_0\nu^3 + {\mathcal{O}}(\nu^5) \ . \label{DDeq2.2.13}
\end{eqnarray}
The leading term of the no spin-flip amplitude, $f(0)$, is the
Thomson term already familiar from nonrelativistic
theory\footnote{By comparing with Eq.~(\ref{DDeq2.1.31}) we see
that we have now converted to Heaviside-Lorentz units, i.e.,
$\alpha_{em}=e^2/4\pi = 1/137$ and $r_{cl}=e^2/4\pi M$, here and
in all following sections.}. The term ${\mathcal{O}}(\nu)$
vanishes because of crossing symmetry, and only the term
${\mathcal{O}}(\nu^2)$ contains information on the internal
structure (spectrum and excitation strengths) of the complex
system. In the forward direction this information appears as the
sum of the electric and magnetic dipole polarizabilities. The
higher order terms ${\mathcal{O}}(\nu^4)$ contain contributions of
dipole retardation and higher multipoles, as will be discussed in
Sec.~\ref{sec:physpol}. By comparing with Eq.~(\ref{DDeq2.2.10}), we can
construct all higher coefficients of the low energy expansion (LEX),
Eq.~(\ref{DDeq2.2.12}), from moments of the total cross section. In
particular we obtain Baldin's sum rule \cite{Bal60,Lap63},
\beqn
\label{DDeq2.2.14}
\alpha + \beta = \frac{1}{2\pi^2}\,
\int_{\nu_0}^{\infty}\,\frac{\sigma_T(\nu')}{\nu'^2}
\,d\nu'\ ,
\eeqn
and from the next term of the expansion a relation for dipole
retardation and quadrupole polarizability. In the case of the
spin-flip amplitude $g$, the comparison of
Eqs.~(\ref{DDeq2.2.11}) and (\ref{DDeq2.2.13}) yields the sum rule
of Gerasimov~\cite{Ger65}, Drell and Hearn~\cite{Dre65},
\beqn
\label{DDeq2.2.15}
\frac{\pi e^2\kappa^2_N}{2M^2}=
\int_{\nu_0}^{\infty}\,\frac{\sigma_{3/2}(\nu')
-\sigma_{1/2}(\nu')}{\nu'}\,d\nu' \, \equiv I \, ,
\eeqn
and a value for the forward spin
polarizability~\cite{Gel54,GGT54},
\beqn
\label{DDeq2.2.16}
\gamma_0= \,-\,\frac{1}{4\pi^2}\,\int_{\nu_0}^{\infty}\,
\frac{\sigma_{3/2}(\nu')-\sigma_{1/2}(\nu')}
{\nu'^3}\,d\nu'\ .
\eeqn
\newline
\indent
Baldin's sum rule was recently reevaluated in Ref.~\cite{Bab98}. 
These authors determined the integral by use of
multipole expansions of pion photoproduction in the threshold
region, old and new total photoabsorption cross sections in the
resonance region (200~MeV$ < \nu < 2$~GeV), and a parametrization of
the high energy tail containing a logarithmical divergence of
$\sigma_T$. The result is
\beqn
 \label{DDeq2.2.18}
\alpha^p+\beta^p & = &
(13.69\pm0.14)\cdot10^{-4}~{\rm{fm}}^3\ , \nonumber \\
\alpha^n+\beta^n & = &
(14.40\pm0.66)\cdot10^{-4}~{\rm{fm}}^3\ ,
\eeqn
for proton and neutron, respectively.
\newline
\indent
Due to the $\nu^{-3}$ weighting of the integral, the forward spin
polarizability of the proton can be reasonably well determined by the GDH
experiment at MAMI. The contribution of the range
200~MeV$<\nu<$\,800~MeV is $\gamma_0 = -[1.87 \pm 0.08$ (stat)
$\pm 0.10$ (syst)$]\cdot10^{-4}$~fm$^4$, the threshold region is
estimated to yield $0.90 \cdot 10^{-4}$~fm$^4$, and only
$-0.04\cdot10^{-4}$~fm$^4$ are expected from energies above
800~MeV \cite{Tia02}. The total result is
\be
\label{DDeq2.2.19}
\gamma_0^p = \left[ -1.01 \pm 0.08~(\mathrm{stat}) \pm 0.10~(\mathrm{syst})
\right] \cdot10^{-4}~\rm{fm}^4 \ .
\ee
We postpone a more detailed discussion of the nucleon's
polarizability to Sec.~\ref{sec:3_9} where experimental findings and
theoretical predictions are compared to the results of dispersion
relations.
\newline
\indent 
As we have seen above, the GDH sum rule is based on very
general principles, Lorentz and gauge invariance, unitarity, and
on one weak assumption: the convergence of an unsubtracted dispersion
relation (DR). It
is of course impossible to ever prove the existence of such a sum
rule by experiment. However, the question is legitimate whether or
not the anomalous magnetic moment on the $lhs$ of
Eq.~(\ref{DDeq2.2.15}) is approximately obtained by integrating
the $rhs$ of that equation up to some reasonable energy, say 3 or
50~GeV. The comparison will tell us whether the anomalous magnetic
moment measured as a ground state expectation value, is related to
the degrees of freedom visible to that energy, or whether it is
produced by very short distance and possibly still unknown
phenomena. Concerning the convergence problem, it is interesting
to note that the GDH sum rule was recently evaluated in
QED~\cite{Dic01} for the electron at order $\alpha_{em}^3$ and
shown to agree with the Schwinger correction to the anomalous
magnetic moment, i.e., $\kappa_e = \alpha_{em} / (2 \pi)$. 
This also gives the electromagnetic correction to the sum rule for the
proton, which is of relative order $(\kappa_e / \kappa_N)^2 \sim 10^{-6}$.  
\newline
\indent 
The GDH sum rule predicts that the integral on the $rhs$
of Eq.~(\ref{DDeq2.2.15}) should be $I_p=205~\mu$b for the proton.
The energy range of the MAMI experiment~\cite{Ahr01} contributes
$I_p(200-800~$MeV)~$=[226~\pm~5$~(stat)$~\pm~12$ (syst)$]~\mu$b. The
preliminary results of the GDH experiment at ELSA~\cite{Hel02}
shows a positive contribution in the range of the 3rd resonance
region, with a maximum value of $\sigma_{3/2}-\sigma_{1/2}\approx
100\mu$b, but only very small contributions at the higher energies
with a possible cross-over to negative values at
$\nu\gtrsim1.8$~GeV. At high $\nu$, above the resonance region,
one usually invokes Regge phenomenology to argue that the integral
converges \cite{Bass97,Bass99}. In particular, for the isovector
channel $\sigma_{1/2} - \sigma_{3/2} \to \nu^{\alpha_1 - 1}$ at
large $\nu$, with $-0.5 \lesssim \alpha_1 \lesssim 0$ being the
intercept of the $a_1(1260)$ meson Regge trajectory. For the
isoscalar channel, Regge theory predicts a behavior 
corresponding to $\alpha_1 \simeq - 0.5$, which is the intercept of the
isoscalar $f_1(1285)$ and $f_1(1420)$ Regge trajectories. However,
these assumptions should be tested experimentally. The approved
experiment SLAC E-159 \cite{E159} will measure the helicity
difference absorption cross section $\sigma_{3/2} - \sigma_{1/2}$
for protons and neutrons in the photon energy range 5 GeV $< \nu <
40$~GeV. This will be the first measurement of $\sigma_{3/2} -
\sigma_{1/2}$ above the resonance region, to test the convergence
of the GDH sum rule and to provide a baseline for our
understanding of soft Regge physics in the spin-dependent forward
Compton amplitude.
\newline
\indent 
According to the latest MAID analysis~\cite{Tia02} the threshold
region yields $I_p$(thr - 200 MeV)\,= -27.5~$\mu$b, with a sign
opposite to the resonance region, because pion S-wave production
contributes to $\sigma_{1/2}$ only. Combining this threshold
contribution with the MAMI value (between 200 and 800~MeV), the
MAID analysis from 800~MeV to 1.66~GeV, and including model estimates for the
$\pi\pi$, $\eta$ and $K$ production channels,  
one obtains an integral value from threshold to 1.66~GeV of \cite{Tia02}~:
\be
\label{eq:ipwl2}
I_p \,(W < 2~\mathrm{GeV}) \,=\, 
\left[ 241 \pm 5~(\mathrm{stat}) \pm 12~(\mathrm{syst}) \pm 7~(\mathrm{model}) 
\right] \,\mu b \, .
\ee 
The quoted model error is essentially due to uncertainties in the
helicity structure of the $\pi\pi$ and $K$ channels.
Based on Regge extrapolations and fits to DIS, the
asymptotic contribution ($\nu > 1.66$~GeV) has been estimated to be 
($-26 \pm 7)~\mu$b in Ref.~\cite{Bia99}, whereas Ref.~\cite{Sim01} 
estimated this to be ($-13 \pm 2)~\mu$b. 
We take the average of both estimates to be ($-20 \pm 9)~\mu$b as 
a range which covers the theoretical uncertainty in the evaluations of 
this asymptotic contribution.  
Putting all contributions together, the result for the integral $I$ of
Eq.~(\ref{DDeq2.2.15}) is
\be
\label{DDeq2.2.20}
I_p \,=\, \left[ 221 \pm 5~(\mathrm{stat}) \pm 12~(\mathrm{syst}) 
\pm 11~(\mathrm{model}) \right] \,\mu b 
\;\approx\; I_p (\rm{sum \ rule}) = 204.8 \, \mu b \ ,
\ee
where the systematical and model errors of different contributions
have been added in quadrature. 
Assuming that the size of the high-energy contribution 
for the estimate of Eq.~(\ref{DDeq2.2.20}) is confirmed by
the SLAC E-159 experiment in the near future, one can
conclude that the GDH sum rule seems to work for the proton.
Unfortunately, the experimental situation is much less clear in the case of the
neutron, for which the sum rule predicts
\be
\label{DDeq2.2.21}
I_n (\rm{sum\ rule}) = 233.2\, \mu b\ .
\ee
From present knowledge of the pion photoproduction multipoles and
models of heavier mass intermediate state, one obtains the estimate
$I_n = \left[ 147~(\pi) \,+\, 55~(\pi\pi)\,-\, 6~(\eta) \right]~\mu$b~$\approx 
196~\mu$b~\cite{Tia02}, from the contributions of the $\pi$, $\pi\pi$
and $\eta$ production channels, thereby assuming the same two-pion
contribution as in the case of the proton.
This estimate for $I_n$ falls short of the sum rule value by about 15 \%. 
Given the model assumptions and the uncertainties in the present data,
one can certainly not conclude that the neutron sum rule is
violated. Possible sources of the discrepancy may be 
a neglect of final state interaction for pion
production off the ``neutron target'' deuteron, 
the helicity structure of two-pion production, 
or the asymptotic contribution, which still remain to be investigated ? 
We shall return to this point in the following Section
when discussing so-called generalized GDH integrals for virtual
photon scattering. In any case, the outcome of the planned
experiments of the GDH collaboration~\cite{GDHneutron} for the
neutron will be of extreme interest.

\subsection{Forward dispersion relations in doubly virtual Compton
scattering (VVCS)}

\begin{figure}[h]
\epsfysize=12.cm
\centerline{\epsffile{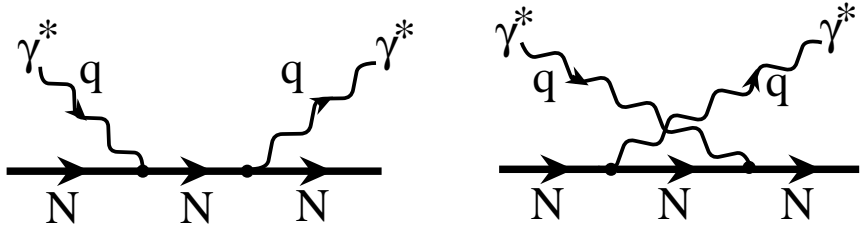}}
\vspace{-8.cm}
\caption{Born diagrams for the doubly virtual Compton scattering
(VVCS) process.}
\label{fig:born_vvcs}
\end{figure}

In this section we consider the forward scattering of a virtual
photon with space-like four-momentum $q$, i.e., $q^2=q^2_0-{\bq}^2
=-Q^2<0$. The first stage of this process, the absorption of the
virtual photon, is related to inclusive electroproduction,
$e+N\rightarrow e'+N'+$ anything, where $e(e^{'})$ and $N(N^{'})$ are
electrons and nucleons, respectively, in the initial (final)
state. The kinematics of the electron is traditionally described
in the $lab$ frame (rest frame of $N$), with $E$ and $E^{'}$ the
initial and final energy of the electron, respectively, and
$\theta$ the scattering angle. This defines the kinematical values
of the emitted photon in terms of four-momentum transfer $Q$ and
energy transfer $\nu$,
\beqn
Q^{2} = 4 EE^{'} \sin^{2} \frac{\theta}{2} , \, \nu = E
-E^{'}\ ,
\label{DDeq2.3.1}
\eeqn
and the $lab$ photon momentum $|{\bq}_{\rm{\sc{lab}}}|= \sqrt{Q^2
+ \nu^2}$. In the $c.m.$ frame of the hadronic intermediate state,
the four-momentum of the virtual photon is $q = (\omega,
{\bq}_{\rm{\sc{cm}}})$ with
\beqn
\omega = \frac{M \nu - Q^2}{W}, \quad
{\bq}_{\mbox{\sc{cm}}} = \frac{M}{W} \,
{\bq}_{\mbox{\sc{lab}}}\ ,
\label{DDeq2.3.2}
\eeqn
where $W$ is the total energy in the hadronic $c.m.$ frame. 
We further introduce the Mandelstam variable $s$  
and the Bjorken variable $x$,

\beqn
s  = 2M \nu + M^{2}-Q^{2}= W^{2}\ ,
\hspace{2cm}
x = \frac{Q^2}{2M \nu}\ .
\label{DDeq2.3.3} 
\eeqn
The virtual photon spectrum is
normalized according to Hand's definition~\cite{Han63} by the
``equivalent photon energy'',
\beqn
K = K_H = \nu (1-x) = \frac{W^{2}-M^{2}}{2M}\ .
\label{DDeq2.3.5}
\eeqn
An alternate choice would be to use Gilman's
definition~\cite{Gil68}, $K_G = |{\bq}_{\mbox{\sc{lab}}}|$.
\newline
\indent The inclusive inelastic cross section may be written in
terms of a virtual photon flux factor $\Gamma_V$ and four partial
cross sections~\cite{Dre01},
\beqn
\frac{d\sigma}{d\Omega\ dE'} = \Gamma_V \sigma
(\nu,Q^2)\ , \label{DDeq2.3.6}
\eeqn
\beqn
\sigma =
\sigma_T+\epsilon\sigma_L+hP_x\sqrt{2\epsilon(1-\epsilon)}\
         \sigma'_{LT}+hP_z\sqrt{1-\epsilon^2}\ \sigma'_{TT}\ ,
\label{DDeq2.3.7}
\eeqn
with the photon polarization
\beqn
\epsilon = \frac{1}{1+2 (1+ \nu^2/Q^2) \tan^2
\theta/2}\ ,
\label{DDeq2.3.8}
\eeqn
and the flux factor
\beqn
\Gamma_V = \frac{\alpha_{em}}{2\pi^2}\ \frac{E'}{E}\ \frac{K}{Q^2}\
\frac{1}{1-\epsilon}\ .
 \label{DDeq2.3.9}
 \eeqn
In addition to the transverse cross section $\sigma_T$ and
$\sigma'_{TT}$ of Eqs.~(\ref{DDeq2.2.4}) and (\ref{DDeq2.2.5}),
the longitudinal polarization of the virtual photon gives rise to
a longitudinal
cross section $\sigma_L$ and a longitudinal-transverse interference
$\sigma'_{LT}$. The two spin-flip (interference)
cross sections can only be
measured by a double-polarization experiment, with $h = \pm 1$
referring to the two helicity states of the (relativistic)
electron, and $P_z$ and $P_x$ the components of the target
polarization in the direction of the virtual photon momentum
${\bq}_{\mbox{\sc{lab}}}$ and perpendicular to that direction in
the scattering plane of the electron. In the following we shall
change the sign of the two spin-flip
cross sections in comparison with Ref.~\cite{Dre01}, i.e.,
introduce the sign convention used in DIS,
\begin{eqnarray}
\label{DDeq2.3.10}
\sigma_{TT}=-\sigma'_{TT}\quad \mathrm{and}
\quad \sigma_{LT} = -\sigma'_{LT}\ .
\end{eqnarray}
\indent
The partial cross sections are related to the quark structure functions
as follows~\cite{Dre01}
\footnote{We note at this point that the factor $\gamma^2$ in the
denominator of $\sigma_L$ is missing in Ref.~\cite{Dre01}.} :
\begin{eqnarray}
\sigma_T& =& \frac{4\pi^2\alpha_{em}}{M K}\ F_1\ ,
\nonumber \\
\sigma_L& =& \frac{4\pi^2\alpha_{em}}{K}\left[ \frac{1+ \gamma^2}
{\gamma^2} \frac{F_2}{\nu} - \frac{F_1}{M}\right]\ ,
\nonumber \\
\sigma_{TT}& =& \frac{4\pi^2\alpha_{em}}{M K} \left(g_1-\gamma^2\,
g_2 \right)\ ,
\nonumber\\
\sigma_{LT}&=& \frac{4\pi^2\alpha_{em}}{M K}\,\gamma\,\left(g_1+g_2\right)\ ,
\label{DDeq2.3.11}
\end{eqnarray}
with the ratio $\gamma = Q/\nu$. 
The helicity cross sections are then given by
\begin{eqnarray}
\label{DDeq2.3.12}
\sigma_{1/2} & = & \frac{4 \pi^2 \alpha_{em}}{M K} \,
\left(F_1 + g_1 - \gamma^2 g_2\right), \nonumber \\
\sigma_{3/2} & = & \frac{4 \pi^2 \alpha_{em}}{M K} \,
\left(F_1 - g_1 + \gamma^2 g_2\right)\ .
\end{eqnarray}
Due to the longitudinal degree of freedom, the virtual photon has
a third polarization vector ${\bvare}_0$ in addition to the
transverse polarization vectors ${\bvare}_{\pm}$ defined in
Eq.~(\ref{DDeq2.2.1}). A convenient definition of this four-vector
is
\beqn
\label{DDeq2.3.13}
\varepsilon_0 = \frac{1}{Q}(|{\bq}|,\,0,\,0,\,q_0)\ ,
\eeqn
where we have chosen the z-axis in the direction of the photon
propagation,
\beqn
\label{DDeq2.3.14}
q = (q_0,\,0,\,0,\,|{\bq}|)\ .
\eeqn
All 3 polarization vectors and the photon
momentum are orthogonal (in the Lorentz metrics!),
\beqn
\label{DDeq2.3.15}
\varepsilon_m\cdot q=0\
,\quad\varepsilon_m^{\ast}\cdot\varepsilon_{m'}=(-1)^m \, \delta_{mm'}\ ,
 \quad {\mbox{for}} \quad m, m' =0,\,\pm1\ .
\eeqn
The invariant matrix element for the absorption of a photon with
helicity $m$ is
\beqn
\label{DDeq2.3.16}
{\mathcal{M}}_m\sim\varepsilon_m\cdot\langle J\rangle\ ,
\eeqn
where $J$ is the hadronic transition current, which is gauge
invariant,
\beqn
\label{DDeq2.3.17}
q\cdot \langle J\rangle = q_0 \langle\rho\rangle -
{\bq}\cdot\langle{\mathbf{j}}\rangle = 0\ .
\eeqn
Being Lorentz invariant, the matrix element ${\mathcal{M}}_m$ can be
evaluated in any system of reference, e.g., in the $lab$ frame and
by use of Eq.~(\ref{DDeq2.3.17}),
\beqn
\label{DDeq2.3.18}
{\mathcal{M}}_0\sim\frac{1}{Q}
(|{\bq}_{\mbox{\sc{lab}}}|\langle\rho\rangle-\nu\langle j_z\rangle) =
\frac{Q}{|{\bq}_{\mbox{\sc{lab}}}|}\langle\rho\rangle =
\frac{Q}{\nu}\langle j_z\rangle\ .
\eeqn
The VVCS amplitude for forward scattering takes the
form (as a $2 \times 2$ matrix in nucleon spinor space)~:
\begin{eqnarray}
\label{DDeq2.3.19}
T(\nu,\,Q^2,\,\theta=0)& \;=\; &
{\bvare}\,'^{\ast}\cdot{\bvare} \, f_T(\nu,\,Q^2) \;+\;
f_L(\nu,\,Q^2) \nonumber\\
&+& \;i{\bsig}\cdot({\bvare}\,'^{\ast}\times{\bvare}) \,g_{TT}(\nu,\,Q^2)
\;-\; i{\bsig}\cdot[({\bvare}\,'^{\ast}-{\bvare})\times \hat{q} \,]
\,g_{LT}(\nu,\,Q^2) \, ,
\end{eqnarray}
where we have generalized the notation of Eq.~(\ref{DDeq2.2.2}) to the
VVCS case. The optical theorem relates the imaginary parts of the 4
amplitudes in Eq.~(\ref{DDeq2.3.19}) to the 4 partial cross
sections of inclusive scattering,
\begin{eqnarray}
\label{DDeq2.3.20}
{\mbox{Im}}\ f_T(\nu,\,Q^2) & = &
\frac{K}{4\pi}\sigma_T(\nu,\,Q^2) \, , \nonumber \\
{\mbox{Im}}\ f_L(\nu,\,Q^2) & = &
\frac{K}{4\pi}\sigma_L(\nu,\,Q^2) \, , \nonumber \\
{\mbox{Im}}\ g_{TT}(\nu,\,Q^2) & = &
\frac{K}{4\pi}\sigma_{TT}(\nu,\,Q^2) \, , \nonumber \\
{\mbox{Im}}\ g_{LT}(\nu,\,Q^2) & = &
\frac{K}{4\pi}\sigma_{LT}(\nu,\,Q^2) \ .
\end{eqnarray}
We note that products $K\,\sigma_T$ etc. are independent of
the choice of $K$, because they are directly proportional to
the measured cross section (see Eqs.~(\ref{DDeq2.3.6}) and
(\ref{DDeq2.3.9})). Of course, the natural choice at this point
would be $K=K_G=|{\bq}_{\mbox{\sc{lab}}}|$, because we
expect the photon three-momentum on the $rhs$ of
Eq.~(\ref{DDeq2.3.20}). However, we shall later evaluate the cross
sections by a multipole decomposition in the $c.m.$ frame for which
$K=K_H$ is the standard choice.
\newline
\indent
The imaginary parts of the scattering amplitudes,
Eqs.~(\ref{DDeq2.3.20}), get contributions from both
elastic scattering at
$\nu_B=Q^2/2M$ and inelastic
processes above pion threshold, for
$\nu>\nu_0=m_{\pi}+(m_{\pi}^2+Q^2)/2M$. The elastic contributions can be calculated
from the direct and crossed Born diagrams of
Fig.~\ref{fig:born_vvcs}, where the electromagnetic vertex
for the transition $\gamma^* (q) + N(p) \to N(p + q)$ is given by
\beqn
\Gamma^\mu \;=\; F_D(Q^2) \, \gamma^\mu \;+\;
F_P(Q^2) \, i \sigma^{\mu \nu} \frac{q_\nu}{2 M} \; ,
\label{eq:bornvvcs}
\eeqn
with $F_D$ and $F_P$ the nucleon Dirac and Pauli form factors,
respectively.
The choice of the electromagnetic vertex according to
Eq.~(\ref{eq:bornvvcs}) ensures gauge invariance when calculating the
Born contribution to the VVCS amplitude, and yields~:
\begin{eqnarray}
\label{DDeq2.3.21}
f_T^{\mbox{\sc{Born}}} (\nu,\,Q^2) & = &
-\frac{\alpha_{em}}{M}
\left (F_D^2 + \frac{\nu_B^2}{\nu^2-\nu_B^2+i \varepsilon}
\,G_M^2\right ) \, ,
\nonumber \\
f_L^{\mbox{\sc{Born}}} (\nu,\,Q^2) & = &
-\frac{\alpha_{em}Q^2}{4M^3}
\left (F_P^2 + \frac{4M^2}{\nu^2-\nu_B^2+i \varepsilon}
\,G_E^2\right ) \, ,
\nonumber \\
g_{TT}^{\mbox{\sc{Born}}} (\nu,\,Q^2) & = &
-\frac{\alpha_{em}\nu}{2M^2}
\left (F_P^2 + \frac{Q^2}{\nu^2-\nu_B^2+i \varepsilon}
\,G_M^2\right ) \, ,
\nonumber \\
g_{LT}^{\mbox{\sc{Born}}} (\nu,\,Q^2) & = &
\frac{\alpha_{em}Q}{2M^2}
\left (F_DF_P - \frac{Q^2}{\nu^2-\nu_B^2+i \varepsilon}
\,G_EG_M \right ) \, .
\end{eqnarray}
The electric $(G_E)$ and magnetic $(G_M)$
Sachs form factors are related to the Dirac $(F_D)$ and Pauli
$(F_P)$ form factors by
\beqn \label{DDeq2.3.22} G_E\,(Q^2) = F_D\,(Q^2)-\tau\,F_P(Q^2)\ ,
 \quad G_M\,(Q^2) = F_D\,(Q^2)+F_P(Q^2)\ , \eeqn
with $\tau=Q^2/4M^2$, and are normalized to
\beqn
\label{DDeq2.3.23}
G_E(0)=e_N\ ,\quad G_M(0)=e_N+\kappa_N=\mu_N\ ,
\eeqn
where $e_N,\ \kappa_N$, and $\mu_N$ are the charge (in units of
$e$), the anomalous and the total magnetic moments (in units of
$e/2M$) of the respective nucleon. 
We have split the elastic
contributions of Eq.~(\ref{DDeq2.3.21}) into a real contribution
(terms in $F_D$ and $F_P$) and a complex contribution (terms in
$G_E$ and $G_M$). The latter terms have a structure like the
susceptibility of Eq.~(\ref{DDeq2.1.5}) and fulfill a dispersion
relation by themselves. By use of Eqs.~(\ref{DDeq2.3.11}),
(\ref{DDeq2.3.20}), and (\ref{DDeq2.3.21}), the imaginary parts of
the Born amplitudes can be related to the elastic contributions of
the quark structure functions and to the form factors,
\begin{eqnarray}
\label{DDeq2.3.24}
\frac{4M}{e^2}\,{\mbox{Im}}\ f_T^{\mbox{\sc{Born}}} & = &
F_1^{\mbox{\sc{el}}} = \frac{1}{2}\,G_M^2\,\delta(1-x) \ ,
\nonumber \\
\frac{4M}{e^2}\,{\mbox{Im}}\ f_L^{\mbox{\sc{Born}}} & = &
\frac{Q^2+4M^2}{2Q^2}\,F_2^{\mbox{\sc{el}}}-F_1^{\mbox{\sc{el}}}=
\frac{2M^2}{Q^2}\,G_E^2\,\delta(1-x) \ ,
\nonumber \\
\frac{4M}{e^2}\,{\mbox{Im}}\ g_{TT}^{\mbox{\sc{Born}}} & = &
g_1^{\mbox{\sc{el}}}-\frac{4M^2}{Q^2}\,g_2^{\mbox{\sc{el}}}=
\frac{1}{2}\,G_M^2\,\delta(1-x) \ ,
\nonumber \\
\frac{4M}{e^2}\,{\mbox{Im}}\ g_{LT}^{\mbox{\sc{Born}}} & = &
\frac{2M}{Q}\,(g_1^{\mbox{\sc{el}}}+g_2^{\mbox{\sc{el}}})=
\frac{M}{Q}\,G_EG_M\,\delta(1-x) \ .
\end{eqnarray}
These equations describe the imaginary parts of the scattering
amplitudes in the physical region at $x=1$ or $\nu=\nu_B$. The
continuation of the amplitudes to negative or complex arguments
follows from crossing symmetry (see, e.g., Eq.~(\ref{DDeq2.2.3}))
and analyticity (see Eq.~(\ref{DDeq2.1.16})).
\newline
\indent
According to Eqs.~(\ref{DDeq2.2.12}) and (\ref{DDeq2.2.13}),
the low energy theorem for real photons asserts that the leading and
next-to-leading order terms in an expansion in $\nu$ are completely
determined by the pole singularities of the Born terms. However, in
the case of virtual photons the limit $\nu\rightarrow0$ has to be
performed with care~\cite{Ji93}, because
\beqn
\label{DDeq2.3.25}
\lim_{\nu\to 0}\ \lim_{Q^2\to 0}\ f(\nu,\,Q^2)\neq
\lim_{Q^2\to 0}\ \lim_{\nu\to 0}\ f(\nu,\,Q^2)\ .
\eeqn
If we choose $Q^2=0$ right away, we reproduce the results of real
Compton scattering, Eqs.~(\ref{DDeq2.2.12}) and
(\ref{DDeq2.2.13}), for $f(\nu)=f_T(\nu,\,Q^2=0)$ and
$g(\nu)=g_{TT}(\nu,\,Q^2=0)$, while $f_L$ and $g_{LT}$ vanish
because of the longitudinal currents involved. On the other hand,
if we choose $Q^2$ finite and let $\nu$ go to zero, the result is
quite different. In particular
\be
\label{DDeq2.3.25a}
f^{\rm{\scriptsize{Born}}}_T(\nu,Q^2=0) =
-\frac{\alpha_{em}}{M}\,e_N^2+\mathcal{O}(\nu^2)\,
\ee
while
\be
\label{DDeq2.3.25b}
f^{\rm{\scriptsize{Born}}}_T(\nu=0,Q^2) =
\frac{\alpha_{em}}{M}\,\kappa_N(2e_N+\kappa_N)+\mathcal{O}(Q^2)\, .
\ee
The surprising result is that a long-wave real photon couples to a
Dirac (point) particle, while a long-wave virtual photon couples
only to a particle with an anomalous magnetic moment, i.e., a
particle with internal structure. The inelastic contributions, on the
other hand, are independent of the order of the limits. 
\newline
\indent
It is now straightforward to construct the full VVCS amplitudes by
dispersion relations in $\nu$ at $Q^2=\,$const.
For the amplitude $f_T$ (which is even in $\nu$),
we shall need a subtracted DR as in the case
of Eq.~(\ref{DDeq2.2.8}),
\begin{eqnarray}
\label{eq:ftdr1} {\mbox{Re}}\ f_T(\nu,\,Q^2) \;=\; {\mbox{Re}}\
f_T(0,\,Q^2) \,+\, \frac{2\nu^2}{\pi}\,{\mathcal{P}}\,
\int_{0}^{\infty}\,\frac {{\rm{Im}}\ f_T(\nu',\,Q^2)} {\nu'
(\nu'^2-\nu^2)}\,d\nu'\ .
\end{eqnarray}
The integral in Eq.~(\ref{eq:ftdr1}) gets contributions from both the
elastic cross section (nucleon pole) at $\nu' = \nu_B$ and from the inelastic
continuum for $\nu' > \nu_0$ ~:
\begin{eqnarray}
\label{eq:ftdr}
{\mbox{Re}}\ f_T(\nu,\,Q^2) & \;=\; &
{\mbox{Re}}\ f_T^{\mbox{\sc{pole}}}(\nu,\,Q^2)
\;+\; \bigg[ {\mbox{Re}}\ f_T(0,\,Q^2) \,-\,
{\mbox{Re}}\ f_T^{\mbox{\sc{pole}}}(0,\,Q^2) \bigg]
\nonumber \\
&+& \frac{\nu^2}{2\pi^2}\,{\mathcal{P}}\,
\int_{\nu_0}^{\infty}\,\frac
{K(\nu',\,Q^2)\sigma_T(\nu',\,Q^2)}
{\nu' (\nu'^2-\nu^2)}\,d\nu'\ .
\end{eqnarray}
In the case of $K=K_H(\nu,\,Q^2)=\nu(1-x)$, the
dispersion integral is of the same form as in
Eq.~(\ref{DDeq2.2.8}) except for a factor $(1-x)$ typical for that
choice of $K$. The pole contribution which enters in
Eq.~(\ref{eq:ftdr}) can be read off Eq.~(\ref{DDeq2.3.21}),
\begin{eqnarray}
\label{eq:ftpole}
{\mbox{Re}}\ f_T^{\mbox{\sc{pole}}} (\nu,\,Q^2) & = &
-\frac{\alpha_{em}}{M}
\frac{\nu_B^2}{\nu^2-\nu_B^2} \,G_M^2(Q^2) \, .
\end{eqnarray}
The function $f_T(\nu,\,Q^2) - f_T^{\mbox{\sc{pole}}}(\nu,\,Q^2)$,
i.e., excluding the nucleon pole term, is continuous in $\nu$.
Therefore, one may perform a low energy expansion in $\nu$,
\begin{eqnarray}
\label{eq:ftlex}
&&{\mbox{Re}}\ f_T(\nu,\,Q^2) \,-\,
{\mbox{Re}}\ f_T^{\mbox{\sc{pole}}}(\nu,\,Q^2) \;=\; \nonumber\\
&&\left[ {\mbox{Re}}\ f_T(0,\,Q^2) \,-\,
{\mbox{Re}}\ f_T^{\mbox{\sc{pole}}}(0,\,Q^2) \right] \;+\;
\left( \alpha(Q^2) + \beta(Q^2) \right) \; \nu^2 \;+\;
{\mathcal{O}}(\nu^4) \, ,
\end{eqnarray}
where the term in ${\mathcal{O}}(\nu^2)$ generalizes the definition
of the sum of electric and magnetic polarizabilities at finite $Q^2$.
Comparing Eqs.~(\ref{eq:ftdr}) and (\ref{eq:ftlex}), 
one obtains the generalization of
Baldin's sum rule to virtual photons,
\begin{eqnarray}
\label{eq:alphapbeta}
\alpha(Q^2) + \beta(Q^2) & \;=\; & \frac{1}{2\pi^2}\,
\int_{\nu_0}^{\infty}\,\frac{K(\nu, \, Q^2)}{\nu} \,
\frac{\sigma_T(\nu,\,Q^2)}{\nu^2}\,d\nu \, ,
\nonumber \\
&\;=\; & \frac{e^2 M}{\pi \, Q^4}\,\int_{0}^{x_0}\,
2x\,F_1(x,\,Q^2)\,dx \, ,
\end{eqnarray}
where in the last line we have expressed the integral in terms of
the nucleon structure function $F_1$ using Eq.~(\ref{DDeq2.3.11}).
The Callan-Gross relation~\cite{Cal68} implies that in the limit of
large $Q^2$ the integrand $2x\,F_1(x,\,Q^2)\rightarrow
F_2(x,\,Q^2)$, i.e., the generalized Baldin sum rule measures the
second moment of $F_1$ and, asymptotically, the first moment of
$F_2$. We can also define the resonance contribution to $\alpha +
\beta$ through the integral
\begin{eqnarray}
\label{eq:alphapbetares}
\alpha_{res}(Q^2) + \beta_{res}(Q^2)
\;=\; \frac{e^2 M}{\pi \, Q^4}\,\int_{x_{res}}^{x_0}\,
2x\,F_1(x,\,Q^2)\,dx \, ,
\end{eqnarray}
where $x_{res}$ corresponds with $W = 2$~GeV.
\newline
\indent
In Fig.~\ref{fig:alphapbeta_vvcs}, we show the $Q^2$ dependence of
$\alpha + \beta$ and compare a resonance estimate with the evaluation
for $Q^2 > 1$~GeV$^2$ obtained from the DIS structure function $F_1$,
using the MRST01 parametrization~\cite{MRST01}. For the resonance
estimate we use the MAID model~\cite{Dre99} for the one-pion channel
and include an estimate for the $\eta$ and $\pi \pi$ channels according to
Ref.~\cite{Dre01}. One sees that at $Q^2 = 0$, the one-pion
channel alone gives about 85 \% of Baldin's sum rule. Including the estimate
for the $\eta$ and $\pi \pi$ channels, one nearly saturates Baldin's sum
rule. Going to $Q^2$ larger than 1~GeV$^2$, we also show
the sum rule estimate of Eq.~(\ref{eq:alphapbetares}) obtained from 
DIS by including only the range $W < 2$~GeV. The comparison of
this result with the resonance estimate of MAID shows that the MAID
model nicely reproduces the $Q^2$ dependence of $\sigma_T$ for $W < 2$~GeV.
By comparing the full DIS estimate with the contribution from
$W~<~2$~GeV, one notices that the sum rule value for
$\alpha + \beta$ at $Q^2 \lesssim 1$~GeV$^2$ is mainly saturated
by the resonance contribution, whereas for $Q^2 \gtrsim 2$~GeV$^2$,
the non-resonance contribution ( $W~>~2$~GeV ) dominates the
sum rule. Therefore, around $Q^2 \simeq 1 - 2$~GeV$^2$, a transition
occurs from a resonance dominated description to a partonic description.
Such a transition was already noticed in Refs.~\cite{Ede98,DKKPT} 
where a resonance estimate for $\alpha + \beta$ was compared with the DIS
estimate, giving qualitatively similar results as shown here 
\footnote{Note however that in Refs.~\cite{Ede98,DKKPT}, the  
virtual photon flux differs from Eq.~(\ref{eq:alphapbetares}).}.
\begin{figure}[h]
\vspace{-1.cm}
\epsfysize=11.5cm
\centerline{\epsffile{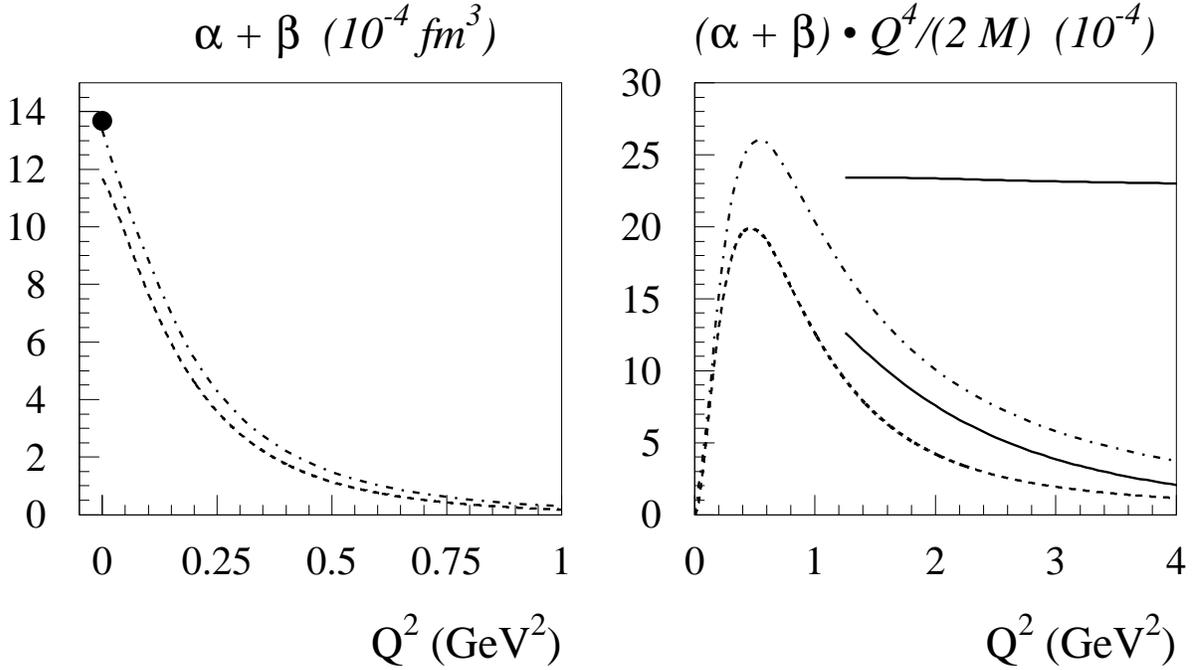}}
\vspace{-.5cm}
\caption{$Q^2$ dependence of the polarizability
$\alpha + \beta$  (left) and
$(\alpha + \beta) \cdot Q^4/(2 M)$ (right) for the proton, as
given by Eq.~(\ref{eq:alphapbeta}). The dashed (dashed-dotted) curves
represent the MAID estimate~\cite{Dre99,Dre01} for the
$\pi$ ($\pi + \eta + \pi\pi$) channels.
The upper solid curve is the evaluation using the DIS structure function
$F_1$~\cite{MRST01}. The lower solid curve is the evaluation
for the resonance region ($W < 2$ GeV) using the same DIS structure function.
The solid circle at $Q^2$ = 0 corresponds to the
Baldin sum rule~\cite{Bab98}.}
\label{fig:alphapbeta_vvcs}
\end{figure}
\newline
\indent
As in the case of $f_T$, 
also the longitudinal amplitude $f_L$ (which is also even in
$\nu$), should obey a subtracted DR~:
\begin{eqnarray}
\label{DDeq2.3.32}
{\mbox{Re}}\ f_L\,(\nu,\,Q^2) &\;=\;&
{\mbox{Re}}\ f_L^{\mbox{\sc{pole}}}(\nu,\,Q^2)
\;+\; \bigg[ {\mbox{Re}}\ f_L(0,\,Q^2) \,-\,
{\mbox{Re}}\ f_L^{\mbox{\sc{pole}}}(0,\,Q^2) \bigg]
\nonumber \\
\;&+&\;  \frac{\nu^2}{2\pi^2}\,{\mathcal{P}}\,\int_{\nu_0}^{\infty}\,
\frac{K(\nu',\,Q^2)\,\sigma_L\,(\nu',\,Q^2)}
{\nu'\,(\nu'^2-\nu^2)}\,d\nu'\ ,
\end{eqnarray}
where the pole part to $f_L$ can be read off 
Eq.~(\ref{DDeq2.3.21}),
\begin{eqnarray}
\label{eq:flpole}
{\mbox{Re}}\ f_L^{\mbox{\sc{pole}}} (\nu,\,Q^2) & = &
-\frac{\alpha_{em} \, Q^2}{M}
\frac{1}{\nu^2-\nu_B^2} \, G_E^2(Q^2) \, .
\end{eqnarray}
Analogously to Eq.~(\ref{eq:ftlex}),
one may again perform a low energy expansion for the non-pole (or
inelastic) contribution to the function $f_L(\nu,\,Q^2)$, defining a
longitudinal polarizability $\alpha_L(Q^2)$ as the coefficient of the
$\nu^2$ dependent term.
Equation~(\ref{DDeq2.3.32}) then yields a sum rule for this polarizability~:
\begin{eqnarray}
\label{eq:alphal}
\alpha_L\,(Q^2) &\;=\;& \frac{1}{2\pi^2}\,
\int_{\nu_0}^{\infty}\,\frac{K(\nu, \, Q^2)}{\nu} \,
\frac{\sigma_{L}\,(\nu,\,Q^2)}{\nu^2}\,d\nu\ , \nonumber\\
&\;=\;& \frac{e^2 \, 4M^3}{\pi \, Q^6}\,\int_{0}^{x_0}\,dx \,
\left\{ \frac{Q^2}{4 M^2} \,
\left[ F_2\,(x,\,Q^2) \,-\, 2 x \, F_1\,(x,\,Q^2) \right]
\,+\, x^2 \, F_2\,(x,\,Q^2) \right\} .
\end{eqnarray}
In the last line we used Eq.~(\ref{DDeq2.3.11}) to express $\alpha_L$
in terms of the first moment of $F_L \equiv F_2 - 2 x F_1$ and
the third moment of $F_2$.
Comparing Eqs.~(\ref{eq:alphapbeta}) and (\ref{eq:alphal}), one sees
that at large $Q^2$, where $F_1, F_2$ and $Q^2 F_L$ are $Q^2$
independent (modulo logarithmic scaling violations), the ratio
$\alpha_L / (\alpha + \beta) \sim 1/Q^2$. The quantity $\alpha_L$ is
therefore a measure of higher twist (i.e. twist-4) matrix elements.
\newline
\indent
In Fig.~\ref{fig:alphal_vvcs}, we show the $Q^2$ dependence for
$\alpha_L$ and compare the MAID model (for the one-pion channel)
with the DIS evaluation of Eq.~(\ref{eq:alphal}), using the MRST01
parametrization \cite{MRST01} for $F_2$ and $F_L$. By confronting the full DIS
estimate with the contribution from the range $W < 2$~GeV, one first
notices that around $Q^2 \simeq 1 - 2$~GeV$^2$, a transition
occurs from a resonance dominated towards a partonic description, as
is also seen for $\alpha + \beta$ in Fig.~\ref{fig:alphapbeta_vvcs}.
Furthermore, by comparing the MAID model with the DIS evaluation of
Eq.~(\ref{eq:alphal}) in the range $W < 2$~GeV, one notices that,
in contrast to the case of $\alpha + \beta$, the MAID model clearly
underestimates $\alpha_L$. This points to a lack of longitudinal
strength in the phenomenological model, which is to be addressed in future
analyses.
\newline
\indent
Similar as in the case of real photons (Baldin sum rule !), the
generalized polarizabilities can be, in principle, constructed
directly from the experimental data. However, this requires a
longitudinal-transverse separation of the cross sections at constant
$Q^2$ over a large energy range. 
\begin{figure}[h]
\vspace{-1.cm}
\epsfysize=11.5cm
\centerline{\epsffile{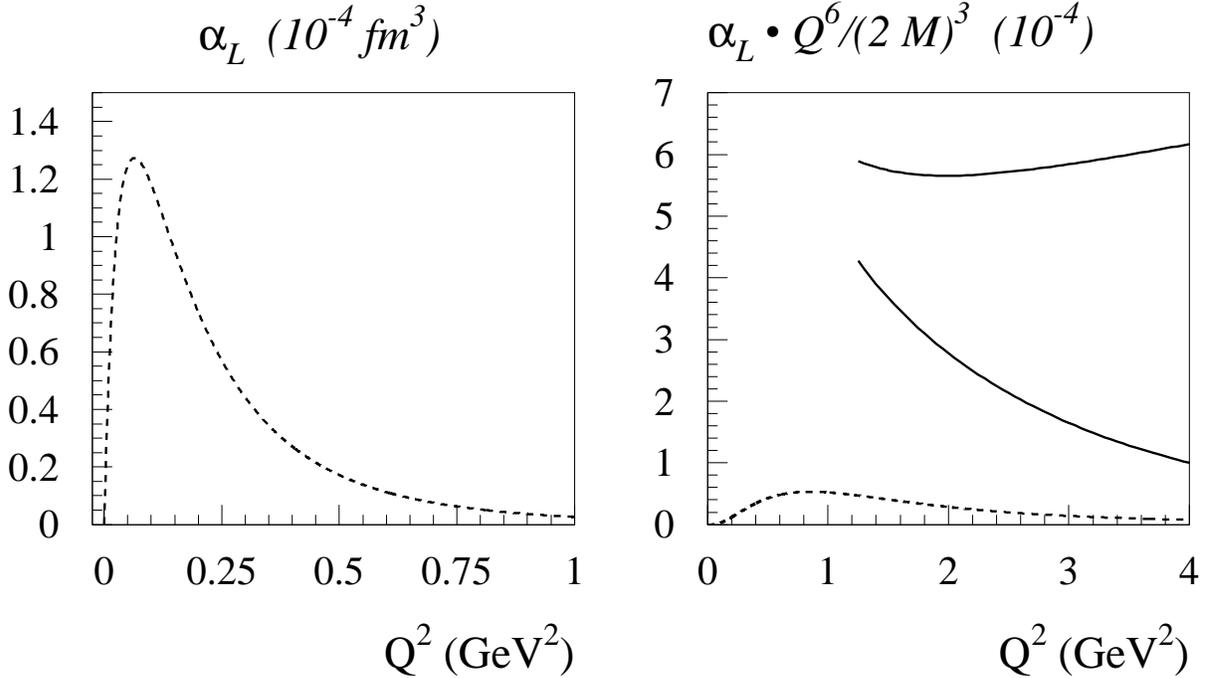}}
\vspace{-.5cm}
\caption{$Q^2$ dependence of the polarizability $\alpha_L$ (left)
and $\alpha_L \cdot Q^6/(2 M)^3$ (right) for the proton,
as given by Eq.~(\ref{eq:alphal}).
The dashed curve represents the MAID estimate~\cite{Dre99,Dre01}
for the one-pion channel.
The upper solid curve is the evaluation using the DIS structure
functions $F_2$ and $F_L$ \cite{MRST01}.
The lower solid curve is the evaluation
for the resonance region ($W < 2$ GeV) using the same DIS structure functions.}
\label{fig:alphal_vvcs}
\end{figure}
\newline
\indent
We next turn to the sum rules for the spin dependent VVCS amplitudes
(see also Ref.~\cite{Ji01} where a nice review of generalized sum rules for
spin dependent nucleon structure functions has been given).
Assuming an appropriate high-energy behavior, the spin-flip amplitude
$g_{TT}$ (which is odd in $\nu$) satisfies
an unsubtracted DR as in Eq.~(\ref{DDeq2.2.9}),
\beqn
\label{DDeq2.3.28}
{\mbox{Re}}\ g_{TT}\,(\nu,\,Q^2) = \frac{2\nu}{\pi}\,
{\mathcal{P}}\,\int_{0}^{\infty}\,\frac
{{\mbox{Im}}\ g_{TT}\,(\nu',\,Q^2)}{\nu'^2-\nu^2} \, d\nu'\ .
\eeqn
Assuming that the integral of Eq.~(\ref{DDeq2.3.28}) converges,
one can separate the contributions from the elastic cross section
at $\nu'=\nu_B$ and the inelastic continuum for $\nu'>\nu_0$, and
by use of Eq.~(\ref{DDeq2.3.20}), one obtains~:
\begin{eqnarray}
\label{DDeq2.3.29}
{\mbox{Re}}\ g_{TT}\,(\nu,\,Q^2) \;=\;
{\mbox{Re}}\ g_{TT}^{\mbox{\sc{pole}}}(\nu,\,Q^2)
\;+\;\frac{\nu}{2\pi^2}\,{\mathcal{P}}\,
\int_{\nu_0}^{\infty}\,\frac{K(\nu',\,Q^2)\,\sigma_{TT}(\nu',\,Q^2)}
{\nu'^2-\nu^2} \, d\nu' \, ,
\end{eqnarray}
where the pole part is given by Eq.~(\ref{DDeq2.3.21}) as~:
\begin{eqnarray}
\label{eq:gttpole}
{\mbox{Re}}\ g_{TT}^{\mbox{\sc{pole}}} (\nu,\,Q^2) & = &
-\frac{\alpha_{em} \, \nu}{2 \, M^2}
\frac{Q^2}{\nu^2-\nu_B^2} \,G_M^2(Q^2) \, .
\end{eqnarray}
Performing next a low energy expansion (LEX) for the non-pole contribution to
$g_{TT}(\nu, Q^2)$, we obtain~:
\begin{eqnarray}
\label{eq:gttlex}
{\mbox{Re}}\ g_{TT}(\nu,\,Q^2) \,-\,
{\mbox{Re}}\ g_{TT}^{\mbox{\sc{pole}}}(\nu,\,Q^2) \;=\;
\left(\frac{2 \, \alpha_{em} }{M^2} \right) \, I_A(Q^2) \; \nu
\;+\; \gamma_0(Q^2) \; \nu^3 \;+\; {\mathcal{O}}(\nu^5) \, .
\end{eqnarray}
For the ${\mathcal{O}}(\nu)$ term, Eq.~(\ref{DDeq2.3.29}) yields a
generalization of the GDH sum rule~:
\beqn
\label{DDeq2.3.30}
I_A(Q^2) &\;=\;&
\frac{M^2}{\pi \, e^2}\,
\int_{\nu_0}^{\infty}\, \frac{K(\nu, \, Q^2)}{\nu} \,
\frac{\sigma_{TT}\,(\nu,\,Q^2)}{\nu}\,d\nu \, , \nonumber\\
&\;=\;& \frac{2 \, M^2}{Q^2}\,
\int_{0}^{x_0}\,dx \, \left\{ g_1\,(x,\,Q^2)
\,-\, \frac{4 M^2}{Q^2} \, x^2 \, g_2\,(x,\,Q^2) \right\} \, ,
\eeqn
where the integral $I_A\,(Q^2)$ has been introduced in
Ref.~\cite{Dre01}. At $Q^2 = 0$, one recovers the GDH sum rule of
Eq.~(\ref{DDeq2.2.15}) as $I_A(0) = - \kappa_N^2 / 4$.
However, it has to be realized that several definitions have been
given how to generalize the integral to finite $Q^2$~\cite{Dre01}.
The definition $I_A$ of Eq.~(\ref{DDeq2.3.30}) has the advantage
that the (arbitrary) factor $K$ in the photon flux disappears (see
the discussion after Eq.~(\ref{DDeq2.3.20})). In other definitions
the factor $K/\nu$ in Eq.~(\ref{DDeq2.3.30}) is simply replaced by
1, which formally makes the integral look like the GDH integral
for real photons, Eq.~(\ref{DDeq2.2.15}). Unfortunately, these
integrals now depend on the definition of $K$ (see
Eq.~(\ref{DDeq2.3.5}). In the following we call these integrals
$I_B$ (Gilman's definition) and $I_C$ (Hand's definition), and
refer the reader to Ref.~\cite{Dre01} for the expressions
analogous to Eq.~(\ref{DDeq2.3.30}) and further details.
We will discuss the ${\mathcal{O}}(\nu)$ term in 
Eq.~(\ref{eq:gttlex}) and the first moment of $g_1$ in detail further
on, and turn first to the ${\mathcal{O}}(\nu^3)$ term.
\newline
\indent
From the ${\mathcal{O}}(\nu^3)$ term of Eq.~(\ref{eq:gttlex}),
one obtains a generalization of the forward spin polarizability,
\beqn
\label{eq:gammao}
\gamma_0\,(Q^2) &\;=\;& \frac{1}{2\pi^2}\,
\int_{\nu_0}^{\infty}\, \frac{K(\nu, \, Q^2)}{\nu} \,
\frac{\sigma_{TT}\,(\nu,\,Q^2)}{\nu^3}\,d\nu \, , \nonumber \\
&\;=\;& \frac{e^2 \, 4M^2}{\pi \, Q^6}\,\int_{0}^{x_0}\,dx \, x^2 \,
\left\{ g_1\,(x,\,Q^2)
\,-\, \frac{4 M^2}{Q^2} \, x^2 \, g_2\,(x,\,Q^2) \right\} \, .
\eeqn
At large $Q^2$, the term proportional to $g_2$ in
Eq.~(\ref{eq:gammao}) can be dropped 
and $\gamma_0$ is then proportional to the third moment of $g_1$.
\newline
\indent
In Fig.~\ref{fig:gammao_vvcs}, we show the $Q^2$ dependence of
$\gamma_0$ and compare the resonance estimate from MAID to the
evaluation with the DIS structure function $g_1$ for
$Q^2 > 1$~GeV$^2$. For the structure function $g_1$, we use the recent
fit performed in~\cite{Blum02}, which also provides $1 \sigma$
error bands for this distribution, allowing us to determine the
experimental error on $\gamma_0$, as shown by the
shaded bands in Fig.~\ref{fig:gammao_vvcs}. At low $Q^2$, one sees
that the estimate for the one-pion channel completely dominates
$\gamma_0$ and reproduces well its measured value at $Q^2$ = 0.
At $Q^2 > 2$~GeV$^2$, the MAID model ($\pi + \eta + \pi\pi$ channels)
is also in good agreement with the DIS evaluation of the $W < 2$~GeV
range in the integral Eq.~(\ref{eq:gammao}) for
$\gamma_0$. Furthermore, comparing the full DIS estimate with the
contribution from the range $W < 2$~GeV, we once more observe the
gradual transition from the resonance dominated to the partonic
region. Around $Q^2$ = 4 GeV$^2$, the $W < 2$~GeV region contributes
about 30~\% to $\gamma_0$, whereas for $\alpha + \beta$ at the same
$Q^2$, this contribution is below 10~\%. This difference can be understood by
comparing the sum rule Eq.~(\ref{eq:gammao}) for $\gamma_0$ with
Eq.~(\ref{eq:alphapbeta}) for $\alpha + \beta$.
From this comparison, one notices that the
sum rule for $\gamma_0$ invokes one additional power of $\nu$ in the
denominator, giving higher weight to the resonance region
as compared with $\alpha + \beta$.
\begin{figure}[h]
\vspace{-1.4cm}
\epsfysize=11.5cm
\centerline{\epsffile{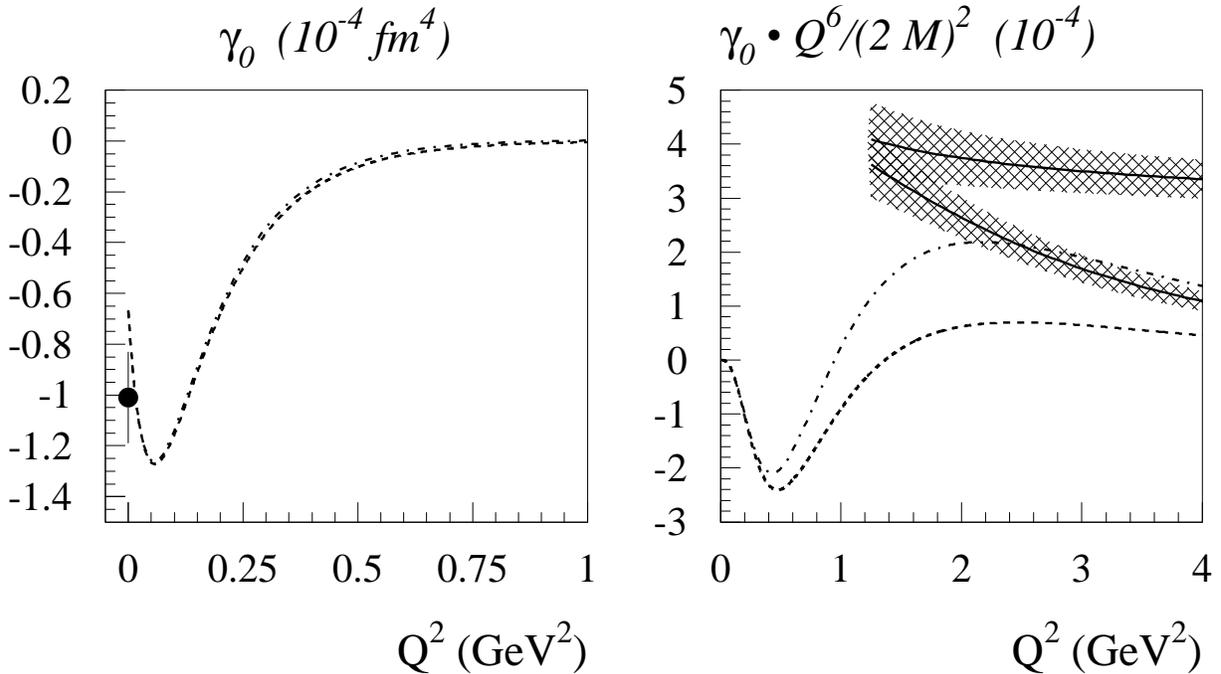}}
\vspace{-.6cm}
\caption{$Q^2$ dependence of the polarizability
$\gamma_0$  (left) and
$\gamma_0 \cdot Q^6/(2 M)^2$ (right) for the proton, as
given by Eq.~(\ref{eq:gammao}).
The dashed (dashed-dotted) curves
represent the MAID estimate~\cite{Dre99,Dre01}
 for the $\pi$ ($\pi + \eta + \pi\pi$) channels.
The upper solid curve is the evaluation using the DIS structure function
$g_1^p$ \cite{Blum02}. The lower solid curve is the evaluation
for the resonance region ($W < 2$ GeV) using the DIS structure function.
The shaded bands represent the corresponding error estimates
as given by Ref.~\cite{Blum02}.
The solid circle at $Q^2$ = 0 corresponds to the
evaluation of Eq.~(\ref{DDeq2.2.19}).}
\label{fig:gammao_vvcs}
\end{figure}
\newline
\indent
We next turn to the amplitude $g_{LT}(\nu, Q^2)$, which is even in $\nu$.
Assuming an unsubtracted DR exists for the amplitude
$g_{LT}$, it takes the form
\begin{eqnarray}
{\mbox{Re}}\ g_{LT}(\nu,\,Q^2) & = &
{\mbox{Re}}\ g_{LT}^{\mbox{\sc{pole}}}(\nu,\,Q^2) \;+\;
\frac{1}{2\pi^2}\,{\mathcal{P}}\,\int_{\nu_0}^{\infty}\,
\frac{\nu'K\,(\nu'\, ,Q^2)\,\sigma_{LT}(\nu'\, ,Q^2)}{(\nu'^2-\nu^2)}\,d\nu'
\, , 
\label{DDeq2.3.35}
\end{eqnarray}
where the pole part is given by Eq.~(\ref{DDeq2.3.21}) as~:
\begin{eqnarray}
\label{eq:gltpole}
{\mbox{Re}}\ g_{LT}^{\mbox{\sc{pole}}} (\nu,\,Q^2) & = &
\, - \, \frac{\alpha_{em} \, Q}{2 \, M^2}
\frac{Q^2}{\nu^2-\nu_B^2} \,G_E(Q^2) \, G_M(Q^2) \, .
\end{eqnarray}
One sees that for the unsubtracted dispersion integral of
Eq.~(\ref{DDeq2.3.35}) to converge, the cross section
$\sigma_{LT}(\nu, Q^2)$ should drop {\it faster} than 1 / $\nu$ at large
$\nu$.
One can then perform a low energy expansion for the non-pole contribution to
$g_{LT}(\nu, Q^2)$, as~:
\begin{eqnarray}
{\mbox{Re}}\ g_{LT}(\nu,\,Q^2) -
{\mbox{Re}}\ g_{LT}^{\mbox{\sc{pole}}}(\nu,\,Q^2) \,=\,
\left(\frac{2 \, \alpha_{em} }{M^2} \right) Q \, I_3(Q^2)
\,+\, Q \, \delta_{LT}(Q^2) \, \nu^2 \,+\, {\mathcal{O}}(\nu^4) \, ,
\nonumber \\
\label{eq:gltlex}
\end{eqnarray}
where $I_3(Q^2)$ has been introduced in Ref.~\cite{Dre01} as
\beqn
\label{eq:i3int}
I_3(Q^2) &\; = \;&
\frac{M^2}{\pi \,e^2}\,
\int_{\nu_0}^{\infty}\,\frac{K(\nu \, , Q^2)}{\nu} \,
\frac{1}{Q} \, \sigma_{LT}\,(\nu,\,Q^2)\,d\nu\  \nonumber\\
&\;=\;& \frac{2 \, M^2}{Q^2}\,\int_{0}^{x_0}\,dx \,
\left\{ g_1\,(x,\,Q^2) \,+\, g_2\,(x,\,Q^2) \right\} \, .
\eeqn
\indent
For the ${\mathcal{O}}(\nu^2)$ term of Eq.~(\ref{eq:gltlex}),
one obtains a generalized longitudinal-transverse polarizability,
\beqn
\label{eq:deltalt}
\delta_{LT}\,(Q^2) &\;=\;& \frac{1}{2\pi^2}\,
\int_{\nu_0}^{\infty}\,\frac{K(\nu, \, Q^2)}{\nu} \,
\frac{\sigma_{LT}(\nu\, ,Q^2)}{Q\,\nu^2}\,d\nu  \nonumber \\
&\;=\;& \frac{e^2 \, 4M^2}{\pi \, Q^6}\,\int_{0}^{x_0}\,dx \, x^2 \,
\left\{ g_1\,(x,\,Q^2) \,+\, g_2\,(x,\,Q^2) \right\} \, .
\eeqn
This function is finite in the limit $Q^2\rightarrow 0$, and
can be evaluated safely on the basis of dispersion
relations. We note that in Ref.~\cite{Dre01}, the
quantity $\delta_0$ differs by the factor $(1-x)$ in the
integrand.
At large $Q^2$, $\delta_{LT}$ is proportional to the third
moment of the transverse spin structure function $g_T \equiv g_1 + g_2$.
In this limit, Wandzura and Wilczek~\cite{WW77} have shown that
when neglecting dynamical (twist-3) quark-gluon correlations, the
transverse spin structure function $g_T$ can be expressed in terms of
the twist-2 spin structure function $g_1$ as~:
\beqn
\label{eq:ww}
g_1\,(x, Q^2) \,+\, g_2\,(x, Q^2) \;=\;
\int_{x}^{1}\, dy \; \frac{g_{1}\,(y,\,Q^2)}{y} \, .
\eeqn
Recent experimental data from SLAC~\cite{Ant99,E155X} for the spin structure
function $g_2$ show that the measured value of $g_2(x, Q^2)$
(in the range $0.02 \leq x \leq 0.8$  and 1~GeV$^2 \leq Q^2 \leq 30$ GeV$^2$)
is consistent with the Wandzura-Wilczek (WW) relation of Eq.~(\ref{eq:ww}).
One can therefore evaluate the {\it rhs} of Eq.~(\ref{eq:deltalt}),
to good approximation,
by calculating the third moment of both sides of Eq.~(\ref{eq:ww}).
By changing the integration variables $(x, y) \to (z, y)$
with $x = z \cdot y$, one obtains~:
\beqn
\label{eq:deltaltww}
\int_0^1 \, dx \, x^2 \; \int_{x}^{1}\, dy \, \frac{g_{1}\,(y,\,Q^2)}{y}
\;=\; \frac{1}{3} \, \int_{0}^{1}\, dy \, y^2 \, g_{1}\,(y,\,Q^2) \, .
\eeqn
Combining Eqs.~(\ref{eq:gammao}) and (\ref{eq:deltalt})
with Eq.~(\ref{eq:deltaltww}) and using the WW relation, we may relate the generalized spin
polarizabilities $\delta_{LT}(Q^2)$ and $\gamma_0(Q^2)$, at large $Q^2$~:
\footnote{Note that for $Q^2 \to \infty$, one can again neglect the
elastic contribution and make the replacement
$\int_{0}^{x_0} \to \int_{0}^{1}$ in Eq.~(\ref{eq:deltalt}).}
\beqn
\label{eq:deltaltww2}
\delta_{LT}(Q^2) \rightarrow \frac{1}{3} \; \gamma_0(Q^2) \, , \hspace{1cm}
Q^2 \to \infty \, .
\eeqn
In Fig.~\ref{fig:deltalt_vvcs}, the $Q^2$ dependence of the
polarizability $\delta_{LT}$ is shown both for
the MAID model (for the one-pion channel)
and for the DIS evaluation of Eq.~(\ref{eq:deltaltww2}).
Comparing the MAID model with the DIS evaluation for the range $W < 2$~GeV,
one notices that the MAID model underestimates $\delta_{LT}$,
similarly as was seen in Fig.~\ref{fig:alphal_vvcs} for $\alpha_L$.
As the polarizability $\delta_{LT}$ involves a longitudinal amplitude,
this may again point to a lack of longitudinal strength in the MAID model.
\begin{figure}[h]
\vspace{-1.4cm}
\epsfysize=11.5cm
\centerline{\epsffile{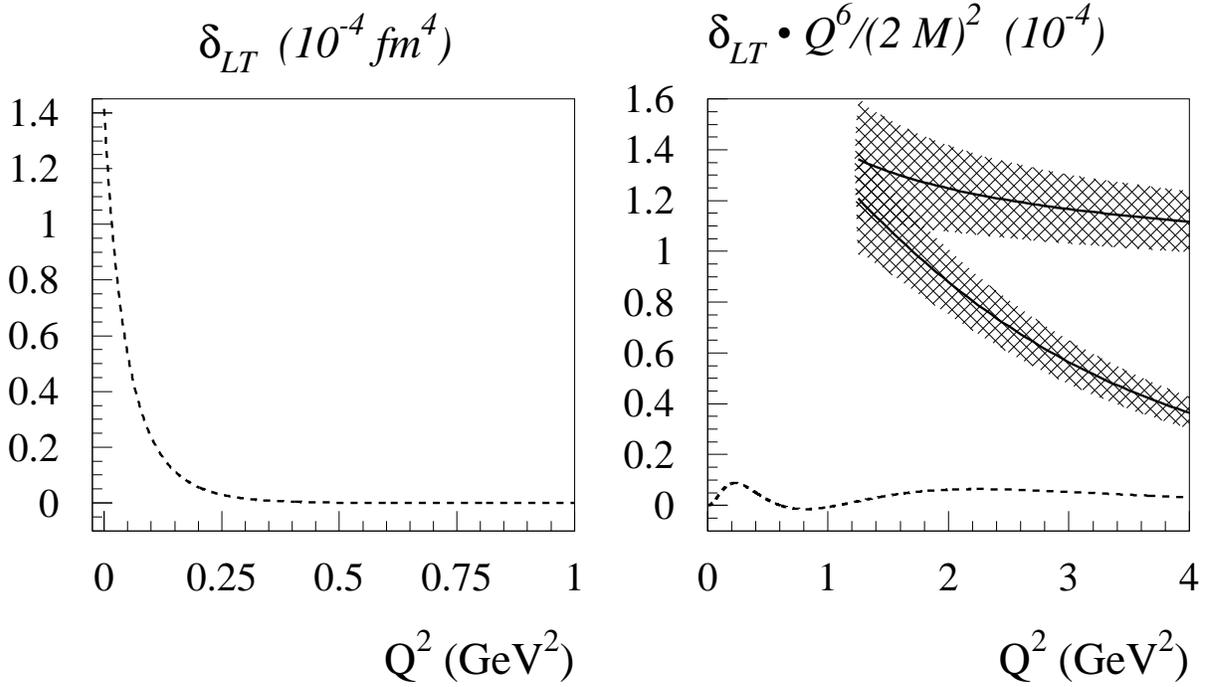}}
\vspace{-.6cm}
\caption{$Q^2$ dependence of the polarizability
$\delta_{LT}(Q^2)$ (left) and
$\delta_{LT}(Q^2) \cdot Q^6/(2 M)^2$ (right) for the proton, as
given by Eq.~(\ref{eq:deltalt}). The dashed curve represents the
MAID estimate for the one-pion channel~\cite{Dre99}.
The upper solid curve is the evaluation using the DIS structure function
$g_T = g_1 + g_2$ \cite{Blum02}. The lower solid curve is the evaluation
for the resonance region ($W < 2$ GeV) using the DIS structure function.
The shaded bands represent the corresponding error estimates
as given by Ref.~\cite{Blum02}.}
\label{fig:deltalt_vvcs}
\end{figure}
\newline
\indent
In order to construct the VVCS amplitudes
which are in one-to-one correspondence with the quark
structure functions, it is useful to cast Eq.~(\ref{DDeq2.3.19})
into a covariant form,
\begin{eqnarray}
\label{DDeq2.3.48}
T(\nu,\,Q^2,\,\theta=0) \;=\;
\varepsilon_{\mu}'^{\ast}\varepsilon_{\nu}
&&\left \{ \left( -g^{\mu\nu}+\frac{q^{\mu}q^{\nu}}{q^2}\right)
T_1(\nu,\,Q^2) \right .\nonumber \\
&& \left . +\frac{1}{p\cdot q} \left(p^{\mu}-\frac{p\cdot
q}{q^2}\,q^{\mu}\right) \left(p^{\nu}-\frac{p\cdot
q}{q^2}\, q^{\nu} \right) T_2(\nu,\,Q^2) \right .\nonumber \\
&& \left . +\frac{i}{M}\,\epsilon^{\mu\nu\alpha\beta}\,q_{\alpha}
s_{\beta}\, S_1(\nu,\,Q^2) \right . \nonumber \\
&& \left . + \frac{i}{M^3}\,\epsilon^{\mu\nu\alpha\beta}\,q_{\alpha}
(p\cdot q\ s_{\beta}-s\cdot q\ p_{\beta})\, S_2(\nu,\,Q^2) \right
\}\, ,
\end{eqnarray}
where $\epsilon_{0123} = +1$, and 
$s^\alpha$ is the nucleon covariant spin vector satisfying
$s \cdot p$ = 0, $s^2$ = -1. With the definition of
Eq.~(\ref{DDeq2.3.48}), all four VVCS amplitudes $T_1, T_2, S_1$ and
$S_2$ have the same dimension of mass. Furthermore,
the 4 new structure functions are related to the previously introduced
VVCS amplitudes of Eq.~(\ref{DDeq2.3.19}) as follows~:
\begin{eqnarray}
\label{DDeq2.3.49}
T_1(\nu,\,Q^2) & = & f_T (\nu,\,Q^2) \, , \\
\label{DDeq2.3.50}
T_2(\nu,\,Q^2) & = & \frac{\nu}{M}\ \frac{Q^2}{\nu^2+Q^2}\,
\left(f_T(\nu,\,Q^2)+f_L(\nu,\,Q^2) \right) \, , \\
\label{DDeq2.3.51}
S_1(\nu,\,Q^2) & = &
\frac{\nu\,M}{\nu^2+Q^2}\left(g_{TT}(\nu,\,Q^2) +
\frac{Q}{\nu}\,g_{LT}(\nu,\,Q^2)\right) \, , \\
\label{DDeq2.3.52}
S_2(\nu,\,Q^2) & = & -
\frac{M^2}{\nu^2+Q^2}\left(g_{TT}(\nu,\,Q^2) -
\frac{\nu}{Q}\,g_{LT}(\nu,\,Q^2)\right) \, .
\end{eqnarray}
The Born contributions to these functions can be expressed in
terms of the form factors by use of Eq.~(\ref{DDeq2.3.21}) 
as follows~:
\begin{eqnarray}
\label{eq:bornt12s12}
T_1^{\mbox{\sc{Born}}} (\nu,\,Q^2) & = &
-\frac{\alpha_{em}}{M}
\left (F_D^2 + \frac{\nu_B^2}{\nu^2-\nu_B^2+i \varepsilon}
\,G_M^2\right ) \, ,
\nonumber \\
T_2^{\mbox{\sc{Born}}} (\nu,\,Q^2) & = &
-\frac{\alpha_{em} \nu}{M^2} \, 
\frac{Q^2}{\nu^2-\nu_B^2+i \varepsilon} \,
\left (F_D^2 + \tau \, F_P^2 \right ) \, ,
\nonumber \\
S_1^{\mbox{\sc{Born}}} (\nu,\,Q^2) & = &
-\frac{\alpha_{em}}{2M}
\left (F_P^2 + \frac{Q^2}{\nu^2-\nu_B^2+i \varepsilon}
\,F_D (F_D + F_P) \right ) \, ,
\nonumber \\
S_2^{\mbox{\sc{Born}}} (\nu,\,Q^2) & = &
\frac{\alpha_{em}}{2}
\frac{\nu}{\nu^2-\nu_B^2+i \varepsilon}
\,F_P (F_D + F_P) \, ,
\end{eqnarray}
One sees that the pole singularities appearing at $\nu=\pm
i\,Q$, due to the denominators 
in Eqs.~(\ref{DDeq2.3.50})-(\ref{DDeq2.3.52}), are 
actually canceled by a corresponding zero in the numerator of
the Born terms. The imaginary parts of the inelastic contributions
follow from Eq.~(\ref{DDeq2.3.20}),
\begin{eqnarray}
\label{DDeq2.3.53}
{\rm Im}\ T_1 & = & \frac{K}{4\pi}\,\sigma_T = \frac{e^2}{4M}\,F_1
\ ,\\
\label{DDeq2.3.54}
{\rm Im}\ T_2 & = & \frac{\nu}{M}\,\frac{Q^2}{\nu^2+Q^2}\,
\frac{K}{4\pi}\,(\sigma_T+\sigma_L) = \frac{e^2}{4M}\,F_2 \ ,\\
\label{DDeq2.3.55}
{\rm Im}\ S_1 & = & \frac{\nu\, M}{\nu^2+Q^2}\,\frac{K}{4\pi}\,
\left(\sigma_{TT}+\frac{Q}{\nu}\,\sigma_{LT}\right )=
\frac{e^2}{4M}\,\frac{M}{\nu}\,g_1 := \frac{e^2}{4M}\,G_1\ ,\\
\label{DDeq2.3.56}
{\rm Im}\ S_2 & = & -\frac{M^2}{\nu^2+Q^2}\,\frac{K}{4\pi}\,
\left(\sigma_{TT}-\frac{\nu}{Q}\,\sigma_{LT}\right )=
\frac{e^2}{4M}\,\frac{M^2}{\nu^2}\,g_2 :=  \frac{e^2}{4M}\,G_2\ .
\end{eqnarray}
In order to cancel the singularities at $\nu=\pm i\,Q$, the
following relations should be fulfilled if the partial cross
sections are continued into the complex $\nu$-plane:
\begin{eqnarray}
\label{DDeq2.3.57}
\sigma_T\,(iQ,\,Q^2) &\;=\;& -\sigma_L \,(iQ,\,Q^2) \, , \nonumber \\
\sigma_{TT}\,(iQ,\,Q^2) &\;=\;& i\,\sigma_{LT} \,(iQ,\,Q^2)\ .
\end{eqnarray}
These relations can be verified by realizing that the
singularities at $Q^2+\nu^2=0$ correspond to the Siegert limit,
${\bq}_{\rm\sc{lab}}\rightarrow 0$, which also implies
${\bq}_{\rm\sc{cm}}\rightarrow 0$.
Furthermore, all multipoles vanish in that limit, except for the
(unretarded) dipole amplitudes. In the case of one-pion production
as presented in Ref.~\cite{Dre01}, these are the amplitudes $E_{0+}$ and
$L_{0+}$ of the transverse and longitudinal currents,
respectively, which become equal in the Siegert limit. The
relations of Eq.~(\ref{DDeq2.3.57}) then follow straightforwardly.
\newline
\indent
We next discuss dispersion relations for the spin dependent 
amplitudes $S_1$ and $S_2$.
The spin-dependent VVCS amplitude $S_1$ is even in $\nu$,
and an unsubtracted DR reads
\beqn
\label{DDeq2.3.62}
{\rm Re}\ S_1(\nu,\,Q^2) 
\,=\, {\rm Re}\ S_1^{{\rm\sc{pole}}}\,+\, 
\frac{2}{\pi}\,{\mathcal{P}}\,\int_{\nu_0}^{\infty}
\frac{\nu'\,{\rm Im}\ S_1(\nu',\,Q^2)}{\nu'^2 - \nu^2} \,d\nu' \, ,
\eeqn
where the pole part ${\rm Re}\ S_1^{{\rm\sc{pole}}}$ is obtained from   
Eq.~(\ref{eq:bornt12s12}) as~: 
\beqn
{\rm Re}\ S_1^{\mbox{\sc{pole}}} (\nu,\,Q^2) & = &
-\frac{\alpha_{em}}{2M}
\frac{Q^2}{\nu^2-\nu_B^2}\,F_D(Q^2) \, \left(F_D(Q^2) + F_P(Q^2) \right ) \, ,
\eeqn
\newline
\indent
We can next perform a low-energy expansion for
$S_1(\nu,\,Q^2) -  S_1^{{\rm\sc{pole}}}(\nu,\,Q^2)$ as~:
\begin{eqnarray}
&&\hspace{-.25cm}{\mbox{Re}}\ S_1(\nu,\,Q^2) -
{\mbox{Re}}\ S_1^{\mbox{\sc{pole}}}(\nu,\,Q^2) \,=\, \nonumber \\
&&\hspace{-.25cm}
\left(\frac{2 \, \alpha_{em} }{M} \right) \, I_1(Q^2)
\,+\, \left[ \left(\frac{2 \, \alpha_{em} }{M} \right) 
{1 \over {Q^2}} \left( I_A(Q^2) - I_1(Q^2) \right)
\,+\, M \delta_{LT}(Q^2) \right] \, \nu^2  \,+\, {\mathcal{O}}(\nu^4) ,
\label{eq:s1lex}
\end{eqnarray}
where the leading term in $\nu^0$ follows
from Eq.~(\ref{DDeq2.3.62}) as~:
\beqn
I_1(Q^2) &\;\equiv\;& \frac{2M^2}{Q^2}\int_0^{x_0}
g_1(x,\,Q^2)\,dx \nonumber \\
&\;=\;& \frac{M^2}{\pi \,e^2}\,
\int_{\nu_0}^{\infty}\,\frac{K(\nu, Q^2)}{(\nu^2 + Q^2)}
\left\{\sigma_{TT}\,(\nu,\,Q^2) \,+\,
\frac{Q}{\nu} \, \sigma_{LT}\,(\nu,\,Q^2) \right\} \,d\nu \, ,
\label{eq:I1} 
\eeqn
which reduces to the GDH sum rule at $Q^2=0$, as 
$I_1(0) = - \kappa_N^2 / 4$. By using Eqs.~(\ref{DDeq2.3.30}), 
(\ref{eq:deltalt}) and (\ref{eq:I1}), one can verify that the term in 
$\nu^2$ in Eq.~(\ref{eq:s1lex})
can be expressed in terms of $I_A, I_1$ and $\delta_{LT}$.
\newline
\indent
At large $Q^2$, $I_1(Q^2)$ has the limit 
\beqn
\label{eq:Aqhighq}
I_1(Q^2)\rightarrow\frac{2 \, M^2}{Q^2} \; \Gamma_1(Q^2)\, , \hspace{1cm}
Q^2 \to \infty \, ,
\eeqn
with \footnote{At $Q^2 \to \infty$ , one
can replace $\int_{0}^{x_0} \to \int_{0}^{1}$,
because the elastic contribution to $\Gamma_1$ vanishes like $Q^{-8}$
and is therefore highly suppressed.}~:
\beqn \Gamma_1(Q^2) \equiv
\int_{0}^{1}\,dx \, g_1\,(x,\,Q^2) \, .
\eeqn
For the first moment
$\Gamma_1$, a next-to-leading order (NLO) QCD fit to all available
DIS data for $g_1^p (g_1^n)$ has been performed in
Ref.~\cite{Ant00}, yielding the values at $Q^2 = 5 \, \mathrm{GeV}^2$~:
\beqn
\Gamma_1^p &\;=\;& 0.118 \pm 0.004 \pm 0.007 \; , \nonumber \\
\Gamma_1^n &\;=\;& -0.058 \pm 0.005 \pm 0.008 \; , \nonumber \\
\Gamma_1^p - \Gamma_1^n &\;=\;& 0.176 \pm 0.003 \pm 0.007 \; .
\label{eq:disg1}
\eeqn
For the isovector combination $\Gamma_1^p -
\Gamma_1^n$, the Bjorken sum rule \cite{Bj66} predicts~:
\beqn
\label{eq:bjsr}
\Gamma_1^p - \Gamma_1^n
\,\to\, \frac{1}{6} \, g_A \, = \, 0.211 \pm 0.001 \, ,
\hspace{1cm} Q^2 \to \infty \, ,
\eeqn
where $g_A$ is the
axial-vector weak coupling constant. The inclusion of QCD
corrections up to order $\alpha_s^3$ yields~\cite{Lar91}~:
\beqn
\label{eq:bjsr2}
\hspace{-.75cm}
\Gamma_1^p - \Gamma_1^n =
\frac{1}{6} \, g_A
\left\{ 1 - \left( \frac{\alpha_s(Q^2)}{\pi} \right)
- 3.5833 \, \left( \frac{\alpha_s(Q^2)}{\pi} \right)^2
- 20.2153 \, \left( \frac{\alpha_s(Q^2)}{\pi} \right)^3
\right\}.
\eeqn
When evaluating Eq.~(\ref{eq:bjsr2}) using 3 light quark flavors
in $\alpha_s$ and fixing $\alpha_s(M_Z^2)$ at 0.114, one obtains
\cite{Ant00}~:
\beqn
\label{eq:bj5}
\Gamma_1^p - \Gamma_1^n \;=\; 0.182 \pm 0.005 \, ,
\hspace{1cm} \mathrm{at} \;\;\; Q^2 = 5 \; \mathrm{GeV}^2 \; .
\eeqn
One sees that the experimental value of Eq.~(\ref{eq:disg1}) is in
good agreement with the Bjorken sum rule value of Eq.~(\ref{eq:bj5}). 
\newline
\indent
In Fig.~\ref{fig:i1p}, we show the $Q^2$ dependence of
$I_1$ for the proton and compare the MAID estimate with the DIS
evaluation for $Q^2 > 1 $~GeV$^2$, using the parametrization of
Ref.~\cite{Blum02} for $g_1$. One immediately sees that
the integral $I_1^p$ has to undergo a
sign change from the large negative GDH sum rule value at $Q^2 = 0$ to
the positive value at large $Q^2$ as extracted from DIS. Recent data
from SLAC \cite{Abe97} and JLab/CLAS \cite{CLAS02} 
cover the intermediate $Q^2$ range. 
In particular, the JLab/CLAS data, 
which extend downwards to $Q^2 \simeq
0.15 $~GeV$^2$, clearly confirm this sign change in the sum rule, which
occurs around $Q^2 \simeq 0.25 $~GeV$^2$. The resonance estimate of
MAID, including $\pi + \eta + \pi\pi$ channels also displays such a
sign change. Given some uncertainties in the evaluation of the $\pi\pi$
channels and higher continua, the calculation qualitatively
reproduces the trend of the data for the $W < 2 $~GeV contribution
to $I_1^p$. At larger $Q^2$, one again
notices the gradual transition from a resonance dominated to a
partonic description. For example, at $Q^2 = 2 $~GeV$^2$, the
$W < 2 $~GeV region amounts to only to 20 \% of the total sum rule value
for $I_1^p$. To gain an understanding of this gradual transition in 
the integral $I_1^p$, it was proposed in \cite{Ans89} to
parametrize the $Q^2$ dependence through a vector meson dominance type model. 
This model was refined in Refs.~\cite{Bur92,Bur93} by adding explicit resonance
contributions which are important at low $Q^2$ as discussed above and  
lead to the following phenomenological parametrization~:
\begin{equation}
I_1^{p, n}(Q^2) \,=\,  I_{1, res}^{p, n}(Q^2) \,+\, 
2 M^2 \, \Gamma_{1, as}^{p, n} \left[ {1 \over (Q^2 + \mu^2)} - 
{{c^{p,n} \, \mu^2} \over (Q^2 + \mu^2)^2} \right] \, ,
\label{eq:bi1}
\end{equation}
where $I_{1, res}^{p, n}(Q^2)$ is the resonance contribution to $I_1^{p,n}$, 
$\Gamma_{1, as}^{p, n}$ are the asymptotic values for the first moments of
$g_1$, and the scale $\mu$ was assumed to be the vector meson mass 
\cite{Ans89}, i.e., $\mu = m_\rho$. 
Furthermore, the parameter $c^{p,n}$ in Eq.~(\ref{eq:bi1}) was chosen as
\begin{equation}
c^{p, n} \,=\, 1 \,+\, {{\mu^2} \over {2 M^2}} \,
{1 \over \Gamma_{1, as}^{p, n}} \left[ {{\kappa^2} \over 4} \,+\, 
I_{1, res}^{p, n}(0) \right] \, ,
\label{eq:bi2}
\end{equation}
so as to reproduce the sum rule at $Q^2$ = 0.
In Fig.~\ref{fig:i1p}, we use the parametrization of 
Eq.~(\ref{eq:bi1}), but take the recent experimental value 
of Eq.~(\ref{eq:disg1}) for $\Gamma_{1, as}^p$. Furthermore, we
use as input for the resonance contribution 
at the real photon point $I_{1, res}^{p}(0)$ (corresponding with $W < 2$~GeV) 
the experimental value from Ref.~\cite{Ahr01}~: 
$I_{1, res}^{p}(0) = -0.95$ (open diamond in Fig.~\ref{fig:i1p}). 
For the $Q^2$ dependence of the resonance contribution, we take the
MAID estimate~\cite{Dre99,Dre01} rescaled to the experimental value 
$I_{1, res}^{p}(0)$ at the real photon point. 
It is seen from Fig.~\ref{fig:i1p} that the resulting 
calculation (shown by the thin solid curve) gives a rather good
description of the sign change occurring in $I_1^p$ at 
$Q^2 \simeq 0.25$~GeV$^2$. 
%
\begin{figure}[h]
\vspace{-.25cm}
\epsfysize=15.cm
\centerline{\epsffile{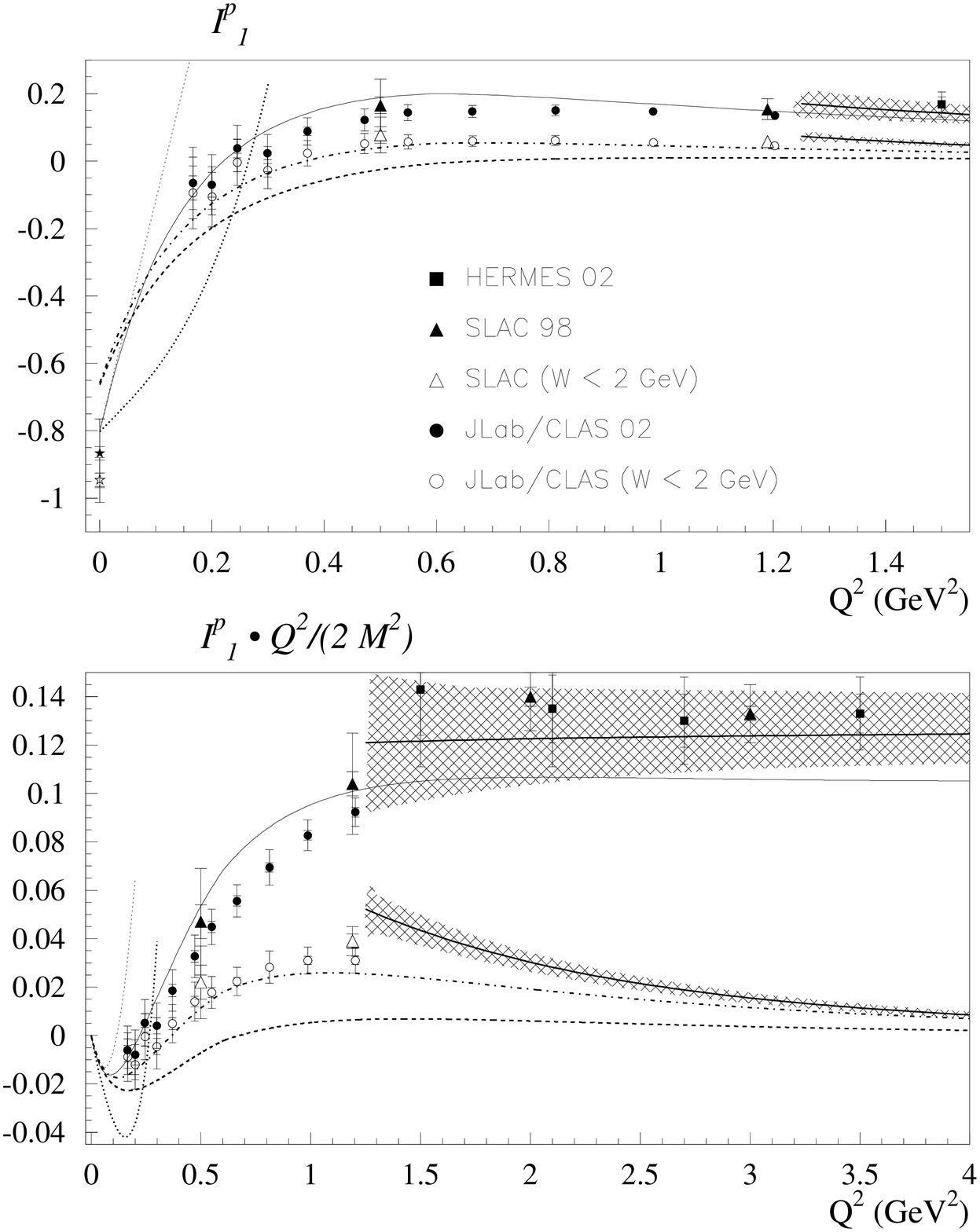}}
\vspace{-.5cm}
\caption{$Q^2$ dependence of the integral $I_1^p$ (upper panel) and
 $I_1^p \cdot Q^2/(2 M^2)$ (lower panel) for the proton, as
given by Eq.~(\ref{eq:I1}).
The dashed (dashed-dotted) curve
represent the MAID estimate~\cite{Dre99,Dre01}
for the $\pi$ ($\pi + \eta + \pi\pi$) channels.
The thin solid curve, covering the whole $Q^2$ range,  
is the parametrization of Eq.~(\ref{eq:bi1})
evaluated as described in the text.
The upper thin dotted curves are the ${\mathcal O}(p^4)$ 
HBChPT results of Ref.~\cite{JiK00a}, whereas the lower thick dotted
curves are the corresponding ${\mathcal O}(p^4)$ relativistic BChPT results 
of Ref.~\cite{Ber02}.
The upper thick solid curve at $Q^2 \geq 1.25$~GeV$^2$ 
is the evaluation using the DIS structure function $g_1^p$ 
of Ref.~\cite{Blum02}. The lower thick solid curve in the same range 
is the evaluation for the resonance region ($W < 2$ GeV) 
using the DIS structure function. The shaded bands around the thick
solid curves represent 
the corresponding error estimates as given by Ref.~\cite{Blum02}.
The open star at $Q^2$ = 0 corresponds with the MAMI data
\cite{Ahr00} combined with the estimate for the non-measured 
contribution in the range $W <$ 2 GeV, as given by Eq.~(\ref{eq:ipwl2}). 
The solid star is the total value of Eq.~(\ref{DDeq2.2.20}), 
which includes the estimate for $W >$ 2 GeV.
The SLAC data are from Ref.~\cite{Abe97}, 
the HERMES data are from Ref.~\cite{Air02}, 
and the preliminary JLab/CLAS data are from
Ref.~\cite{CLAS02} (inner error bars are statistical errors only, outer
error bars include systematical errors).
}
\label{fig:i1p}
\end{figure}
\newline
\indent
The following Fig.~\ref{fig:ian} displays the results for the
generalized GDH integral for the neutron, as derived from the
$^3$He data of the Hall A Collaboration at JLab~\cite{Ama02} and
corrected for nuclear effects according to the procedure of
Ref.~\cite{Cio97}. Recently, also the generalized GDH integral for the
deuteron has been measured by the Clas Collaboration 
at JLab~\cite{Kuhn93,Yun03}, 
and will provide a cross-check for the extraction of the generalized 
GDH integral for the neutron.
The comparison of the existing neutron  
data with the MAID results in Fig.~\ref{fig:ian} shows the 
same problem as already discussed for real photons: The
helicity difference in the low-energy region is not properly
described by the existing phase shift analyses. However, the
strong curvature at $Q^2\approx 0.1$~GeV$^2$ agrees nicely with the
predictions. 
In Fig.~\ref{fig:ian}, we also show the corresponding parametrization of 
Eq.~(\ref{eq:bi1}) for $I_A$, by using the value of Eq.~(\ref{eq:disg1}) 
for $\Gamma_{1, as}^n$ and the MAID estimate for the
resonance contribution. It is seen that 
the resulting calculation gives a rather good
description for the generalized GDH integral $I_A$ for the neutron. 
\begin{figure}[h]
\vspace{-1.cm}
\epsfysize=9.25cm
\centerline{\epsffile{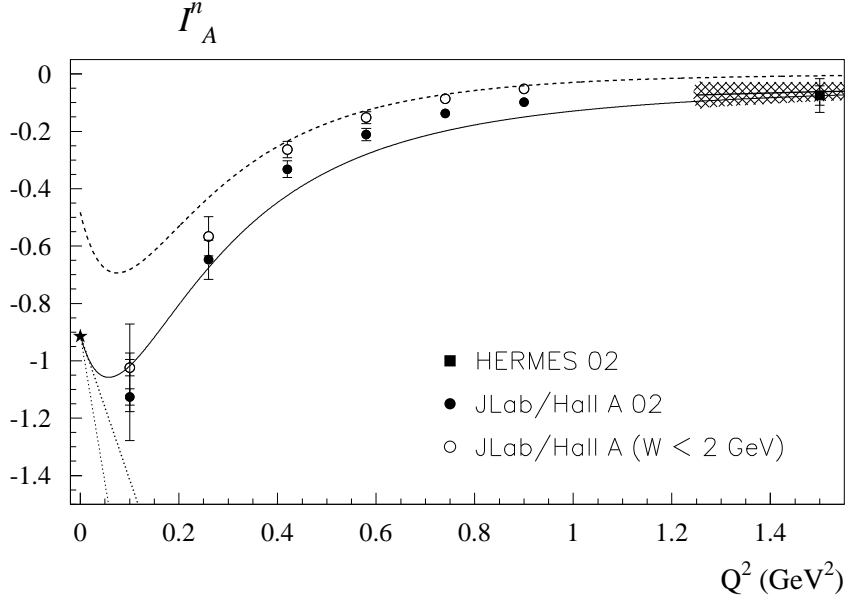}}
\vspace{-.5cm}
\caption{The generalized GDH integral $I_A(Q^2)$ vs. $Q^2$ for the neutron.
The dashed curve represents the MAID estimate~\cite{Dre99,Dre01}
for the $\pi$ channel. 
The thin solid curve is the parametrization of Eq.~(\ref{eq:bi2}),
using MAID to calculate the resonance contribution. 
The lower thin dotted curve is the ${\mathcal O}(p^4)$ 
HBChPT results of Ref.~\cite{JiK00a}, whereas the upper thick dotted
curve is the corresponding ${\mathcal O}(p^4)$ relativistic BChPT results 
of Ref.~\cite{Ber02}.
The thick solid curve for $Q^2 \geq 1.25$~GeV$^2$ 
is the evaluation using the DIS structure function
$g_1^n$, and the shaded band represents the corresponding error estimate
as given by Ref.~\cite{Blum02}.
The $^3$He data of the Hall A Collaboration at JLab~\cite{Ama02} 
are corrected for nuclear effects according to Ref.~\cite{Cio97}. 
The HERMES data are from Ref.~\cite{Air02}. 
For both data sets : inner error bars are statistical errors only, outer
error bars include systematical errors.
The GDH sum rule value is indicated by the star. }
\label{fig:ian}
\end{figure}
\newline
\indent
In Figs.~\ref{fig:i1p} and \ref{fig:ian}, 
we also present the heavy baryon chiral perturbation theory (HBChPT) 
calculation to ${\mathcal O}(p^4)$ of Ref.~\cite{JiK00a}, as well as the 
relativistic baryon ChPT (relativistic BChPT) calculation to  ${\mathcal
O}(p^4)$ of Ref.~\cite{Ber02}.
From the comparison of both the HBChPT and relativistic BChPT calculations 
to the individual proton and neutron generalized GDH integrals, 
one sees that the chiral expansion may only be applied in 
a very limited range of $Q^2 \lesssim 0.05$~GeV$^2$.
This can be understood from the phenomenological calculations discussed above, 
where it became obvious that the GDH integrals for proton and neutron at 
small $Q^2$ are dominated by the $\Delta(1232)$ resonance
contribution. However in the p - n difference, the $\Delta(1232)$
contribution and other isospin 3/2 resonances drop out. Therefore, 
it was noted in Ref.~\cite{Bur01} that the
HBChPT expansion may be applied in a larger $Q^2$ range for the
difference $I_1^p - I_1^n$.
In Fig.~\ref{fig:i1pmn}, we display the $Q^2$ dependence of
the proton - neutron difference $I_1^p - I_1^n$. 
It is indeed seen that the $Q^2$ dependence of the ChPT calculations,
in particular the HBChPT calculation,  
is much less steep for the p - n difference and follows the 
phenomenological estimate over a larger $Q^2$ range. 
Therefore this opens up the possibility, as discussed in 
Ref.~\cite{Bur01}, 
to extend the $Q^2$ range of the ChPT calculation upwards in $Q^2$. 
On the other hand, the extension of the operator
product expansion for $\Gamma_1^p - \Gamma_1^n$ 
to a value around $Q^2 \simeq 0.5$~GeV$^2$ 
requires the control of higher twist terms, which lattice QCD
estimates show to be rather small \cite{Sch00}. 
This may open the possibility to bridge the gap 
between the low and high $Q^2$ regimes, at least for this particular
observable.
\begin{figure}[h]
\vspace{-1.25cm}
\epsfysize=10.cm
\centerline{\epsffile{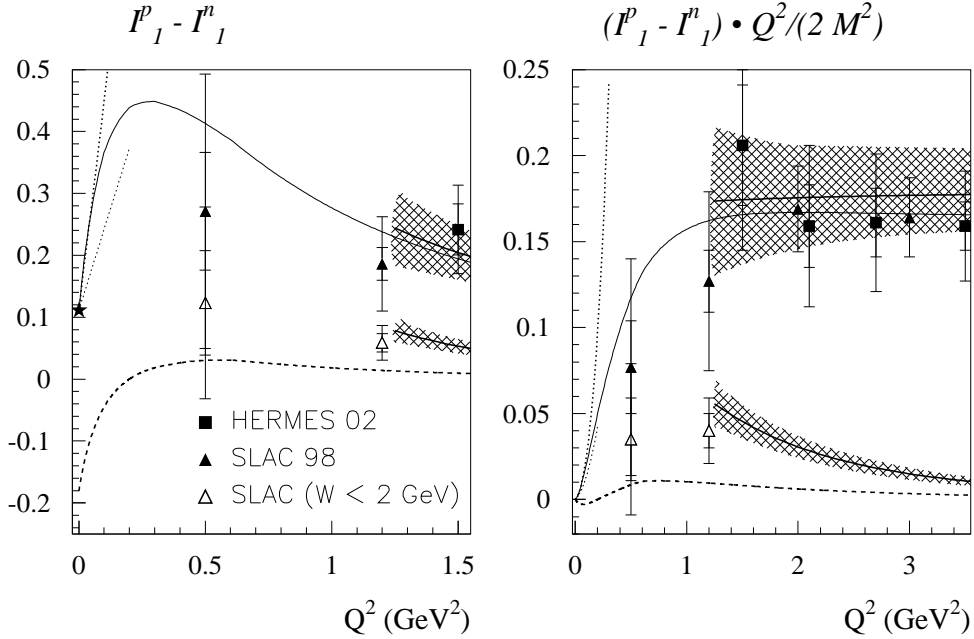}}
\vspace{-.5cm}
\caption{$Q^2$ dependence of the integral $I_1^p - I_1^n$ (left) and
$(I_1^p - I_1^n) \cdot Q^2/(2 M^2)$ (right) for the
proton - neutron difference, as
given by Eq.~(\ref{eq:I1}). The dashed curve is the
MAID estimate~\cite{Dre99} for the one-pion channel.
The thin solid curve is the parametrization of Eq.~(\ref{eq:bi2}),
using MAID to calculate the resonance contribution. 
The lower thin dotted curves are the ${\mathcal O}(p^4)$ 
HBChPT results of Ref.~\cite{JiK00a}, whereas the upper thick dotted
curves are the corresponding ${\mathcal O}(p^4)$ 
relativistic BChPT results of Ref.~\cite{Ber02}.
The upper thick solid curve for $Q^2 \geq 1.25$~GeV$^2$ 
is the evaluation using the DIS structure function
$g_1^p - g_1^n$~\cite{Blum02}, whereas 
the lower thick solid curve is the evaluation
for the resonance region ($W < 2$ GeV) using the DIS structure function.
The shaded bands represent the corresponding error estimates
as given by Ref.~\cite{Blum02}.
The SLAC data are from Ref.~\cite{Abe97}, 
and the HERMES data are from Ref.~\cite{Air02}  
(for both data sets : inner error bars are statistical errors only, outer
error bars include systematical errors).
The GDH sum rule value is indicated by the star.}
\label{fig:i1pmn}
\end{figure}
\newline
\indent
The second spin-dependent VVCS amplitude $S_2$ is odd in $\nu$,
which leads to the unsubtracted DR
\begin{eqnarray}
\label{DDeq2.3.64}
{\rm Re}\ S_2(\nu,\,Q^2) 
\,&=&\, \frac{2\nu}{\pi} \,{\mathcal{P}}\, \int_{0}^{\infty}
\frac{ {\rm Im}\ S_2(\nu',\,Q^2)}{\nu'^2-\nu^2} \,d\nu'\ \nonumber \\
\,&=&\, {\rm Re}\ S_2^{{\rm\sc{pole}}}
\,+\, \frac{2\nu}{\pi}  \,{\mathcal{P}}\, \int_{\nu_0}^{\infty}
\frac{ {\rm Im}\ S_2(\nu',\,Q^2)}{\nu'^2-\nu^2} \,d\nu' \, ,
\end{eqnarray}
where the pole part ${\rm Re}\ S_2^{{\rm\sc{pole}}}$ is obtained from  
Eq.~(\ref{eq:bornt12s12}) as~:
\beqn
{\rm Re}\ S_2^{\mbox{\sc{pole}}} (\nu,\,Q^2) & = &
\frac{\alpha_{em}}{2}
\frac{\nu}{\nu^2-\nu_B^2}\,F_P(Q^2) \, \left(F_D(Q^2) + F_P(Q^2) \right ) \, ,
\eeqn
\newline
\indent
Assuming further that the high-energy behavior of $S_2$ is given by
\beqn
\label{eq:highs2}
S_2(\nu, Q^2) \to \nu^{\alpha_2} \, ,
\hspace{1cm} \mathrm{for} \;\; \nu \to \infty \, ,
\hspace{1cm} \mathrm{with} \;\; \alpha_2 < -1 \, ,
\eeqn
one can also write down an unsubtracted dispersion relation for
the amplitude $\nu \, S_2$ (which is even in $\nu$),
\begin{eqnarray}
\label{DDeq2.3.67}
{\rm Re}\ (\nu \, S_2(\nu,\,Q^2)) 
\,&=&\, \frac{2}{\pi} \,{\mathcal{P}}\, \int_{0}^{\infty}
\frac{\nu'^2 {\rm Im}\ S_2(\nu,\,Q^2)}{\nu'^2-\nu^2}\,d\nu'\ \nonumber \\
\,&=&\, {\rm Re}\ (\nu \, S_2)^{{\rm\sc{pole}}}
\,+\, \frac{2}{\pi} \,{\mathcal{P}}\, \int_{\nu_0}^{\infty}
\frac{\nu'^2 {\rm Im}\ S_2(\nu,\,Q^2)}{\nu'^2-\nu^2}\,d\nu'\, ,
\end{eqnarray}
where the pole part is obtained from Eq.~(\ref{eq:bornt12s12}) as~:
\beqn
{\rm Re}\ (\nu \, S_2(\nu,\,Q^2))^{\mbox{\sc{pole}}}  & = &
\frac{\alpha_{em}}{2}
\frac{\nu_B^2}{\nu^2-\nu_B^2}\,F_P(Q^2) \, \left(F_D(Q^2) + F_P(Q^2)
\right ) \, .
\label{eq:nus2pole}
\eeqn
If we subtract Eq.~(\ref{DDeq2.3.67}) from Eq.~(\ref{DDeq2.3.64})
multiplied by $\nu$, we obtain the ``superconvergence relation''
(for {\it any} value of $Q^2$),
\beqn
\label{eq:drbcs2}
0 =\int_0^{\infty} {\rm Im}\ S_2(\nu,\,Q^2)\,d\nu\ ,
\eeqn
i.e., the pole contribution and the inelastic contribution to that
integral should cancel. Equation~(\ref{eq:drbcs2}) is known as the
Burkhardt-Cottingham (BC) sum rule \cite{BC70}.
When Eq.~(\ref{eq:drbcs2}) is expressed in terms of the nucleon structure
function $g_2(x, Q^2$), the BC sum rule implies the vanishing of the
first moment of $g_2$, i.e.,
\beqn
\label{eq:drbcg2}
0 =
\int_{0}^{1}\, dx \, g_{2}\,(x,\,Q^2) \, ,
\eeqn
and the convergence condition of Eq.~(\ref{eq:highs2}) leads to~:
\beqn
\label{eq:highg2}
g_2(x, Q^2) \to x^{\tilde \alpha_2} \, ,
\hspace{1cm} \mathrm{for} \;\; x \to 0 \, ,
\hspace{1cm} \mathrm{with} \;\; \tilde \alpha_2 > -1 \, .
\eeqn
Separating the elastic and inelastic contributions in Eq.~(\ref{eq:drbcg2})
and using Eq.~(\ref{DDeq2.3.24}), we may express the BC sum rule
for {\it any} value of $Q^2$ as~:
\beqn
\label{DDeq2.3.46}
I_2(Q^2) \;\equiv\; \frac{2M^2}{Q^2}\int_0^{x_0}g_2(x,\,Q^2)\,dx \;=\;
\frac{1}{4} \, F_P(Q^2) \, \left( F_D(Q^2) + F_P(Q^2) \right) \, .
\eeqn
Alternatively the BC sum rule can be written in terms of
the Sachs form factors and the absorption cross
sections, i.e., by
\beqn
\label{eq:drbcg4}
I_2(Q^2)&\;=\;&\frac{M^2}{\pi \, e^2 }\,
\int_{\nu_0}^{\infty}\,\frac{K(\nu,\,Q^2)}{\nu^2 + Q^2}
\, \left\{ \,- \sigma_{TT}(\nu,\,Q^2) \,+\,
\frac{\nu}{Q} \, \sigma_{LT}(\nu,\,Q^2) \, \right\} \, d \nu 
\nonumber \\
&\;=\;& \frac{1}{4}\ \frac{G_M(Q^2)(G_M(Q^2)-G_E(Q^2))}{1+\tau}\, .
\eeqn
\newline
\indent
Performing a low energy expansion 
for $(\nu S_2) - (\nu S_2)^{{\rm\sc{pole}}}$, 
we obtain from Eq.~(\ref{DDeq2.3.67})
\footnote{Note that the relation of Ref.~\cite{Dre01}, i.e., 
$I_A'(0) - I_1'(0) = M^2 / (2 \, \alpha_{em}) \, 
\cdot \left( \gamma_0(0) - \delta_{LT}(0) \right)$ ensures that the 
$\nu^4$ term in $\nu S_2$ has no singularity at $Q^2 = 0$.}~:
\begin{eqnarray}
&&{\mbox{Re}}\ \nu \, S_2(\nu,\,Q^2) -
{\mbox{Re}}\ (\nu \, S_2(\nu,\,Q^2))^{\mbox{\sc{pole}}} \,=\,\nonumber \\
&& \left(2 \, \alpha_{em} \right) \, I_2(Q^2)
\,-\, \left(2 \, \alpha_{em} \right) 
{1 \over {Q^2}} \left( I_A(Q^2) - I_1(Q^2) \right) \, \nu^2  
\nonumber \\
&&+\, {1 \over Q^2} \left[ \left(2 \, \alpha_{em} \right) 
{1 \over {Q^2}} \left( I_A(Q^2) - I_1(Q^2) \right)
\,+\, M^2 \, \left( \delta_{LT}(Q^2) - \gamma_0(Q^2) \right) \right] \, \nu^4  
\,+\, {\mathcal{O}}(\nu^6) ,
\label{eq:s2lex}
\end{eqnarray}
in terms of the integrals $I_2, I_1, I_A$ and spin polarizabilities 
$\gamma_0$ and $\delta_{LT}$ introduced before. 
For the Born contribution, we obtain from 
Eqs.~(\ref{eq:bornt12s12}) and (\ref{eq:nus2pole}) that 
\beqn
{\rm Re}\ (\nu \, S_2(\nu,\,Q^2))^{\mbox{\sc{Born}}}  \,-\,
{\rm Re}\ (\nu \, S_2(\nu,\,Q^2))^{\mbox{\sc{pole}}}  \,=\,
\frac{\alpha_{em}}{2}
\,F_P(Q^2) \, \left(F_D(Q^2) + F_P(Q^2) \right ) ,
\eeqn
yielding
\beqn
I_2^{\mbox{\sc{Born}}} (Q^2) \,=\,
\frac{1}{4}
\,F_P(Q^2) \, \left(F_D(Q^2) + F_P(Q^2) \right ) \, .
\eeqn
It is interesting to note that the Born contribution to $(\nu S_2)$
leads exactly to the BC sum rule value of Eq.~(\ref{DDeq2.3.46}).
Furthermore, the Born contribution also leads to 
$I_1^{\mbox{\sc{Born}}} (Q^2) = I_A^{\mbox{\sc{Born}}} (Q^2) 
= - \,F_P^2(Q^2) \,/\, 4$. 
\newline
\indent
By comparing Eqs.~(\ref{eq:i3int}), (\ref{eq:I1}) 
and (\ref{eq:drbcg4}) one obtains that $I_3$, defined by 
Eq.~(\ref{eq:gltlex}), can be expressed as~:
\beqn 
I_3(Q^2) \,=\, I_1(Q^2) \,+\, I_2(Q^2) \, . 
\eeqn
If the BC sum rule holds at $Q^2$ = 0, 
one obtains $I_3(0) = e_N \kappa_N / 4$. 
\newline
\indent
The BC sum rule has been shown to be satisfied in the case of quantum
electrodynamics by a calculation in lowest order of 
$\alpha_{em}$ \cite{Tsai75}.
In perturbative QCD, the BC sum rule was calculated for a quark
target to first order in $\alpha_s$ and also shown to hold
\cite{Alt94}.
Furthermore, it is interesting to note that
the validity of the Wandzura-Wilczek relation of
Eq.~(\ref{eq:ww}) for the transverse spin structure function $g_1 + g_2$
implies that the BC sum rule is satisfied. Indeed one directly
obtains Eq.~(\ref{eq:drbcg2}) by integrating Eq.~(\ref{eq:ww}). 
\newline
\indent
For a nucleon target, the BC sum rule has recently been evaluated at
small $Q^2$ in HBChPT at order ${\mathcal{O}}(p^4)$ \cite{KSV02}, and
has also been shown to hold to this order. 
\newline
\indent
In Fig.~\ref{fig:I2p}, we show the MAID model prediction
for $I_2(Q^2)$ of the proton and compare it with the BC sum rule
value.
It is obvious from Fig.~\ref{fig:I2p} that at small $Q^2$, 
the one-pion channel nearly saturates the BC sum rule prediction. 
At intermediate values of $Q^2$, the MAID calculation
starts to fall short of the sum rule, because the $\pi\pi$ channels and
higher continua become increasingly important.
One also notices that the HBChPT result at order 
${\mathcal{O}}(p^4)$ \cite{KSV02} for the first moment of $g_2$ 
remains close to the phenomenological sum rule evaluation, in the range up to
$Q^2 \simeq 0.3$~GeV$^2$. 
For the higher moments of $g_2$, it was shown in Ref.~\cite{KSV02}
that the $\Delta$-contribution is very small, so that the moments 
of $g_2$ seem to be a promising observable to bridge the gap between
the HBChPT description at the lower $Q^2$ and the perturbative QCD
result at the larger $Q^2$.
In the large $Q^2$ region, the first moment of $g_2$
was recently evaluated by the E155 Collaboration~\cite{E155X}
at $Q^2$ = 5 GeV$^2$, and the integral of $g_2$ over
the range $0.02 \leq x \leq 0.8$ was found to be $-0.044 \pm 0.008 \pm 0.003$.
Although this value differs significantly from zero, it does not
represent a conclusive test of the BC sum rule, because 
the behavior of $g_2$ is still unknown in the small $x$ region, 
which remains to be explored by future experiments. 
\begin{figure}[h]
\vspace{-1.7cm}
\epsfysize=9.8cm
\centerline{\epsffile{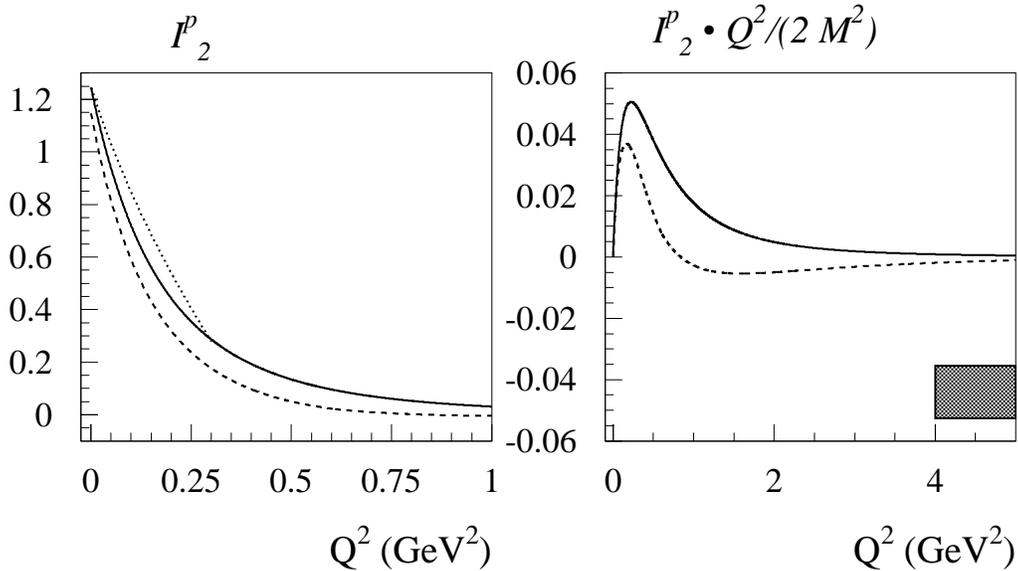}}
\vspace{-.6cm}
\caption{$Q^2$ dependence of the integral $I_2$ (left) and
$I_2 \cdot Q^2/(2 M^2)$ (right) for the proton, as
given by Eq.~(\ref{DDeq2.3.46}). The dashed curve represents the
MAID estimate~\cite{Dre99} for the one-pion channel. 
The dotted curve is the HBChPT result 
at order ${\mathcal{O}}(p^4)$ \cite{KSV02}.
The solid curve is the Burkhardt-Cottingham sum rule 
({\it rhs} of Eq.~(\ref{DDeq2.3.46})), using the dipole
parametrization for $G_M^p$ and
the parametrization for $G_E^p / G_M^p$ following
from the recent JLab data~\cite{Jon00,Gay02}.
The shaded band represents the evaluation 
using the recent SLAC E155 data for $g_2$
integrated over the range $0.02 \leq x \leq 0.8$~\cite{E155X}.}
\label{fig:I2p}
\end{figure}

\section{Dispersion relations in real Compton scattering (RCS)}
\label{sec:rcs}


\subsection{Introduction}

As we have seen in the previous section, forward photon scattering
is closely related to properties of the excitation spectrum of the
probed system. By use of dispersion relations (DRs) it becomes possible
to set up sum rules on the basis of general principles and to
determine certain combinations of polarizabilities from the
knowledge of the absorption cross sections alone. In the following
we shall discuss the general case of RCS and set up dispersion
relations valid for all angles.

\begin{figure}[h]
\vspace{-.25cm} \epsfxsize=13cm
\centerline{\epsffile{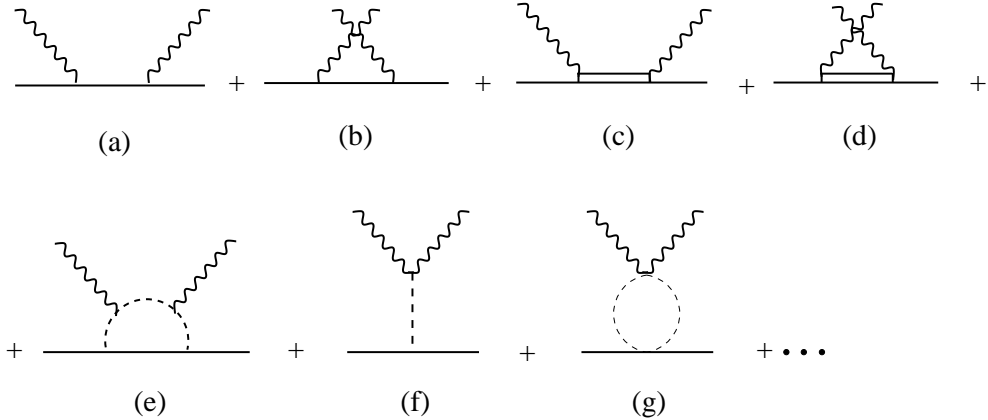}} \caption[]{Some typical
intermediate states contributing to Compton scattering off the nucleon. 
Upper row: The direct (a) and crossed (b) Born diagrams with intermediate
nucleons, a typical resonance excitation in the $s$-channel (c) and
its crossed version (d). Lower row: Typical mesonic contributions
with photon scattering off an intermediate pion (e), the pion pole
diagram (f) and a correlated two-pion exchange such
as the ``$\sigma$ meson'' (g).} 
\label{fig:graphs_rcs}
\end{figure}

\begin{figure}[h]
\epsfxsize=11.5cm
\centerline{\epsffile{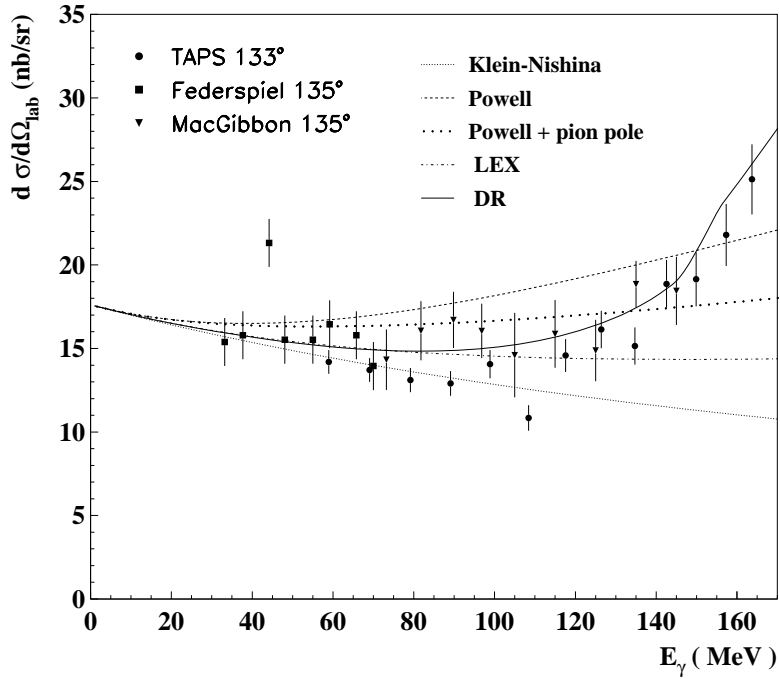}} 
\vspace{-.5cm}
\caption[]{
Differential cross section for Compton scattering off the proton
as a function of the $lab$ photon energy $E_\gamma$ and at fixed
scattering angle $\theta_{\mathrm{lab}}=135^{{\rm o}}$. The curves show the
full cross section from fixed-$t$ subtracted dispersion relations (solid), the
Klein-Nishina cross section (small dots), the Powell cross section
(dashed), the Powell plus $\pi^0$ pole cross section (large
dots), and the low energy expansion (LEX) including also the
leading order contributions from the scalar polarizabilities
(dashed-dotted).} \label{fig:rcs_lex}
\end{figure}

Some typical processes contributing to RCS are shown in
Fig.~\ref{fig:graphs_rcs}. When the nucleon is taken as a
structureless Dirac particle, only the nucleon pole terms
contribute. These are diagrams (a) and (b) for the $s$ and $u$
channels, respectively. The differential cross section for this
situation, first obtained by Klein and Nishina~\cite{Kle29} in
1929, is shown in Fig.~\ref{fig:rcs_lex}. The inclusion of the
anomalous magnetic moment leads to a far more complicated
result corresponding to the Powell cross section~\cite{Pow49}
in Fig.~\ref{fig:rcs_lex}. If we add the pion pole term,
Fig.~\ref{fig:graphs_rcs} (f), the cross section drops one third
 to the original result of Klein and Nishina. This term is, of
course, due to the decay $\pi^0\rightarrow\gamma+\gamma$, and
therefore directly related to the axial anomaly, 
derived on general grounds as Wess-Zumino-Witten term~\cite{WZW}. 
The pion pole term is often referred to as triangle anomaly, because the vertex
$\pi\gamma\gamma$ can be resolved into a triangular quark loop, a
diagram not allowed in a classical theory and only appearing due to
the renormalization process of quantum field theory. As we see
from the figure, the pion pole term gives a considerable
contribution for backward scattering, its effect is sometimes
included in the backward spin polarizability $\gamma_{\pi}$ (the
index $\pi$ stands for $\theta=180^{\circ}$!), though from the
standpoint of dispersion relations it should be considered as a
pole term like the  nucleon pole terms. 
Except for the diagrams (a), (b), and (f),
all other and higher
diagrams in Fig.~\ref{fig:graphs_rcs} have no pole structure, but
correspond to excited states in $s$-, $u$- or $t$-channel processes. As
such they lead to dispersive contributions whose lowest terms are
given by the six leading polarizabilities of RCS on the nucleon. 
The result of a calculation taking account only of the electric and magnetic
dipole polarizabilities is labeled LEX in Fig.~\ref{fig:rcs_lex}.
It is obvious that the low energy expansion is correct only up
to about 80~MeV, in a region where the ``world data'' scatter and
give only limited information on the polarizabilities. Therefore,
the analysis of the modern data has been based on dispersion theory
whose results are labeled by DR in the figure. Clearly the higher
order terms become more and more important with increasing photon
energies, particularly after crossing the pion threshold (seen as
a kink at about 150~MeV) with a sharp rise if the energy increases
further towards the $\Delta (1232)$ resonance.

\subsection{Kinematics}

\begin{figure}[h]
\epsfxsize=8cm
\centerline{\epsffile{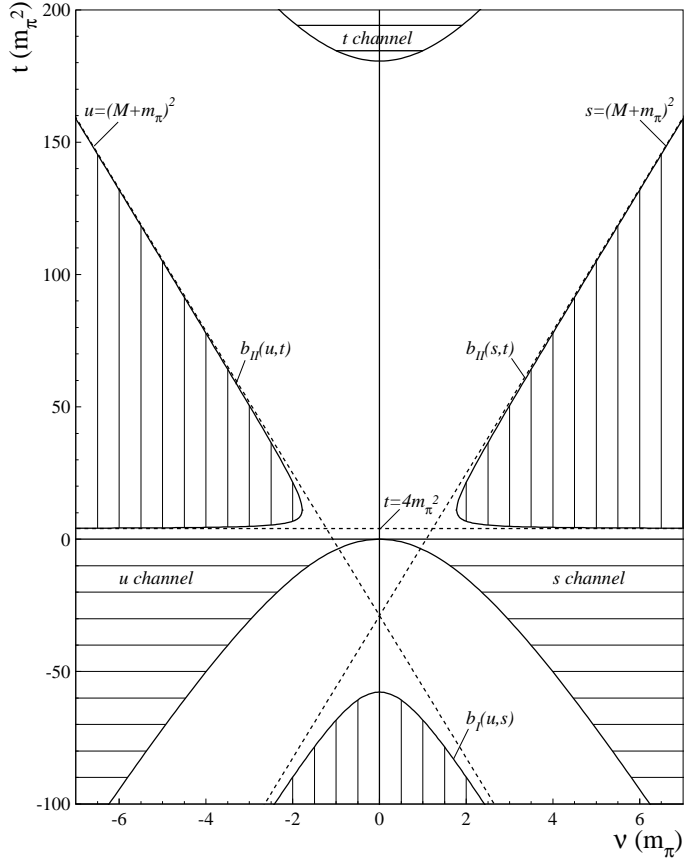}}
\caption[]{The Mandelstam plane for real Compton scattering. The
physical regions are horizontally hatched. The spectral regions (with
boundaries $b_I$ and $b_{II}$) are vertically hatched.}
\label{fig:mandelstam_rcs}
\end{figure}

\indent
Assuming invariance under parity, charge conjugation and time reversal
symmetry, the general amplitude for Compton scattering can be
expressed by 6 independent structure functions $A_i(\nu,t)$,
$i=1,...,6$
~\cite{Lvo97}.
These structure functions depend on two Lorentz
invariant variables, e.g., $\nu$ and $t$ as defined in the following.
Denoting the momenta of the initial state photon and proton by $q$ and
$p$ respectively, and with corresponding final state momenta
 $q'$ and $p'$, the familiar Mandelstam variables are
\begin{equation}
s=(q+p)^2\ ,\ \ t=(q-q')^2\ ,\ \ u=(q-p')^2\ ,
\label{eq3.1.1}
\end{equation}
with the constraint $s+t+u=2M^2$.
The variable $\nu$ is defined by
\begin{equation}
\nu=\frac{s-u}{4M}\ .
\label{eq3.1.2}
\end{equation}
The orthogonal coordinates of the Mandelstam plane, $\nu$ and $t$,
are related to the initial ($E_{\gamma}$) and final
($E_{\gamma}'$) photon $lab$ energies, and to the $lab$ scattering angle
$\theta_{\mathrm{lab}}$ by
\beqn
\label{eq3.1.3}
t &  = & - 4E_{\gamma}E_{\gamma}'\sin^2 \frac{\theta_{\mathrm{lab}}}{2} =
- 2M (E_{\gamma}-E_{\gamma}')\ , \nonumber \\
\nu & = & E_{\gamma} + \frac{t}{4M} = \frac{1}{2}
 (E_{\gamma}+E_{\gamma}')\ .
\eeqn
The physical regions of the Mandelstam plane are shown in
Fig.~\ref{fig:mandelstam_rcs} by the horizontally hatched areas.
The vertically hatched areas
are the spectral regions discussed in detail in  App. A of
Ref.~\cite{DGP99}. The boundaries of the physical regions in the
$s,\ u$ and $t$ channels are determined by the zeros of the Kibble
function
\be
\label{DDeq3.1.4}
\Phi(s,t,u) = t(us-M^4)=0\ .
\ee
In particular the RCS experiment takes place in the $s$-channel
region, limited by the line $t=0$ (forward scattering, $\theta=0^{\circ}$)
and the lower right part of the hyperbola $us=M^4$ (backward scattering,
$\theta=180^{\circ})$. The $u$-channel region is obtained by
crossing $(\nu\rightarrow-\nu)$, and the $t$-channel region in
the upper part of Fig.~\ref{fig:mandelstam_rcs} corresponds to
the process $\gamma+\gamma\rightarrow N+\bar{N}$ and requires a
value of $t\ge4M^2$.

\subsection{Invariant amplitudes and nucleon polarizabilities}

\indent
The invariant Compton tensor can be constructed as
\be
\label{DDeq3.2.1}
T_{fi} = \varepsilon_{\mu} \, \varepsilon_{\nu}'^{\ast} \,
\bar{u} \, (p',\lambda'_N) H^{\mu\nu} \, u(p,\lambda_N)\ ,
\ee
where $\varepsilon$ and $\varepsilon'$ are the polarization
vectors of the incoming and outgoing photon, respectively, as
defined in Eq.~(\ref{DDeq2.2.1}), $u$ and $\bar u$ are the nucleon
spinors, and $\lambda_N$ ($\lambda'_N$) are the nucleon helicities
in the initial (final) states respectively.
The Compton tensor $H^{\mu\nu}$ can be built from the
four-momentum vectors and Dirac matrices as follows~\cite{Pra58}:
\beqn
\label{DDeq3.2.2}
H^{\mu\nu} & = & -\frac{P'^{\mu}P'^{\nu}}{P'^2}\,(T_1+T_2 \, \kdagger
                 )
                 -\frac{N^{\mu}N^{\nu}}{N^2}\,(T_3+T_4 \, \kdagger)
                 \nonumber
                 \\
             &&  +i\,\frac{P'^{\mu} N^{\nu} - P'^{\nu} N^{\mu}}
                 {P'^2K^2}\,\gamma_5 \, T_5
                 +i\,\frac{P'^{\mu} N^{\nu} + P'^{\nu} N^{\mu}}
                 {P'^2K^2}\,\gamma_5 \,\kdagger \,T_6\ ,
\eeqn
where
\be
\begin{array}{lcl}
P^{\mu} =  {\textstyle\frac{1}{2}}\,(p+p')^{\mu} \quad & ,
\quad &
K^{\mu} =  {\textstyle\frac{1}{2}}\,(q+q')^{\mu}\ , \nonumber\\
Q^{\mu} =  {\textstyle\frac{1}{2}}\,(q'-q)^{\mu} \quad & ,
\quad &
P'^{\mu} = P^{\mu}-\frac{(P\cdot K)}{K^2}\,K^{\mu}\ , \nonumber \\
N^{\mu} \equiv
\varepsilon^{\mu\nu\alpha\beta}P'_{\nu}Q_{\alpha}K_{\beta}\ ,
\nonumber
\end{array}
\label{DDeq3.2.3}
\ee
with $\varepsilon^{0123}=-1$.
The six tensorial objects in Eq.~(\ref{DDeq3.2.2}) form a complete
 basis, and
the amplitudes $T_1,\ldots T_6$ of Prange are scalar functions of
$\nu$ and $t$ containing the nucleon dynamics. Unfortunately, the
Prange amplitudes have singularities in the forward and backward
directions leading to linear dependencies at these points
(kinematical constraints). L'vov~\cite{Lvo81} has therefore
proposed a different tensor basis, resulting in the set of
amplitudes
\be
\begin{array}{lcl}
A_1 =  {\textstyle\frac{1}{t}}\,[T_1+T_3+\nu(T_2+T_4)] \quad & ,
\quad &
A_2 =  {\textstyle\frac{1}{t}}\, [2T_5+\nu(T_2+T_4)] \ , \nonumber\\
A_3 = \frac{M^2}{M^4-su}[T_1-T_3-{\textstyle\frac{t}{4\nu}}(T_2-T_4)]
\quad & , \quad &
A_4 =\frac{M^2}{M^4-su}[2MT_6-{\textstyle\frac{t}{4\nu}}(T_2-T_4)]\
,\nonumber \\
A_5 = {\textstyle\frac{1}{4\nu}} [T_2+T_4] \quad & , \quad &
A_6 = {\textstyle\frac{1}{4\nu}} [T_2-T_4] \ .
\end{array}
\label{DDeq3.2.4}
\ee
These L'vov amplitudes have no kinematical constraints and are
symmetrical under crossing,
\be
A_i(-\nu,t) = A_i(\nu,t)\ ,\quad i=1,\ldots 6\ .
\label{DDeq3.2.5}
\ee
In the spirit of dispersion relations we build the invariant
amplitudes by adding the pole contributions of
Fig.~\ref{fig:graphs_rcs} (a), (b) and (f),
 and an integral over the spectrum of excited
intermediate states. Furthermore, we define the polarizabilities 
by subtracting the nucleon pole contributions $A_i^B$ 
from the amplitudes and introduce the quantities
\footnote{Alternatively, the polarizabilities can be defined by 
also subtracting the $\pi^0$ pole contribution in the case of the
amplitude $A_2$.}
\begin{equation}
A_i^{NB}(\nu,t) = A_i(\nu,t) \;-\; A_i^B(\nu,t)\ .
\label{eq3.2.1}
\end{equation}
The polarizabilities are related to these functions and their
derivatives at the origin of the Mandelstam plane, $\nu=t=0$,
\be
\label{eq3.2.2}
a_i\equiv A_i^{NB} (0,0)\,,\quad a_{i,\nu} = \left(\frac{\partial
A_i^{NB}}{\partial\nu}\right)_{\nu=t=0}\,,\quad
a_{i,t} = \left(\frac{\partial
A_i^{NB}}{\partial t}\right)_{\nu=t=0}\ .
\ee
For the spin-independent (scalar) polarizabilities $\alpha=\alpha_{E1}$ and
$\beta=\beta_{M1}$, one finds the two combinations
\begin{eqnarray}
\label{eq:alphaplusbeta}
\alpha_{E1} + \beta_{M1} \;&=&\; - {1 \over {2 \pi}} \; (a_3 \;+\; a_6)\;, \\
\alpha_{E1} - \beta_{M1} \;&=&\; - {1 \over {2 \pi}} \; a_1\;,
\label{eq:nospinpol0}
\end{eqnarray}
related to forward and backward Compton scattering respectively.
The 4 spin-dependent (vector) polarizabilities $\gamma_1$ to
$\gamma_4$ of Ragusa~\cite{Rag93}, and the multipole spin
polarizabilities $\gamma_{E1E1}$, 
$\gamma_{M1M1}$, $\gamma_{M1E2}$, $\gamma_{E1M2}$
of Ref.~\cite{Bab98a} (see Section~\ref{sec:physpol}), are defined by~:
\begin{eqnarray}
&&\hspace{-.75cm} \gamma_0 \;\equiv\; \gamma_1 - \gamma_2 - 2 \gamma_4 
\;=\; - \gamma_{E1E1} - \gamma_{M1M1} - \gamma_{M1E2} - \gamma_{E1M2} 
\;=\; {1 \over {2 \pi} M} \; a_4\;, \label{eq23} \\
&&\hspace{-.75cm} \gamma_{13} \;\equiv\; \gamma_1 + 2 \gamma_3 
\;=\; - \gamma_{E1E1} + \gamma_{E1M2} 
\;=\; -\;{1 \over {4 \pi} M} \; (a_5 \;+\; a_6)\;, \label{eq24} \\
&&\hspace{-.75cm} \gamma_{14} \;\equiv\; \gamma_1 - 2 \gamma_4 
\;=\; - \gamma_{E1E1} - 2 \gamma_{M1M1} - \gamma_{E1M2} 
\;=\; {1 \over {4 \pi} M} \; (2 \, a_4 \;+\; a_5 \;-\; a_6)\;, \label{eq25} \\
&&\hspace{-.75cm} \gamma_\pi \;\equiv\; \gamma_1 + \gamma_2 + 2 \gamma_4 
\;=\; - \gamma_{E1E1} + \gamma_{M1M1} + \gamma_{M1E2} - \gamma_{E1M2} 
\;=\; -\;{1 \over {2 \pi} M} \; (a_2 \;+\; a_5),
\label{eq:spinpol0}
\end{eqnarray}
where $\gamma_0$ and $\gamma_\pi$ are the spin polarizabilities in
the forward and backward directions respectively. Since the
$\pi^0$ pole (see Fig.~\ref{fig:graphs_rcs} (f)) contributes to
$A_2$ only, the combinations $\gamma_0$, $\gamma_{13}$ and
$\gamma_{14}$ of Eqs.~(\ref{eq23})-(\ref{eq25}) are independent of
the pole term, and only the backward spin polarizability
$\gamma_\pi$ is affected by this term.

\subsection{RCS data for the proton and extraction of proton polarizabilities}
\label{sec:rcsproton}

A pioneering experiment in Compton scattering off the proton was
performed by Gol'danski {\it et al.}~\cite{Gol60} in 1960. Their
result for the electric polarizability was $\alpha_{E1} = 9\pm 2$,
with a large uncertainty in the normalization of the cross section
giving rise to an additional systematical error of $\pm5$. We note
that here and in the following all scalar polarizabilities are
given in units of $10^{-4}$~fm$^3$. The next effort to determine
the polarizabilities is due to the group of Baranov~\cite{Bar74}.
The data were taken with a bremsstrahlung beam with photon
energies up to 100~MeV, and the polarizabilities were obtained by
a fit to the low-energy expansion (LEX). However, such energies
are outside the range of the LEX. A later reevaluation by use of
dispersion relations~\cite{Mac95} lead to center values of
$\alpha_{E1} \approx 12$ and $\beta_{M1}\approx -6$, far outside
the range of Baldin's sum rule and more recent results for the
magnetic polarizability $\beta_{M1}$. In any case these findings were
much to the surprise of everybody, because the spin flip
transition from the nucleon to the dominant $\Delta$(1232)
resonance was expected to provide a large paramagnetic
contribution of order $\beta{\mbox{\scriptsize{para}}}\approx10$.
The first modern experiment was performed at Illinois in
1991~\cite{Fed91}. It was done with a tagged photon beam, thereby
improving the capability to measure absolute cross sections, and
in the region of energies between 32 and 72~MeV where the LEX was
applicable. Unfortunately, by the same token the cross sections
were small with the consequence of large error bars. The
experiment was repeated by the Saskatoon-Illinois group at higher
energies above~\cite{Hal93} and below~\cite{Mac95} the pion
threshold, and evaluated in the framework of dispersion relations
with much improved results on the polarizabilities. These results
were confirmed, within the error bars, by the Brookhaven group
working with photons produced by laser backscattering from a
high-energy electron beam~\cite{Ton98}. Even more precise data
were recently obtained by the A2 collaboration at MAMI, using the
TAPS setup at energies below pion threshold~\cite{Olm01}. The
results of these modern experiments are compiled in
Table~\ref{tab3.6.1}.
\begin{table}[h]
\begin{center}
\caption{Values for the scalar polarizabilities of the proton as
obtained from the modern experiments. 
\label{tab3.6.1}}
\begin{tabular}{lllll}
Data set        & Energies    & Angles         & $\qquad
\alpha_{E1}+\beta_{M1}$ &
$\qquad \alpha_{E1}-\beta_{M1}$ \\
& (MeV)       & (degree) &\qquad  ($10^{-4}$ fm$^3$)& \qquad ($10^{-4}$ fm$^3$)\\
\hline
Illinois 1991~\cite{Fed91}   & 32-72  & 60, 135 &
$15.8\pm4.5\pm0.1$ & $11.9\pm5.3\pm0.2$ \\
Saskatoon 1993~\cite{Hal93}  & 149-286 & 24-135 &
$12.1\pm1.7\pm0.9$ & $7.9\pm1.4\pm2.0$ \\
Saskatoon 1995~\cite{Mac95}  & 70-148  & 90, 135 &
$15.0\pm3.1\pm0.4$ & $10.8\pm1.8\pm1.0$ \\
LEGS 1998~\cite{Ton98}  & 33-309   & 70-130 & $13.23\pm
0.86^{+0.20}_{-0.49}$ & $10.11\pm1.74^{+1.22}_{-0.86}$ \\
MAMI/TAPS 2001~\cite{Olm01} & 55-165   & 59-155 &
$13.1\pm0.6\pm0.8$  & $10.7\pm0.6\pm0.8$
\end{tabular}
\end{center}
\end{table}
\newline
\indent
A fit to all modern low-energy data constrained by the sum rule
relation $\alpha_{E1}+\beta_{M1} = 13.8\pm0.4$ leads to the
results~\cite{Olm01}:
\beqn
\label{DDeq3.6.0a}
\alpha_{E1} & = & 12.1  \pm
0.3(\rm{stat}) \mp 0.4(\rm{syst}) \pm 0.3(\rm{mod})\ ,
\nonumber \\
\beta_{M1} & = & 1.6 \pm 0.4(\rm{stat}) \pm 0.4(\rm{syst}) \pm
0.4(\rm{mod})\ ,
\eeqn
the errors denoting the statistical, systematical and
model-dependent errors, in order. This new global average
confirms, beyond any doubt, the dominance of the electric
polarizability $\alpha_{E1}$ and the tiny value of the magnetic
polarizability $\beta_{M1}$, which has to come about by a cancellation
of the large paramagnetic contribution of the $N\Delta$ spin-flip
transition with a nearly equally strong diamagnetic term.
\newline
\indent
Much less is known about the spin polarizabilities of the proton,
except for the forward spin polarizability 
$\gamma_0= \gamma_1-\gamma_2-2\gamma_4 = [-1.01\pm 0.08 \rm{(stat)} \pm 0.1
\rm{(syst)}] \cdot10^{-4}~\rm{fm}^4$, which
is determined by the GDH experiment at MAMI and dispersion
relations according to Eq.~(\ref{DDeq2.2.19}). However, the only
other combination for which there exists experimental information
is the backward spin polarizability $\gamma_{\pi} =
\gamma_1+\gamma_2+2\gamma_4$. Dispersive contributions from the
$s$-channel integral have been found to be positive and in the range
of $5\lesssim \gamma_{\pi}\,(\rm{disp})\lesssim 10$ (here and in
the following in units of $10^{-4}$~fm$^4$). 
In addition to this dispersive part, a large contribution 
comes from the $t$-channel $\pi^0$ exchange, 
$\gamma_\pi ({\pi^0}$-pole) $\simeq -46.7$ 
(see Eqs.~(\ref{eq:piopole})-(\ref{eq:pi0gagacoup})), 
giving a total result of $-42 <\gamma_\pi < -37.$
These theoretical
predictions have been challenged by a first experimental value
presented by the LEGS group~\cite{Ton98} who found from a combined
analysis of pion photoproduction and Compton scattering:
\be
\label{eq3.6.4}
\gamma_{\pi} = -27.1\pm2.2\,(\rm{stat+syst})
^{+2.8}_{-2.4}\,(\rm{mod})\ ,
 \ee
where the first error combines statistical and systematical
uncertainties, and the second one represents the model error.
However, there is now contradicting evidence from recent MAMI data
obtained both at low energies~\cite{Olm01} and in the region of
the $\Delta$ resonance (\cite{Gal01}, \cite{Wol01}, \cite{Cam02}).
Though these new results vary somewhat depending on the subsets of
data in different energy regions and on the input of the
underlying dispersion analysis, they are well in the range of the
expectations from both dispersion theory and chiral perturbation
theory. Typical values are:
\be
\label{eq3.6.5}
\gamma_{\pi} =  \left \{
\begin{array}{ll}
-36.1\pm2.1(\rm{stat+syst})\pm0.8(\rm{mod}) & \quad \mbox{\cite{Olm01}} \\
 -37.9\pm0.6(\rm{stat+syst})\pm3.5(\rm{mod})
& \quad \mbox{\cite{Gal01}}\ ,
\end{array} \right .
\ee
where the statistical plus systematical error dominates in the low
energy region while the model dependency gives rise to a large
uncertainty for the experiments in the $\Delta$ region.
\newline
\indent
The predictions for the other spin and higher order
polarizabilities from dispersion analysis and ChPT will be
compared in the following chapters. Unfortunately, these
polarizabilities are all small and can hardly be deduced without
dedicated polarization studies. This will require a new generation
of experiments with polarized beams, polarized targets, and recoil
polarimetry.

\subsection{Extraction of neutron polarizabilities}

The experimental situation concerning the polarizabilities of the
neutron is still quite unsatisfactory. The electric polarizability
$\alpha_{E1}^n$ can in principle be measured by scattering low
energy neutrons off the Coulomb field of a heavy nucleus, while
the magnetic polarizability $\beta_{M1}^n$ remains essentially
unconstrained. This technique seemed to be very promising until
the beginning of the 90's, when Schmiedmayer {\it et
al.}~\cite{Sch91} published a value of
$\alpha_{E1}^n=12.6\pm1.5\,(\rm{stat})\pm2.0\,(\rm{syst})$,
obtained by the scattering neutrons with energies 50~eV$\le E_n\le
50$~keV off a $^{208}$Pb target. Shortly later Nikolenko and
Popov~\cite{Nik92} argued that the errors were underestimated by a
factor of 5. These findings were confirmed by a similar
experiment~\cite{Koe95} resulting in $\alpha_{E1}^n=0\pm5$, and by
a further analysis of the systematical errors~\cite{Eni97} leading
to the estimate $7\lesssim\alpha_{E1}^n\lesssim19$.
\newline
\indent
The two remaining methods to measure $\alpha_{E1}^n$ are
quasi-free Compton scattering off a bound neutron, or elastic
scattering from the deuteron. The first experiment on quasi-free
Compton scattering by a neutron bound in the deuteron was
performed by Rose {\it et al.}~\cite{Ros90}. Interpreted in
conjunction with Baldin's sum rule, the result is
$0<\alpha_{E1}^n<14$ with a mean value $\alpha_{E1}^n\approx10.7$.
The small sensitivity of the experiment follows from the fact that
Thomson scattering vanishes for the neutron, and therefore also
the important interference between the Thomson term and the
leading non-Born amplitude (present in the LEX of the proton!) is
absent for the neutron. It was therefore proposed to measure
$\alpha_{E1}^n-\beta_{M1}^n$ at photon energies in the $\Delta$
region and at backward angles. Of course, the analysis will
strongly depend on final-state interactions and two-body currents.
The quality of the analysis can be tested, to some extent, by also
measuring the polarizabilities of the bound proton. Such results
obtained by the TAPS Collaboration at MAMI were quite
promising~\cite{Wis99}, $\alpha_{E1}^p-\beta_{M1}^p=10.3\pm1.7\
(\rm{stat\ +\ syst})\pm1.1$ (mod). The experiment was then
extended to the neutron by the CATS/SENECA
Collaboration~\cite{Kos02}. Data were collected with both a
deuterium and a hydrogen target and analyzed within the framework
of Levchuk {\it et al.}~\cite{Lev00} by use of different
parametrizations of pion photoproduction multipoles and
nucleon-nucleon interactions. The agreement between the
polarizabilities of free and bound protons was again quite
satisfactory, and the final result for the (bound) neutron was
\be \label{eq3.7.1} \alpha_{E1}^n-\beta_{M1}^n = 9.8\pm3.6\,
(\rm{stat})^{+2.1}_{-1.1}\,(\rm{syst})\pm2.2\,(\rm{mod})\ . \ee
The quasi-free scattering cross section obtained at MAMI is in
good agreement with an earlier datum of a Saskatoon
group~\cite{Kol00} measured at 247~MeV and
$\theta_\mathrm{lab}=135^{\circ}$. From the ratio between the neutron and
the proton results this group derived a most probable value of
$\alpha_{E1}^n-\beta_{M1}^n=12$, however with a very large error
bar.
The comparison between proton and neutron shows that there is no
significant isovector contribution in the scalar polarizabilities
of the nucleon.
\newline
\indent
The second type of experiment, $\gamma d \to \gamma d$, has been
performed at SAL \cite{Horn00} and at MAX-lab \cite{Lundin02}. 
An analysis with the formalism of Ref.~\cite{Lev00} gave the 
results~:
\begin{eqnarray} 
\label{hornlundin} 
\alpha_{E1}^n-\beta_{M1}^n \;&=&\; -4.8 \pm 3.9 \hspace{1cm} 
\mbox{\cite{Horn00}} \, , \nonumber \\
\;&=&\; +3.2 \pm 3.1 \hspace{1cm} 
\mbox{\cite{Lundin02}} \, ,  
\end{eqnarray}
leading to values compatible with zero. 
By comparing the two methods to extract neutron polarizabilities from
deuteron experiments, 
we observe a clear tendancy that elastic Compton scattering leads to
smaller values than those extracted from quasi-free scattering, which
remains to be studied by future investigations.


\subsection{Unsubtracted fixed-$t$ dispersion relations}
\label{sec:3_6}

\indent The invariant amplitudes $A_i$ are free of kinematical
singularities and constraints, and obey the crossing symmetry
Eq.~(\ref{DDeq3.2.5}). Assuming further analyticity and an
appropriate high-energy behavior, these amplitudes fulfill
unsubtracted DRs at fixed $t$,
\begin{equation}
\mathrm{Re} A_i(\nu, t) \;=\; A_i^B(\nu, t) \;+\; {2 \over \pi}
\; {\mathcal P} \int_{\nu_{0}}^{+ \infty} d\nu' \; {{\nu' \;
\mathrm{Im}_s A_i(\nu',t)} \over {\nu'^2 - \nu^2}}\;,
\label{eq:unsuba}
\end{equation}
where $A_i^B$ are the Born (nucleon pole) contributions as in App.
A of Ref.~\cite{Lvo97}, $\mathrm{Im}_s A_i$ the discontinuities
across the $s$-channel cuts of the Compton process and $\nu_{0} =
m_\pi + (m_\pi^2 + t/2)/(2 M)$. However, such unsubtracted DRs
require that at high energies ($\nu \rightarrow \infty$) the
amplitudes $\mathrm{Im}_s A_i(\nu,t)$ drop fast enough so that
the integral of Eq.~(\ref{eq:unsuba}) is convergent and the
contribution from the semi-circle at infinity can be neglected.
For real Compton scattering, Regge theory predicts the following
high-energy behavior for $\nu \rightarrow \infty$ and fixed
$t$~\cite{Lvo97}:
\be \label{eq:rcsregge1}
\begin{array}{lcr}
A_{1}, A_{2} \;\sim\; \nu^{\alpha_M(t)} \quad  & , \quad &
\left(A_3 + A_6 \right) \;\sim\; \nu^{\alpha_P(t) -
2} \;, \nonumber \\
 A_{3}, A_{5} \;\sim\; \nu^{\alpha_M(t) - 2}
\quad & , \quad &  A_{4} \;\sim\; \nu^{\alpha_M(t) - 3} \;,
\end{array}
\ee
where $\alpha_M(t) \lesssim 0.5$ (for $t \leq$ 0) is a meson Regge
trajectory, and where $\alpha_P(t)$ is the Pomeron trajectory
which has an intercept $\alpha_P(0) \approx $ 1.08. Note that the
Pomeron dominates the high energy behavior of the combination of
$A_3 + A_6$. From the asymptotic behavior of
Eq.~(\ref{eq:rcsregge1}), it follows that for RCS unsubtracted
dispersion relations do not exist for the amplitudes $A_1$ and
$A_2$. The reason for the divergence of the unsubtracted integrals
is essentially given by fixed poles in the $t$-channel, notably
the exchange of the neutral pion (for $A_2$) and of a somewhat
fictitious $\sigma$-meson (for $A_1$) with a mass of about 600~MeV
and a large width, which models the two-pion continuum with the
quantum numbers $I=J=0$. In order to obtain useful results for
these two amplitudes, L'vov {\it et al.}~\cite{Lvo97} proposed to
close the contour of the integral in Eq.~(\ref{eq:unsuba}) by a
semi-circle of finite radius $\nu_{max}$ (instead of the usually
assumed infinite radius!) in the complex plane, i.e.  the real
parts of $A_1$ and $A_2$ are calculated from the decomposition
\begin{equation}
\mathrm{Re} A_i(\nu, t) \;=\; A_i^B(\nu, t) \;+\; A_i^{int}(\nu,
t) \;+\; A_i^{as}(\nu, t) \;, \label{eq:aintas}
\end{equation}
with $A_i^{int}$ the $s$-channel integral from pion threshold
$\nu_{0}$ to a finite upper limit $\nu_{max}$, and an `asymptotic
contribution' $A_i^{as}$ representing the contribution along the
finite semi-circle of radius $\nu_{max}$ in the complex plane. In
the actual calculations, the $s$-channel integral is typically
evaluated up to a maximum photon energy $E_\gamma = \nu_{max} -
t/(4 M) \approx 1.5$~GeV, for which the imaginary part of the
amplitudes can be expressed through unitarity by the meson
photoproduction amplitudes (mainly 1$\pi$ and 2$\pi$
photoproduction) taken from experiment.
\newline
\indent
All contributions from higher energies are then absorbed in the
asymptotic term, which is replaced by a finite number of energy
independent poles in the $t$ channel. In particular the asymptotic
part of $A_1$ is parametrized by the exchange of a scalar particle
in the $t$ channel, i.e. an effective ``$\sigma$
meson''~\cite{Lvo97},
\begin{equation}
A_1^{as}(\nu, t) \approx A_1^{\sigma}(t) \;=\; {{F_{\sigma \gamma
\gamma} \; g_{\sigma NN}} \over {t - m_\sigma^2}} \;,
\label{eq:a1sigma}
\end{equation}
where $m_\sigma$ is the $\sigma$ mass, and $g_{\sigma NN}$ and
$F_{\sigma \gamma \gamma}$ are the couplings of the $\sigma$  to
nucleons and photons respectively. In Ref.~\cite{Lvo97}, the
product of the $\sigma$ couplings in the numerator of
Eq.~(\ref{eq:a1sigma}) is used as a fit parameter, which
determines the value of $\alpha - \beta$ through
Eq.~(\ref{eq:nospinpol0}).
\newline
\indent In a similar way, the asymptotic part of $A_2$ is
described  by the $\pi^0$ $t$-channel pole~:
\begin{equation}
A_2^{\pi^0}(0, t) \;=\; {{F_{\pi^0 \gamma \gamma} \; g_{\pi NN}}
\over {t - m_\pi^2}} \;. \label{eq:piopole}
\end{equation}
The coupling $F_{\pi^0 \gamma \gamma}$ is determined through the
$\pi^0 \rightarrow \gamma \gamma$ decay as
\begin{equation}
\Gamma\left( \pi^0 \rightarrow \gamma \gamma\right) \;=\; {1 \over
{64 \, \pi} } \, m_{\pi^0}^3 \, F_{\pi^0 \gamma \gamma}^2 \;.
\end{equation}
Using $\Gamma\left( \pi^0 \rightarrow \gamma \gamma\right)$ =
7.74~eV \cite{PDG98}, one obtains
\begin{equation}
F_{\pi^0 \gamma \gamma} = - \frac{e^2}{4\pi^2 f_\pi}= -0.0252 \,
\, \mbox{GeV}^{-1}, \label{eq:pi0gagacoup}
\end{equation}
where $f_\pi = 92.4 $ MeV  is the pion-decay constant and the sign
is in accordance with the $\pi^0 \gamma \gamma$ coupling in the
chiral limit, given by the Wess-Zumino-Witten effective chiral
Lagrangian~\cite{WZW}. With the $\pi NN$ coupling constant taken
from Ref.~\cite{Pav99}, $g_{\pi NN}^2/4 \pi$ = 13.73, the product
of the couplings in Eq.~(\ref{eq:piopole}) takes the value
$F_{\pi^0 \gamma \gamma} \, g_{\pi NN} \approx$ -0.331 GeV$^{-1}$, 
leading to a value of -46.7 for the pion pole contribution to 
$\gamma_\pi$. On the other hand, the effective chiral Lagrangian
yields the value -43.5~\cite{HHKK98}. 
\newline
\indent This procedure is relatively safe for $A_2$ because of the
dominance of the $\pi^0$ pole or triangle anomaly, which is well
established both experimentally and on general grounds as
Wess-Zumino-Witten term. However, it introduces a considerable
model-dependence in the case of $A_1$. Though $\sigma$ mesons have
been repeatedly reported in the past, their properties were never
clearly established. Therefore, this particle should be
interpreted as a parametrization of the $I=J=0$ part of the
two-pion spectrum, which shows up differently in different
experiments and hence has been reported with varying masses and
widths.


\subsection{Subtracted fixed-$t$ dispersion relations}
\label{sec:fixedt}

\indent As has been stated in the previous section, unsubtracted
DRs do not converge for the amplitudes $A_1$ and $A_2$. Moreover,
the amplitude $A_3$ converges only slowly, and in practice has to
be fixed by Baldin's sum rule. In order to avoid the convergence
problems and the phenomenology necessary to determine the
asymptotic contributions, it was suggested to consider DRs at fixed
$t$ that are once subtracted at $\nu=0$~\cite{DGP99},
\begin{equation}
\mathrm{Re} A_i(\nu, t) \;=\; A_i^B(\nu, t) \;+\; \left[ A_i(0,
t) - A_i^B(0, t) \right] \;+\;{2 \over \pi} \;\nu^2\; {\mathcal P}
\int_{\nu_0}^{+ \infty} d\nu' \; {{\; \mathrm{Im}_s A_i(\nu',t)}
\over {\nu' \; (\nu'^2 - \nu^2)}}\;. \label{eq:sub}
\end{equation}
These subtracted DRs should converge for all 6 invariant amplitudes
due to the two additional powers of $\nu'$ in the denominator, and
they are essentially saturated by the $\pi N$ intermediate states
as will be shown later. In other words, the lesser known
contributions of two and more pions as well as higher continua are
small and may be treated reliably by simple models.
\newline
\indent The price to pay for this alternative is the appearance of
the subtraction functions $A_i(\nu=0, t)$, which have to be
determined at some small (negative) value of $t$. We do this by
setting up once-subtracted DRs, this time in the variable $t$,
\begin{eqnarray}
A_i(0, t) \,-\, A_i^B(0, t) \,&=&\,
\left[ A_i(0, 0) \,-\, A_i^B(0, 0)
\right] \;+\;
\left[ A_i^{t-pole}(0, t) \,-\, A_i^{t-pole}(0, 0) \right] \nonumber\\
& +&\,{t \over \pi} \, \int_{(2 m_\pi)^2}^{+ \infty} dt' \;
{{\mathrm{Im}_t A_i(0,t')} \over {t' \; (t' - t)}} \,-\,{t \over
\pi} \, \int_{- \infty}^{-2 m_{\pi}^2 - 4 M m_\pi} dt' \;
{{\mathrm{Im}_t A_i(0,t')} \over {t' \; (t' - t)}} \;,
\label{eq:subt}
\end{eqnarray}
where $A_i^{t-pole}(0, t)$ represents the contribution of poles in
the $t$ channel, in particular of the $\pi^0$ pole in the case of
$A_2$, which is given by Eq.~(\ref{eq:piopole}).
\newline
\indent 
To evaluate the dispersion integrals,
the imaginary part due to $s$-channel cuts in Eq.~(\ref{eq:sub}) 
is determined, through unitarity relation, from the scattering amplitudes of photoproduction on the nucleon.
Due to the  energy denominator $1/\nu'(\nu'^2-\nu^2)$ in the subtracted 
dispersion integrals, the most important contribution is from the $\pi
N$ intermediate states, while mechanisms involving more pions or
heavier mesons in the intermediate states are largely suppressed.  
In our calculation, we evaluate the $\pi N$ contribution using the multipole 
amplitudes from the analysis of
Hanstein, Drechsel and Tiator (HDT)~\cite{HDT} at energies $\nu\le
500$~MeV and at the higher energies up to $\nu \simeq 1.5$ GeV 
we take as input the SAID multipoles (SP02K solution)~\cite{said}.  
The expansion of
$\mbox{Im}_s A_i$ into this set of multipoles is truncated at a
maximum angular momentum $j_{max}=l\pm 1/2=7/2,$ with the exception of
the energy range in the unphysical region   
where we use $j_{max}=3/2$.  
The higher partial waves with $j\ge j_{max}+1$ are
evaluated analytically in the one-pion exchange (OPE) approximation.
The relevant formulas to implement the calculation are reported in
App. B and C of Ref.~\cite{Lvo97}.
The multipion intermediate states are approximated by the inelastic
decay channels of the $\pi N$ resonances.  
Since a multipole analysis is not yet available for the two-pion channel,  
we assume that this inelastic contribution follows the helicity structure
of the one-pion photoproduction amplitudes. In this approximation, we
first calculate the resonant part of the pion photoproduction multipoles
using the Breit-Wigner parametrization of Ref.~\cite{said}, which is then
scaled by a suitable factor to include the inelastic decays of the
resonances. 
It was found, however, that in the subtracted dispersion relation formalism, 
the sensitivity to the multipion channels is very small and that subtracted 
dispersion relations are essentially saturated at $\nu\simeq 0.4$ GeV.
\newline
\indent
The imaginary part in the $t$-channel integral from $4 m_\pi^2
\rightarrow + \infty$ in Eq.~(\ref{eq:subt}) is saturated by the
possible intermediate states for the $t$-channel process (see, for example,
Fig.~\ref{fig:graphs_rcs} (e) and (g)), which lead to cuts along
the positive $t$-axis. For values of $t$ below the $K \bar K$
threshold, the $t$-channel discontinuity is essentially saturated
by $\pi \pi$ intermediate states.  
As a consequence, the dependence of the subtraction functions on the momentum 
transfer $t$ can be calculated by including the experimental information on 
the $t$-channel
 process through $\pi\pi$ intermediate states as 
$\gamma\gamma\rightarrow\pi\pi\rightarrow N\bar{N}$.
In Ref~\cite{DGP99}, a unitarized amplitude for the 
$\gamma\gamma\rightarrow\pi\pi$ subprocess was constructed, and 
a good description of the available data was found.
This information is then combined with the $\pi\pi\rightarrow N\bar{N}$ 
amplitudes determined from dispersion theory by analytical continuation of 
$\pi N$ scattering amplitudes~\cite{Hoe83}.
In practice, the upper limit of integration along the positive-$t$ cut
is $t=0.78 $ GeV$^2$, corresponding to the highest $t$ value for which the 
$\pi\pi\rightarrow N\bar{N}$ amplitudes are given in Ref.~\cite{Hoe83}.
In App.~\ref{sec:app1}, we show in detail how the discontinuities 
$\mathrm{Im}_t A_i$ of the invariant amplitudes $A_i (i=1,...,6)$ in the 
$t$-channel ($\gamma\gamma\rightarrow N\bar{N}$) can be 
expressed in terms of the corresponding $\gamma\gamma\rightarrow \pi\pi$ and 
$\pi\pi\rightarrow N\bar{N}$ amplitudes.
\newline
\indent
The second integral in Eq.~(\ref{eq:subt})
extends from $-\infty$ to $a=-2\, (m_{\pi}^2 + 2 M m_\pi) \approx -
0.56$~GeV$^2$. As we are interested in evaluating
Eq.~(\ref{eq:subt}) for small (negative) values of $t$ ($|t|\ll
|a|$), the integral from $- \infty$ to $a$ will be highly
suppressed by the denominator of the subtracted DRs, resulting in a small 
contribution.
This contribution is estimated by saturation
with the $\Delta$-resonance
and non-resonant $\pi N$ intermediate states.
In particular, we calculate the non-resonant $\pi N$ contribution to the
Compton amplitudes through unitarity relation from the OPE and 
nucleon-pole pion-photoproduction amplitudes,
while we consider the amplitudes corresponding to diagrams (c) and (d) of  
Fig.~\ref{fig:graphs_rcs} for the $\Delta$-resonance excitation .
Finally, the corresponding contributions to the discontinuities of the 
invariant amplitudes $A_i$ at $\nu=0$ and negative $t$ are obtained by 
analytical continuation in the unphysical region.
We estimated the total uncertainty resulting from the negative $t$ integral
and the two-pion contributions, by calculating the cross sections 
with and without the negative $t$ integral and the two-pion contributions.
The total difference can be estimated of the order of 3-5 \%, as
long as we restrict ourselves to the calculation of observables up to
the $\Delta$-resonance region.
\newline
\indent 
Once the $t$ dependence of the subtraction functions $A_i(0,t)$ 
is determined, the subtraction constants $A_i(0,
0)$ remain to be fixed. Although all 6 subtraction constants $a_1$
to $ a_6$ could be used as fit parameters, we shall restrict the
fit to the parameters $a_1$ and $a_2$, or equivalently to
$\alpha_{E1} - \beta_{M1}$ and $\gamma_\pi$.  The subtraction
constants $a_4, a_5$ and $a_6$ will be calculated through an
unsubtracted sum rule, as derived from Eq.~(\ref{eq:unsuba}),
\begin{equation}
a_{4, 5, 6} \;=\; {2 \over \pi} \; \int_{\nu_0}^{+ \infty}
d\nu' \; {{ \mathrm{Im}_s A_{4, 5, 6}(\nu',t = 0)} \over
{\nu'}}\;. \label{eq:a4a5a6}
\end{equation}
The remaining subtraction constant $a_3$, which is related to
$\alpha_{E1} + \beta_{M1}$ through Eq.~(\ref{eq:alphaplusbeta}),
will be fixed through Baldin's sum rule.


\subsection{Hyperbolic (fixed-angle) dispersion relations}
\label{sec:hypdr}

\begin{figure}[h]
\vspace{-.25cm} 
\epsfxsize=12.cm
\centerline{\epsffile{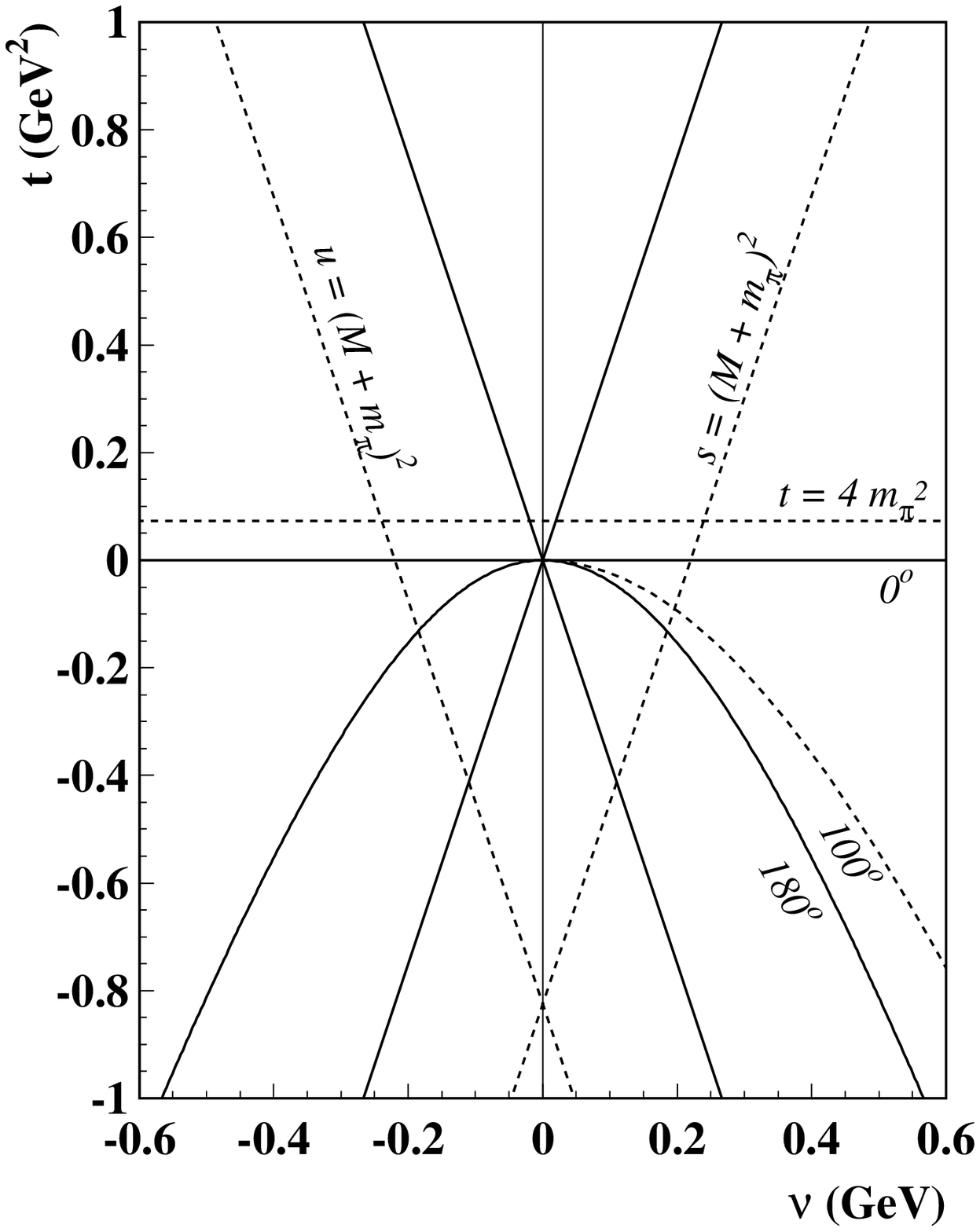}} 
\vspace{-.5cm}
\caption[]{Integration
paths in the $s$-channel region of the
 Mandelstam plane for RCS at fixed $lab$ angle.}
\label{fig:mandelstam_hyp1}
\end{figure}

\begin{figure}[h]
\vspace{-.25cm} \epsfxsize=7.5cm
\centerline{\epsffile{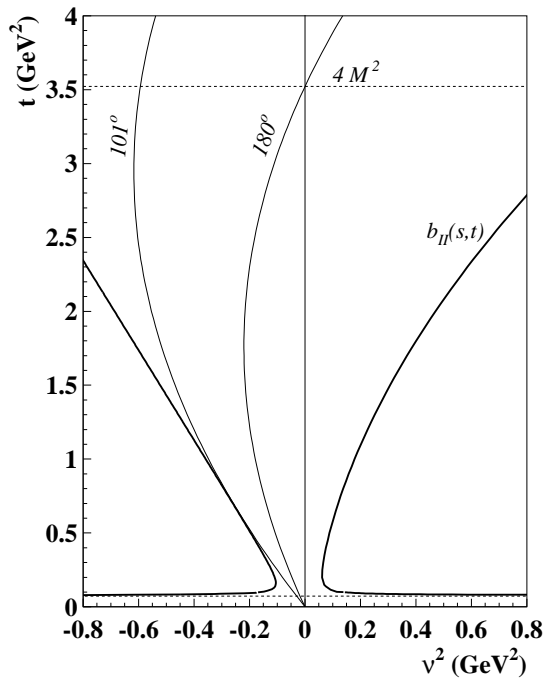}} 
\vspace{-.5cm}
\caption[]{Fixed lab
angle integration paths for RCS in the $\nu^2-t$ plane.
The right thick line corresponds to the spectral function 
$b_{II}(s,t)$  which determines the semi-minor axis of the ellipse of 
convergence,
while the left thick line follows from the semi-major axis of the 
ellipse of convergence and gives the boundary of convergence 
at $\nu^2<0$.}
\label{fig:mandelstam_hyp2}
\end{figure}

\begin{figure}[t]
\vspace{-.25cm}
\epsfxsize=14cm
\centerline{\epsffile{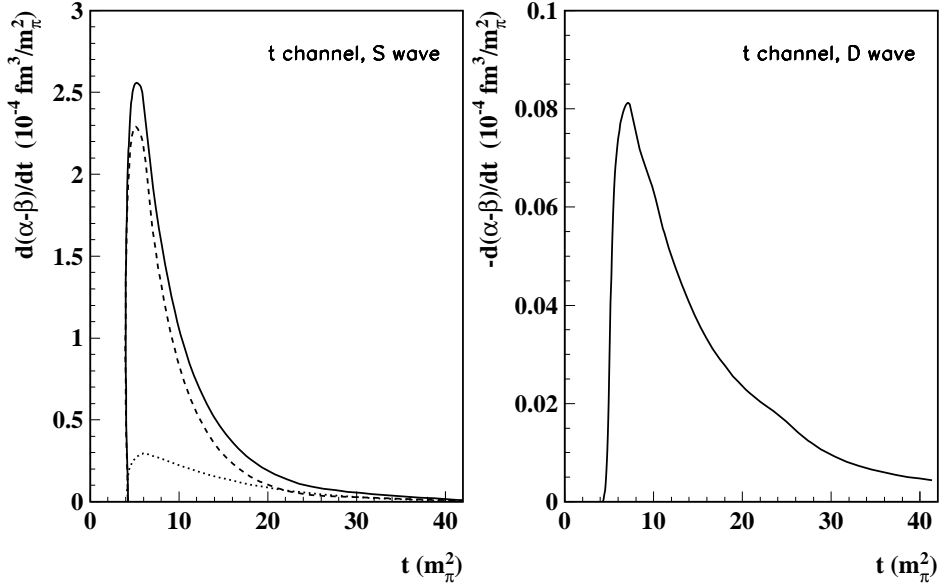}} 
\vspace{-.25cm}
\caption[]{The
integrand for the $t$-channel contribution from S-waves
  (left panel) and D-waves (right panel) 
to the polarizability $\alpha_{E1}-\beta_{M1}$. Dashed curve:
contribution from the $\pi^+ \pi^-$ channel, dotted curve:
contribution from the $\pi^0 \pi^0$ channel, solid curve: full
result, sum of charged and neutral channels. }
\label{fig:alpha_minus_beta_integrand}
\end{figure}

As we have seen in the previous sections, DRs at constant $t$ have
the shortcoming that the dispersion integrals get contributions
from the unphysical region between the boundaries of the physical
$s$ and $u$ channel regions. Though in principle the integrand in
this region can be constructed by extrapolating a partial wave
expansion of the Compton amplitudes, the calculation is limited in
practice to low partial waves. In order to improve the convergence
for larger values of $t$, fixed-angle DRs have been
proposed~\cite{Ber74} and applied to Compton
scattering~\cite{Hol94,Lvo99}. In particular for
$\theta_\mathrm{lab}=180^{\circ},$ the path of integration runs along the
lower boundary of the $s$-channel region (see Fig.
\ref{fig:mandelstam_hyp1}) from the origin of the Mandelstam plane
to infinity (``$s$-channel contribution''), complemented by a path
of integration in the upper half-plane (``$t$-channel
contribution''). The $s-u$ crossing symmetric hyperbolic
integration paths are given by~:
\begin{equation}
(s-a)(u-a)=b \qquad b=(a-M^2)^2 \, ,
\end{equation}
where $a$ is in one-to-one correspondence with the {\it lab}
and {\it c.m.} scattering angles as~:
\begin{equation}
a \,=\, - M^2 \, {{1 + \cos \theta_\mathrm{lab}} \over {1 - \cos
\theta_\mathrm{lab}}} \, ,
\hspace{2cm}a \,=\, - s \, {{1 + \cos \theta_\mathrm{cm}} \over {1 - \cos
\theta_\mathrm{cm}}} \, .
\end{equation}
A few contours corresponding with fixed values of $a$ are shown in
Fig.~\ref{fig:mandelstam_hyp1}.
Along such a path at fixed $a$, one can write down a dispersion
integral as
\begin{eqnarray}
{\rm Re} A_i(s,t,a)&=&A_i^{B}(s,t,a)+A_i^{t-pole}(s,t,a)\nonumber\\
& &\nonumber\\
& +&\frac{1}{\pi}\int_{(M+m_\pi)^2}^{\infty} {\rm d}s'\,\, {\rm
Im}_s A_i(s',\tilde t,a)
\left[\frac{1}{s'-s}+\frac{1}{s'-u}-\frac{1}{s'-a}\right]\nonumber\\
& &\nonumber\\
&+&\frac{1}{\pi}\int_{4m_\pi^2}^{\infty}{\rm d} t'\,\, \frac{{\rm
Im}_t A_i(\tilde s,t',a)}{t'-t}\ , \label{eq:hypdr1}
\end{eqnarray}
where the discontinuity in the $s$ channel ${\rm Im}_s A_i(s',\tilde t,a)$
is evaluated along the hyperbola given by
\begin{equation}
(s'-a)(u'-a)=b, \qquad s'+\tilde t+ u'=2M^2
\, ,
\end{equation}
and $\mathrm{Im}_t A_i(\tilde s,t',a)$ runs along the path defined by the 
hyperbola
\begin{equation}
(\tilde s-a)(\tilde u-a)=b, \qquad \tilde s+ t'+ \tilde u=2M^2
\, .
\end{equation}
The integrals in Eq.~(\ref{eq:hypdr1}) have a similar form as in
the case of subtracted DRs (Sec.~\ref{sec:fixedt}) except that the
individual partial waves are multiplied with different kinematical
factors depending on the angle. Though the problems of the partial
wave expansion are now cured in the lower half plane (for $t<0$), the
integration in the upper half plane (for $t>0$) still runs through the
unphysical region $4m_{\pi}^2\le t<4M^2$. 
In the latter region the $t$-channel partial wave expansion of ${\rm
Im}_t A_i$ convergences if $-0.594 \; \mbox{GeV}^2\le a \le 0$, corresponding to 
$ 101^{{\rm o}} \le \theta_\mathrm{lab} \le 180^{{\rm o}}$.
In fact, a simultaneous investigation of the $s-$ and $t-$channel Lehmann 
ellipses leads to the result that the convergence of the partial wave 
expansion is limited at positive $\nu^2$ by the spectral function 
$b_{II}(s,t)$, and at negative $\nu^2$ by the left thick line shown in 
Fig.~\ref{fig:mandelstam_hyp2} which follows from the semi-major axis 
of the ellipse of convergence~\cite{Hoe83}.
\newline
\indent
Summing up we find that DRs at $t=$ const are perfect for $t=0$,
i.e., $\theta_\mathrm{lab}=0$, and run into problems with increasingly
negative $t$ values, particularly at backwards angles, while DRs at
$\theta_\mathrm{lab}=$~const are best at $\theta_\mathrm{lab}=180^{\circ}$ and
loose accuracy with decreasing angle. Therefore, the two
techniques nicely complement each other.
\newline
\indent Holstein and Nathan~\cite{Hol94} investigated backward DRs
($\theta_\mathrm{lab}=180^{\circ}$) in order to get rigorous bounds for
the backward scalar polarizability, $\alpha_{E1}-\beta_{M1}$, and
to connect the polarizabilities of nucleons and pions. From the
$s$-channel integral they found for the proton
$(\alpha_{E1}-\beta_{M1})_s = 4.8-10.8\pm3.0$,
 where the numbers on the $rhs$ refer to one-pion
intermediate states with parity change, one-pion states without
parity change, and the error due to the unknown multipole
structure of heavier intermediate states. For the $t$-channel the
result was $(\alpha_{E1}-\beta_{M1})_t=10.3-1.7$, the first
contribution due to S-wave $\pi\pi$ states, the much smaller
second number from D-waves. Correlations are found to play an
important role for the $\pi\pi$ S-waves. While the pion
polarizability increases the Born contribution from 16.1 to 19.1,
the hadronic interaction decreases this value to 10.3 as shown
above.
Their total sum of the $s$- and $t$-channel contributions is still
considerably smaller than the nowadays accepted experimental
value, but the interesting finding of this investigation was the
partial resolution of the phenomenological $\sigma$ meson by the
$\pi\pi$ continuum. 
\newline
\indent
The same technique was later applied by L'vov
and Nathan~\cite{Lvo99} to study the puzzle of the large
dispersive contributions to the backward spin polarizability as
deduced by the LEGS experiment~\cite{Ton98}. Their final result
was $\gamma_{\pi} = - 39.5\pm2.4$, obtained by adding the
contributions of the $\pi^0$ pole $(-45.0\pm1.6)$, the $s$-channel
$\pi N$ states $(7.31\pm1.8)$ and small contributions of $\pi\pi
N$ states as well as $\eta$ and $\eta'$ in the $t$-channel.
\newline
\indent
Our results for the proton polarizabilities from unsubtracted
fixed-angle DRs are presented in the following three tables. We
choose 6 convenient combinations of the polarizabilities as seen
in Table~\ref{fixed_angle} where the total results are given. For
the reason outlined before, the fixed-angle results deteriorate
with decreasing angle, and therefore only the range $100^{{\rm
o}}\le \theta_\mathrm{lab}\le180^{{\rm o}}$ is shown. The contributions
of the $s$- and $t$-channel paths are listed separately in
Tables~\ref{fixed_angle_s} and \ref{fixed_angle_t}, respectively.
\begin{table}[h]
\caption[]{The proton polarizabilities evaluated by unsubtracted
fixed-angle dispersion relations at different $lab$ scattering
angles. See Eqs.~(\ref{eq:alphaplusbeta})-(\ref{eq:spinpol0}) for definitions.
The values are given in units of $10^{-4}$ fm$^3$ for the scalar 
polarizabilities, and $10^{-4}$ fm$^4$ for the spin polarizabilities.
\label{fixed_angle}} \vspace{0.35 truecm}
\begin{center}
\begin{tabular}{||c|c|c|c|c|c|c||}
\hline\hline $ \theta_\mathrm{lab}$ & $\alpha_{E1} - \beta_{M1}$ &
$\alpha_{E1} + \beta_{M1}$ & $\gamma_{\pi}^{disp}$ & $\gamma_{13}$
& $\gamma_{14}$ & $\gamma_{0}$
\\
\hline $ 180^{{\rm o}}$ & 10.89 & 10.80 & 7.79 & 4.32 & -2.36 &
-1.07
\\
\hline $ 140^{{\rm o}}$ & 10.58 & 10.93 & 7.62 & 4.33 & -2.35 &
-1.09
\\
\hline $ 100^{{\rm o}}$ & 9.36 & 11.41 & 6.93 & 4.40 & -2.28 &
-1.14
\\
\hline\hline
\end{tabular}
\end{center}
\end{table}
\newline
\indent
In general we expect that the forward polarizabilities,
$\alpha_{E1} + \beta_{M1}$ and $\gamma_{0}$, are described best by
forward DRs, i.e., the values at $\theta_\mathrm{lab}=0^{{\rm o}}$ (last
line of Table~\ref{fixed_angle_s}).
In comparing with the Baldin sum rule, Eq.~(\ref{DDeq2.2.18}), it
becomes obvious that $\alpha_{E1} + \beta_{M1}$ is not yet
saturated at the upper limit of our integration, $\nu=1.5$~GeV.
However, $\gamma_{0}$ is in good agreement with the experimental
analysis, Eq.~(\ref{DDeq2.2.19}), because of the better
convergence of the integral Eq.~(\ref{DDeq2.2.16}). The backward
polarizabilities, on the other hand, should be evaluated by paths
along backward angles. From Table~\ref{fixed_angle_s} we find
indeed that the $s$-channel contribution for $\alpha_{E1} -
\beta_{M1}$ and $\gamma_{\pi}^{disp}$ is pretty stable for $
\theta_\mathrm{lab}\gtrsim100^{{\rm o}}$.
\begin{table}[h]
\caption[]{ The $s$-channel contribution to the proton
polarizabilities of Table~\ref{fixed_angle}, first integral on the
$rhs$ of Eq.~(\ref{eq:hypdr1}).
Note that the values for $\theta_{\mathrm{lab}}=0^{\mathrm{o}}$ are identical with the
results of fixed-$t$ dispersion relations at $t=0$. 
Units as in Table~\ref{fixed_angle}.
\label{fixed_angle_s}}
\vspace{0.35 truecm}
\begin{center}
\begin{tabular}{||c|c|c|c|c|c|c||}
\hline\hline $ \theta_\mathrm{lab}$ & $\alpha_{E1} - \beta_{M1}$ &
$\alpha_{E1} + \beta_{M1}$ & $\gamma_{\pi}^{disp}$ & $\gamma_{13}$
& $\gamma_{14}$& $\gamma_{0}$
\\
\hline $ 180^{{\rm o}}$ & -5.56 & 7.52 & 7.71 & 2.75 & -2.70 &
-1.07
\\
\hline $ 140^{{\rm o}}$ & -5.63 & 7.65 & 7.70 & 2.75 & -2.70 &
-1.09
\\
\hline $ 100^{{\rm o}}$ & -5.76 & 8.13 & 7.71 & 2.82 & -2.62 &
-1.14
\\
\hline $ 60^{{\rm o}}$ & -5.76 & 9.25 & 7.94 & 3.19 & -2.31 &
-1.07
\\
\hline $ 20^{{\rm o}}$ & -5.49 & 11.29 & 8.89 & 4.01 & -1.68 &
-0.82
\\
\hline $ 0^{{\rm o}}$ &-5.30 & 11.94 & 9.29 & 4.28 & -1.54 & -0.75
\\
\hline\hline
\end{tabular}
\end{center}
\end{table}%
\begin{table}[h]
\caption[]{ The $t$-channel contribution to the proton
polarizabilities of Table~\ref{fixed_angle},
 last integral on the $rhs$ of Eq.~(\ref{eq:hypdr1}).
Units as in Table~\ref{fixed_angle}.
\label{fixed_angle_t}} \vspace{0.35 truecm}
\begin{center}
\begin{tabular}{||c|c|c|c|c|c|c||}
\hline\hline $ \theta_\mathrm{lab}$ & $\alpha_{E1} - \beta_{M1}$ &
$\alpha_{E1} + \beta_{M1}$ & $\gamma_{\pi}^{disp}$ & $\gamma_{13}$
& $\gamma_{14}$ & $\gamma_{0}$
\\
\hline $ 180^{{\rm o}}$ & 16.46 & 3.28 & 0.08 & 1.57 & 0.34 & 0
\\
\hline $ 140^{{\rm o}}$ & 16.20 & 3.28 & -0.08 & 1.57 & 0.34 & 0
\\
\hline $ 100^{{\rm o}}$ & 15.11 & 3.28 & -0.78 & 1.57 & 0.34 & 0
\\
\hline\hline
\end{tabular}
\end{center}
\end{table}
\newline
\indent
However, the astounding fact is the large $t$-channel contribution
for $\alpha_{E1} - \beta_{M1}$ (see Table~\ref{fixed_angle_t}).
From the last line in Table~\ref{fixed_angle_s}, one sees that the
$s$-channel integral up to $\nu_{max} = 1.5$~GeV yields only a small
contribution of about +3.3 to the electric polarizability. It is
remarkable to observe that the bulk contribution resides at energies
beyond 1.5~GeV. The bad convergence of the $s$-channel integral is
related to a strong concentration of the spectral strength in the
$t$-channel, close to two-pion threshold. This effect is clearly
reflected by the large $t$-channel contribution of about +9.9 to
$\alpha_{E1}$ (see Table~\ref{fixed_angle}, first line). 
The integrand for the $t$-channel integral is shown in
Fig.~\ref{fig:alpha_minus_beta_integrand} for the $\pi^+\pi^-$ and
$\pi^0\pi^0$ channel and for the sum of both channels.
The maximum of the integral is at $t\approx 0.09$~GeV$^2$, and displays a
long tail reaching out to higher values of $t$.
It is obvious that
this contribution contains the phenomenological $\sigma$ meson,
which has to be introduced to describe the data in the framework
of unsubtracted DRs at constant $t$. Our best value for $\alpha_{E1}
- \beta_{M1}$ comes from backward angles with an error estimated
from the stability in the region $ 140^{{\rm
o}}\le\theta_\mathrm{lab}\le180^{{\rm o}}$,
\be
\alpha_{E1} - \beta_{M1}=10.7\pm 0.2\ .
\ee
In Tables~\ref{fixed_angle}-\ref{fixed_angle_t}
only the dispersive contribution to the backward spin polarizability
has been listed.
If we add the large
$\pi^0$ pole contribution (see
Eqs.~(\ref{eq:piopole})-(\ref{eq:pi0gagacoup})),
we obtain
\be
\gamma_{\pi} = -38.8\pm1.8\ ,
\ee
the largest error being due to the value of the $\pi^0$ pole
contribution.
\newline
\indent
According to Table~\ref{fixed_angle_t}, the $t$-channel
contributions for the remaining combinations $\gamma_{13}$ and
$\gamma_{14}$ are very stable while the $s$-channel results depend
on the path of integration. 
In the case of the polarizability $\gamma_{13}$,
the backward value and the forward value agree within 1$\%$  (see 
Table~\ref{fixed_angle} and the last line of Table~\ref{fixed_angle_s}).
 However, the forward value of $\gamma_{14}=-1.54$
differs from its backward value substantially.
Since this polarizability contains the amplitude $a_4,$ which in turn 
is related to the forward polarizability $\gamma_0$, we expect that
the forward value is more realistic.
\newline
\indent
In summary, we obtain the following results~:
\begin{eqnarray}
\gamma_{13} \,&=&\, 4.30 \pm 0.02 \ , \nonumber \\
\gamma_{14} \,&=&\, -1.95 \pm 0.41 \ , \nonumber \\
\gamma_0 \,&=&\, -0.91 \pm  0.16 \ ,
\end{eqnarray}
where the central value and the errors are derived by combining
forward (last line of Table~\ref{fixed_angle_s}) 
and backward (first line of Table~\ref{fixed_angle}) DRs.
\newline
\indent
In order to improve the convergence,
we shall also consider  hyperbolic DRs that are once subtracted at $s=u=M^2,$ 

\begin{eqnarray}
{\rm Re} A_i(s,t,a)& = &A_i^{B}(s,t,a)
+[A_i(M^2,0,a)-A_i^{B}(M^2,0,a)]\nonumber\\
& &\nonumber\\
&+&\frac{1}{\pi}\int_{(M+m_\pi)^2}^{\infty} {\rm d}s'\,\,
{\rm Im}_s A_i(s',\tilde t,a)
\left[\frac{(s-M^2)}{(s'-s)(s'-M^2)}+\frac{(u-M^2)}{(s'-u)(s'-M^2)}
\right]
\nonumber\\
& &\nonumber\\
&+&[A_i^{t-pole}(t)-A_i^{t-pole}(0)]
+\frac{t}{\pi}\int_{4m_\pi^2}^{\infty}{\rm d} t'\,\,
\frac{{\rm Im}_t A_i(\tilde s,t',a)}{t'(t'-t)} \ .
\label{eq:hyp_sub}
\end{eqnarray}
In addition to the better convergence in the $s$- and $t$-channel integrals,
the subtraction at $s=u=M^2$ allows us to pursue a similar strategy as in 
the case of fixed-$t$ subtracted DRs. We note in fact that the subtraction 
constants in Eq.~(\ref{eq:hyp_sub}) are again related to the polarizabilities, 
i.e. $a_i=[A_i(M^2,0,a)-A_i^{B}(M^2,0,a)]$ independent of the 
value of $a$.


\subsection{Comparison of different dispersion relation
approaches to RCS data}
\label{sec:3_9} 

\begin{figure}[h]
\vspace{-.25cm} \epsfxsize=12cm
\centerline{\epsffile{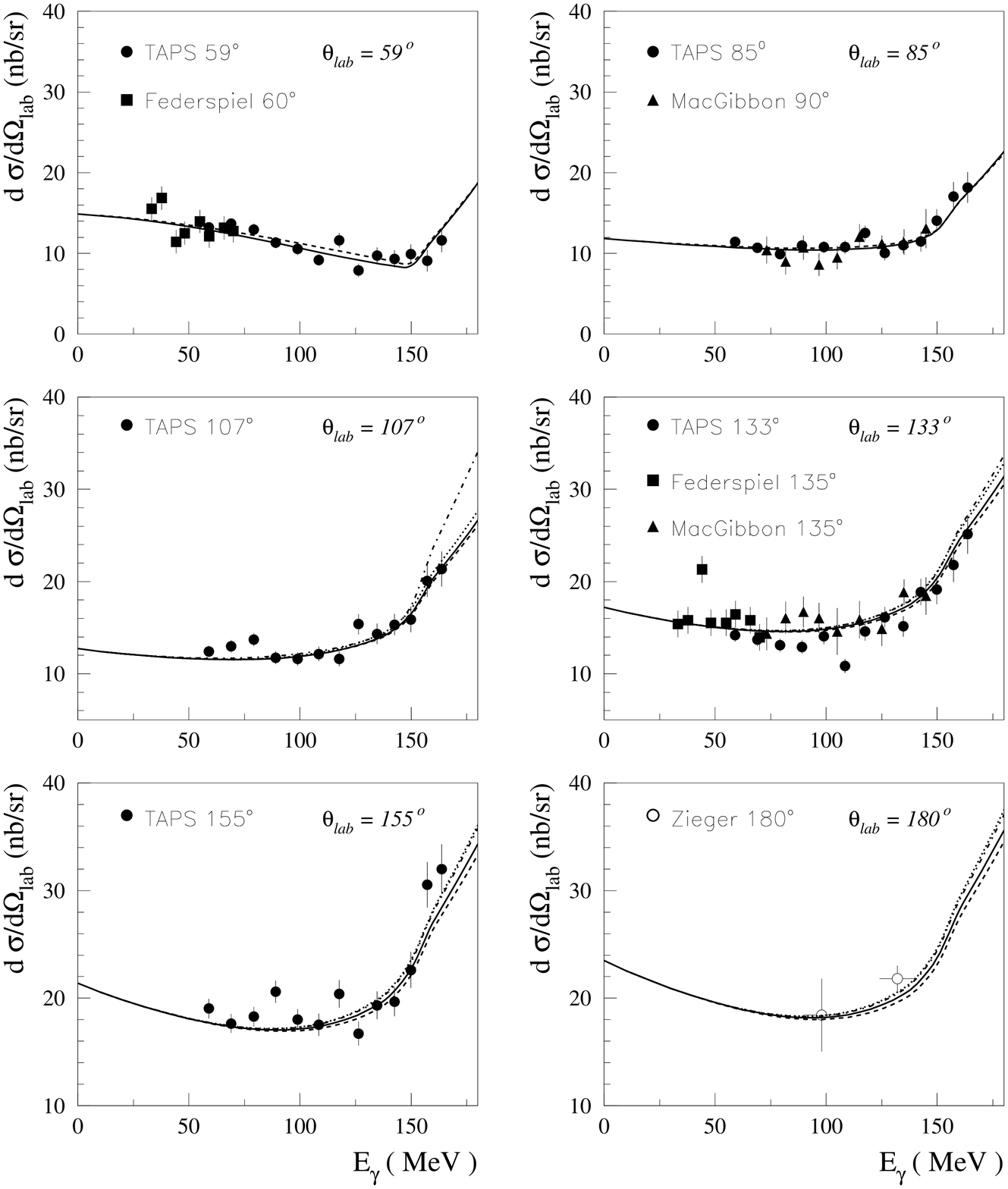}}
\vspace{-.5cm}
\caption[]{ Differential cross section for Compton
scattering off the proton as a function of the $lab$ photon energy
$E_\gamma$ and at different scattering angles. Full curves:
results from fixed-$t$ subtracted DRs, dashed curves: fixed-$t$
unsubtracted DRs, dotted curves: hyperbolic subtracted DRs,
dashed-dotted curves: hyperbolic unsubtracted DRs. All result are
shown for fixed values of $\alpha_{E1}+\beta_{M1} = 13.8$,
$\alpha_{E1}-\beta_{M1} = 10$, and $\gamma_\pi=-37.$ 
The experimental data are from Ref.~\cite{Olm01} (full circles),
 Ref.~\cite{Fed91} (diamonds), Ref.~\cite{Mac95} (triangles), 
and Ref.~\cite{Zieger} (open circles). 
}
\label{fig:rcs_taps_data}
\end{figure}

\begin{figure}[tb]
\vspace{-.5cm} 
\epsfxsize=9.75cm
\centerline{\epsffile{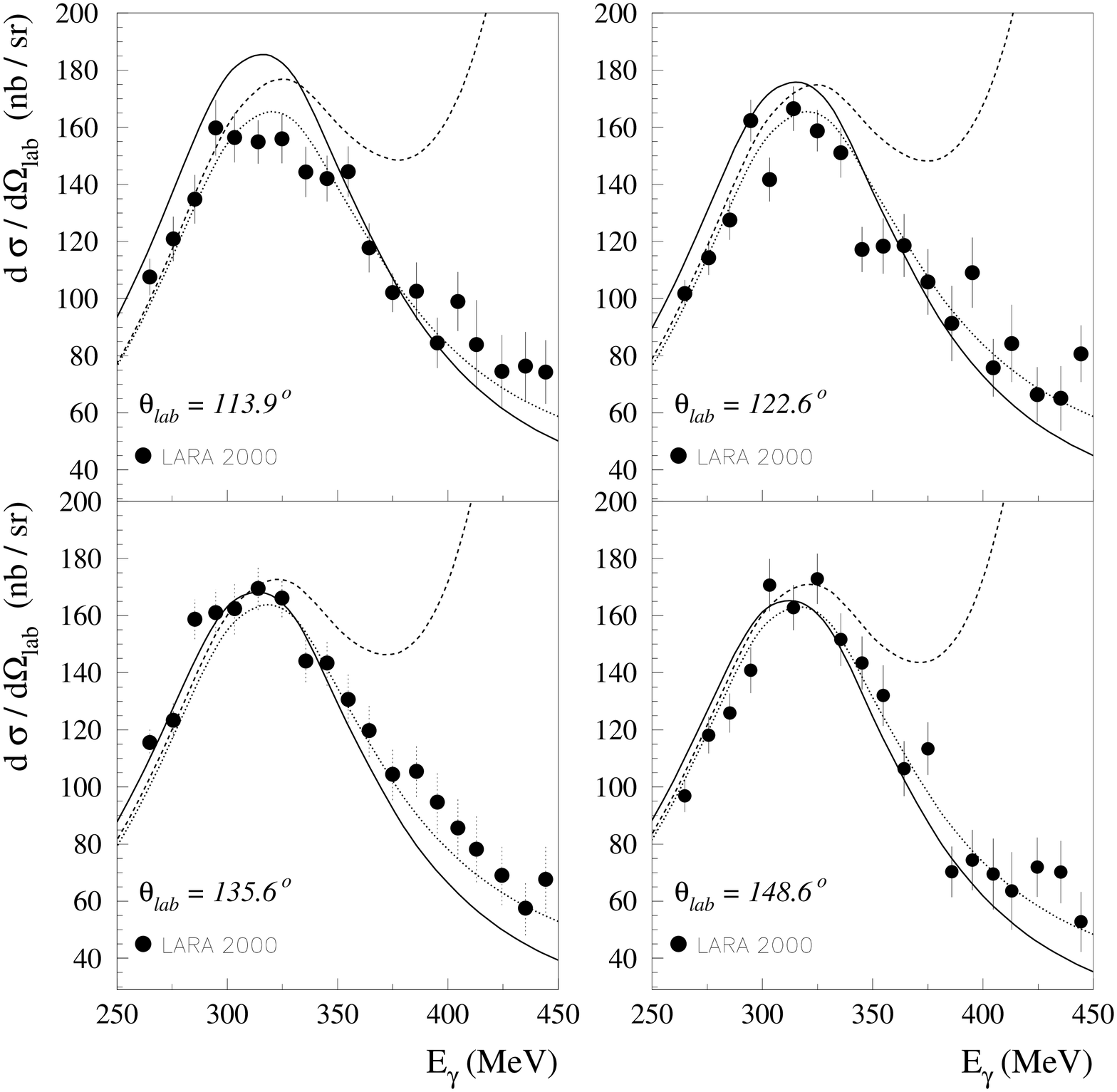}} 
\vspace{-.25cm}
\caption[]{
Differential cross section for Compton scattering off the proton
as a function of the $lab$ photon energy $E_\gamma$ and at
different scattering angles. Full curves: results from hyperbolic
unsubtracted DRs, dashed curves: fixed-$t$ subtracted DRs, dotted
curves: fixed-$t$ unsubtracted DRs. All results are shown for fixed
values of $\alpha_{E1}+\beta_{M1} = 14.05$,
$\alpha_{E1}-\beta_{M1} = 10$, and $\gamma_\pi=-38.$
The experimental data are from Ref.~\cite{Wol01}.
\label{fig:comparison_lara_1} }
\end{figure}

\begin{figure}[t!] 
\epsfxsize=9.5cm
\centerline{\epsffile{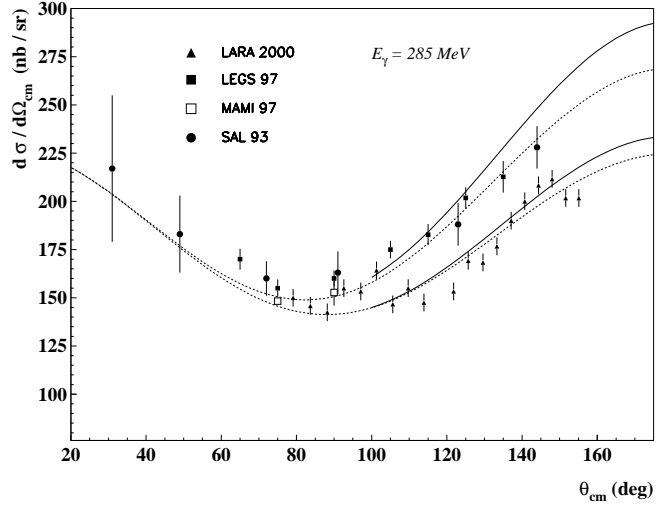}} 
\vspace{-.5cm}
\caption[]{The angular distribution 
at fixed photon $lab$ energy $E_{\gamma}=285$~MeV.
The results are displayed for $\alpha_{E1} +\beta_{M1} =14.05$,
$\alpha_{E1}-\beta_{M1} =10$ and different values of the backward
spin polarizability.
The dashed and solid curves are the results  
from fixed-$t$ subtracted DR and hyperbolic subtracted DR, respectively,
for $\gamma_\pi=-27$ (pair of upper curves) 
and $\gamma_\pi=-38$ (pair of lower curves).
The results from hyperbolic DRs are shown at backward angles, 
$\theta_{\mathrm{cm}}>100^{\mathrm{o}}$.
The experimental data are from Refs.~\cite{Wol01} (triangles),
\cite{Ton98} (full diamonds),
 \cite{Huen97} (open diamonds), and \cite{Hal93} (circles).
} \label{fig:rcs_eg285}
\end{figure}

\begin{figure}[h!]
\vspace{-1.25cm} 
\epsfxsize=9.25cm 
\centerline{\epsffile{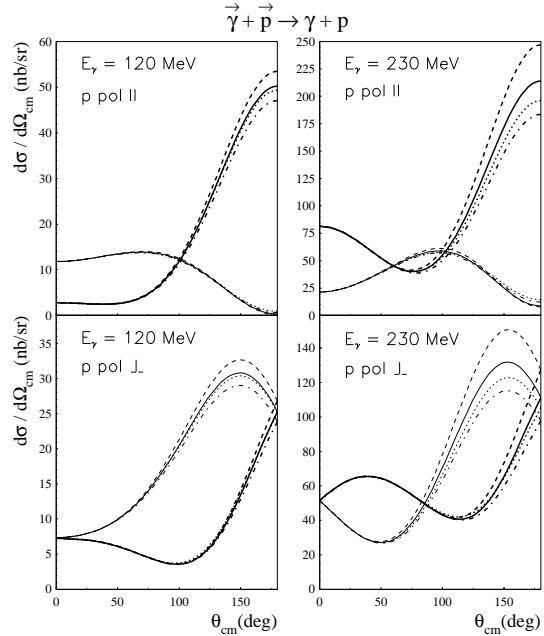}} 
\vspace{-3.cm} 
\caption[]{ Double polarization differential cross
sections for Compton scattering off the proton, with circularly
polarized photon and target proton polarized along the photon
direction (upper panels) or perpendicular to the photon direction
and in the plane (lower panels). The thick and thin curves
correspond to a proton polarization along the positive and
negative directions, respectively. The results of the dispersion
calculation at fixed-$t$ are for fixed $\alpha_{E1}+\beta_{M1}=13.8$,
fixed $\alpha_{E1}-\beta_{M1}=10,$ 
and $\gamma_\pi=-32$ (full curves), 
$\gamma_\pi=-27$ (dashed curves), and $\gamma_\pi=-37$
(dashed-dotted curves). We also show the result for  
$\alpha_{E1}+\beta_{M1}=13.8$,
$\alpha_{E1}-\beta_{M1}=8$ and $\gamma_\pi=-37$ (dotted curves). }
\label{fig:doublepol_circ}
\end{figure}

\begin{figure}[H]
\vspace{-1.3cm} 
\epsfxsize=9.5cm 
\centerline{\epsffile{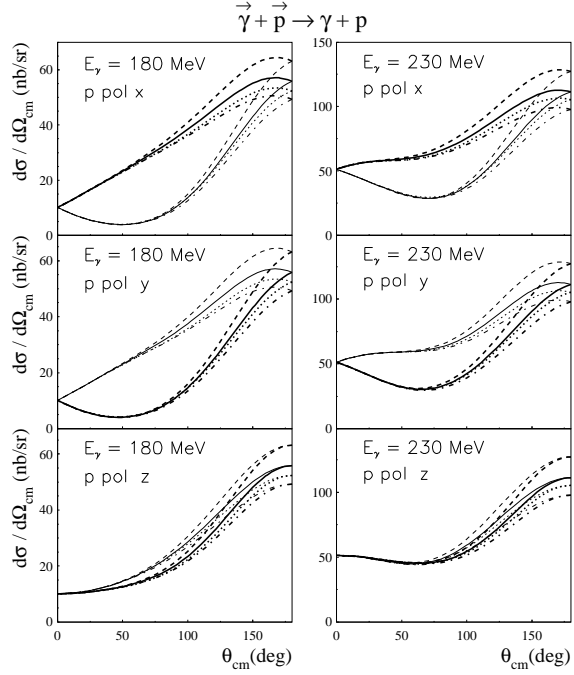}} 
\vspace{-.3cm} 
\caption[]{ Double polarization differential cross
sections for Compton scattering off the proton, with linearly polarized
photon and target proton polarized parallel or perpendicular to the 
scattering plane.
Upper panels: target polarized along the $x$ direction in the scattering plane
and photon with linear polarization at an angle $\Phi=+45^{\mathrm{o}}$
 (thick curves) and $\Phi=-45^{\mathrm{o}}$ (thin curves)
 with respect to the scattering plane.
Middle panels: target polarized along the $y$ direction
perpendicular to the scattering plane
and linearly polarized photon parallel
 (thick curves) and perpendicular (thin curves) to the scattering plane.
Lower panels: target polarized along the $z$ direction in the scattering plane 
and photon with linear polarization at an angle $\Phi=+45^{\mathrm{o}}$
 (thick curves) and $\Phi=-45^{\mathrm{o}}$ (thin curves)
 with respect to the scattering plane.
 The results of the dispersion
calculation at fixed-$t$ are for fixed $\alpha_{E1}+\beta_{M1}=13.8$,
fixed $\alpha_{E1}-\beta_{M1}=10,$ 
and $\gamma_\pi=-32$ (full curves), 
$\gamma_\pi=-27$ (dashed curves), and $\gamma_\pi=-37$
(dashed-dotted curves). 
We also show the result for  
$\alpha_{E1}+\beta_{M1}=13.8$,
$\alpha_{E1}-\beta_{M1}=8$ and $\gamma_\pi=-37$ (dotted curves). }
\label{fig:doublepol_long}
\end{figure}

In this subsection we compare the results from fixed-$t$ and
hyperbolic DRs, in both their subtracted and unsubtracted
versions, with some selected experimental data.
Fig.~\ref{fig:rcs_taps_data} shows the differential cross section
in the low-energy region for various $lab$ angles, obtained at
fixed values of $\alpha_{E1},\ \beta_{M1},$ and $\gamma_{\pi}$. The
results from subtracted and unsubtracted fixed-$t$ DRs (full and
dashed curves) are nearly identical except for extreme backward
scattering. Hyperbolic DRs, on
the other hand, can only be trusted in the backward hemisphere.
The unsubtracted version (dashed-dotted curve) clearly fails at
$\theta_{\mathrm{lab}}=107^{\circ}$ and above pion threshold, and of course
even more so at smaller scattering angles. However, it is
extremely satisfying that in all other cases the four different
approaches agree within the experimental error bars. We can
therefore conclude that the analysis of the low-energy data is
well under control and that quite reliable values can be extracted
for the polarizabilities, in particular $\alpha_{E1},\ \beta_{M1},$
and $\gamma_{\pi}$, which have a large influence on the low-energy
cross sections.
\newline
\noindent
In Table~\ref{fit_no_hallin_fixedt}, we show the results from subtracted 
fixed-$t$ DRs for the  fit of the 
polarizabilities $\alpha_{E1},\ \beta_{M1},$ and $\gamma_{\pi}$ 
to the modern low-energy data.
In particular, we analyzed both the set of recent data from the  
TAPS~\cite{Olm01} experiment and the full set of low-energy data of 
Refs.~\cite{Mac95,Fed91,Olm01,Zieger}.
For the fitting procedure we used the standard $\chi^2$ 
minimization~\footnote{The program package MINUIT from the CERNlib was used.}
by following two different strategies:
1) the spin polarizability, $\gamma_\pi$, the polarizability difference,
$\alpha_{E1}-\beta_{M1}$, and the polarizability sum, 
$\alpha_{E1}+\beta_{M1},$ were all used as independent free parameters;
2) $\gamma_\pi$ and $\alpha_{E1}-\beta_{M1}$ were varied in the fit, while
 $\alpha_{E1}+\beta_{M1}$ was constrained by Baldin's sum rule, using as 
additional datum point the result from the recent re-evaluation of 
Ref.~\cite{Olm01}, $\alpha_{E1}+\beta_{M1} = 13.8\pm 0.4.$ 
We note that the two fit procedures give consistent results  
within the error bars.
Only the value for the polarizability sum $\alpha_{E1}+\beta_{M1}$ 
 obtained from the fit is slightly underestimated with respect to the 
expected value from the Baldin sum rule. However, we note that the set of 
the fitted data mainly covers the backward-angle region 
where cross sections are quite insensitive to the scalar 
polarizability sum and mainly depend on the correlated effects of
$\gamma_\pi$ and $\alpha_{E1}-\beta_{M1}$.
\newline
\noindent
\begin{table}[h]
\caption[]{
 \label{fit_no_hallin_fixedt}
The polarizabilities $\alpha_{E1}-\beta_{M1}$ and $\gamma_\pi$ as obtained by 
fitting the differential cross sections from different experiments with 
fixed-$t$ subtracted DRs: 
``TAPS'' refers to the data from Ref.~\cite{Olm01} (65 data points) fitted by 
 using the definition of $\chi^2$ in Eq.~(\ref{eq:chi_squared}),
 and ``global'' denotes the values of the fit to 
the set of data from Refs.~\cite{Mac95,Fed91,Olm01,Zieger}
(a total of 101 data points) with the $\chi^2$-function of 
Eq.~(\ref{eq:chi_squared_ext}).
The term {\it fixed} denotes that 
$\alpha_{E1}+\beta_{M1}=13.8 \pm 0.4 $ is included as a 
constraint, while {\it free} indicates that this combination is also 
a free parameter.
The first error band is statistical and the second one is systematic.}
\vspace{0.5 truecm}
\begin{center}
\begin{tabular}{c|c|c|c}
\hline\hline 
      & $\alpha_{E1} + \beta_{M1}$ 
      & $13.8\pm 0.4$  ({\it fixed})    
      & $12.6 \pm 1.0 \pm 1.0 $ ({\it free})   \\
TAPS  & $\alpha_{E1}-\beta_{M1}$ 
      & $11.2\pm 1.2\pm 1.9 $ 
      & $11.4\pm 1.3\pm 1.7 $ \\   
      & $\gamma_\pi$ 
      & $-35.7\pm 3.9 \pm 0.6$ 
      & $-35.6\pm 2.1 \pm 0.4 $ \\
      & $\chi^2_{\mathrm{red}} $
      & $82.1/(66-2)=1.3$
      & $80.6/(65-3)=1.3$  
\\
\hline\hline  
      & $\alpha_{E1} + \beta_{M1}$   
      & $13.8 \pm 0.4$ ({\it fixed}) 
      &  $13.2 \pm 0.9 \pm 0.7$ ({\it free})\\
global  & $\alpha_{E1}-\beta_{M1}$ 
        & $11.3\pm 1.1 \pm 2.7 $ 
        & $11.1 \pm 1.1 \pm 0.8$   \\
        & $\gamma_\pi$ 
        &  $-35.9 \pm 1.8 \pm 3.2$ 
	&  $-36.0\pm1.8\pm 3.2$\\
      & $\chi^2_{\mathrm{red}}$ 
      &$116.0/(102-7)=1.2$ 
      &$115.7/(101-8)=1.2$        
\\
\hline\hline 
\end{tabular}
\end{center}
\end{table}
\begin{table}[h]
\caption[]{
The polarizabilities $\alpha_{E1}-\beta_{M1}$ and $\gamma_\pi$ as 
obtained by the fit of Ref.~\cite{Olm01}  with fixed-$t$ unsubtracted DR 
to  the differential cross sections from different experiments~: 
``TAPS'' refers to the data from Ref.~\cite{Olm01} (65 data points) fitted by 
using the definition of $\chi^2$ in Eq.~(\ref{eq:chi_squared}),
 and ``global'' denotes the values of the fit to 
the set of data from Refs.~\cite{Mac95,Fed91,Olm01,Zieger} 
(a total of 101 data points) with the $\chi^2$-function of 
Eq.~(\ref{eq:chi_squared_ext}).
The first error band is statistical and the second one is systematic.
The terms {\it fixed} and {\it free} 
are defined in Table~\ref{fit_no_hallin_fixedt}.
 \label{fit_no_hallin_fixedt_olmos}}
\vspace{0.5 truecm}
\begin{center}
\begin{tabular}{c|c|c}
\hline\hline 
      & $\alpha_{E1} + \beta_{M1}$ 
      &  ({\it free})    
        \\
      & $\alpha_{E1}$ 
        & $12.2\pm 0.8 \mp 1.4$ 
       \\    
TAPS  & $\beta_{M1}$ 
      & $0.8\pm0.9\pm0.5 $ 
       \\
      & $\gamma_\pi$ 
      & $ -35.9\pm2.3 \mp 0.4 $ 
       \\
      & $\chi^2_{\mathrm{red}} $
      & $80.6/(65-3)=1.3$
\\
\hline\hline  
      & $\alpha_{E1} + \beta_{M1}$  
      &  ({\it fixed})     
 \\
       & $\alpha_{E1}$ 
        & $12.4\pm0.6\mp0.5$ 
      \\
global   & $\beta_{M1}$ 
        & $1.4\pm0.7\pm 0.4 $    \\
        & $\gamma_\pi$ 
        & $-36.1\pm2.1\mp0.4 $    \\
      & $\chi^2_{\mathrm{red}}$ 
      &$108.4/(102-7)=1.1$        
\\
\hline\hline 
\end{tabular}
\end{center}
\end{table}
\newline
\indent
In the fit procedure to the TAPS data, 
we used the standard definition of $\chi^2$, i.e.,
\begin{equation}
\chi^2=\sum \left[\left(
\frac{(\sigma_{\mathrm{exp}}-\sigma_{\mathrm{theo}})}
{\Delta\sigma}\right)^2\right],
\label{eq:chi_squared}
\end{equation}
where $\sigma_{\mathrm{exp}}$ are the experimental and  
$\sigma_{\mathrm{theo}}$ the calculated cross-sections, and ${\Delta\sigma}$
are the experimental error bars. 
In Eq.~(\ref{eq:chi_squared}), the experimental errors were estimated 
according to Ref.~\cite{Olm01}, by adding in quadrature the statistical errors
 and the ``random systematic errors'' which were estimated,
from  uncertainties in the experimental geometry 
and from the statistics of the 
simulation, to  be equal to 
$\pm 5\%$ of the measured cross sections.
These statistical errors, including the random systematic uncertainties,
determine the first error bar in the values of the 
polarizabilities reported as TAPS-fit values in 
Table~\ref{fit_no_hallin_fixedt}.
The second error bar in these fitted values corresponds to systematic 
uncertainties. 
These systematic error bars were obtained by rescaling 
the differential cross section by $\pm 3\%$, assuming
that the systematic uncertainties in the data are mainly due to errors in 
the normalization of the measured cross sections.
\newline
\noindent
The global fit to the different data sets of 
Refs.~\cite{Mac95,Fed91,Olm01,Zieger} was performed by 
using a different $\chi^2$-function, namely
\begin{equation}
\chi^2=\sum \left[\left(
\frac{(N\sigma_{\mathrm{exp}}-\sigma_{\mathrm{theo}})}
{N\Delta\sigma}\right)^2\right]
+\left(\frac{N-1}{\Delta\sigma_{\mathrm{sys}}}\right)^2,
\label{eq:chi_squared_ext}
\end{equation}
where $N$ is a normalization parameter used to change the normalization for 
each data set within its systematic errors $\Delta\sigma_{\mathrm{sys}}$ 
taken equal to $\pm 3\%$ of the measured cross sections.
According to Refs.~\cite{Olm01} and~\cite{Mac95},
the minimization of this extended $\chi^2$-function
was performed by taking the polarizabilities and the normalization constants
for each data set as free parameters.
The resulting uncertainties in the fitted values of the 
polarizabilities include contributions from both the statistical and the 
systematic errors. The purely statistical contribution to these error bars 
was obtained by fitting the data with fixed values of the 
normalization constants. On the other hand, the net systematic error bars 
were derived by assuming that the total uncertainty is the result of the sum 
in quadrature of the statistical and systematic contributions.
These statistical and systematic errors are given by the first and second 
error bar, respectively, in the values of the polarizabilities quoted as 
global-fit values in Table.~\ref{fit_no_hallin_fixedt}. 
%
\newline
\noindent
In Table~\ref{fit_no_hallin_fixedt_olmos} we also show the results of 
Ref.~\cite{Olm01} from a fit within fixed-$t$ unsubtracted DRs.
We note that all the analyses, fixed-$t$ subtracted, fixed-$t$ unsubtracted 
and subtracted hyperbolic DRs, 
are in quite good agreement, giving us confidence that the model dependence 
of polarizability extraction is well under control.
\newline
The analysis is more model dependent when we move towards higher energies. As
shown in Fig.~\ref{fig:comparison_lara_1}, the results of
subtracted fixed-$t$ DRs have serious numerical problems at energies
above the $\Delta$(1232). The reason for this failure is, however,
not the high energy region itself, which is strongly suppressed by
the denominator in Eq.~(\ref{eq:sub}). Quite on the contrary, the
denominator creates the problem near the lower limit $\nu_0$ of
the integral, which extends into the region where the amplitude
has to be constructed by a continuation of the partial wave series
into the unphysical region (the area between the line
$s=(M+m_{\pi})^2$ and the $s$-channel region in
Fig.~\ref{fig:mandelstam_rcs}). Moreover, there are some
systematic differences between fixed-$t$ and hyperbolic DRs at these
higher energies. The fact that the data seem to favor fixed-$t$ DRs
is not surprising: The calculations are performed with
polarizabilities essentially derived by this method. Since these
data are taken at backward angles, hyperbolic DRs should in fact
be quite appropriate. As can be seen from the following
Fig.~\ref{fig:comparison_lara_sub_uns}, subtracted and
unsubtracted hyperbolic DRs agree quite nicely, with the exception of the
lowest scattering angle where the subtraction is necessary.
\begin{figure}[h!]
\vspace{-.6cm} \epsfxsize=9.75cm
\centerline{\epsffile{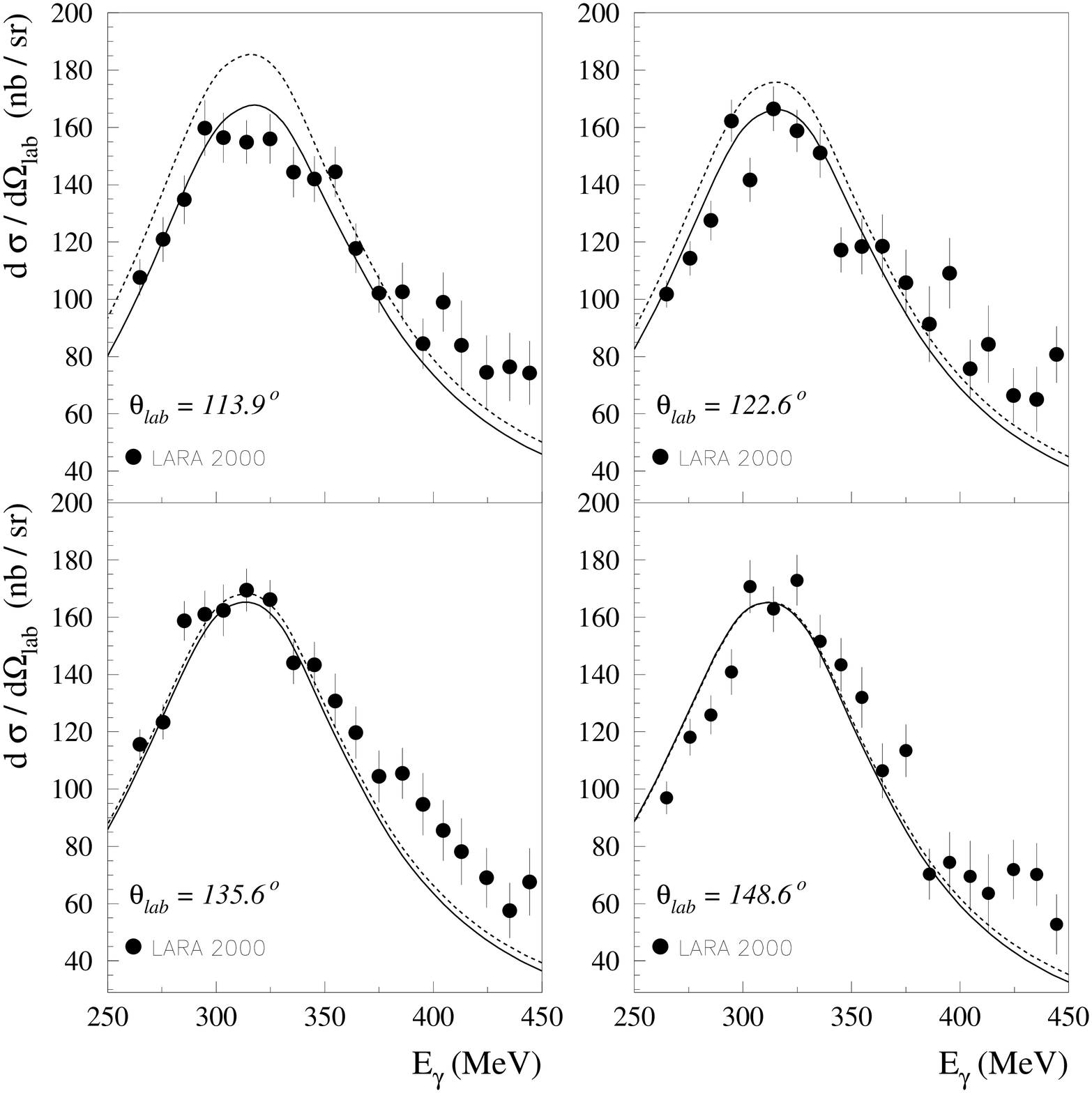}}
\vspace{-.25cm}
\caption[]{ Differential cross section for Compton scattering off
the proton as a function of the $lab$ photon energy $E_\gamma$ and
at different scattering angles. Full curves: results from
hyperbolic subtracted DRs, dashed curves: hyperbolic unsubtracted
DRs. All results are shown for fixed values of
$\alpha_{E1}+\beta_{M1} = 14.05$, $\alpha_{E1}-\beta_{M1} = 10$,
and $\gamma_\pi=-38.$ 
The experimental data are from Ref.~\cite{Wol01}.} 
\label{fig:comparison_lara_sub_uns}
\end{figure}
\newline
\noindent
Figure~\ref{fig:rcs_eg285} shows the angular distribution at an
energy somewhat below the $\Delta$(1232), which turns out to be
quite sensitive to the backward spin polarizability. Our
calculations confirm the finding of Ref.~\cite{Wol01}:  The
value of $\gamma_{\pi}$ derived by the LEGS
collaboration~\cite{Ton98} is related to the fact that the LEGS and SAL
data lie systematically above the recent MAMI results. This
difference can be partly compensated by a small change of the
$M_{1+}$ multipole, e.g., a 2 \% increase of $M_{1+}$ raises the cross
section by nearly 10 \%. 
However, the backward-forward asymmetry can not be changed that way 
but requires a strong variation of $\gamma_{\pi}$, in addition of the
effect of the (known) $E_{1+} / M_{1+}$ ratio.
\newline
\indent
In order to get new and independent information on the spin
polarizabilities, it will be necessary to perform double
polarization experiments. 
Figure~\ref{fig:doublepol_circ} shows the differential cross sections
for circularly polarized photons and target polarized perpendicular or
parallel to the photon beam.
Both for parallel and
perpendicular polarization, a spin-flip of the target proton
changes the cross section by large factors. The sensitivity to the
backward spin polarizability turns out to be largest at the higher
energies and for circularly polarized photons hitting protons with
polarizations perpendicular to the photon beam. It is also
demonstrated in Fig.~\ref{fig:doublepol_circ} that even an
unreasonably large 20~\% decrease of $\alpha_{E1}-\beta_{M1}$ can only 
simulate a
change in $\gamma_{\pi}$ of about 2-3 units, making this an ideally
suited observable to access $\gamma_\pi$.
\newline
\noindent
In the case of linearly polarized photons, one can access three 
additional independent observables.
In particular, we can classify these polarization observables
by assuming the $xz$ plane as the photon scattering plane, 
with the quantization axis along the direction of the incoming photon momentum,
and denoting with $\Phi$ the angle between the polarization vector of the 
photon and the $x$ axis.
With respect to this frame, one can measure the cross sections with 
the target polarized along the $x$ or $z$ direction and 
the photon polarization at $\Phi=\pm 45^{\mathrm{o}}$, 
and the cross sections with the target polarization perpendicular to the 
scattering plane and the photon polarization parallel ($\Phi=0^{\mathrm{o}}$)
or perpendicular ($\Phi=90^{\mathrm{o}}$) to the scattering plane.
The results from fixed-$t$ subtracted DRs for these observables
 are displayed in Fig.~\ref{fig:doublepol_long}, and show similar sensitivity
to the $\gamma_\pi$ and $\alpha_{E1}-\beta_{M1}$ polarizabilities
as in the case of double polarization observables
with circularly polarized photons.

\clearpage


\subsection{Physics content of the nucleon polarizabilities}
\label{sec:physpol}

The physical content of the polarizabilities can be visualized
best by effective multipole interactions for the coupling of the
electric $(\vec{E})$ and magnetic $(\vec{H})$ fields of the photon
with the internal structure of the nucleon~\cite{Bab98a,Hol00},

\be
\label{eq3.8.1}
H_{\rm{eff}} = -4\pi
\sum_{n=1}\left(\tilde{H}_{\rm{eff}}^{(2n)} +
\tilde{H}_{\rm{eff}}^{(2n+1)}\right) \ ,
\ee
where the even and odd upper indices refer to scalar and vector
polarizabilities, respectively. In particular, the lowest scalar
polarizabilities are contained in
\beqn
\tilde{H}_{\rm{eff}}^{(2)} & = &
{\textstyle\frac{1}{2}}\,\alpha_{E1}\,\vec{E}\,^2 +
{\textstyle\frac{1}{2}}\,\beta_{M1}\,\vec{H}\,^2\ , \nonumber \\
\tilde{H}_{\rm{eff}}^{(4)} & = &
{\textstyle\frac{1}{2}}\,\alpha_{E1,\nu}\,\dot{\vec{E}}\,^2 +
{\textstyle\frac{1}{2}}\,\beta_{M1,\nu}\,\dot{\vec{H}}\,^2
+ {\textstyle\frac{1}{12}}\,\alpha_{E2}\,E_{ij}^2+
{\textstyle\frac{1}{12}}\,\beta_{M2}\,H_{ij}^2\ .
\label{eq3.8.2}
\eeqn

The leading term contains the (static!) electric and magnetic
dipole polarizabilities, $\alpha=\alpha_{E1}$ and
$\beta=\beta_{M1}$. In the subleading term there appear two
derivatives of the fields with regard to either time or space,
$\dot{\vec{E}} = \partial_t\vec{E}$ and
$E_{ij}=\frac{1}{2}(\nabla_iE_j+\nabla_jE_i)$ respectively.
Applied to a plane wave photon, the subleading term is therefore
$\mathcal{O}(\omega^2)$ relative to the leading one. The terms in
$\alpha_{E1,\nu}$ and $\beta_{E1,\nu}$ are, of course, retardation
or dispersive corrections to the respective leading order dipole
polarizabilities, while $\alpha_{E2}$ and $\beta_{E2}$ are the
electric and magnetic quadrupole polarizabilities. Combining the
static dipole polarizabilities with all terms in the sum with time
derivatives only, we obtain the ``dynamical dipole
polarizabilities'' $\alpha_{E1}(\omega)$ and $\beta_{M1}(\omega)$.
The terms involving the gradients build up higher
polarizabilities, at fourth order the (static) electric $(\alpha_{E2})$ and
magnetic $(\beta_{M2})$ quadrupole polarizabilities. In a similar notation the
lowest vector or spin polarizabilities are defined by
\beqn
\tilde{H}_{\rm{eff}}^{(3)} & = &
{\textstyle\frac{1}{2}}\gamma_{E1E1}
\,\vec{\sigma}\cdot(\vec{E}\times\dot{\vec{E}})
+
{\textstyle\frac{1}{2}}\gamma_{M1M1}\,
\vec{\sigma}\cdot(\vec{H}\times\dot{\vec{H}})
\nonumber \\ && - \gamma_{M1E2}\, E_{ij}\,\sigma_iH_j -
\gamma_{E1M2}\, H_{ij}\,\sigma_iE_j\ , \\
\tilde{H}_{\rm{eff}}^{(5)} & = &
{\textstyle\frac{1}{2}}\gamma_{E1E1,\nu}\,\vec{\sigma}\cdot(\dot{\vec{E}}\times\ddot{\vec{E}})
+
{\textstyle\frac{1}{2}}\gamma_{M1M1,\nu}\,\vec{\sigma}\cdot(\dot{\vec{H}}\times\ddot{\vec{H}})
\nonumber \\ && - \gamma_{M1E2,\nu}\,
\dot{E}_{ij}\,\sigma_i\dot{H}_j
   - \gamma_{E1M2,\nu}\,\dot{H}_{ij}\,\sigma_i\dot{E}_j
\nonumber \\ && -
2\gamma_{E2E2}\,\epsilon_{ijk}\,\sigma_iE_{jl}\dot{E}_{kl}
   - 2\gamma_{M2M2}\,\epsilon_{ijk}\,\sigma_iH_{jl}\dot{H}_{kl}
\nonumber \\ && + 3\gamma_{M2E3}\,\sigma_i\,E_{ijk}H_{jk}
   - 3\gamma_{E2M3}\,\sigma_i\,H_{ijk}E_{jk}\ ,
\label{eq3.8.3}
\eeqn
where
\be
E_{ijk}  =  {\textstyle\frac{1}{3}}\,(\nabla_i\nabla_jE_k+
\nabla_i\nabla_kE_j + \nabla_j\nabla_kE_i)  -
{\textstyle\frac{1}{15}}\, (\delta_{ij}\Delta E_k + \delta_{jk}\Delta E_i +
\delta_{ik}\Delta E_j)\ . \nonumber
\label{eq3.8.4}
\ee
\indent
As in the spin-averaged case, four of the terms in the
$\mathcal{O}(\omega^5)$ polarizabilities are simply dispersive
corrections to the $\mathcal{O}(\omega^3)$ expressions. All
polarizabilities defined above can be related to the multipole
expansions given in Ref.~\cite{Bab98a}. In terms of the standard
notation of spherical tensors, the polarizabilities correspond to
the following coupling of electromagnetic transition operators:
\beqn
\alpha_{EL}\sim[E_L\times E_L]^0\ \quad &,& \quad
\beta_{ML}\sim[M_L\times M_L]^0\ , \nonumber \\
\gamma_{ELEL}\sim[E_L\times E_L]^1\quad &,& \quad
\gamma_{MLML}\sim[M_L\times M_L]^1\ ,  \\
\gamma_{M(L-1)EL}\sim[M_{L-1}\times E_L]^1\quad &,& \quad
\gamma_{E(L-1)ML}\sim[E_{L-1}\times M_L]^1 \ . \nonumber
\label{eq3.8.5}
\eeqn
The higher order polarizabilities given above are uniquely defined
by the quantities $a_i,\ a_{i,\nu}$ and $a_{i,t}$ of Eq.~(\ref{eq3.2.2}),
as discussed in detail in Refs.~\cite{Bab98a} and~\cite{Hol00}.
In particular we find for the leading terms the relations
\begin{eqnarray}
\begin{array}{lcl}
\alpha_{E1} = - {\textstyle\frac{1}{4\pi}}\,(a_1+a_3+a_6), \  & 
  &
\beta_{M1} = + {\textstyle\frac{1}{4\pi}}\,(a_1-a_3-a_6)\ , \\
\alpha_{E2} = - {\textstyle\frac{3}{\pi}}\,(a_{1,t}+a_{3,t}+a_{6,t}),
\ &  \ &
\beta_{M2} = +
{\textstyle\frac{3}{\pi}}\,(a_{1,t}-a_{3,t}-a_{6,t})\ , \\
\gamma_{E1E1} = {\textstyle\frac{1}{8\pi M}}\,
(a_2-a_4+2a_5+a_6) \  ,&    &
\gamma_{M1M1} =  - {\textstyle\frac{1}{8\pi M}}\,
(a_2+a_4+2a_5-a_6)\ , \\
\gamma_{M1E2} = {\textstyle\frac{1}{8\pi M}}\,
(a_2+a_4+a_6) \ , &   &
\gamma_{E1M2} =  {\textstyle\frac{1}{8\pi M}}\,
(a_2-a_4-a_6)\ , \\
\gamma_{E2E2} = {\textstyle\frac{1}{24\pi M}}\,
(a_{2,t}-a_{4,t}+3a_{5,t}+2a_{6,t}) \,
\  &   &
\gamma_{M2M2} = {\textstyle\frac{1}{24\pi M}}\,
(-a_{2,t}-a_{4,t}-3a_{5,t}+2a_{6,t}), \\
\gamma_{M2E3} = - {\textstyle\frac{1}{12\pi M}}\,
(a_{2,t}+a_{4,t}+a_{6,t})
\ , &    &
\gamma_{E2M3} = - {\textstyle\frac{1}{12\pi M}}\,
(-a_{2,t}+a_{4,t}+a_{6,t}) \ .
\end{array}\nonumber\\
\label{eq3.8.6}
\end{eqnarray}
where we neglected recoil contributions of $\mathcal{O}(M^{-2}).$
For details see Ref.~\cite{Hol00}.
In terms of Ragusa's polarizabilities $\gamma_i$ one has
\be
\begin{array}{lcl}
\gamma_{E1E1} = -\gamma_1-\gamma_3 \ ,&  \  &
\gamma_{M1M1} =  \gamma_4\ ,\\
\gamma_{M1E2} = \gamma_2+\gamma_4\ ,  &  \  &
\gamma_{E1M2} = \gamma_3\ . 
\end{array}
\label{eq:ragusa_mult}
\ee
\indent
With these definitions we can now complete the expansion of the
forward scattering amplitudes, Eqs.~(\ref{DDeq2.2.12}) and (\ref{DDeq2.2.13}),
to the next order:
\beqn
f(\nu) & = & - \frac{e^2e_N^2}{4\pi M}+
(\alpha_{E1}(\nu)+\beta_{M1}(\nu))\,\nu^2 +
{\textstyle\frac{1}{12}}\,(\alpha_{E2}(\nu)+
\beta_{M2}(\nu))\,\nu^4+\mathcal{O}(\nu^6)\ , \\
g(\nu) & = & - \frac{e^2\kappa_N^2}{8\pi M^2}\,\nu +
\gamma_0(\nu)\nu^3 + \tilde{\gamma}_0(\nu)\nu^5+\mathcal{O}(\nu^7) \ .
\label{eq3.8.7}
\eeqn
with
$\alpha_{EL}(\nu)=\alpha_{EL}+\alpha_{EL,\nu}\nu^2+\mathcal{O}(\nu^4)$
and similarly
for the magnetic terms. In the spin-flip amplitude we have defined
\beqn
\gamma_0(\nu) & = & - (\gamma_{E1E1}(\nu) + \gamma_{M1M1}(\nu) +
\gamma_{M1E2}(\nu) + \gamma_{E1M2}(\nu))\ , \\
\tilde{\gamma}_0(\nu) & = & -
(\gamma_{E2E2}(\nu) + \gamma_{M2M2}(\nu) + {\textstyle\frac{8}{5}}
\gamma_{M2E3}(\nu)+ {\textstyle\frac{8}{5}}
\gamma_{E2M3}(\nu))\ .
\label{eq3.8.8}
\eeqn
We repeat that all combinations of polarizabilities appearing in
the forward direction can be evaluated safely on the basis of DRs.
However, the individual polarizabilities suffer from the
non-convergence of the unsubtracted DRs for $A_1$ and $A_2$, and the
bad convergence of $A_3$. In the following section we shall
compare the predictions of DRs, ChPT and some QCD-motivated models
amongst each other and with the available experimental data.
\begin{figure}[h]
\vspace{-0.1cm}
\epsfxsize=7.cm
\centerline{\epsffile{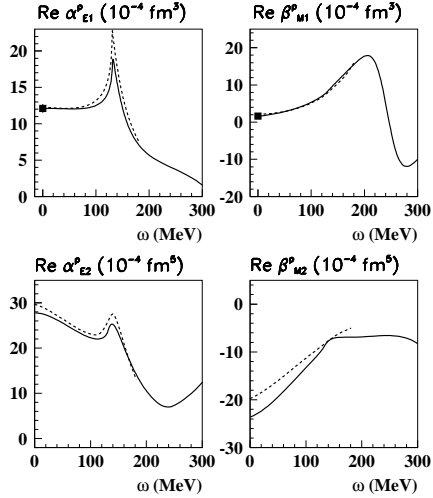}} 
\vspace{-.25cm}
\caption[]{
The real part of the proton
polarizabilities $\alpha_{E1},\ \beta_{M1}$ (upper panels)
and $\alpha_{E2},\ \beta_{M2}$ (lower panels) as function of the
photon {\it c.m.} energy $\omega$.
Full curves: results from fixed-$t$ subtracted
dispersion relations.
Dashed curves: predictions in leading order HBChPT from Ref.~\cite{HGHP02}
for the isoscalar contribution to the dynamical polarizabilities
up to $\omega = 170$ MeV.
The diamonds are the experimental values for the dipole static 
polarizabilities~\cite{Olm01} which are used to fit low-energy constants. 
} \label{fig:dyn_polarizab_proton}
\end{figure}


The imaginary parts of the dynamical polarizabilities are determined
from the scattering amplitudes of photoproduction on the nucleon by
the unitarity relation.
If we take into account only the contribution from one-pion
intermediate states, the unitarity relations take the following simple
form~\cite{Bab98a}~: 
\begin{eqnarray}
\label{eq:im_dyn_pol} {\rm Im}\, \alpha_{E1} (\omega) &=&
\frac{k_{\pi}}{\omega^2}\sum_c (2|E^{(c)}_{2-}|^2+
|E^{(c)}_{0+}|^2
),\nonumber\\
{\rm Im}\, \beta_{M1} (\omega) &=& \frac{k_{\pi}}{\omega^2}\sum_c
(2|M^{(c)}_{1+}|^2+ |M^{(c)}_{1-}|^2
),\nonumber\\
{\rm Im}\, \alpha_{E2} (\omega) &=& \frac{k_{\pi}}{\omega^4} 36
\sum_c (3|E^{(c)}_{3-}|^2+ |E^{(c)}_{1+}|^2
),\nonumber\\
{\rm Im}\, \beta_{M2} (\omega) &=& \frac{k_{\pi}}{\omega^4} 36
\sum_c (3|M^{(c)}_{2+}|^2+ |M^{(c)}_{2-}|^2 ),
\end{eqnarray}
where $k_{\pi}$ is the pion momentum, 
$\omega$ the photon {\it c.m.} energy, and 
$E_{l\pm}^c$ and  $M_{l\pm}^c$ are pion photoproduction multipoles
which are summed over the different isotopic or charge channels.
\newline
\noindent
The real part of these amplitudes, calculated both in
 dispersion theory and HBChPT~\cite{Gri02,HGHP02}, is displayed in
 Fig.~\ref{fig:dyn_polarizab_proton}.
The dynamical polarizabilities allow for a very detailed study of the 
internal degrees of freedom.
For example, $\alpha_{E1}$ and $\alpha_{E2}$ clearly show cusp effects
 due to the opening of the pion threshold, and $\beta_{M1}$ exhibits
 the $\Delta$-resonance structure, with the real part passing through
 zero at the resonance position.
The HBChPT calculation nicely reproduces the results of DRs.
  
\subsection{DR predictions for nucleon polarizabilities and comparison
with theory}
\label{sec:polmod}

In a nonrelativistic model like the constituent quark model (CQM),
the scalar dipole polarizabilities can be expressed by
\begin{eqnarray}
\label{DDeq3.9.1}
\alpha_{E1} &=& 2\alpha_{em} \sum_{n\ne0} \frac{|\langle n| d_z|0
\rangle|^2} {E_n-E_0} + \frac{\alpha_{em}}{3M}
\langle 0|\sum_i e_i\, r_i^2|0\rangle\ ,\\
\label{DDeq3.9.2}
\beta_{M1} &=& 2\alpha_{em} \sum_{n\ne0} \frac{|\langle n| \mu_z|0
\rangle|^2} {E_n-E_0} - \frac{\alpha_{em}}{2M}
\langle 0|d^2+\sum_i d_i^2|0\rangle\ ,
 \end{eqnarray}
where ${\bd} = \sum {\bd}_i = \sum e_i{\br}_i$ and
${\bmu} = \sum {\bmu}_i = \sum \frac{e_i}{2m_i}{\bsig}_i$
are the electric and magnetic dipole operators in the $c.m.$ frame
of the nucleon.
For simplicity the quark masses are taken as
$m_i={\textstyle{\frac{1}{3}}}M$, and the quark charges $e_i$ are
in units of $e$. The terms ${\mathcal{O}}(M^{-1})$ in 
Eqs.~(\ref{DDeq3.9.1},\ref{DDeq3.9.2}) 
are retardation or recoil terms, which are
small corrections in atomic physics but actually quite sizeable
for the quark dynamics of the nucleon. Clearly the first term on
the $rhs$ of both equations is positive, because the dipole matrix
elements appear squared and the excitation energy $E_n-E_0$ is
positive. The higher order terms ${\mathcal{O}}(M^{-1})$, however,
are positive for $\alpha_{E1}$ but negative for $\beta_{M1}$. In
the case of the magnetic polarizability, the leading term
describes the paramagnetism which is essentially due to the
spin-flip transition from the nucleon to the $\Delta$ (1232),
while the subleading term represent Langevin's diamagnetism. The
simple CQM with an oscillator potential connects the $rms$ radius
$\langle r^2\rangle^{1/2}$ with the oscillator frequency,
$\omega_0=3/(M \langle r^2\rangle)$, and yields~\cite{Pet81}
\be
\label{DDeq3.9.3}
\alpha_{E1} = (2\alpha_{em}) / (M\,\omega_0^2)
+ {\mathcal{O}}(M^{-2})\ .
\ee
Unfortunately, it is not possible to describe both size and
excitation energy in this model. If we use the proper size, say
the electric Sachs radius of the proton, $\langle r^2\rangle=\langle
r^2\rangle_E^p$, $\alpha_{E1}$ is grossly overestimated with a
value of about 40. On the other hand, the correct excitation energy for the
dominant dipole mode N$^{\ast}$ (1520), leads to a value much too
small, $\alpha_{E1}\approx3.5$. Concerning the magnetic
polarizability, the magnetic dipole transition to the $\Delta$
(1232) yields a large paramagnetic value,
$\beta^{\Delta}_{M1}\approx12$, which is somewhat reduced by the
diamagnetic terms.
\newline
\indent
The fact that we underestimate $\alpha_{E1}$ if using the
excitation energy of the N$^{\ast}$ (1520) is easily understood:
The energy denominator in Eq.~(\ref{DDeq3.9.1}) has been taken to
be nearly 600~MeV, while electric dipole absorption due to pion
S-wave production already takes place at much smaller energies.
The strong dependence of $\alpha_{E1}$ on the size of  the oscillator parameter
can of course be used to get close to the experimental numbers,
and indeed reasonable results were obtained using the MIT bag
model~\cite{Hec81,Sch84}. However, it was also early recognized
that no complete picture of the nucleon can emerge without
including the pion cloud. In fact, a detailed study of the
polarizabilities in a chiral quark model showed that for a
reasonable quark core radius of 0.6~fm, the pion cloud
contributions are clearly dominant~\cite{Wei85}.
\newline
\indent
Systematic calculations of pion cloud effects became possible with the
development of chiral perturbation theory (ChPT), an expansion in
the external momenta and the pion mass (``$p$ expansion''). The
first calculation of Compton scattering in that scheme was
performed by Bernard, Kaiser and Mei{\ss}ner in 1991~\cite{BKM91}.
Keeping only the leading term in $1 / m_\pi$, they found the following
simple relation at order $p^3$ (one-loop calculation)
\be
\label{DDeq3.9.4}
\alpha_{E1} = 10\beta_{M1} = \frac
{5\alpha_{em}g_A^2}{96\pi f_{\pi}^2m_{\pi}} = 12.2\ ,
\ee
in
remarkable agreement with experiment. The calculation was later
repeated in heavy baryon ChPT, which allows for a consistent
chiral power counting, and extended to ${\mathcal{O}}(p^4)$
yielding~\cite{BKSM93} 
\be 
\label{DDeq3.9.5} 
\alpha_{E1}^p = 10.5\pm2.0\quad , \quad \beta_{M1}^p = 3.5\pm3.6 \ . 
\ee
The error bars for these values indicate that several low-energy
constants were determined by resonance saturation, e.g., by
putting in phenomenological information about the $\Delta$ (1232)
resonance. Since this resonance lies close, it may not be
justified to ``freeze'' the degrees of freedom of this near-by
resonance. It is for this reason that the ``small scale
expansion'' (SSE) was proposed which includes the excitation
energy of the $\Delta$ (1232) as an additional expansion parameter
(``$\varepsilon$ expansion''). Unfortunately, at ${\mathcal{O}}(\varepsilon^3)$
the ``dynamical'' $\Delta$~\cite{HHK97,HHKK98} 
increases the polarizabilities to values far above the data~\cite{HHKK98},
\be
\label{DDeq3.9.6}
\alpha_{E1}^p =16.4\quad {\rm{and}} \quad
\beta_{M1}^p = 9.1\ .
\ee
Since large loop corrections are expected at
${\mathcal{O}}(\varepsilon^4)$, a calculation to this order might
remedy the situation. Elsewise, one would have to shift the
problem to large contributions of counterterms, thus loosing the
predictive power.
\newline
\indent
The comparison between the predictions for the scalar polarizabilities in 
heavy baryon ChPT and in fixed-$t$ DR is given in 
Table ~\ref{fixed_t_scalar_p}.
The differences between the DR analyses of Ref.~\cite{Hol00} (HDPV) and 
Ref.~\cite{Bab98a} (BGLMN) can be explained by  different inputs for the 
one-pion multipoles (in Ref.~\cite{Bab98a} the solution of SAID-SP97K was 
used) and  different approximations for the multipion channels.
In Ref.~\cite{Bab98a}, in addition to the parametrization of the 
resonant contribution of the inelastic channels mentioned in 
Sec.~\ref{sec:fixedt}, the  nonresonant contribution to the two-pion 
photoproduction channel was modeled by calculating 
the OPE diagram of the $\gamma N\rightarrow \pi\Delta$ reaction.
The difference between the data and the model for two-pion photoproduction 
consisting of resonant mechanism plus the OPE diagram for the nonresonant 
mechanism, was then fitted and attributed to a phenomenological, 
nonresonant $\gamma N\rightarrow \pi \Delta$ S-wave correction term. 
The effect of the multipion channels can be seen mainly in the 
sum $\alpha_{E1}+\beta_{M1},$ which, within the  BGLMN 
analysis, approximately reproduces 
the value of Baldin's sum rule as given in  Eq.~(\ref{DDeq2.2.18}). 
\newline
\indent
Furthermore, in Table~\ref{fixed_t_scalar_p}, we also show the results
for $\alpha_{E1}$ and $\beta_{M1}$ obtained in the 
dressed K-matrix model of Ref.~\cite{KS01}, which turn out to be 
quite close to the DR results. 
\begin{table}[h]
\caption[]{ Theoretical predictions for scalar polarizabilities of
the proton:
to  ${\mathcal{O}}(p^3)$   
in HBChPT~\cite{BKM91},
 to $ {\mathcal{O}}(\varepsilon^3)$
in the small scale expansion~\cite{HHKK98},
in the fixed-$t$ dispersion relation analyses
of Ref.~\cite{Hol00} (HDPV) and Ref.~\cite{Bab98a} (BGLMN), 
and in the dressed K-matrix model of Ref.~\cite{KS01} (KS).
In the DR calculations
$ \alpha_{E1}-\beta_{M1}=10.0$ is used as input.
The values are given in units of $10^{-4}$ fm$^3$ for the dipole 
polarizabilities and in units of $10^{-4}$ fm$^5$ for the quadrupole
polarizabilities.
 \label{fixed_t_scalar_p}}
\vspace{0.5 truecm}
\begin{center}
\begin{tabular}{||c|c|c|c|c|c|c||}
\hline\hline 
 & ${\mathcal{O}}(p^3)$ &$ {\mathcal{O}}(\varepsilon^3)$ &
HDPV& BGLMN & KS
\\
\hline 
$ \alpha_{E1}$  & 13.6 & 16.4  & 11.0 & 11.9 & 12.1
\\
\hline 
$ \beta_{M1}$   & 1.4  &  9.1  & 1.0  & 1.9 & 2.4 
\\
\hline\hline  
$ \alpha_{E2}$  &  22.1 & 26.2 & 28.8  & 27.5 &
\\
\hline 
$ \beta_{M2}$   & -9.5  &-12.3 & -23.7 & -22.4 &
\\
\hline\hline
\end{tabular}
\end{center}
\end{table}
\newline
\indent
The spin polarizabilities were calculated within the HBChPT
approach in Ref.~\cite{BKM95}. Taking out a common factor
$C=\alpha_{em}g_A/(4\pi^2f_{\pi}^2m_{\pi}^2)$, Ragusa's
polarizabilities at ${\mathcal{O}}(p^3)$ read
\be
\label{DDeq3.9.7} \{ \gamma_1,\gamma_2,\gamma_3,\gamma_4\} =
C\left\{ -1+\frac{g_A}{6},\, 0+
\frac{g_A}{12},\,\frac{1}{2}+\frac{g_A}{24},\,
-\frac{1}{2}-\frac{g_A}{24}\right\}\ , \ee where the first term
for each $\gamma_i$ is the contribution of the $t$-channel $\pi^0$
pole term, and the second one the dispersive contribution.
Clearly, the pole term is the dominant feature except for the case
of $\gamma_2$. Whether or not the pole term should be included or
dropped in the definition of the spin polarizabilities is an open
discussion, though from the standpoint of DRs the pole terms and
dispersion integrals are clearly separated.
\newline
\indent
In Table~\ref{fixed_t_spin_p}, 
we compare the proton results from heavy baryon ChPT, 
fixed-$t$ DR and hyperbolic DR analyses for the dispersive contribution to 
the spin polarizabilities of Eq.~(\ref{eq3.8.6}).
The agreement between the different DR results is quite satisfactory 
in all cases, and the spread among the different DR values can be seen
as the best possible error estimate of such calculations to date. 
Let us also notice that the dressed K-matrix model of Ref.~\cite{KS01} 
also yields values which are quite close to the DR analysis, 
except for $\gamma_{E1 M2}$ which comes out much larger 
in absolute value in the dressed K-matrix model and is responsible for
the too large and positive value obtained in that model for $\gamma_0$
compared to experiment.   
One also sees from Table~\ref{fixed_t_spin_p} that 
the ChPT predictions disagree in some cases, 
both among each other and with the DR results. 
It is obvious that the reason for these problems deserves further study.
\newline
\indent 
In the following we shall discuss the forward spin
polarizabilities $\gamma_0$.
 As is obvious from Eqs.~(\ref{eq:ragusa_mult}) and (\ref{DDeq3.9.7}), 
the $\pi^0$ pole term cancels
in the forward direction, and in agreement with forward dispersion
relations, Eq.~(\ref{DDeq2.2.16}), only excited intermediate
states contribute. Two recent calculations of $\gamma_0$ at
${\mathcal{O}}(p^4)$ yield the following result~\cite{Kum00,JiK00}
\be \label{DDeq3.9.8} \gamma_0 = \frac{g_A}{6}\,C
\left(1-\frac{\pi m_{\pi}}{8M} (21+3\kappa_p-2\kappa_n)\right) =
4.5-8.4 = -3.9\ . 
\ee
In another independent investigation, Gellas {\it et
al.}~\cite{Gel00} arrived at a value $\gamma_0=-1$, close to the
experimental value $\gamma_0 = (-1.01 \pm 0.08 \pm 0.10) $
of Eq.~(\ref{DDeq2.2.19}).
 However, the
apparent discrepancy within the HBChPT calculations 
is not related with any differences
concerning the observables, but merely a matter of definition of a
polarizability~\cite{Bir01,Gel01}. For comparison, the SSE at
${\mathcal{O}}(\varepsilon^3)$ predicts~\cite{HHKK98}
\be
\label{DDeq3.9.9} \gamma_0 = 4.6(\pi N) - 2.4 (\Delta) -
0.2(\pi\Delta) = 2.0\ , \ee the individual contributions being due
to $\pi N$ loops, $\Delta$ poles, and $\pi\Delta$ loops.
\begin{table}[h]
\caption[]{ Theoretical predictions for the dispersive contribution to 
spin polarizabilities of
the proton:
to  ${\mathcal{O}}(p^3)$ in HBChPT~\cite{HHKK98}, 
to  ${\mathcal{O}}(p^4)$   
in HBChPT from the two derivations of Refs.~\cite{Gel01} and ~\cite{Kum00},
 to $ {\mathcal{O}}(\varepsilon^3)$
in the small scale expansion~\cite{Gel01},
 in fixed-$t$ dispersion relation analysis
of Ref.~\cite{Hol00} (HDPV) and Ref.~\cite{Bab98a} (BGLMN), and in our 
calculation with hyperbolic dispersion relations (HYP. DR) at 
$\theta_{\mathrm{lab}}=180^{\mathrm{o}}$. 
Furthermore, the column (KS) gives the results in the dressed K-matrix
model of Ref.~\cite{KS01}.
The values are given in units of $10^{-4}$ fm$^4$ for the lower order 
polarizabilities and in units of $10^{-4}$ fm$^6$ for the higher order 
polarizabilities. 
\newline
$^*$ : Ref.~\cite{Hemgdh} has suggested that a contribution of +2.5
 from the $\Delta$-pole is still missing.  
 \label{fixed_t_spin_p}}
\vspace{0.5 truecm}
\begin{center}
\begin{tabular}{||c|c|c|c|c|c|c|c|c||}
\hline\hline 
& ${\mathcal{O}}(p^3)$ & ${\mathcal{O}}(p^4)~\mbox{\cite{Gel01}}$ & $ {\mathcal{O}}(\varepsilon^3)$ 
 & ${\mathcal{O}}(p^4)~\mbox{\cite{Kum00}}$ & HDPV  & BGLMN & HYP. DR & KS
\\
\hline 
$\gamma_{E1E1}$  & -5.7 & -1.8  & -5.4  & -1.4 & -4.3 &-3.4 & -3.8 & -5.0 
\\
\hline 
$ \gamma_{M1M1}$  & -1.1 & 0.4$^*$  & 1.4  & 3.3  & 2.9 & 2.7 & 2.9 & 3.4
\\
\hline  
$ \gamma_{E1M2}$ & 1.1  & 0.7  &  1.0  & 0.2  & -0.01  & 0.3 & 0.5 & -1.8
\\
\hline 
$ \gamma_{M1E2}$  & 1.1 & 1.8  &  1.0  & 1.8  &  2.1  & 1.9 & 1.6 & 1.1
\\
\hline\hline 
$ \gamma_0$  & 4.6  & -1.1   & 2.0   & -3.9  & -0.7   &-1.5  & -1.1 & 2.4\\
\hline
$ \gamma_\pi^{disp}$ & 4.6  & 3.3   &6.8    & 6.3  & 9.3  & 7.8 &7.8 & 11.4 \\
\hline\hline 
$ \gamma_{E2E2}$  & -0.4 & 0.08  &-0.28  &      & -0.16  & &     &
\\
\hline $ \gamma_{M2M2}$  & -0.03 & 0.06  &-0.03  &      &-0.09   & &    &
\\
\hline $ \gamma_{E2M3}$  & 0.11 &0.03  &0.11   &      &0.08    &    & &
\\
\hline $ \gamma_{M2E3}$  & 0.11 &0.03  &0.11   &      &0.06    &  & & 
\\
\hline\hline
\end{tabular}
\end{center}
\end{table}
\newline
\indent
The corresponding results for the neutron scalar and spin
polarizabilities are shown in Tables~\ref{fixed_t_scalar_n}
and~\ref{fixed_t_spin_n}. In the case of the scalar polarizabilities, 
the difference between proton and neutron dispersive results is quite small. 
This is in qualitative agreement with the ChPT calculations, in which
the isovector effects appear only at the fourth order. At this order
however, unknown low-energy constants enter the scalar
polarizabilities. 
For the lower-order spin polarizabilities of the neutron, 
shown in Table~\ref{fixed_t_spin_n}, it is amusing to note that the 
${\mathcal{O}}(p^4)$ HBChPT predictions of Ref.~\cite{Kum00} 
are quite close to the hyperbolic DR results.
The higher-order spin polarizabilities of the proton and neutron 
are predicted to be equal, because only the  
isoscalar contribution to the $t$-channel structure constants $a_{i,t}$ 
of Eq.~(\ref{eq3.8.6}) was taken into account.
\newline
\indent
Finally, we also like to mention that very recently the first lattice
QCD calculations of hadron electric and magnetic polarizabilities were
reported~\cite{latticepol}. While these initial results look quite
promising, it is too premature to make a quantitative comparison with
experiment at this time, 
because the current lattice calculations were performed for
rather large values of the quark mass (corresponding with pion masses
$\gtrsim 500$~MeV). However, in the near future such calculations can
be envisaged for values of $m_\pi$ down to about 300~MeV. Furthermore,
in this range one may make use of ChPT
results which calculate the dependence of the polarizabilities on
$m_\pi$. This opens up the prospect to extrapolate the lattice results 
downwards in $m_\pi$,  and bridge the gap in $m_\pi$ 
between the existing lattice calculations and the chiral limit. 
Such a study is very worthwhile to investigate in a future work. 
\begin{table}[h]
\caption[]{ 
Same as Table~\ref{fixed_t_scalar_p}, but for the scalar
polarizabilities of the neutron.
In the DR calculations 
$ \alpha_{E1}-\beta_{M1}=11.5$ is used as input.
 \label{fixed_t_scalar_n}}
\vspace{0.5 truecm}
\begin{center}
\begin{tabular}{||c|c|c|c|c|c|c||}
\hline\hline 
 & ${\mathcal{O}}(p^3)$ &$ {\mathcal{O}}(\varepsilon^3)$ &
HDPV& BGLMN & KS
\\
\hline 
$ \alpha_{E1}$  & 13.6 & 16.4  & 12.3 & 13.3 & 12.7 
\\
\hline 
$ \beta_{M1}$   & 1.4  &  9.1  & 0.8  & 1.8  & 1.8
\\
\hline\hline  
$ \alpha_{E2}$  &  22.1 & 26.2 & 28.8  & 27.2 &
\\
\hline 
$ \beta_{M2}$   & -9.5  &-12.3 & -23.7 & -23.5 &
\\
\hline\hline
\end{tabular}
\end{center}
\end{table}

\begin{table}[h]
\caption[]{
Same as Table~\ref{fixed_t_spin_p}, but for the spin polarizabilities
of the neutron.
\newline
$^*$ : Ref.~\cite{Hemgdh} has suggested that a contribution of +2.5
 from the $\Delta$-pole is still missing.  
 \label{fixed_t_spin_n}}
\vspace{0.5 truecm}
\begin{center}
\begin{tabular}{||c|c|c|c|c|c|c|c|c||}
\hline\hline 
& ${\mathcal{O}}(p^3)$ & ${\mathcal{O}}(p^4)~\mbox{\cite{Gel01}}$ & $ {\mathcal{O}}(\varepsilon^3)$ 
 & ${\mathcal{O}}(p^4)~\mbox{\cite{Kum00}}$ & HDPV  & BGLMN & HYP. DR & KS
\\
\hline 
$ \gamma_{E1E1}$ & -5.7 & -4.2 & -5.4  & -4.2 & -5.9   &-5.6 & -4.7 & -4.8
\\
\hline 
$ \gamma_{M1M1}$  & -1.1 & 0.4$^*$  & 1.4  & 2.3  & 3.8  & 3.8 &2.8 & 3.5
\\
\hline  
$ \gamma_{E1M2}$ & 1.1 & 0.5  &  1.0  & 0.4  & -0.9  & -0.7 & 0.4 & -1.8
\\
\hline 
$ \gamma_{M1E2}$  & 1.1 & 2.2  & 1.0  & 2.2  &  3.1  & 2.9 & 2.0 & 1.1
\\
\hline\hline 
$ \gamma_0$  &4.6   &1.1    &2.0    &-0.7   & -0.07   &-0.4  & -0.5 & 2.0\\
\hline
$ \gamma_\pi^{disp}$ & 4.6  &6.3  & 6.8  & 8.3  &  13.7  & 13.0 & 9.2 & 11.2 \\
\hline\hline $ \gamma_{E2E2}$  & -0.4 & 0.08  &-0.28  &  & -0.16  & & &    
\\
\hline $ \gamma_{M2M2}$  & -0.03 & 0.06  &-0.03  &      &-0.09   &   &  &
\\
\hline $ \gamma_{E2M3}$  & 0.11 &0.03  &0.11   &      &0.08    &   &  &
\\
\hline $ \gamma_{M2E3}$  & 0.11 &0.03  &0.11   &      &0.06    &   &  &
\\
\hline\hline
\end{tabular}
\end{center}
\end{table}

\section{Dispersion relations in virtual Compton scattering (VCS)}
\label{sec:vcs}

\subsection{Introduction}

In this section, we discuss dispersion relations for the virtual
Compton scattering (VCS) process. In this process, denoted as 
$\gamma^* + p \to \gamma + p$, a spacelike virtual photon ($\gamma^*$)
interacts with a nucleon (we consider a proton in all of the
following as experiments are only performed for a proton target so far) 
and a real photon ($\gamma$) is produced. 
At low energies, this real photon plays the role of an 
applied quasi-static electromagnetic field, and the VCS process
measures the response of the nucleon to this applied field. 
In the real Compton scattering process discussed in 
Sec. \ref{sec:rcs}, this response is characterized by global 
nucleon structure constants such as the nucleon 
dipole and higher order polarizabilities. 
In contrast, for the VCS process,   
the virtuality of the initial photon can be dialed so as to map out 
the spatial distribution of these nucleon polarizabilities, giving
access to so-called generalized polarizabilities (GPs). 
\newline
\indent
First unpolarized VCS observables have been obtained from experiments 
at the MAMI accelerator \cite{Roc00} at a virtuality $Q^2$ = 0.33 GeV$^2$, 
and recently at JLab \cite{Fon02} at higher virtualities, 
$1 < Q^2 < 2~$GeV$^2$. 
Both experiments measured two combinations of GPs. 
Further experimental programs are underway 
at the intermediate energy electron accelerators 
(MIT-Bates \cite{Mis97}, MAMI \cite{dHo01}, and JLab \cite{Hyd02}) 
to measure both unpolarized and polarized VCS observables. 
\newline
\indent
VCS experiments at low outgoing photon energies 
can be analyzed in terms of low-energy expansions (LEXs), proposed in
Ref.~\cite{Gui95}. 
In the LEX, only the leading term (in the energy of the produced real photon) 
of the response to the quasi-constant electromagnetic field, 
due to the internal structure of the system, is taken into account. 
This leading term depends linearly on the GPs.  
As the sensitivity of the VCS cross sections to the GPs 
grows with the photon energy, it is
advantageous to go to higher photon energies, provided one can keep the
theoretical uncertainties under control when approaching and crossing the pion
threshold. The situation can be compared to RCS as described in
Sec. \ref{sec:rcs}, where it was shown that  
one uses a dispersion relation formalism 
to extract the polarizabilities at energies 
above pion threshold, with generally larger effects on the
observables. 
\newline
\indent
In this section, we describe the application of
a dispersion relation formalism to the VCS reaction with the aim to
extract GPs from VCS experiments over a larger energy range.  
We will also review the present status and future prospects 
of VCS experiments and describe the physics contained in the GPs. 
For more details on VCS, see also the reviews of Refs.~\cite{Gui98,Vdh00}.

\subsection{Kinematics and invariant amplitudes}

\begin{figure}[ht]
\epsfxsize=8.5cm
\centerline{\epsffile{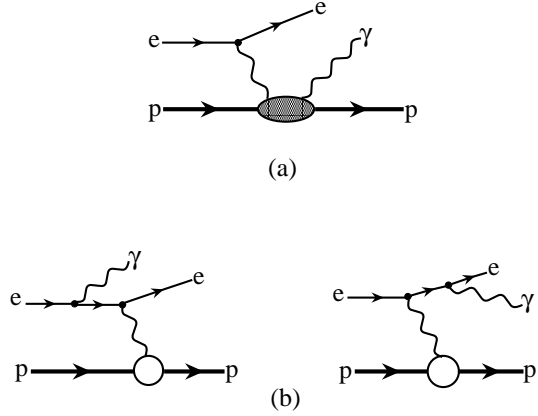}}
\vspace{-.75cm}
\caption[]{(a) FVCS process, (b) BH process.}
\label{fig:diagrams}
\end{figure}

The VCS process on the proton is accessed through 
the $e p \to e p \gamma$ reaction.
In this process, the final photon can be emitted
either by the proton, which is referred to as the fully virtual
Compton scattering (FVCS) process, or by
the lepton, which is referred to as the Bethe-Heitler (BH) process.
This is shown graphically in Fig.~\ref{fig:diagrams},
leading to the amplitude $T^{ee'\gamma}$ of the $e p \to e p \gamma$
reaction as the coherent sum of the BH and the FVCS process~:
\begin{equation}
T^{ee'\gamma}=T^{BH}+T^{FVCS}.
\end{equation}
The BH amplitude $T^{BH}$ is exactly calculable from QED if one knows
the nucleon electromagnetic form factors. The FVCS amplitude
$T^{FVCS}$ contains, in the one-photon exchange approximation, the VCS
subprocess $\gamma^* p \to \gamma p$. We refer to Ref.~\cite{Gui98}
where the explicit expression of the BH amplitude is given,
and where the construction of the
FVCS amplitude from the $\gamma^* p \to \gamma p$ process is discussed.

\begin{figure}[h]
\vspace{-.25cm}
\epsfxsize=6.cm
\centerline{\epsffile{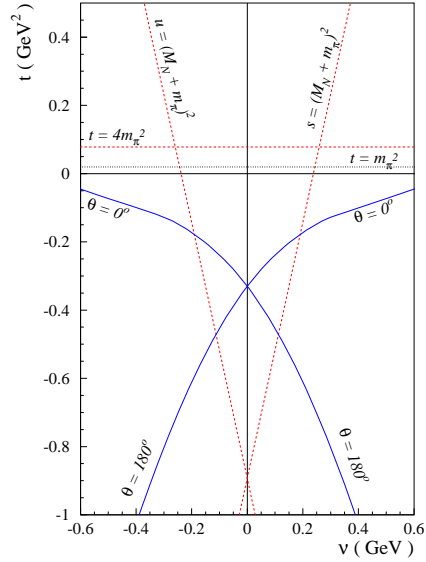}}
\caption[]{The Mandelstam plane for virtual Compton scattering at
  $Q^2$ = 0.33 GeV$^2$. The boundaries of the physical $s$-channel
  region are $\theta = 0^{\mathrm{o}}$ and
$\theta = 180^{\mathrm{o}}$ for $\nu > 0$, the $u$-channel region
  is obtained by crossing, $\nu \to - \nu$.
The curves for $\theta = 0^{\mathrm{o}}$ and
$\theta = 180^{\mathrm{o}}$ intersect at $\nu = 0$, $t = - Q^2$,
which is the point where the generalized polarizabilities are defined.}
\label{fig:mandelstam}
\end{figure}

\indent
We characterize the four-vectors of the virtual (real) photon in the VCS
process $\gamma^* p \to \gamma p$
by $q$ ($q'$) respectively, and the four-momenta of initial (final)
nucleons by $p$ ($p'$) respectively.
In the VCS process, the initial photon is spacelike and we denote its
virtuality in the usual way by $q^2$ = - $Q^2$.
Besides $Q^2$, the VCS process can be described by the
Mandelstam invariants
\begin{equation}
s = (q + p)^2, \hspace{.5cm}
t = (q - q')^2, \hspace{.5cm}
u = (q - p')^2\;,
\end{equation}
with the constraint
\begin{equation}
s + t + u = 2 M^2 - Q^2 \;,
\end{equation}
where $M$ denotes the nucleon mass.
We furthermore introduce the variable $\nu$, which changes sign under
$s \leftrightarrow u$ crossing :
\begin{equation}
\nu = {{s - u} \over {4 M}} \,=\, 
E_\gamma^{lab} \,+\, {1 \over {4 M}} \left( t - Q^2 \right) \; ,
\end{equation}
where $E_\gamma^{lab}$ is the virtual photon
energy in the {\it lab} frame.
In the following, we choose $Q^2$, $\nu$ and $t$ as the independent
variables to describe the VCS process. In Fig.~\ref{fig:mandelstam},
we show the Mandelstam plane for the VCS process at a fixed value
of $Q^2$ = 0.33 GeV$^2$, at which the experiment of Ref.~\cite{Roc00} 
was performed.

\indent
The VCS helicity amplitudes can be written as
\begin{equation}
T_{\lambda' \lambda_N';  \lambda \lambda_N} =\, - e^2
\varepsilon_\mu(q, \lambda) \, \varepsilon^{'*}_\nu(q', \lambda')
\, \bar u(p', \lambda_N') \, {\mathcal M}^{\mu \nu} \,u(p,
\lambda_N)\ , \label{eq:matrixele}
\end{equation}
where the polarization four-vectors of the virtual (real) photons
are denoted by $\varepsilon$ ($\varepsilon^{'}$), and their
helicities by $\lambda$ ($\lambda'$), with $\lambda = 0, \pm 1$
and $\lambda' = \pm 1$. The nucleon helicities are $\lambda_N,
\lambda_N' = \pm 1/2$, and $u, \bar u$ are the nucleon spinors.
The VCS tensor ${\mathcal
M}^{\mu\nu}$ in Eq.~(\ref{eq:matrixele}) can be decomposed into a
Born (B) and a non-Born part (NB) :
\begin{equation}
{\mathcal M}^{\mu \nu} \;=\; {\mathcal M}^{\mu \nu}_{B} \,+\,
{\mathcal M}^{\mu \nu}_{NB} \;.
\label{eq:vcsbnb}
\end{equation}
In the Born process, the virtual photon is
absorbed on a nucleon and the intermediate state remains a nucleon,
whereas the non-Born process contains all nucleon excitations
and meson-loop contributions.
The separation between Born and non-Born parts is performed in the
same way as described in Ref.~\cite{Gui95}, to which we refer for
details (see also Ref.~\cite{SKK96}).
\newline
One can proceed by parametrizing the VCS tensor of Eq.~(\ref{eq:vcsbnb})
in terms of 12 independent amplitudes.
In Ref.~\cite{DKK98}, a gauge-invariant tensor basis was found 
so that the resulting
non-Born invariant amplitudes are free of kinematical singularities and
constraints, which is an important property when setting up a
dispersion relation formalism.
This tensor takes the form
\begin{equation}
{\mathcal M}^{\mu \nu} \;=\; \sum_{i = 1}^{12}
\; f_i(Q^2, \nu, t) \, \rho^{\mu \nu}_i \;,
\label{eq:nonborn}
\end{equation}
where the 12 independent tensors $\rho^{\mu \nu}_i$
are given in App.~B.
The corresponding 12 amplitudes $f_i$ are expressed
in terms of the invariants $Q^2$, $\nu$ and $t$. 
The tensor basis $\rho^{\mu \nu}_i$ is chosen
 such that the resulting invariant amplitudes
$f_i$ are either even or odd under crossing, which leads to the
following
symmetry relations for the $f_i$ at the real photon point :
\begin{eqnarray}
&&f_i \left( 0, \nu, t \right) \,=\,
+ \, f_i \left( 0, - \nu, t \right),
\hspace{.2cm} (i = 1, 2, 6, 11) \;, \nonumber \\
&&f_i \left( 0, \nu, t \right) \,=\,
- \, f_i \left( 0, - \nu, t \right),
\hspace{.2cm} (i = 4, 7, 9, 10) \;,
\label{eq:phocross}
\end{eqnarray}
while the amplitudes $f_3$, $f_5$, $f_8$, $f_{12}$
do not contribute at this point, because
the corresponding tensors vanish in the limit $Q^2 \rightarrow 0$.
\newline
\indent
Nucleon crossing combined with charge conjugation provides the
more general constraints on the $f_i$ at arbitrary virtuality
$Q^2$ :
\begin{eqnarray}
f_i \left( Q^2, \nu, t \right) \,&=&\,
+ \, f_i \left( Q^2, - \nu, t \right),
\hspace{.1cm} (i = 1, 2, 5, 6, 7, 9, 11, 12), \hspace{.1cm} \nonumber \\
f_i \left( Q^2, \nu, t \right) \,&=&\,
- \, f_i \left( Q^2, - \nu, t \right),
\hspace{.1cm} (i = 3, 4, 8, 10). \hspace{.2cm}
\label{eq:nucross}
\end{eqnarray}
When using dispersion relations, it will be convenient to work with
12 amplitudes that are all even in $\nu$. This is achieved by
introducing the amplitudes $F_i$ ($i$ = 1,...,12) as follows :
\begin{eqnarray}
F_i \left( Q^2, \nu, t \right) \,&=&\, f_i \left( Q^2, \nu, t \right),
\hspace{.2cm} (i = 1, 2, 5, 6, 7, 9, 11, 12)\;, \nonumber\\
F_i \left( Q^2, \nu, t \right) \,&=&\,
{1 \over \nu} \, f_i \left( Q^2, \nu, t \right),
\hspace{.2cm} (i = 3, 4, 8, 10)\;,
\label{eq:fampl}
\end{eqnarray}
satisfying $F_i \left( Q^2, -\nu, t \right)
= F_i\left( Q^2, \nu, t \right)$ for
$i$ = 1,...,12.
As the non-Born invariant amplitudes
$f^{NB}_{3, 4, 8, 10} \thicksim \nu$ for $\nu \rightarrow 0$,
the definition of Eq.~(\ref{eq:fampl}) ensures that also all the
non-Born amplitudes 
$F_i^{NB}$ ($i$ = 1,...,12) are free from kinematical singularities.
The results for the Born amplitudes $F_i^B$ are listed in
App.~B of Ref.~\cite{Pas01}.
\newline
\indent
From Eqs.~(\ref{eq:phocross}) and (\ref{eq:nucross}), one furthermore finds that
$F_7$ and $F_9$ vanish at the real photon point.
Since 4 of the tensors also vanish in the limit $Q^2 \rightarrow 0$, only
the six amplitudes $F_1$, $F_2$, $F_4$, $F_6$, $F_{10}$
and $F_{11}$ enter in real Compton scattering (RCS). These 6
amplitudes are related
to the RCS amplitudes of Eq.~(\ref{DDeq3.2.4}) by
\begin{eqnarray}
- e^2 \, F_1 \,&=& - A_1 \,-\, \left( {{t - 4 M^2} \over {4 M^2}}
\right) \, A_3 \,+\, {{\nu^2} \over {M^2}} \, A_4 \,+\, A_6 \;, \nonumber\\
- e^2 \, F_2 \,&=& - {1 \over {2 M^2}} \, \left[ A_3 \,+\, A_6
  \,-\, {t \over {4 M^2}} \, A_4 \right] \;, \nonumber\\
- e^2 \, F_4 \,&=& {1 \over {2 M^2}} \, A_4 \;, \nonumber\\
- e^2 \, F_6 \,&=& {1 \over {4 M^2}} \,
\left[ - \left( {{t - 4 M^2} \over {4 M^2}} \right) \, A_4
\,+\, A_6 \right]  \;, \nonumber\\
- e^2 \, F_{10} \,&=& - {1 \over {2 M}} \, \left[ A_5 - A_6
\right]\;, \nonumber\\
- e^2 \, F_{11} \,&=& - {1 \over 4 M} \, \left[ A_2 \,-\,
{{t - 4 M^2 + 4 \nu^2} \over {4 M^2}} \, A_4 \,+\, A_6 \right], \hspace{.4cm}
\label{eq:fversusa}
\end{eqnarray}
where the charge factor $- e^2$ appears explicitly on the {\it
lhs} of Eq.~(\ref{eq:fversusa}), because this factor is included
in the usual definition of the $A_i$.


\subsection{Definitions of nucleon generalized polarizabilities}

The behavior of the non-Born VCS tensor
${\mathcal M}^{\mu \nu}_{NB}$ of Eq.~(\ref{eq:nonborn})
at low energy ($\rmqp \equiv |{\bq}_{cm}^{\; '}| \to 0$)
but at arbitrary three-momentum $\rmq \,\equiv\, |{\bq}_{cm}\,|$
of the virtual photon, can be parametrized by 6 generalized
polarizabilities (GPs), which will be
denoted by $P^{(\mathcal{M}' \, L',\, \mathcal{M} \,L)S}(\rmq)$
\cite{Gui95,DKK98,DKM97}.
In this notation, $\mathcal{M}$ ($\mathcal{M}'$) refers to the
electric $(E)$, magnetic $(M)$ or longitudinal $(L)$ nature of the initial
(final) photon, $L$ ($L'$) represents the angular momentum of the
initial (final) photon, and $S$ differentiates between the
spin-flip ($S=1$) and non spin-flip ($S=0$)
character of the transition at the nucleon side. 
Assuming that the emitted real photons have low energies, we may use
the dipole approximation ($L' = 1$). For a dipole transition in the
final state, angular momentum and parity conservation lead to 10 GPs
\cite{Gui95}. Furthermore, it has been shown~\cite{DKK98} that nucleon crossing
combined with charge conjugation symmetry of the VCS amplitudes
provide 4 additional constraints among the 10 GPS. 
A convenient choice for the 6 independent GPs appearing in that approximation 
has been proposed in Ref.~\cite{Gui98}~:
\begin{eqnarray}
&&P^{(L1,L1)0}(\rmq),\; P^{(M1,M1)0}(\rmq), \;
\label{eq:defgpunpol} \\
&&P^{(L1,L1)1}(\rmq),\; P^{(M1,M1)1}(\rmq),\;
P^{(M1,L2)1}(\rmq),\; P^{(L1,M2)1}(\rmq). \hspace{.5cm}
\label{eq:defgppol}
\end{eqnarray}
We note at this point that the difference between the transverse electric and
longitudinal transitions is of higher order in $\rmq$, which explains why
the electric multipoles can be replaced by the longitudinal ones
in the above equations.
\newline
\indent
In the limit $\rmq\rightarrow 0$ one finds the following
relations between the VCS and RCS polarizabilities~\cite{DKK98}~:
\be
\begin{array}{lcl}
P^{(L1,L1)0}(0)=-{{4 \pi} \over {e^2}}\, \sqrt{\frac{2}{3}} \,
\alpha \quad & , \quad
& P^{(M1,M1)0}(0)=-{{4 \pi} \over {e^2}}\, \sqrt{\frac{8}{3}} \,
\beta \;, \nonumber \\
P^{(L1,M2)1}(0)=-{{4 \pi} \over {e^2}}\, \frac{\sqrt{2}}{3} \,
\gamma_3 \quad &, \quad
& P^{(M1,L2)1}(0)=-{{4 \pi} \over {e^2}}\, \frac{2 \sqrt{2}}{3 \sqrt{3}} \,
\left( \gamma_2 + \gamma_4 \right) \;, \nonumber \\
P^{(L1,L1)1}(0)= 0 \quad &, \quad &P^{(M1,M1)1}(0)= 0 \; .
\label{eq_3_47}
\end{array}
\ee
\indent
In terms of invariants, the limit $\rmqp \to 0$ at finite
three-momentum $\rmq$ of the virtual photon corresponds to
$\nu \to 0$ and $t \to -Q^2$ at finite $Q^2$.
One can therefore express the GPs in terms of
the VCS invariant amplitudes $F_i$ at the point $\nu = 0, t = -Q^2$ for
finite $Q^2$, for which we introduce the shorthand~:
\begin{equation}
\bar F_i(Q^2) \;\equiv\; F_i^{NB} \left(Q^2, \nu = 0, t = - Q^2 \right) \;.
\label{eq:fbardef}
\end{equation}
The relations between the GPs and the $\bar F_i(Q^2)$ can be found in
Ref.~\cite{DKK98}.
\newline
\indent
Analogously to the sum rules which we discussed in Sec.~\ref{sec:rcs} 
for the nucleon polarizabilities at $Q^2$ = 0, we now
turn to dispersion relations for the GPs. 
From the high-energy behavior of the amplitudes $F_i$, 
it was found in Ref.~\cite{Pas01} 
that the unsubtracted DRs do not exist for the amplitudes
$F_1$ and $F_5$, but can be written down for the other amplitudes.
Therefore, unsubtracted DRs for the GPs will hold for those GPs which do not
depend on the two amplitudes $F_1$ and $F_5$.
However, the amplitude $F_5$ can appear in the combination $F_5 + 4 \, F_{11}$,
because this combination has a high-energy behavior
leading to a convergent integral \cite{Pas01,Pas00}.
Among the six GPs we find four combinations that do not depend on
$F_1$ and $F_5$~:
\begin{eqnarray}
P^{\left(L 1, L 1\right)0} &&+\, {1 \over 2}
P^{\left(M 1, M 1\right)0} =
{{-2} \over {\sqrt{3}}}\left( {{E + M} \over
    E}\right)^{1/2} M \tilde q_0 \,
\left\{ {{\rmq^2} \over {\tilde q_0^2}}\, \bar F_2 +
\left( 2 \, \bar F_6 + \bar F_9 \right) - \bar F_{12} \right\},
\label{eq:gpdisp1} \\
P^{\left(L 1, L 1\right)1} &&=
{1 \over {3 \sqrt{2}}}\,\left( {{E + M} \over E}\right)^{1/2}
\,\tilde q_0\,
\left\{ \left( \bar F_5 + \bar F_7 + 4\, \bar F_{11} \right)
+ 4 \, M \, \bar F_{12} \right\},
\label{eq:gpdisp2} \\
P^{\left(L 1, M 2\right)1} &&-\, {1 \over {\sqrt{2} \, \tilde q_0}}
P^{\left(M 1, M 1\right)1} =
{1 \over {3}} \left( {{E + M} \over E}\right)^{1/2}
{{M \, \tilde q_0} \over {\rmq^2}} \nonumber\\
\times &&
\left\{ \left( \bar F_5 + \bar F_7 + 4\, \bar F_{11} \right)
+ 4 \, M \left( 2 \, \bar F_6 + \bar F_9 \right) \right\},
\label{eq:gpdisp3} \\
P^{\left(L 1, M 2\right)1} &&+ \,
{{\sqrt{3}} \over {2}}  P^{\left(M 1, L 2\right)1} =
{1 \over {6}} \left( {{E + M} \over E}\right)^{1/2} \,
{{\tilde q_0} \over {\rmq^2}} \nonumber \\
\times&&
\left\{ \tilde q_0 \left( \bar F_5 + \bar F_7 + 4\, \bar F_{11} \right)
+ 8 \, M^2 \left( 2 \, \bar F_6 + \bar F_9 \right) \right\}, \hspace{.5cm}
\label{eq:gpdisp4}
\end{eqnarray}
where $E = \sqrt{\rmq^2 + M^2}$ denotes the initial proton {\it c.m.} 
energy and $\tilde q_0 = M - E$ the virtual photon {\it c.m.} 
energy in the limit $\rmqp$ = 0. For small values of $\rmq$,
we observe the relation $\tilde q_0 \approx - \, \rmq^2 / (2 M)$.
Furthermore, in the limit $\rmqp$ = 0, the value of $Q^2$ is
always understood as being $\tilde Q^2 \equiv \rmq^2 - \tilde q_0^2$,
which we denote by $Q^2$ for simplicity of the notation.


\subsection{Fixed-$t$ dispersion relations}
\label{sec:vcsfixedt}

With the choice of the tensor basis of App.~B,
and taking account of the crossing relation Eq.~(\ref{eq:nucross}),
the resulting non-Born VCS invariant amplitudes $F_i$ ($i$ = 1,...,12)
are free of kinematical singularities and constraints,
and even in $\nu$, i.e., $F_i(Q^2,\nu, t) \,=\, F_i(Q^2,-\nu, t)$.
Assuming further analyticity and
an appropriate high-energy behavior, these amplitudes fulfill
unsubtracted dispersion relations
\footnote{As a historical remark we note that 
dispersion relations have been considered for the first time for the
virtual Compton scattering process in~\cite{Ber61}. This work
considered however a different set of amplitudes as discussed here. 
To avoid numerical artefacts due to kinematical singularities we will
only consider here DRs in the amplitudes $F_i$ which
are free from such singularities and constraints.} 
with respect to the variable $\nu$ at
fixed $t$ and fixed virtuality $Q^2$,
\begin{eqnarray}
\hspace{-0.5cm}
\mathrm{Re} F_i^{NB}(Q^2, \nu, t) \;=\;
 F_i^{pole}(Q^2, \nu, t) \,-\,  F_i^{B}(Q^2, \nu, t) \,+\,
{2 \over \pi} \; {\mathcal P} \int_{\nu_{0}}^{+ \infty} d\nu' \;
{{\nu' \; \mathrm{Im}_s F_i(Q^2, \nu',t)} \over {\nu'^2 - \nu^2}} \; ,
\nonumber \\
\label{eq:unsub}
\end{eqnarray}
where we explicitly indicate that the {\it lhs} of Eq.~(\ref{eq:unsub})
represents the non-Born ($NB$) parts of the amplitudes. 
In Eq.~(\ref{eq:unsub}), $F_i^{B}$ is defined as in the discussion 
following Eq.~(\ref{eq:vcsbnb}), 
whereas $F_i^{pole}$ represents the nucleon pole contribution
(i.e. energy factors in the numerators are evaluated at the pole
position\footnote{Note that of the twelve VCS amplitudes $F_i$, only
for the amplitudes $F_1$, and a combination of $F_5$ and $F_{11}$ 
there is a difference between the Born and pole parts.}). 
Furthermore in Eq.~(\ref{eq:unsub}), 
$\mathrm{Im}_s F_i$ are the discontinuities across the $s$-channel cuts of the
VCS process, starting at the pion production threshold,
which is the first inelastic channel,
i.e.,  $\nu_{0} = m_\pi + (m_\pi^2 + t/2 + Q^2/2)/(2 M)$.
\newline
\indent
Besides the absorptive singularities due to
physical intermediate states which contribute
to the {\it rhs} of dispersion integrals as Eq.~(\ref{eq:unsub}),
one might wonder if other singularities exist giving rise to
imaginary parts. Such additional singularities could come from
so-called anomalous thresholds~\cite{Bjo65,Pil79}, which
arise when a hadron is a loosely bound system of other hadronic
constituents which can go on-shell (such as is the case of a nucleus
in terms of its nucleon constituents), leading to so-called triangular
singularities. It was shown that in the case of strong
confinement within QCD, the quark-gluon structure of hadrons
does not give rise to additional anomalous thresholds
\cite{Jaf92,Oeh95}, and the quark singularities are turned into hadron
singularities described through an effective field theory.
Therefore, the only anomalous thresholds arise for those
hadrons which are loosely bound composite systems of other
hadrons (e.g., the $\Sigma$ particle in terms of $\Lambda$ and $\pi$).
For the nucleon case, such anomalous thresholds are
absent, and the imaginary parts entering the dispersion integrals
of Eq.~(\ref{eq:unsub}) are calculated from
absorptive singularities due to $\pi N$, $\pi \pi N$,
... physical intermediate states.
\newline
\indent
The assumption that unsubtracted dispersion relations
as in Eq.~(\ref{eq:unsub}) hold, requires
that at high energies ($\nu \rightarrow \infty$
at fixed $t$ and fixed $Q^2$) the
amplitudes $\mathrm{Im}_s F_i(Q^2,\nu,t)$ ($i$ = 1,...,12)
drop fast enough so that the
integrals of Eq.~(\ref{eq:unsub}) are convergent and the contribution from
the semi-circle at infinity can be neglected.
The high-energy behavior of the amplitudes $F_i$ was investigated 
in~\cite{Pas01} by considering 
the Regge limit ($\nu \rightarrow \infty$, at fixed $t$ and fixed
$Q^2$) of the VCS helicity amplitudes. 
As mentioned above, it follows 
from this analysis that for the amplitudes $F_1$ and
$F_5$, an unsubtracted dispersion integral 
does not exist, whereas the other ten VCS amplitudes can
be evaluated through unsubtracted dispersion integrals as in
Eq.~(\ref{eq:unsub}).
\newline
\indent
Having specified the VCS invariant amplitudes and their high energy
behavior, we are now ready to set up the DR formalism. 
The difference between Born and pole terms in Eq.~(\ref{eq:unsub})
vanishes for the four combinations of GPs on the {\it lhs} of
Eqs.~(\ref{eq:gpdisp1} - \ref{eq:gpdisp4}). They can be directly evaluated by
unsubtracted DRs through the following integrals
for the corresponding $\bar F_i(Q^2)$~:
\begin{equation}
\bar F_i(Q^2) \;=\;
{2 \over \pi} \; \int_{\nu_{0}}^{+ \infty} d\nu' \;
{{\mathrm{Im}_s F_i(Q^2, \nu',t = - Q^2)} \over {\nu'}}\;.
\label{eq:sumrule}
\end{equation}
\indent
We will next discuss in Sec.~\ref{vcs_schannel} how the $s$-channel
dispersion integrals of Eqs.~(\ref{eq:unsub}) 
and (\ref{eq:sumrule}) are evaluated.
In particular, unitarity will allow us to express the imaginary parts
of the VCS amplitudes in terms of
$\pi N$, $\pi \pi N$,... intermediate states.
Subsequently, we will show in Sec.~\ref{asympt} how to deal with the
remaining two VCS invariant amplitudes for which one cannot write
down unsubtracted DRs.

\subsubsection{s-channel dispersion integrals}
\label{vcs_schannel}

The imaginary parts of the amplitudes $F_i$ in Eq.~(\ref{eq:unsub}) are
obtained through the imaginary part of the VCS helicity amplitudes
$T_{\lambda' \lambda_N';  \lambda \lambda_N}$
defined in Eqs.~(\ref{eq:matrixele}) and (\ref{eq:nonborn}).
The VCS helicity amplitudes can be expressed by the $F_i$
in a straightforward manner, even though the calculation is cumbersome.
The main difficulty, however, is the inversion of the relation
between the two sets of amplitudes,
i.e., to express the twelve amplitudes $F_i$
in terms of the twelve independent helicity amplitudes.
This problem has been solved in Refs.~\cite{Pas01,Gor02} 
in two different ways. Firstly,
the inversion was performed numerically by applying different
algorithms. Secondly, an explicit analytical inversion was found 
as detailed in Ref.~\cite{Gor02}. The two different methods allow us to
cross-check the results.
\newline
\indent
Having expressed the amplitudes $F_i$ in terms of the helicity
amplitudes, the latter are determined by using unitarity.
Denoting the VCS helicity amplitudes by $T_{fi}$, the
unitarity relation takes the generic form
\begin{equation}
2 \, \mathrm{Im}_s \, T_{fi} = \sum_{X} (2\pi)^4 \delta^4 (P_X - P_i)
T_{Xf}^{\dagger} T_{Xi}\,,
\label{eq:schunit}
\end{equation}
where the sum runs over all possible intermediate states $X$.
Here we are mainly interested in VCS through the
$\Delta(1232)$-resonance region.
Therefore, we restrict ourselves to the dominant contribution by
only taking account of the $\pi N$ $s$-channel intermediate states.
If one wants to extend the dispersion formalism to higher energies, 
the influence of additional channels, like the $\pi \pi N$
intermediate states has to be addressed. 
The helicity amplitudes for $\pi N$ intermediate states 
are expressed in terms of pion photo- and electroproduction
multipoles as specified in App.~C.4 of Ref~\cite{Pas01}.
The calculations are performed by use of 
the phenomenological MAID analysis~\cite{Dre99},
which contains both resonant and non-resonant pion production mechanisms. 
This state-of-the-art analysis is based on the existing pion photo-
and electroproduction data. A direct evaluation of the {\it rhs} of
Eq.~(\ref{eq:schunit}) is not possible due to an incomplete coverage
of the phase space with the present data sets. 
Therefore, a phenomenological analysis is needed to fully calculate
the dispersive input. 

\subsubsection{Asymptotic parts and dispersive contributions beyond $\pi N$}
\label{asympt}

To evaluate the VCS amplitudes $F_1$ and $F_5$ in an unsubtracted
DR framework, we proceed as
in the case of RCS~\cite{Lvo97}. This amounts to perform the
unsubtracted dispersion integrals of Eq.~(\ref{eq:unsub}) for $F_1$ and $F_5$
along the real $\nu$-axis in the range
$-\nu_{max}\leq\nu\leq+\nu_{max}$,
and to close the contour by a semi-circle with radius $\nu_{max}$
in the upper half of the complex $\nu$-plane, with the result
\begin{equation}
\hspace{-0.5cm}
\mathrm{Re} F_{i}^{NB}(Q^2, \nu, t) =
F_i^{pole}(Q^2, \nu, t) -  F_i^{B}(Q^2, \nu, t) +
F_{i}^{int}(Q^2, \nu, t) + F_{i}^{as}(Q^2, \nu, t) ,
\label{eq:unsub2}
\end{equation}
for ($i = 1, 5$),
where the integral contributions $F_{i}^{int}$ (for $i = 1, 5$) are given by
\begin{equation}
F_{i}^{int}(Q^2, \nu, t) \;=\;
{2 \over \pi} \; {\mathcal P} \int_{\nu_{0}}^{\nu_{max}} d\nu' \;
{{\nu' \; \mathrm{Im}_s F_{i}(Q^2, \nu',t)} \over {\nu'^2 - \nu^2}}\;,
\label{eq:unsub3}
\end{equation}
and with the contributions of the semi-circle of radius $\nu_{max}$
identified with the asymptotic contributions
($F_1^{as}$, $F_5^{as}$).
\newline
\indent
Evidently, the separation between asymptotic and integral contributions
in Eq.~(\ref{eq:unsub2}) is specified by the value of $\nu_{max}$.
The total result for $F_i^{NB}$ is formally independent
of the specific value of $\nu_{max}$.
In practice, however, $\nu_{max}$ is chosen to be not too large so
that one can evaluate the dispersive integrals of
Eq.~(\ref{eq:unsub3}) from threshold up to $\nu_{max}$ sufficiently accurate. 
As we are mainly interested here in a description of VCS up to
$\Delta(1232)$-resonance energies,
the dispersion integrals are saturated by their $\pi N$
contribution and we choose $\nu_{max} = 1.5$ GeV. 
In the following, we denote this contribution by $F_i^{\pi N}$. 
Furthermore, the remainder is estimated by an energy-independent
function, which parametrizes the asymptotic contribution 
$F_i^{as}$ due to $t$-channel poles, and which contains 
all dispersive contributions beyond 
the value $\nu_{max} = 1.5$ GeV. We will next discuss the
asymptotic contributions $F_5^{as}$ and $F_1^{as}$.
\newline
\begin{itemize}

\item{The asymptotic contribution $F_5^{as}$} \\

The asymptotic contribution to the amplitude $F_5$ predominantly
results from the $t$-channel $\pi^0$-exchange 
\footnote{ 
As mentioned before, the $\pi^0$-pole only contributes to the amplitudes
$F_5$ and $F_{11}$, but drops out in the combination
$( F_5 + 4 \, F_{11} )$, which therefore has a different high-energy
behavior.}~:
\begin{eqnarray}
F^{as}_{5}(Q^2, \nu, t) \,&\approx&\, F_{5}^{\pi^0}(Q^2, t)
\,=\, -4 \, F_{11}^{\pi^0}(Q^2, t) \,=\,  - {1 \over {M} e^2} \,
{ g_{\pi NN} \; {F_{\pi^0 \gamma \gamma}\left( Q^2 \right)}
\over {t - m_\pi^2}} \, .
\label{eq:piopolevcs}
\end{eqnarray}
For the $Q^2$-dependence of $F_{\pi^0 \gamma \gamma} \left( Q^2 \right)$,
one can use the interpolation formula proposed in ~\cite{Bro81}~:
\begin{equation}
F_{\pi^0 \gamma \gamma} \left( Q^2 \right) \; = \;
{ F_{\pi^0 \gamma \gamma} (0)
\over {1 + Q^2/ (8 \, \pi^2 \, f_\pi^2)}} \, ,
\label{eq:piogagaff}
\end{equation}
where $F_{\pi^0 \gamma \gamma}(0)$ has been given in
Eq.~(\ref{eq:pi0gagacoup}).
Equation~(\ref{eq:piogagaff})
provides a rather good parametrization of the
$\pi^0 \gamma^* \gamma$ form factor data over the whole $Q^2$ range.
\newline
\indent
When fixing the asymptotic contribution $F_5^{as}$ through its
$\pi^0$-pole contribution as in Eq.~(\ref{eq:piopolevcs}),
one can determine one more GP of the nucleon, in addition to the
4 combinations of Eqs.~(\ref{eq:gpdisp1}) - (\ref{eq:gpdisp4}).
In particular, the GP $P^{\left(M 1, M 1\right)1}$ 
which contains $F_5$ can be calculated through
\begin{eqnarray}
P^{\left(M 1, M 1\right)1} \left( Q^2 \right) \,=&&\, -
{{\sqrt{2}} \over {3}}\,\left( {{E + M} \over E}\right)^{1/2}
{{M \, \tilde q_0^2} \over {\rmq^2}} \,
\left\{ \bar F_5 (Q^2) \,+\, \tilde q_0 \, \bar F_{12} (Q^2) \right\}.
\label{eq:gp1111}
\end{eqnarray}
\newline
\item{The asymptotic part and dispersive contributions beyond $\pi N$
to $F_1$} \\
\label{sec:f1}

We next turn to the high-energy contribution to $F_1$.
As in the case of RCS,
the asymptotic contribution to the amplitude $F_1$
originates predominantly from the $t$-channel $\pi \pi$ intermediate states.
In a phenomenological analysis, this continuum
is parametrized through the exchange of a scalar-isoscalar particle in the
$t$-channel, i.e. an effective ``$\sigma$''-meson, as suggested 
in~\cite{Lvo97} and discussed in Sec.~\ref{sec:3_6} for the RCS case.
In this spirit, the difference between $F_1^{NB}$ and its $\pi N$ contribution 
can be parametrized at $Q^2 = 0$ by the energy-independent function~:
\begin{eqnarray}
&&F_1^{NB}(0, \nu, t) \,-\, F_1^{\pi N}(0, \nu, t) \, \approx \,
\left[ F_1^{NB}(0, 0, 0) - F_1^{\pi N}(0, 0, 0) \right]
\, {{1} \over {1 - t/m_\sigma^2}} \;,
\label{eq:f1asrcs}
\end{eqnarray}
where $F_1^{\pi N}$ is evaluated through a
dispersive integral as discussed in Eq.~(\ref{eq:sumrule}), and
 the $\sigma$-meson mass
$m_\sigma$ is a free parameter as in the fixed-$t$ unsubtracted 
RCS dispersion analysis.
A fit to the $t$-dependence of RCS data results in
 $m_\sigma \approx$ 0.6~GeV \cite{Lvo97}.
The value $F_1^{NB}(0, 0, 0)$ is then considered as a remaining global
fit parameter to be extracted from experiment.
It can be expressed physically in terms of the magnetic dipole polarizability
$\beta$~:
\begin{equation}
F_1^{NB}(0, 0, 0) \,=\, {{4 \pi} \over {e^2}} \; \beta  \;.
\label{eq:f1beta}
\end{equation}
The term $F_1^{\pi N}(0, 0, 0)$ in Eq.~(\ref{eq:f1asrcs}),
can be calculated through a dispersion integral and results in the value~:
\begin{equation}
\alpha_{em}\,F_1^{\pi N}(0, 0, 0) \,=\, \beta^{\pi N}
\, = \, 9.1 \;,
\label{eq:betapin}
\end{equation}
in units of $10^{-4}$~fm$^3$. 
From the $\pi N$ contribution $\beta^{\pi N}$ of Eq.~(\ref{eq:betapin}),
and the phenomenological value $\beta$ of Eq.~(\ref{DDeq3.6.0a}),
one obtains the difference
\begin{equation}
\beta - \beta^{\pi N} = - 7.5 \; ,
\label{eq:bmbpin}
\end{equation}
which enters in the {\it rhs} of Eq.~(\ref{eq:f1asrcs}).
As discussed before, the small total value of the 
magnetic polarizability $\beta$ 
comes about by a near cancellation between a large
(positive) paramagnetic contribution ($\beta^{\pi N}$) and a
large (negative) diamagnetic contribution ($\beta - \beta^{\pi N}$),
i.e., the asymptotic part of $F_1$ parametrizes the diamagnetism.
\newline
\indent
Turning next to the $Q^2$ dependence of the asymptotic contribution to
$F_1$, it has been proposed in \cite{Pas01} to parametrize 
this part of the non-Born term $F_1^{NB}(Q^2, \nu, t)$
beyond its $\pi N$ dispersive contribution, by
an energy independent $t$-channel pole of the form~:
\begin{equation}
F_1^{NB}(Q^2, \nu, t) \, - \, F_1^{\pi N}(Q^2, \nu, t) \, \approx \,
{{f(Q^2)} \over {1 - t/m_\sigma^2}} \;.
\label{eq:f1asvcs1}
\end{equation}
The function $f(Q^2)$ in Eq.~(\ref{eq:f1asvcs1}) can be obtained by
evaluating the $lhs$ of Eq.~(\ref{eq:f1asvcs1}) at the point where the
GPs are defined, i.e., $\nu = 0$ and $t = - Q^2$, at finite $Q^2$.
This leads to~:
\begin{equation}
f(Q^2) \, = \,
\left[ \bar F_1(Q^2) - \bar F_1^{\pi N}(Q^2) \right] \;
\left(1 + Q^2/m_\sigma^2 \right) \;,
\label{eq:f1asvcs2}
\end{equation}
where the shorthand $\bar F_1(Q^2)$ is defined in Eq.~(\ref{eq:fbardef}).
Equations (\ref{eq:f1asvcs1}) and (\ref{eq:f1asvcs2}) then lead
to the following expression for the VCS amplitude $F_1^{NB}$~:
\begin{eqnarray}
F_1^{NB}(Q^2, \nu, t) &&\approx F_1^{\pi N}(Q^2, \nu, t)
\,+\, \left[ \bar F_1(Q^2) \,-\, \bar F_1^{\pi N}(Q^2) \right]\,
{{1 + Q^2/m_\sigma^2} \over {1 - t/m_\sigma^2}} , \hspace{.2cm}
\label{eq:f1asvcs}
\end{eqnarray}
where the $\pi N$ contributions $F_1^{\pi N}(Q^2, \nu, t)$ and
$\bar F_1^{\pi N}(Q^2)$ 
are calculated through dispersion integrals as 
given by Eqs.~(\ref{eq:sumrule}) and (\ref{eq:unsub3}) respectively.
Consequently, the only unknown quantity on the {\it rhs} of
Eq.~(\ref{eq:f1asvcs}) is $\bar F_1(Q^2)$, which can be directly used as
a fit parameter at finite $Q^2$.
The quantity $\bar F_1(Q^2)$ can be expressed in terms of
the generalized magnetic polarizability $P^{(M 1, M 1)0}$
of Eq.~(\ref{eq:defgpunpol}) as~\cite{DKK98}~:
\begin{eqnarray}
\bar F_1(Q^2) \,&=&\,
- \sqrt{{3 \over 8}}\,\left( {{2 E} \over {E + M}}\right)^{1/2} \,
P^{\left(M 1, M 1\right)0}(Q^2)
\,\equiv \, {{4 \pi} \over {e^2}} \,
\left( {{2 E} \over {E + M}}\right)^{1/2} \, \beta(Q^2) \, ,
\label{eq:betaq}
\end{eqnarray}
where $\beta(Q^2$) is the generalized magnetic polarizability,
which reduces to the polarizability $\beta$ of RCS at $Q^2$ = 0.
The parametrization of Eq.~(\ref{eq:f1asvcs}) for $F_1$ then permits to
directly extract $\beta(Q^2)$ from VCS observables 
at a fixed $Q^2$.
In the following, we consider a convenient parametrization of the
$Q^2$ dependence of $\beta(Q^2)$ in order to provide predictions
for VCS observables.
For this purpose it was proposed in \cite{Pas01} to use a dipole 
parametrization for the difference $\beta(Q^2) -
\beta^{\pi N}(Q^2)$, which enters in the {\it rhs}
of Eq.~(\ref{eq:f1asvcs}) via Eq.~(\ref{eq:betaq}), as
\footnote{The dipole form displays the $1/Q^4$ behavior at large $Q^2$
as expected from  perturbative QCD.}~:
\begin{equation}
\beta(Q^2) - \beta^{\pi N}(Q^2) \,=\, {{ \left( \beta - \beta^{\pi N}\right) }
\over {\left( 1 + Q^2 / \Lambda^2_{\beta} \right)^2 }} \, ,
\label{eq:gpbetaparam}
\end{equation}
where the RCS value $(\beta - \beta^{\pi N})$ on the {\it rhs} is
given by Eq.~(\ref{eq:bmbpin}). The mass scale $\Lambda_{\beta}$
in Eq.~(\ref{eq:gpbetaparam}) determines the $Q^2$ dependence,  
and hence gives us the information how the diamagnetism
is spatially distributed in the nucleon. Using the dipole
parametrization of Eq.~(\ref{eq:gpbetaparam}), one can extract
$\Lambda_{\beta}$ from a fit to VCS data at different $Q^2$ values, 
and check the parametrization of Eq.~(\ref{eq:gpbetaparam}) for the 
asymptotic contribution to $\beta(Q^2)$. 
\newline
\indent
To have some educated guess on the physical value of $\Lambda_\beta$, we next
discuss two microscopic calculations of the diamagnetic contribution
to the GP $\beta(Q^2)$. The diamagnetism of the nucleon is dominated
by the pion cloud surrounding the nucleon. 
This diamagnetic contribution has been estimated in 
Ref.~\cite{Pas01} through a
DR calculation of the $t$-channel $\pi \pi$ intermediate
state contribution to $F_1$. Such a dispersive estimate
has been discussed before for RCS in Sec. \ref{sec:fixedt},
where it was shown that the asymptotic part of $F_1$ 
(or equivalently $A_1$) can be related to
the $\gamma \gamma \to \pi \pi \to N \bar N$ process. The dominant
contribution is due to the $\pi \pi$ intermediate state with
spin and isospin zero ($I = J = 0$). The generalization to VCS
leads then to the identification of $F_1^{as}$ with
the following unsubtracted DR in $t$ at fixed energy $\nu = 0$~:
\begin{equation}
\bar F_1^{as} (Q^2)\,=\,
\frac{1}{\pi}\int^\infty_{4m_{\pi}^2} dt'\,
\frac{\mathrm{Im}_t F_1 (Q^2,0, t')}{t'+Q^2}\,.
\label{eq:t-DR}
\end{equation}
The imaginary part on the {\it rhs} of Eq.~(\ref{eq:t-DR})
has been evaluated in Ref.~\cite{Pas01} through 
the subprocesses $\gamma^* \gamma\rightarrow \pi\pi$ and $\pi\pi\rightarrow
N{\bar{N}}$. To describe the $Q^2$ dependence of the 
$\gamma^* \gamma\rightarrow \pi\pi$ amplitude, 
which is dominated by the unitarized Born amplitude (on the pion), 
the pion electromagnetic form factor was included.
The result for this dispersive estimate of $\bar F_1^{as}$
through $t$-channel $\pi \pi$ intermediate states 
is shown in Fig.~\ref{fig:f1_asymp}, and compared 
with the corresponding evaluation of Ref.~\cite{Met96} 
in the linear $\sigma$-model (LSM). 
The LSM calculation overestimates the value of
$\bar F_1^{as}(0)$ (or equivalently $\beta_{as}$) by
about 30\% at any realistic value of $m_\sigma$, which is a free
parameter in this calculation. However, as for the dispersive
calculation, it also shows a steep $Q^2$ dependence.

\begin{figure}[ht]
\vspace{-.4cm}
\epsfxsize=7.cm
\centerline{\epsffile{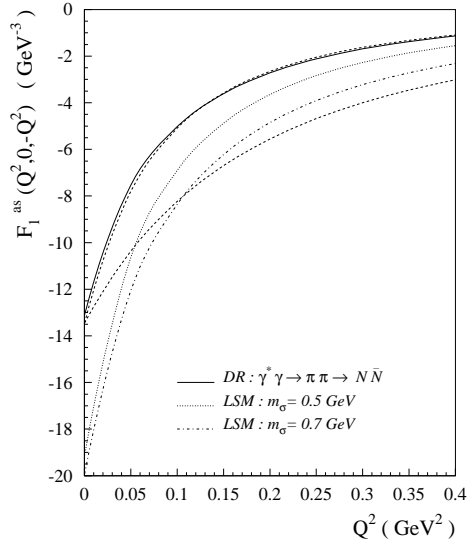}}
\vspace{-.25cm}
\caption[]{Theoretical estimates of the asymptotic contribution
  $F_1^{as}$ : DR calculation~\cite{Pas01}
of the $\gamma^* \gamma \to \pi \pi \to N \bar N$ process
(solid curve); linear $\sigma$-model (LSM)
calculation~\cite{Met96} with $m_\sigma$ = 0.5 GeV
(dotted curve) and $m_\sigma$ = 0.7 GeV (dashed-dotted curve).
The dashed curves are dipole parametrizations
according to Eq.~(\ref{eq:gpbetaparam}), which are fixed
to the phenomenological value at $Q^2$ = 0 and are shown for
two values of the mass-scale, $\Lambda_\beta$ = 0.4 GeV
(upper dashed curve, nearly coinciding with solid curve)
and $\Lambda_\beta$ = 0.6 GeV (lower dashed curve).}
\label{fig:f1_asymp}
\end{figure}
\indent
Furthermore in Fig.~\ref{fig:f1_asymp}, 
the two model calculations discussed above are compared 
with the dipole parametrization 
of Eq.~(\ref{eq:gpbetaparam}) for the two values $\Lambda_\beta$ = 0.4~GeV
and $\Lambda_\beta$ = 0.6 GeV. It is seen that these values are
compatible with the microscopic estimates discussed before.
In particular, the result for $\Lambda_\beta$ = 0.4 GeV is nearly
equivalent to the dispersive estimate of $\pi \pi$
exchange in the $t$-channel. The value of the mass scale
$\Lambda_\beta$ is small compared to the typical scale of
$\Lambda_D \approx$ 0.84 GeV appearing in the nucleon
magnetic (dipole) form factor.
This reflects the fact that diamagnetism has its physical
origin in the pionic degrees of freedom,
i.e., the diamagnetism is situated in the surface and
intermediate region of the nucleon.
\newline
\item{Dispersive contributions beyond $\pi N$ to $F_2$} \\

Though we can write down unsubtracted DRs
for all invariant amplitudes (or combinations of invariant
amplitudes) except for $F_1$ and $F_5$,
one might wonder about the quality of our approximation to saturate the
unsubtracted dispersion integrals by $\pi N$ intermediate states only.
This question is particularly relevant for the
amplitude $F_2$, for which we next investigate the size of
dispersive contributions beyond the $\pi N$ channel.
We start with the case of RCS, where one can
quantify the higher dispersive corrections to $F_2$, because the value of
$F_2^{NB}$ at the real photon point can be expressed exactly
(see Eqs.~(\ref{eq_3_47}) and (\ref{eq:gpdisp1})) in terms of the scalar
polarizability $(\alpha + \beta)$ as
\begin{equation}
F_2^{NB}(0, 0, 0) \,=\, - {{4 \pi} \over {e^2}} \, {1 \over {(2 M)^2}} \,
\left(\alpha + \beta \right) \, .
\label{eq:f2rcs}
\end{equation}
The $\pi N$ dispersive contribution provides the value
\begin{equation}
(\alpha+\beta)^{\pi N} = 11.6 \, ,
\label{eq:apbpin}
\end{equation}
which falls short by about 15 \% compared to the Baldin's sum rule value
of Eq.~(\ref{DDeq2.2.18}). The remaining part originates from higher
dispersive contributions ($\pi \pi N$, ...) to $F_2$.
These higher dispersive contributions could be calculated through
unitarity, by use of Eq.~(\ref{eq:schunit}), similarly to the
$\pi N$ contribution. However, the present data for the production of
those intermediate states (e.g., $\gamma^* N \to \pi \pi N$) are still
too scarce to evaluate the imaginary parts of the VCS amplitude $F_2$
directly. Therefore, we estimate the dispersive contributions
beyond $\pi N$ by an energy-independent constant,
which is fixed to its phenomenological value at $\nu = t = 0$. This yields
at $Q^2 = 0$~: 
\begin{eqnarray}
F_2^{NB}(0, \nu, t) &&\approx F_2^{\pi N}(0, \nu, t)
\,-\, {{4 \pi} \over {e^2}} \, {1 \over {(2 M)^2}} \,
\left[ \, \left(\alpha + \beta \right) \,-\,
\left(\alpha + \beta \right)^{\pi N} \, \right], \hspace{.2cm}
\label{eq:f2approx}
\end{eqnarray}
which is an exact relation at $\nu = t = 0$, the point where
the polarizabilities are defined.
\newline
\indent
The approximation of Eq.~(\ref{eq:f2approx}) to replace
the dispersive contributions beyond $\pi N$ by a
constant can only be valid if one stays below the thresholds for those
higher contributions. Since the next threshold beyond $\pi N$ is $\pi \pi N$,
the approximation of Eq.~(\ref{eq:f2approx}) restricts us in practice
to energies below the $\Delta(1232)$-resonance.
\newline
\indent
We next consider the extension to VCS, and focus our efforts to
describe VCS into the $\Delta(1232)$-resonance region.
Analogously to Eq.~(\ref{eq:f2approx}) for RCS, the
dispersive contributions beyond $\pi N$ are approximated by an
energy-independent constant. This constant is fixed
at arbitrary $Q^2$, $\nu = 0$, and $t = - Q^2$, which is the point
where the GPs are defined. One thus obtains for $F_2^{NB}$  \cite{Pas01}~:
\begin{equation}
F_2^{NB}(Q^2, \nu, t) \approx F_2^{\pi N}(Q^2, \nu, t) \;+\;
\left[ \bar F_2(Q^2) \,-\, \bar F_2^{\pi N}(Q^2) \right] \, ,
\label{eq:f2asvcs}
\end{equation}
where $\bar F_2(Q^2)$ is defined as in Eq.~(\ref{eq:fbardef}), and can be
expressed in terms of GPs through relations as given by 
Eqs.~(\ref{eq:gpdisp1}) - (\ref{eq:gpdisp4}). 
We saturate the 3 combinations of spin
GPs in Eqs.~(\ref{eq:gpdisp2}) - (\ref{eq:gpdisp4}) by their $\pi N$
contribution, and include for
the fourth spin GP of Eq.~(\ref{eq:gp1111}) also the $\pi^0$-pole contribution.
Therefore, we only consider dispersive contributions beyond the $\pi N$
intermediate states for the two scalar GPs, which are therefore
the two fit quantities in the present DR formalism for
VCS. In this way, one can use Eq.~(\ref{eq:gpdisp1}),
to write the difference $\bar F_2(Q^2) - \bar
F_2^{\pi N}(Q^2)$ entering in the {\it rhs} of Eq.~(\ref{eq:f2asvcs})
as~:
\begin{eqnarray}
\bar F_2(Q^2) - \bar F_2^{\pi N}(Q^2) \, &&\approx \,
{{4 \pi} \over {e^2}} \,
\left( {{2 E} \over {E + M}}\right)^{1/2}
\, {{\tilde q_0} \over {\rmq^2}} \, {1 \over {2 M}}
\nonumber \\
&& \times \left\{\;  \left[\alpha(Q^2) \,-\, \alpha^{\pi N}(Q^2)\right]
\,+\,  \left[\beta(Q^2) \,-\, \beta^{\pi N}(Q^2)\right] \; \right\},
\label{eq:alphaq}
\end{eqnarray}
in terms of the generalized magnetic polarizability $\beta(Q^2)$  
of Eq.~(\ref{eq:betaq}), 
and the generalized electric polarizability $\alpha(Q^2$),
which is related to the GP $P^{(L1,L1)0}(Q^2)$ by Eq.~(\ref{eq_3_47}).
\newline
\indent
We stress that Eqs.~(\ref{eq:f1asvcs}) and (\ref{eq:alphaq})
are intended to extract the two GPs $\alpha(Q^2)$ and $\beta(Q^2)$
from VCS observables minimizing the model dependence as much as possible.
As discussed before for $\beta(Q^2)$, we next consider 
a convenient parametrization of the $Q^2$ dependence
of $\alpha(Q^2)$ in order to provide predictions for VCS observables.
For this purpose, a dipole form has been proposed in Ref.~\cite{Pas01} 
for the difference
$\alpha(Q^2) - \alpha^{\pi N}(Q^2)$ which enters in the {\it rhs} of
Eq.~(\ref{eq:alphaq}),
\begin{equation}
\alpha(Q^2) - \alpha^{\pi N}(Q^2) \,=\,
{{\left( \alpha - \alpha^{\pi N} \right)}
\over {\left( 1 + Q^2 / \Lambda^2_{\alpha} \right)^2 }} \, ,
\label{eq:gpalphaparam}
\end{equation}
where the $Q^2$ dependence is governed by the mass scale
$\Lambda_\alpha$, the second free parameter of the DR formalism.
In Eq.~(\ref{eq:gpalphaparam}), the RCS value
\begin{equation}
(\alpha - \alpha^{\pi N}) \,=\,  9.6 \, ,
\end{equation}
is obtained from the phenomenological value of Eq.~(\ref{DDeq3.6.0a}) for
$\alpha$, and from the calculated $\pi N$ contribution,
$\alpha^{\pi N}$ =  2.5.
Using the dipole parametrization of Eq.~(\ref{eq:gpalphaparam}),
one can extract the free parameter $\Lambda_{\alpha}$ from
a fit to VCS data at different $Q^2$ values.

\end{itemize}


\subsection{VCS data for the proton and extraction of generalized
polarizabilities}

\begin{figure}[ht]
\epsfxsize=10cm
\centerline{\epsffile{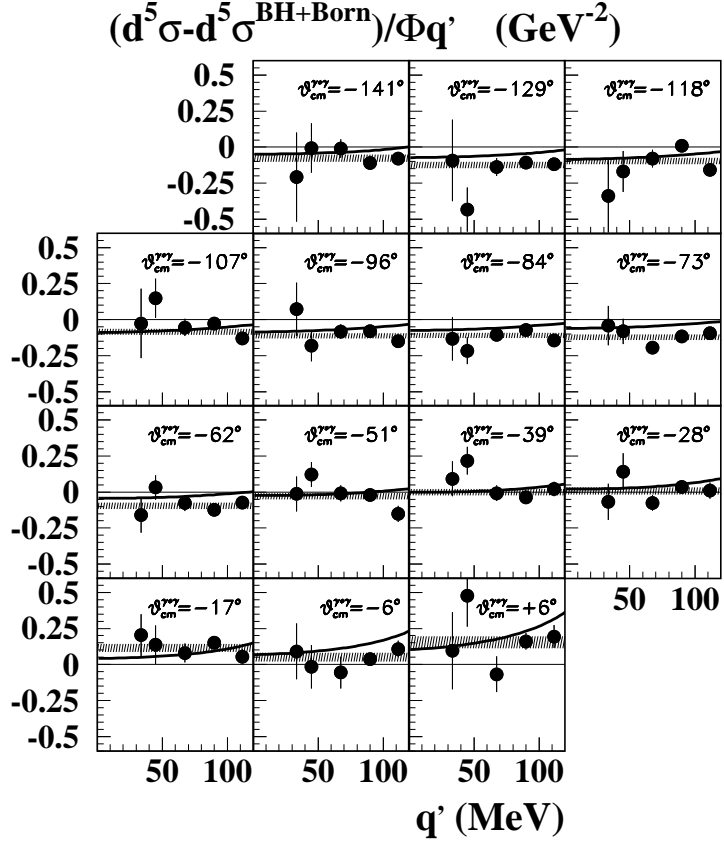}}
\vspace{-.75cm}
\vspace{-1cm}
\caption[]{ $(d^5\sigma-d^5\sigma^{{\rm BH+Born}})/\Phi \rmqp$ 
for the $e p \to e p \gamma$ reaction as function 
of the outgoing-photon energy $\rmqp$ in MAMI kinematics : $\varepsilon
=0.62$, $\rmq = 0.6$ GeV, $\phi = 0^{\mathrm{o}}$, 
and for different photon {\it c.m.} angles 
$\theta^{\gamma*\gamma}_{{\rm cm}}$. 
The data and the shaded bands, representing the best fit to 
the data within the LEX formalism, are from~\cite{Roc00}. 
The solid curves are the DR results taking into 
account the full $\rmqp$ dependence of the non-Born contribution to the cross 
section. The asymptotic contributions are calculated according to  
Eqs.~(\ref{eq:gpbetaparam}) and (\ref{eq:gpalphaparam}), 
with $\Lambda_\beta=0.6$ GeV and $\Lambda_\alpha = 1$ GeV, respectively.}
\label{fig:vcs_mami_spectrum}
\end{figure}

\begin{figure}[ht]
\epsfxsize=8.25cm
\centerline{\epsffile{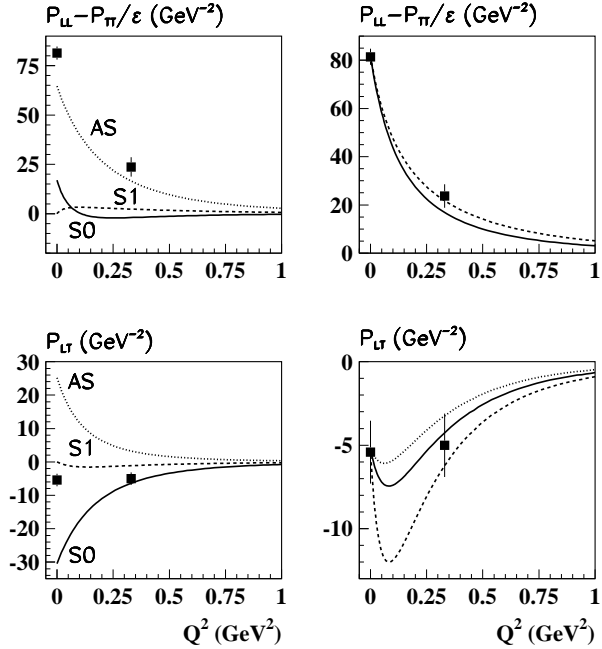}}
\vspace{-.3cm}
\caption[]{Results for the unpolarized structure functions
$P_{LL} - P_{TT}/\varepsilon$ (upper panels),
and $P_{LT}$ (lower panels), for $\varepsilon$ = 0.62.
Upper left panel~: 
dispersive $\pi N$ contribution of the GP $\alpha$ (solid curve, S0), 
dispersive $\pi N$ contribution of the spin-flip GPs (dashed curve, S1), and 
the asymptotic contribution (AS) of $\alpha$ according to 
Eq.~(\ref{eq:gpalphaparam}) with $\Lambda_\alpha=1$ GeV (dotted curve).
Upper right panel~: total result for 
$P_{LL} - P_{TT}/\varepsilon $ 
(sum of the three contributions on the upper left panel) 
for $\Lambda_\alpha=1$ GeV (solid curve) and 
$\Lambda_\alpha=1.4$ GeV (dashed curve).
Lower left panel~: dispersive 
$\pi N$ contribution of the GP $\beta$ (solid curve, S0), contribution 
of the spin-flip GPs (dashed curve, S1), and the asymptotic contribution (AS)
of $\beta$ according to Eq.~(\ref{eq:gpbetaparam}) 
with $\Lambda_\beta=0.6$ GeV (dotted curve). 
Lower right panel~: total result for $P_{LT}$,   
for $\Lambda_\beta=0.7$ GeV (dotted curve), 
$\Lambda_\beta=0.6$ GeV (solid curve), 
and $\Lambda_\beta=0.4$ GeV (dashed curve).
The RCS data are from Ref.~\cite{Olm01}, and the VCS data at $Q^2=0.33
\mbox{\, GeV}^2$ from Ref.~\cite{Roc00}.}
\label{fig:meaning_pol}
\end{figure}

\begin{figure}[ht]
\vspace{-.3cm}
\epsfxsize=11.cm
\centerline{\epsffile{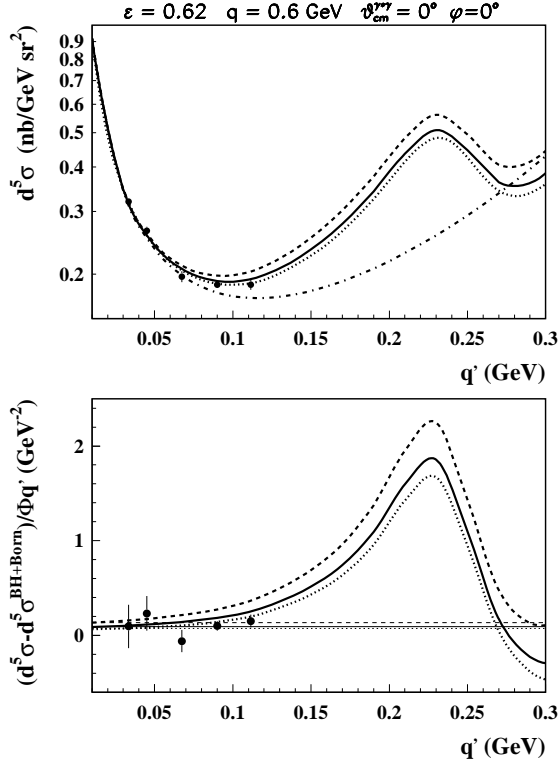}}
\vspace{-.2cm}
\caption[]{
Upper panel: The differential cross section for the reaction 
$e p \to e p \gamma$ as function of the outgoing-photon energy $\rmqp$ 
in MAMI kinematics : $\varepsilon =0.62$, $\rmq = 0.6$ GeV, 
and for $\theta^{\gamma*\gamma}_{{\rm cm}} = 0^{\rm o}$, in plane ($\phi=0^{\rm o}$).
The BH + B contribution is given by the dashed-dotted curve.
Lower panel: 
Results for $(d^5\sigma-d^5\sigma^{{\rm BH+Born}})/\Phi \rmqp$ 
as function of $\rmqp$ .
The total DR results are obtained with the
asymptotic parts of Eqs.~(\ref{eq:gpbetaparam}) and (\ref{eq:gpalphaparam}),
using a fixed value of $\Lambda_\alpha = 1 $ GeV and 
for the same three values of $\Lambda_\beta$ as displayed in the lower
right plot of Fig.~\ref{fig:meaning_pol}, i.e. 
$\Lambda_\beta = 0.7$ GeV (dotted curve), 
$\Lambda_\beta = 0.6$ GeV (solid curve), and 
$\Lambda_\beta = 0.4$ GeV (dashed curve).
In the lower panel, the DR calculations 
taking into account the full energy dependence 
of the non-Born contribution (thick curves) are compared to the 
corresponding results within the LEX formalism (thin horizontal curves).
The data are from Ref.~\cite{Roc00}.}
\label{fig:vcs_mami_thetap0_qp_dep}
\end{figure}

\begin{figure}[ht]
\epsfxsize=7cm
\centerline{\epsffile{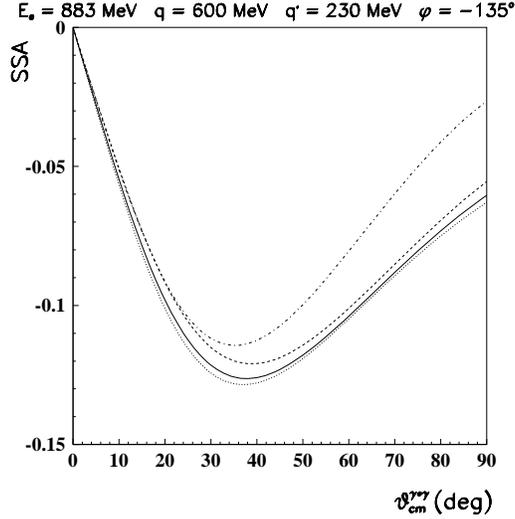}}
\vspace{0cm}
\caption[]{Electron single spin asymmetry (SSA) for VCS in MAMI kinematics
as function of the photon scattering angle.
The full dispersion results are shown for the values: 
$\Lambda_\alpha = 1$ GeV, $\Lambda_\beta = 0.6 $ GeV (solid curve), 
$\Lambda_\alpha = 1$ GeV, $\Lambda_\beta = 0.4 $ GeV (dashed curve),
$\Lambda_\alpha = 1$ GeV, $\Lambda_\beta = 0.7 $ GeV (dotted curve), 
and   $\Lambda_\alpha = 1.4$ GeV, $\Lambda_\beta = 0.4 $ GeV (dashed-dotted curve). }
\label{fig:vcs_ssa}
\end{figure}

\begin{figure}[ht]
\vspace{-.5cm}
\epsfxsize=9.5cm
\centerline{\epsffile{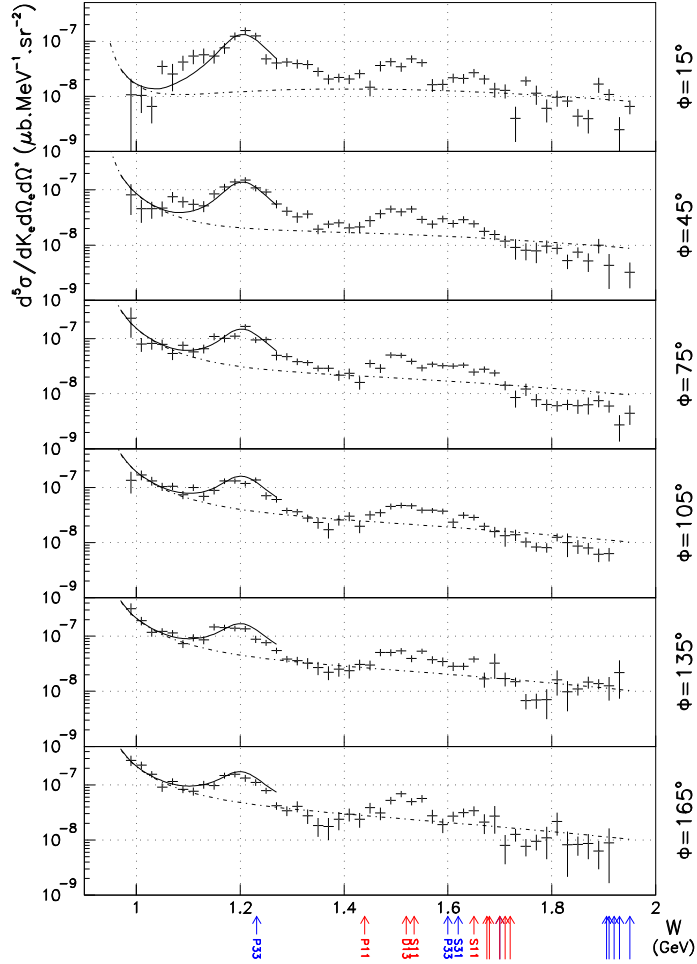}}
\vspace{0cm}
\caption[]{The differential cross sections for the $e p \to e p
\gamma$ reaction as function of the {\it c.m.} energy W in JLab kinematics : 
$E_e$ = 4.032 GeV, $Q^2=1.0$ GeV$^2$, and for fixed scattering angle 
$\theta^{\gamma*\gamma}_{\rm cm} = -167.2 ^{\rm o}$, 
for different out-of-plane angles $\phi$. 
The BH + B contribution is given by the dashed curve.
The total DR result is shown by the solid curve (limited to W $<$ 1.25 GeV) 
for the values $\Lambda_\alpha =1.0 $ GeV and $\Lambda_\beta =0.45$ GeV. 
The data are from Ref.~\cite{Fon02}.}
\label{fig:vcsres_jlab}
\end{figure}

\begin{figure}[ht]
\epsfxsize=11.cm
\centerline{\epsffile{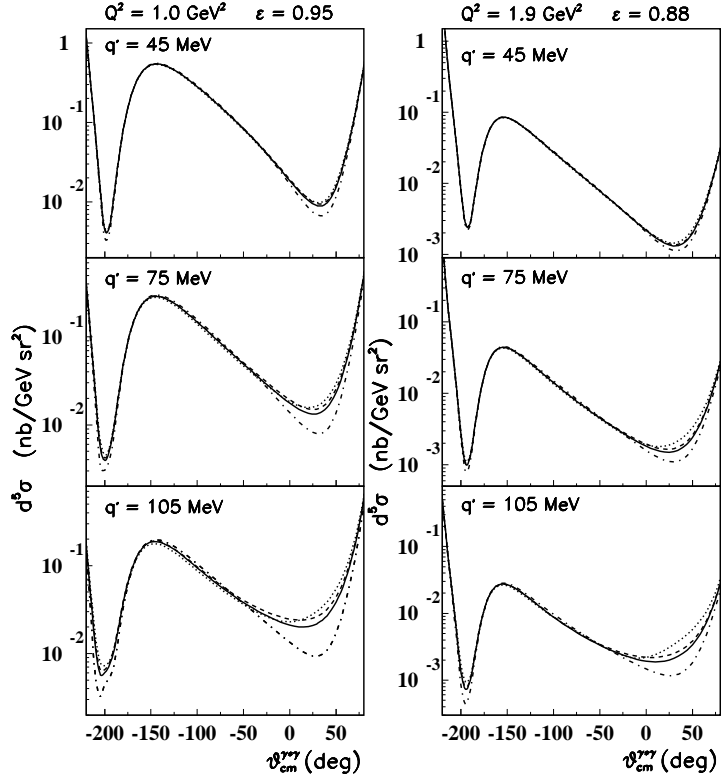}}
\caption[]{
The differential cross section for the reaction $e p \to e p \gamma$ 
as function of the photon scattering angle, at 
different values of the outgoing-photon 
energy in JLab kinematics: 
$Q^2=1$ GeV$^2$ and $\varepsilon= 0.95$ (left panels)
and $Q^2=1.9$ GeV$^2$ and $\varepsilon=0.88$ (right panels). 
The BH + B cross sections are shown by the dashed-dotted curves. 
The DR results are displayed 
with the asymptotic terms parametrized as in 
Eqs.~(\ref{eq:gpalphaparam}) and (\ref{eq:gpbetaparam}), using the
values : 
$\Lambda_\alpha$ = 1~GeV and $\Lambda_\beta$ = 0.6~GeV (solid curves), 
$\Lambda_\alpha$ = 1~GeV and $\Lambda_\beta$ = 0.4~GeV (dashed curves),
$\Lambda_\alpha$ = 1.4~GeV and $\Lambda_\beta$ = 0.6~GeV (dotted curves).}
\label{fig:vcs_jlab_bel_thr}
\end{figure}

\begin{figure}[ht]
\vspace{-3.25cm}
\epsfxsize=12cm
\centerline{\epsffile{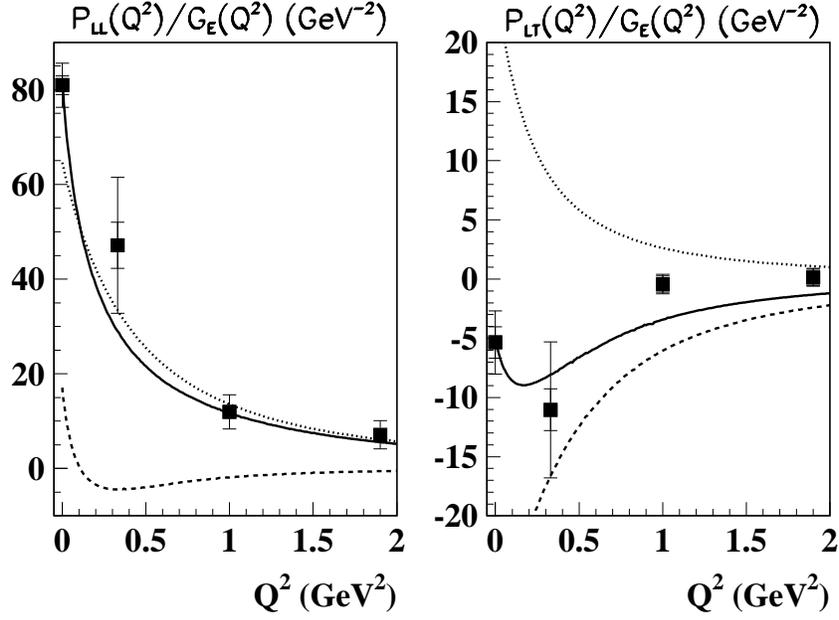}}
\vspace{0cm}
\caption[]{Results for the unpolarized VCS structure functions 
$P_{LL}$ (left panel) and $P_{LT}$ (right panel) divided by the proton 
electric form factor. 
Dashed lines: dispersive $\pi N$ contributions.
Dotted lines: asymptotic contributions calculated according to
Eqs.~(\ref{eq:gpbetaparam}) and (\ref{eq:gpalphaparam}) with 
$\Lambda_\alpha = 0.92 $ GeV (left panel) and   
$\Lambda_\beta = 0.66 $ GeV (right panel). 
Solid curves: total results, 
sum of the dispersive and asymptotic contributions.
The RCS data are from Ref.~\cite{Olm01}, the VCS MAMI data 
at $Q^2 = 0.33$ GeV$^2$ are from Ref.~\cite{Roc00}, and 
the preliminary VCS JLab data at $Q^2 = 1$ GeV$^2$ and 
$Q^2 = 1.9$ GeV$^2$ from Ref.~\cite{Fon02} (inner error bars are statistical
errors only, outer error bars include systematical errors). 
The values for $P_{LL}$ at $Q^2 > 0$ were extracted by use of a
dispersive estimate for the not yet separated $P_{TT}$ contribution.}  
\label{fig:vcs_asy_ff}
\end{figure}

\begin{figure}[ht]
\vspace{-1.cm}
\epsfxsize=8.75cm
\centerline{\epsffile{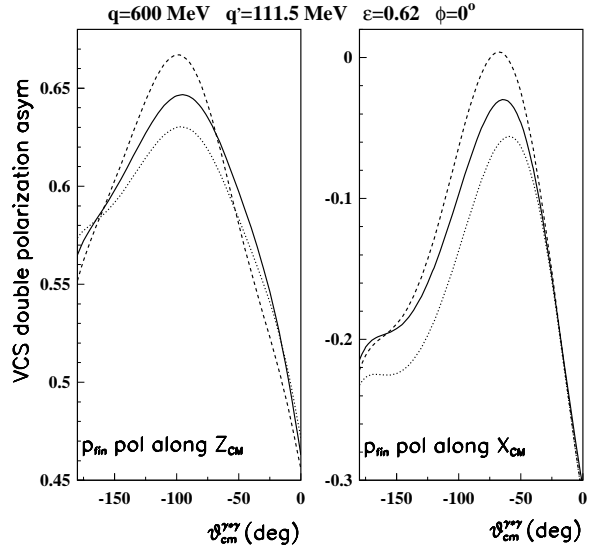}}
\vspace{-.25cm}
\caption[]{VCS double-polarization asymmetry (polarized electron,
recoil proton polarization along either the $z$- or $x$- directions 
in the {\it c.m.} frame) 
in MAMI kinematics as function of the photon scattering angle.
The dotted curves correspond to the BH+B contribution.
The solid curves show the total DR results for the values of  
$\Lambda_\alpha$ = 1~GeV, $\Lambda_\beta$ = 0.6~GeV.
The dashed curves are the HBChPT predictions from~\cite{Hem00}.}
\label{fig:vcs_mami_doublepol}
\end{figure}

Having set up the dispersion formalism for VCS, we now show the
predictions for the different $e p \to e p \gamma$ observables 
for energies up to the $\Delta(1232)$-resonance region. 
The aim of the experiments is to extract the 6 GPs of 
Eqs.~(\ref{eq:defgpunpol}) and (\ref{eq:defgppol}) 
from both unpolarized and polarized
observables. We will compare the DR results, which take account of 
the full dependence of the $e p \to e p \gamma$ observables 
on the energy ($\rmqp$) of the emitted photon, 
with a low-energy expansion (LEX) in $\rmqp$. 
In the LEX of observables, only the first three terms of a Taylor
expansion in $\rmqp$ are taken into account. 
In such an expansion in $\rmqp$, the experimentally extracted 
VCS unpolarized squared amplitude 
$\calm^{\rm exp}$ takes the form \cite{Gui95}~:
\begin{equation}
\calm^{\rm exp}=\frac{\calm^{\rm exp}_{-2}}{\rmqp^2}
+\frac{\calm^{\rm exp}_{-1}}{\rmqp}
+\calm^{\rm exp}_0+O(\rmqp) \, .
\label{eq:unpolsqramp}
\end{equation}
Due to the low energy theorem (LET), the threshold coefficients 
$\calm^{\rm exp}_{-2}$ and $\calm^{\rm exp}_{-1}$ are
known~\cite{Gui95}, and are fully determined from the Bethe-Heitler +
Born (BH + B) amplitudes. 
The information on the GPs is contained in \( \calm^{\rm exp}_0\), 
which contains a part originating from the 
BH+B amplitudes and another one which is a linear combination of the GPs, with
coefficients determined by the kinematics. 
The unpolarized observable $\calm^{\rm exp}_0$ 
can be expressed in terms of 3 structure functions
$P_{LL}(\rmq)$, $P_{TT}(\rmq)$, and $P_{LT}(\rmq)$ by ~\cite{Gui95}~:
\begin{eqnarray}
\hspace{-0.7cm}
\calm^{\rm exp}_0 - \calm^{\rm BH+B}_0 
= 2 K_2  \Bigg\{ 
v_1 \left[ {\varepsilon  P_{LL}(\rmq) - P_{TT}}(\rmq)\right] 
+ \left(v_2-\frac{\qt0}{\rmq}v_3\right)\sqrt {2\varepsilon \left( 
{1+\varepsilon }\right)} P_{LT}(\rmq) \Bigg\}, 
\label{eq:vcsunpol}
\end{eqnarray}
where $K_2$ is a kinematical factor, $\varepsilon$ is the virtual
photon polarization (in the standard notation used in electron
scattering), and $v_1, v_2, v_3$ are kinematical
quantities depending on $\varepsilon$ and $\rmq$  
as well as on the {\it cm} polar and azimuthal angles
($\theta^{\gamma*\gamma}_{{\rm cm}}$  and $\phi$, respectively) of  the
produced real photon (for details see Ref.~\cite{Gui98}). 
The 3 unpolarized observables of Eq.~(\ref{eq:vcsunpol}) 
can be expressed in terms of the 6 GPs as \cite{Gui95,Gui98}~:
\begin{eqnarray}
&&P_{LL} \,=\, - 2\sqrt{6} \, M \, G_E \, P^{\left( {L1,L1} \right)0} \;, 
\label{eq:unpolobsgp1} \\
&&P_{TT} \,=\, - 3 \, G_M \, \frac{\rmq^2}{\qt0} 
\left( P^{(M1,M1)1}\,-\,\sqrt{2} \, \qt0 \, P^{(L1,M2)1} \right) , 
\label{eq:unpolobsgp2} \\
&&P_{LT} \,=\, \sqrt{\frac{3}{2}} \, \frac{M \, \rmq}{Q} \, G_E \, 
P^{(M1,M1)0} \,+\,\frac{3}{2} \, \frac{Q \, \rmq}{\qt0}\,G_M\,
P^{(L1,L1)1}, \hspace{.6cm}
\label{eq:unpolobsgp3}
\end{eqnarray}
where $G_E$ and $G_M$ stand for the electric and magnetic nucleon form factors
$G_E(Q^2)$ and $G_M(Q^2)$, respectively. 
\newline
\indent
The first VCS experiment was performed at MAMI \cite{Roc00} and 
the response functions $P_{LT}$ and $P_{LL}$~-~$P_{TT}/\varepsilon$ 
were extracted at $Q^2$ = 0.33 GeV$^2$ by performing a LEX to these VCS data,
according to Eq.~(\ref{eq:vcsunpol}). To test the validity of such a
LEX, we show in Fig.~\ref{fig:vcs_mami_spectrum} the DR predictions for
the full energy dependence of the non-Born part of the $e p \to e p
\gamma$ cross section in the kinematics of the MAMI experiment
\cite{Roc00}. This energy dependence is compared with the LEX, which
predicts a linear dependence in $\rmqp$ 
for the difference between the experimentally
measured cross section and its BH + B contribution. 
The result of a best fit to the data in the framework of the LEX is
indicated by the horizontal bands in Fig.~\ref{fig:vcs_mami_spectrum} 
for the quantity $(d^5\sigma-d^5\sigma^{{\rm BH+Born}})/\Phi
\rmqp$, where $\Phi$ is a phase space factor defined in Ref.~\cite{Gui95}. 
The fivefold differential cross section $d^5 \sigma$ is differential
with respect to the electron {\it lab} energy and {\it lab} angles and
the proton {\it c.m.} angles, and stands in all of the following for 
$d \sigma\,/\, dk^e_{lab} \, d \Omega^e_{lab} \,d \Omega^p_{c.m.}$. 
It is seen from Fig.~\ref{fig:vcs_mami_spectrum} that the DR results 
predict only a modest additional energy dependence 
up to $\rmqp \simeq$ 0.1 GeV and for most of the photon angles involved, 
and therefore support the LEX analysis of \cite{Roc00}. Only for 
forward angles, $\theta^{\gamma*\gamma}_{{\rm cm}} \approx 0$, 
which is the angular range from which the value of $P_{LT}$ is extracted, 
the DR calculation predicts a stronger energy dependence in the range 
up to $\rmqp \simeq$ 0.1 GeV, as compared to the LEX. 
\newline
\indent
In Fig.~\ref{fig:meaning_pol}, we display the response functions  
$P_{LL} - P_{TT}/\varepsilon$ and $P_{LT}$ at $Q^2$ = 0.33 GeV$^2$, 
which have been extracted from the cross section data 
of Fig.~\ref{fig:vcs_mami_spectrum} \cite{Roc00}, and compare them
with the corresponding DR calculations.
For the electromagnetic form factors in 
Eqs.~(\ref{eq:unpolobsgp1})-(\ref{eq:unpolobsgp3}) we use the H\"ohler parametrization~\cite{Hoe76} as in the analysis of the MAMI experiment~\cite{Roc00}.
In the lower panel of Fig.~\ref{fig:meaning_pol}, the  
$Q^2$-dependence of the VCS response function $P_{LT}$ is displayed, 
which reduces to the magnetic polarizability $\beta$  
at the real photon point. 
At finite $Q^2$, it contains both the scalar GP $\beta(Q^2)$ and  
the spin GP $P^{(L 1, L 1)1}$, as seen from Eq.~(\ref{eq:unpolobsgp3}). 
It is obvious from Fig.~\ref{fig:meaning_pol} that 
the structure function $P_{LT}$ results from a large
dispersive $\pi N$ (paramagnetic) contribution, which is dominated by 
$\Delta(1232)$ resonance excitation, and a large asymptotic 
(diamagnetic) contribution
to $\beta$ with opposite sign, leading to a relatively small net result. 
The asymptotic contribution is shown in 
Fig.~\ref{fig:meaning_pol} with the parametrization of 
Eq.~(\ref{eq:gpbetaparam}) for the values 
$\Lambda_\beta$ = 0.4 and $\Lambda_\beta$ = 0.6 GeV, 
which were also displayed in Fig.~\ref{fig:f1_asymp}.
Due to the large cancellation in $P_{LT}$, its $Q^2$ dependence is
a very sensitive observable to study the interplay of the two mechanisms. 
In particular, one expects a faster fall-off of the asymptotic
contribution with $Q^2$ in comparison to the 
$\pi N$ dispersive contribution, as
discussed before. This is highlighted by the measured value 
of $P_{LT}$ at $Q^2$ = 0.33 GeV$^2$ \cite{Roc00}, 
which is comparable to the value of $P_{LT}$ at $Q^2$ = 0. 
As seen from Fig.~\ref{fig:meaning_pol}, 
this points to an interesting structure in the  
$Q^2$ region around 0.05 - 0.1~GeV$^2$,
where forthcoming data are expected from an experiment at MIT-Bates
\cite{Mis97}.  
\newline
\indent
In the upper panel of Fig.~\ref{fig:meaning_pol}, we show the 
$Q^2$-dependence of the VCS response function 
$P_{LL}$~-~$P_{TT}/\varepsilon$, which reduces 
at the real photon point to the electric polarizability 
$\alpha$. At non-zero $Q^2$, $P_{LL}$ 
is directly proportional to the scalar GP $\alpha(Q^2)$, as 
seen from Eq.~(\ref{eq:unpolobsgp1}), and the response function  
$P_{TT}$ of Eq.~(\ref{eq:unpolobsgp2}) contains only spin GPs.
As is shown by Fig.~\ref{fig:meaning_pol}, the $\pi N$
dispersive contribution to $\alpha$ and to the spin GPs in $P_{TT}$ 
are smaller than the asymptotic contribution to $\alpha$. 
At $Q^2$ = 0, the $\pi N$ dispersive and asymptotic contributions to
$\alpha$ have the same sign and lead to a large value of $\alpha$, 
in contrast to $\beta$ where both contributions 
have opposite sign and largely cancel each other in their sum. 
\newline
\indent
Increasing the energy, we show 
in Fig.~\ref{fig:vcs_mami_thetap0_qp_dep} the DR predictions
for photon energies in the $\Delta(1232)$-resonance region. 
It is seen that the $e p \to e p \gamma$ cross section 
rises strongly when crossing the
pion threshold. In the dispersion relation formalism, which is based
on unitarity and analyticity, the rise of the cross section with $\rmqp$
below pion threshold, due to virtual $\pi N$ intermediate states,
is connected to the strong rise of the cross section with $\rmqp$ 
when a real $\pi N$ intermediate state can be produced. 
It is furthermore seen from Fig.~\ref{fig:vcs_mami_thetap0_qp_dep}
(lower panel) that the region
between pion threshold and the $\Delta$-resonance peak displays 
an enhanced sensitivity to the GPs through the interference with the
rising Compton amplitude due to $\Delta$-resonance excitation. 
For example, at $\rmqp \simeq$ 0.2 GeV, the 
predictions for $P_{LT}$ in the lower right panel of
Fig.~\ref{fig:meaning_pol} for $\Lambda_\beta$ = 0.4
GeV and $\Lambda_\beta$ = 0.6 GeV give a difference of about 20 \%
in the non-Born squared amplitude. In contrast, the LEX prescription 
results in a relative effect for the same two values of $P_{LT}$ 
of about 10\% or less. 
This is similar to the situation discussed in Sec. \ref{sec:rcsproton} 
for RCS, where the region between pion
threshold and the $\Delta$-resonance position 
also provides an enhanced sensitivity to the
polarizabilities and is used to extract those polarizabilities from
data using a DR formalism. 
Therefore, the energy region between pion threshold and the
$\Delta$-resonance seems promising to measure VCS observables 
with an increased sensitivity to the GPs. Such an experiment has 
been proposed at MAMI and is underway~\cite{dHo01}.
\newline
\indent
When crossing the pion threshold, the VCS amplitude acquires
an imaginary part due to the coupling to the $\pi N$
channel. Therefore, single polarization observables become non-zero
above pion threshold. A particularly relevant observable is the
electron single spin asymmetry (SSA), which is obtained by flipping the
electron beam helicity \cite{Gui98}. 
For VCS, this observable is mainly due to the interference of the 
real BH + B amplitude with the imaginary part of the VCS amplitude. 
As the SSA vanishes in-plane, its 
measurement requires an out-of-plane experiment. 
Such experiments have been proposed both at MAMI \cite{dHo01} and 
at MIT-Bates \cite{Kal97}. 
In Fig.~\ref{fig:vcs_ssa}, the SSA is shown for a kinematics in the 
$\Delta(1232)$ region, corresponding with $W \approx 1.2$~GeV.  
The DR calculation firstly shows that the SSA is quite
sizable in the $\Delta(1232)$ region. 
The SSA, which is mainly sensitive to the imaginary part of the VCS amplitude, 
displays only a rather weak dependence on the magnetic GPs $\beta(Q^2)$, 
and shows a modest dependence on $\alpha(Q^2)$.  
Therefore, it provides an excellent cross-check of the 
dispersive input in the DR formalism for VCS, in particular by comparing 
at the same time the pion and photon electroproduction channels 
through the $\Delta$ region. 
\newline
\indent
Going to higher $Q^2$, the VCS process has also been measured at JLab 
and data have been obtained  
both below pion threshold at $Q^2$ = 1 GeV$^2$ \cite{Deg01}, 
at $Q^2$ = 1.9 GeV$^2$ \cite{Jam00}, 
as well as in the resonance region around $Q^2$ = 1 GeV$^2$ \cite{Lav01} 
(see Ref.~\cite{Fon02} for a short review of these JLab data).
\newline
\indent
In Fig.~\ref{fig:vcsres_jlab}, 
we show the results for the $e p \to e p \gamma$ reaction 
in the resonance region at $Q^2 = 1$~GeV$^2$ and at a backward angle. 
These are the first VCS measurements ever performed in the resonance region.
We also display the DR calculations of \cite{Pas01} for the cross section.
The data clearly show the excitation of the $\Delta(1232)$ resonance, 
and display a second and third resonance region, mainly due to the
excitations of the $D_{13}(1520)$ and $F_{15}(1680)$ resonances. 
The DR calculations reproduce well the $\Delta(1232)$ region. 
Due to scarce information for the dispersive input above the 
$\Delta(1232)$ resonance, the DR calculations cannot be extended at
present into the second and third resonance regions. Between pion
threshold and the $\Delta(1232)$ resonance, the calculations show a sizable
sensitivity to the GPs, in particular to $P_{LL}$ 
in this backward angle kinematics, and  
seem very promising to extract information on the electric polarizability.  
The precise extraction of GPs from VCS data at these higher values of $Q^2$,
requires an accurate knowledge of the nucleon 
electromagnetic form factors (FFs) in this region. 
For the proton electromagnetic form factors, we
use the new empirical fit of \cite{Bra01}, 
which includes the recent high accuracy measurements performed at JLab 
for the ratio of proton electric FF $G_E$ to the magnetic FF $G_M$ 
in the $Q^2$ range 0.4 - 5.6 GeV$^2$ \cite{Jon00,Gay02}. 
From Fig.~\ref{fig:vcsres_jlab}, one sees that a good description of
the JLab data is obtained by the values $\Lambda_\alpha =1.0 $ GeV 
and $\Lambda_\beta =0.45$ GeV.
\newline
\indent
Besides the measurement in the resonance region, the 
$e p \to e p \gamma$ reaction has also been measured at 
JLab below pion threshold for three values of the outgoing photon energy  
at $Q^2$ = 1 GeV$^2$ \cite{Deg01}, 
and at $Q^2$ = 1.9 GeV$^2$ \cite{Jam00}.  
For those kinematics, we show in 
Fig.~\ref{fig:vcs_jlab_bel_thr} the differential cross sections
as well as the non-Born effect relative to the BH + B cross section. 
It is seen from 
Fig.~\ref{fig:vcs_jlab_bel_thr} that the sensitivity to the GPs
is largest where the BH + B cross section becomes small, in
particular in the angular region between 0$^\mathrm{o}$ and 50$^\mathrm{o}$. 
In Fig.~\ref{fig:vcs_jlab_bel_thr}, we show the non-Born effect
for different values of the GPs. 
From the JLab data below pion threshold, the 
two unpolarized structure functions $P_{LL} - P_{TT}/\varepsilon$ and $P_{LT}$ 
have been extracted at 
$Q^2$ = 1 GeV$^2$ and at $Q^2$ = 1.9 GeV$^2$ \cite{Fon02}. 
For this extraction below pion threshold, both the LEX and the DR
formalisms can be used. A nice agreement between the results of both
methods for the structure functions was found in Ref.~\cite{Fon02}.
The preliminary results at $Q^2$ = 1 GeV$^2$ and at $Q^2$ = 1.9 GeV$^2$ 
for $P_{LL}$ \footnote{The present experiments, which are
performed at a fixed value of $\varepsilon$ only measure the combination
$P_{LL} - P_{TT}/\varepsilon$. To extract $P_{LL}$ from these data, we 
calculate the relatively small (spin-flip) contribution $P_{TT}$
(shown by the curves labeled S1 on the left panel 
of Fig.~\ref{fig:meaning_pol}) in the DR formalism and subtract it
from the measured value.} 
and $P_{LT}$ are displayed in Fig.~\ref{fig:vcs_asy_ff}, alongside 
the RCS point and the results at $Q^2 = 0.33$ GeV$^2$. By dividing out
the form factor $G_E$, one sees from Eq.~(\ref{eq:unpolobsgp1}) that 
$P_{LL}$ is proportional to the electric GP $\alpha(Q^2)$, whereas $P_{LT}$ 
is proportional to the magnetic GP $\beta(Q^2)$ plus some correction due
to the spin flip GP $P^{(L1,L1)1}$ which turns out to be small in the
DR formalism as discussed further on. 
One sees from Fig.~\ref{fig:vcs_asy_ff}
that the best fit value for 
$\Lambda_\alpha \simeq 0.92$~GeV 
yields an electric polarizability which is dominated by the asymptotic
contribution and has a similar $Q^2$ behavior as the dipole form
factor. However, the best fit value for 
$\Lambda_\beta \simeq 0.66$~GeV is substantially lower, indicating
that the diamagnetism, which is related to pionic degrees of freedom,
drops faster with $Q^2$. One nicely sees that the data confirm the
interplay between para- and dia-magnetism in $\beta$ as function of $Q^2$. 
\newline
\indent
Until now, we discussed only unpolarized VCS observables. 
An unpolarized VCS experiment gives access to only 3 combinations of
the 6 GPs, as given by Eqs.~(\ref{eq:unpolobsgp1}) - (\ref{eq:unpolobsgp3}). 
It was shown in Ref.~\cite{Vdh97} that VCS double
polarization observables with polarized lepton beam and polarized target
(or recoil) nucleon,
will allow us to measure three more combinations of GPs. Therefore a
measurement of unpolarized VCS observables (at different values of 
$\varepsilon$) and of 3 double-polarization observables 
will give the possibility to disentangle all 6 GPs. 
The VCS double polarization observables, which are denoted by 
\(\Delta \calm (h, i) \) for an electron of helicity $h$, are defined as 
the difference of the squared amplitudes for recoil (or target) proton
spin orientation in the direction and opposite to 
the axis $i$ ($i = x, y, z$), where 
the $z$-direction is chosen along the virtual photon momentum 
(see Ref.~\cite{Vdh97} for details). 
In a LEX, this polarized squared amplitude yields~:   
\begin{equation}
\dcalm^{\rm exp}=\frac{\dcalm^{\rm exp}_{-2}}{\rmqp^2}
+\frac{\dcalm^{\rm exp}_{-1}}{\rmqp}+\dcalm^{\rm exp}_0+O(\rmqp) \,. \;\;
\label{eq_3_50}
\end{equation}
Analogous to the unpolarized squared amplitude (\ref{eq:unpolsqramp}), 
the threshold coefficients $\dcalm^{\rm exp}_{-2}$, 
$\dcalm^{\rm exp}_{-1}$ are known due
to the LET. 
It was found in Ref.~\cite{Vdh97} that 
the polarized squared amplitude $\Delta \calm^{\rm exp}_0$ 
can be expressed in terms of three new structure functions
$P_{LT}^z(\rmq)$, $P_{LT}^{'z}(\rmq)$, and $P_{LT}^{'\perp}(\rmq)$. 
These new structure functions are related to the spin GPs according to
Refs.~\cite{Vdh97,Gui98}~: 
\begin{eqnarray}
&&P_{LT}^z \;=\; \frac{3 \, Q \,\rmq}{2 \, \qt0} \, G_M \, P^{(L1,L1)1}
\,-\, \frac{3\, M \, \rmq}{Q} \, G_E \, P^{(M1,M1)1}, \hspace{.4cm}\\
&&P_{LT}^{'z} \;=\; -\frac{3}{2}\, Q \, G_M \, P^{(L1,L1)1}
\,+\, \frac{3 \, M \, \rmq^2}{Q \, \qt0} \, G_E \, P^{(M1,M1)1}, 
\hspace{.4cm}\\
&&P_{LT}^{'\perp} \;=\; \frac{3 \,\rmq \, Q}{2 \, \qt0} \,G_M \, 
\left(P^{(L1,L1)1} \,-\, \sqrt{\frac{3}{2}} \, \qt0 \, P^{(M1,L2)1}\right).
\hspace{.4cm} 
\label{eq:polobsgp}
\end{eqnarray}
While $P_{LT}^z$ and $P_{LT}^{'z}$ can be accessed by in-plane
kinematics ($\phi = 0^{\mathrm{o}}$), the measurement of $P_{LT}^{'\perp}$
requires an out-of-plane experiment. 
\newline
\indent
In Fig.~\ref{fig:vcs_mami_doublepol}, 
we show the dispersion results for the double polarization
observables, with polarized electron and by measuring the recoil
proton polarization either along the virtual photon direction
($z$-direction) or parallel to the reaction plane and perpendicular to
the virtual photon ($x$-direction). The double polarization
asymmetries are quite large (due to a non-vanishing
asymmetry for the BH + B mechanism), but the DR calculations show
only small relative effects due to the spin GPs below pion threshold. 
However, a heavy-baryon chiral perturbation theory (HBChPT) 
calculation to ${\mathcal O}(p^3)$ \cite{Hem00} shows a 
significantly larger effect due to larger values of the spin GPs 
in this calculation, as will be discussed in the next section. 
Although these double polarization observables are tough to measure, 
a first test experiment is already planned at MAMI \cite{dHo01}.


\subsection{Physics content of the nucleon generalized polarizabilities}

\begin{figure}[ht]
\epsfxsize=8.75cm
\centerline{\epsffile{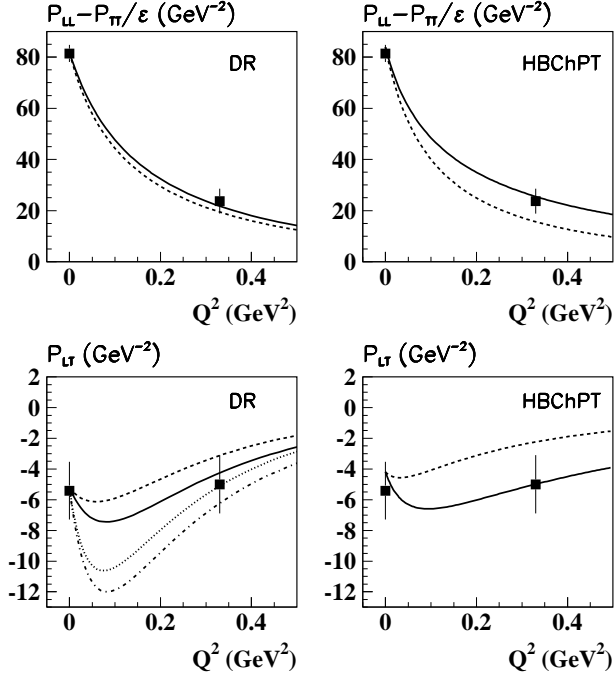}}
\vspace{-0.3cm}
\caption[]{
Comparison between the VCS unpolarized structure functions 
calculated within the DR formalism \cite{Pas01} and 
the ${\mathcal O}(p^3)$ HBChPT~\cite{Hem00,Hem97}.
Upper panels :
results for the $P_{LL} - P_{TT}/\varepsilon$ 
structure function for $\varepsilon = 0.62$ predicted from 
DR (left) and ${\mathcal O}(p^3)$ HBChPT (right).
For the DR predictions, the contribution from the electric GP
$\alpha(Q^2)$ for $\Lambda_\alpha = 1.4$ GeV (dashed curve), is
compared with the sum of the contributions from the scalar 
and spin-flip GPs (solid curve).
Lower panels : results for $P_{LT}$ within DR (left) and 
${\mathcal O}(p^3)$ HBChPT (right).
In the left panel, the contribution from $\beta(Q^2)$   
is shown for the values 
$\Lambda_\beta = 0.6$ GeV (dashed curve) and $\Lambda_\beta = 0.4$ GeV 
(dotted curve).
The total results, sum of the contributions from scalar and spin-flip 
GPs, are shown for $\Lambda_\beta = 0.6$ GeV (solid curve)
and for $\Lambda_\beta = 0.4$ GeV (dashed-dotted curves).
In the right panel, the predictions from HBChPT are shown
for the contribution from $\beta(Q^2)$ alone (dashed curve),
and for the total result (solid curve), which
includes the GP $P^{(L1,L1)1}$. 
The RCS data are from Ref.~\cite{Olm01}, and the VCS data at $Q^2=0.33
\mbox{\, GeV}^2$ from Ref.~\cite{Roc00}.}
\label{fig:plot_compreh}
\end{figure}

\begin{figure}[ht]
\epsfxsize=11cm
\centerline{\epsffile{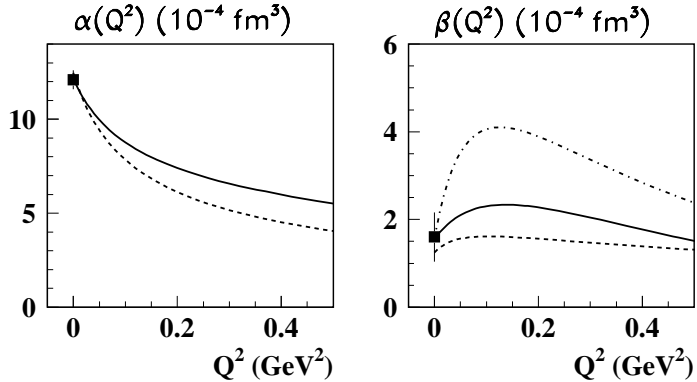}}
\vspace{-5.75cm}
\caption[]{
Left panel : comparison between the results for the electric GP 
$\alpha(Q^2)$ predicted by the DR  formalism for  $\Lambda_\alpha = 1.4$  
GeV (solid curve) and ${\mathcal O}(p^3)$ HBChPT (dashed curve).
Right panel : comparison between the results for the 
magnetic GP $\beta(Q^2)$ predicted by the DR formalism for  
$\Lambda_\beta = 0.6$ GeV (full line) 
and  $\Lambda_\beta = 0.4$ (dashed-dotted line), 
and by ${\mathcal O}(p^3)$ HBChPT (dashed line).}
\label{fig:plot_compreh2}
\end{figure}

\begin{figure}[ht]
\epsfxsize=9.75cm
\centerline{\hspace{1cm}\epsffile{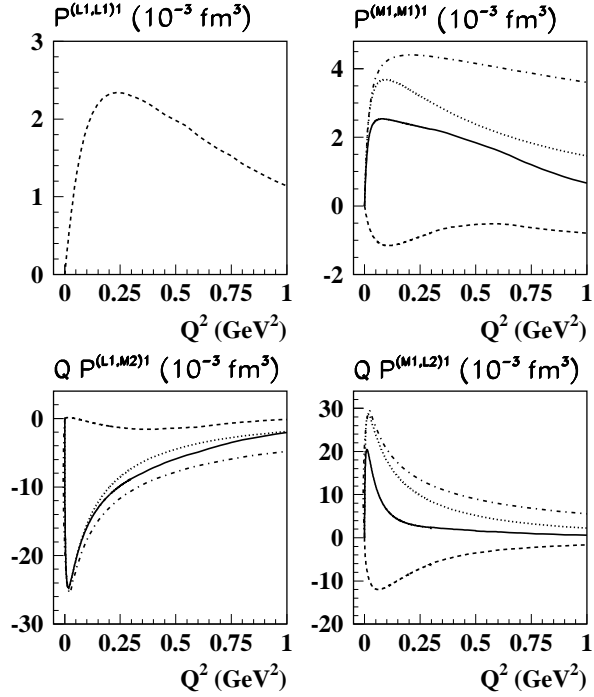}}
\vspace{-.25cm}
\caption[]{$Q^2$-dependence of the spin-flip GPs 
as calculated in Refs.~\cite{Pas01,Pas00}.
The dashed curves correspond to the dispersive $\pi N$ contribution,
the dotted curves show the $\pi^0$-pole contribution, and the
solid curves are the sum of the dispersive and $\pi^0$-pole
contributions.
For comparison, we also show the $\pi^0$-pole contribution when setting
the $\pi^0 \gamma^* \gamma$ form factor equal to 1 (dashed-dotted curves).
Note that $P^{(L1,L1)1}$ has no $\pi^0$-pole contribution.}
\label{fig:polarizab_anom}
\end{figure}

Having discussed the present status of the VCS experiments and the
combinations of GPs which have been extracted from such experiments so
far, we now turn in some more detail to the physics content of these
GPs and compare different model predictions.  
\newline
\indent
We start our discussion with the VCS unpolarized structure
functions as shown in Fig.~\ref{fig:plot_compreh}, and compare the DR
results of Ref.~\cite{Pas01} with the ${\mathcal O}(p^3)$
HBChPT calculations~\cite{Hem00,Hem97}. The DR results have been
shown before in Fig.~\ref{fig:meaning_pol}, where we discussed the
different mass scales parametrizing the asymptotic parts in the GPs 
$\alpha$ and $\beta$. In Fig.~\ref{fig:plot_compreh}, we show in
addition the effect of the spin GPs on these response functions and
compare them with the corresponding calculation in HBChPT. One notices
that the effect of the spin GPs is much smaller in the DR calculation
than in ${\mathcal O}(p^3)$ HBChPT, in particular for the spin GPs
entering $P_{TT}$. The good agreement with the data found in the 
${\mathcal O}(p^3)$ HBChPT calculation is for an important part due to the 
larger size of the spin GPs in this calculation. 
\newline
\indent
The comparison between the spin independent GPs in both calculations
is shown in Fig.~\ref{fig:plot_compreh2}. From this figure, we see
a qualitative agreement between both the DR and the ${\mathcal
O}(p^3)$ HBChPT results. In particular, we see that in both
calculations the $Q^2$ dependence of the electric and magnetic GPs is
quite different. The electric GP shows a rather smooth $Q^2$ behavior,
much as the nucleon electric form factor, whereas the magnetic GP has
a characteristic structure at small $Q^2$. In the DR calculation, this
results due to a cancellation between a large paramagnetic $\Delta$
contribution and a diamagnetic contribution (due to $t$-channel
$\pi\pi$ exchange) which have a different $Q^2$ behavior, as was
already noticed in the early 
effective Lagrangian calculation of Ref.~\cite{Vdh96}.
In the ${\mathcal O}(p^3)$ HBChPT, this structure in $\beta(Q^2)$, at
low $Q^2$, results from $\pi N$ loop effects. 
By Fourier transforming the GPs $\alpha(Q^2)$ 
\footnote{In the notation of Ref.~\cite{LSPUD01}, $\alpha(Q^2)$  
is denoted as the so-called {\it longitudinal} electric GP 
$\alpha_L$, to distinguish it from a higher order (in the
outgoing photon energy) so-called {\it transverse} electric GP 
$\alpha_T$.} 
and $\beta(Q^2)$ in the Breit frame, it was argued in \cite{LSPUD01} 
that one obtains a spatial distribution of the induced 
electric polarization $\alpha(r)$ and magnetization $\beta(r)$ of the
nucleon. The picture which then emerges from the $\pi N$ loop contribution 
in the HBChPT calculation is as expected from a classical
interpretation of diamagnetism. Due to a change in the external
magnetic field, pionic currents start circulating around the nucleon, 
and give rise to an induced magnetization, opposite to the applied
field. This diamagnetic effect leads at distances $r \geq 1/m_\pi$ 
to an negative value for $\beta(r)$, whereas for distances $r \leq 1/m_\pi$ 
the paramagnetism dominates and $\beta(r)$ is positive. 
Therefore, as the momentum transfer $Q^2$ increases, the negative
long-distance contribution to the magnetic GP due to the pion cloud,
no longer contributes and hence $\beta(Q^2)$ increases. This nicely
explains the positive slope of $\beta(Q^2)$ at $Q^2$ = 0 and the 
characteristic turn-over at low $Q^2$ in the HBChPT
calculation as is seen in Fig.~\ref{fig:plot_compreh2}. 
Hence it will be interesting to see the results of a measurement
around $Q^2 \simeq 0.05 - 0.1$~GeV$^2$ for $P_{LT}$ performed 
at MIT-Bates~\cite{Mis97}
to reveal the nature of the diamagnetism in the nucleon. 
\newline
\indent
We next discuss the spin-flip GPs.  
In Fig.~\ref{fig:polarizab_anom}, we show the dispersive and $\pi^0$-pole
contributions to the 4 spin GPs as well as their sum, 
according to the calculations of~\cite{Pas01,Pas00}.
For the presentation, we multiply in Fig.~\ref{fig:polarizab_anom}
the GPs $P^{\left(L 1, M 2\right)1}$ and $P^{\left(M 1, L 2\right)1}$
with $Q$, in order to better compare the $Q^2$ dependence when including
the $\pi^0$-pole contribution, which itself drops very fast with
$Q^2$. The $\pi^0$-pole does
not contribute to the GP $P^{\left(L 1, L 1\right)1}$, but
is seen to dominate the other three spin GPs.
It is however possible to find, besides the
GP $P^{\left(L 1, L 1\right)1}$, the two combinations
given by Eqs.~(\ref{eq:gpdisp3}) and (\ref{eq:gpdisp4})
of the remaining three spin GPs,
for which the $\pi^0$-pole contribution drops out \cite{Pas00}.

\begin{figure}[ht]
\epsfxsize=9.75cm
\centerline{\epsffile{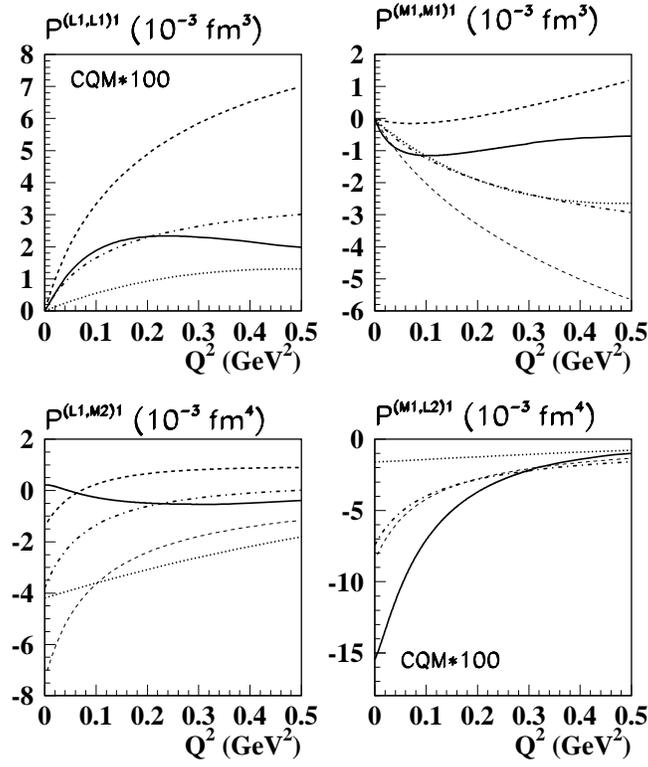}}
\vspace{-.25cm}
\caption[]{Results for the spin-flip GPs excluding the $\pi^0$-pole
  contribution in different model calculations.
The solid curves correspond to the dispersive $\pi N$ contribution 
\cite{Pas01,Pas00}.
The thin dashed curves show the results of 
${\mathcal O}(p^3)$ HBChPT~\cite{Hem00},
whereas the thick dashed curves 
for $P^{(L 1,L 1)1}$, $P^{(M 1,M 1)1}$, and $P^{(L 1,M 2)1}$ 
are the ${\mathcal O}(p^4)$ HBChPT results \cite{Kao02b}. 
The dashed-dotted curves correspond to the predictions of the
linear $\sigma$-model~\cite{Met96}, and the dotted curves are the results of
the nonrelativistic constituent quark model~\cite{PSD01}.
Note that the constituent quark model results (CQM) for $P^{(L 1,L 1)1}$ and
$P^{(M 1,L 2)1}$ are multiplied (for visibility) by a factor 100.}
\label{fig:polarizab_comp}
\end{figure}

\begin{figure}[h]
\epsfxsize=9.cm
\centerline{\epsffile{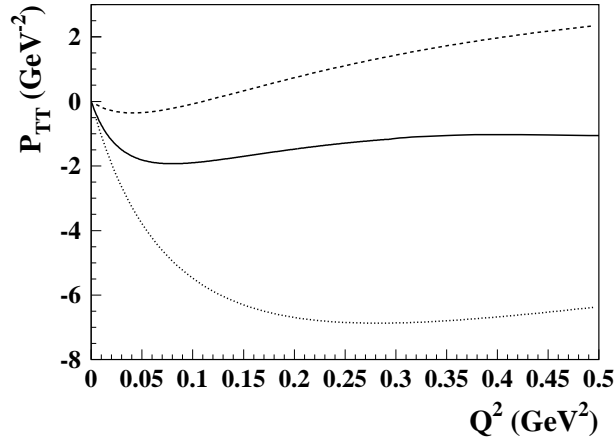}}
\vspace{-0.3cm}
\caption[]{
Results for the VCS structure function $P_{TT}$. 
Dotted curve : ${\mathcal O}(p^3)$ HBChPT~\cite{Hem97};
dashed curve : ${\mathcal O}(p^4)$ HBChPT~\cite{Kao02b}; 
solid curve : dispersive evaluation~\cite{Pas00,Pas01}.
}
\label{fig:ptt}
\end{figure}

\indent
In Fig.~\ref{fig:polarizab_comp} we show the results of the dispersive
contribution to the four spin GPs, and compare them to the results of 
the nonrelativistic constituent quark model~\cite{PSD01},
the HBChPT to ${\mathcal O}(p^3)$~\cite{Hem00,Hem97}, 
the recent HBChPT calculation to ${\mathcal O}(p^4)$~\cite{Kao02b},
and the linear $\sigma$-model~\cite{Met96}.
The constituent quark model (CQM) calculation gives negligibly small
contributions for the GPs $P^{\left(L 1, L 1\right)1}$ and
$P^{\left(M 1, L 2\right)1}$, whereas the GPs
$P^{\left(M 1, M 1\right)1}$ and $P^{\left(L 1, M 2\right)1}$
receive their dominant contribution from the excitation of the
$\Delta (1232)$ ($M1 \to M1$ transition) and the $N^*\ (1520)$
($E1 \to M2$ or $L1 \to M2$ transitions), respectively.
The smallness of $P^{\left(L 1, L 1\right)1}$ and $P^{\left(M 1, L
2\right)1}$ in the CQM can be understood by noting that those two
GPs can be expressed in terms of a GP which involves a transition from 
$L 0$ (Coulomb monopole) to $M1$, through a crossing symmetry relation 
\cite{DKK98} as~:
\begin{equation}
\sqrt{3} \, {{\rmq^2} \over {\tilde q_0}} \, P^{\left(L 1, L 1\right)1} \;=\;
P^{\left(M 1, L 0\right)1} \,+\, 
{{\rmq^2} \over \sqrt{2}} \,P^{\left(M 1, L 2\right)1} \, . 
\label{eq:gpl1l11}
\end{equation}
The GPs on the {\it rhs} of Eq.~(\ref{eq:gpl1l11}) 
encode the response to a static
magnetic dipole field ($M1$) of the nucleon charge density ($L0$) or the 
electric quadrupole density ($L2$). In a non-relativistic CQM
calculation \cite{Gui95,PSD01}, the only response to an applied 
static magnetic field is the alignment of the quark spins, whereas the
charge density or electric quadrupole density remain unchanged. 
Therefore, both GPs $P^{\left(M 1, L 0\right)1}$ and
$P^{\left(M 1, L 2\right)1}$ are vanishingly small in the quark model,
as well as $P^{\left(L 1, L 1\right)1}$ through Eq.~(\ref{eq:gpl1l11}).
Consequently, $P^{\left(M 1, L 0\right)1}$ and
$P^{\left(M 1, L 2\right)1}$ are promising observables to study the
effects of the pion cloud surrounding the nucleon. A large
contribution of pionic effects for these GPs is indeed observed
in the HBChPT and in the linear $\sigma$-model calculations. 
One furthermore notices from Fig.~\ref{fig:polarizab_comp} that 
the ${\mathcal O}(p^3)$ HBChPT predicts a 
rather strong increase with $Q^2$ for the GPs 
$P^{\left(L 1, L 1\right)1}$ and $P^{\left(M 1, M 1\right)1}$.
For $P^{\left(L 1, L 1\right)1}$ this result is confirmed by the 
${\mathcal O}(p^4)$ calculation \cite{Kao02b}. 
For the GP $P^{\left(M 1, M 1\right)1}$, it was found in Ref.~\cite{Kao02b}
that the ${\mathcal O}(p^4)$ calculation gives a large reduction compared to
the ${\mathcal O}(p^3)$ result, and calls the convergence of the
HBChPT result for this observable into question. 
The linear $\sigma$-model, which takes account of part of the higher order
terms of a consistent chiral expansion, in general results in smaller values
for the GPs $P^{\left(L 1, L 1\right)1}$ and $P^{\left(M 1, M 1\right)1}$ 
compared with the corresponding calculations 
to leading order in HBChPT. For the GP $P^{\left(L 1, M 2\right)1}$,
its value at $Q^2$ = 0, which is related to the spin polarizability 
$\gamma_3$ through Eq.~(\ref{eq_3_47}), was reported  
in Sec.~\ref{sec:polmod}. In particular, we can notice that the 
${\mathcal O}(p^4)$ HBChPT result yields a relatively large
correction, bringing it in better agreement with the DR result. 
From Fig.~\ref{fig:polarizab_comp} one notices that the $Q^2$
dependence of the ${\mathcal O}(p^4)$ HBChPT calculation 
for the GP $P^{\left(L 1, M 2\right)1}$ is rather 
weak \cite{Kao02b}, and results in a near constant reduction for this
observable compared to the ${\mathcal O}(p^3)$ calculation.
\newline
\indent
The comparison in Fig.~\ref{fig:polarizab_comp} clearly indicates
that a satisfying theoretical description of the spin-dependent GPs over a
larger range in $Q^2$ is still a challenging task. This calls for 
VCS experiments which are sensitive to the spin-dependent GPs. 
Two types of experiments can be envisaged in this regard. 
Firstly, one notices from Eq.~(\ref{eq:vcsunpol}) that 
an unpolarized VCS experiment at different values of 
$\varepsilon$ (by varying the beam energy) allows one to disentangle the 
response functions $P_{LL}$ and $P_{TT}$. The latter contains the
combination of the spin GPs $P^{\left(M 1, M 1\right)1}$ 
and $P^{\left(L 1, M 2\right)1}$ given by Eq.~(\ref{eq:unpolobsgp2}). 
In Fig.~\ref{fig:ptt}, we show the response function $P_{TT}$ and
compare the DR predictions~\cite{Pas00,Pas01} with the 
${\mathcal O}(p^3)$ HBChPT result~\cite{Hem97} and the 
${\mathcal O}(p^4)$ HBChPT result~\cite{Kao02b}. 
One notices large corrections at ${\mathcal O}(p^4)$ to the HBChPT
result. Therefore, the main difference between the DR result and the 
${\mathcal O}(p^3)$ HBChPT result for the measured response function 
$P_{LL} - P_{TT}/\varepsilon$, as shown in the upper panels of 
Fig.~\ref{fig:plot_compreh}, is largely reduced by the 
${\mathcal O}(p^4)$ HBChPT calculations. It will be very worthwhile to
directly measure the response function $P_{TT}$ which will provide an 
interesting check on our understanding of the 
spin densities of the nucleon, and allow to 
extract the electric polarizability $\alpha(Q^2)$ unambiguously from
the measurement of $P_{LL} - P_{TT}/\varepsilon$.
\newline
\indent
To access the other spin GPs, which do not appear in $P_{TT}$
it was discussed before that one has to
resort to double polarization observables. 
It was shown in Fig.~\ref{fig:vcs_mami_doublepol} that such 
observables are particularly sensitive to the different 
predictions for spin GPs, and are very promising to
measure in the near future \cite{dHo01}, so as 
to sharpen our understanding of the spin
dependent response of the nucleon to an applied electromagnetic field.  


\section{Conclusions and perspectives}
\label{sec:concl}

In this review, we have applied dispersion relations to real
and virtual Compton scattering processes off a nucleon as a powerful
tool to connect different observables and to extract nucleon structure
quantities.
\newline
\indent For forward real Compton scattering, sum rules directly
connect low energy quantities to the polarized or unpolarized
total absorption cross sections. We discussed in some detail the
recent evaluations of the Baldin sum rule and the status of the
GDH sum rule. The latter involves an integral over the helicity
difference cross section $\sigma_{1/2} - \sigma_{3/2}$, for photon
and proton helicities having the same or opposite signs. This
helicity difference cross section for the proton has now been
measured at MAMI and ELSA through the resonance region, up to $W
\lesssim 2.5$~GeV. It displays a region around pion threshold which
is dominated by S-wave pion production, for which $\sigma_{1/2}$
dominates. Furthermore, these data clearly exhibit three resonance
regions with dominance of $\sigma_{3/2}$. By performing the GDH
integral up to $W \lesssim 2$~GeV, one overestimates the sum rule
value for the proton by about 15~\%, indicating that the anomalous
magnetic moment is mostly related to the low-lying degrees of
freedom. A measurement of $\sigma_{1/2} - \sigma_{3/2}$ at
energies up to $W \lesssim 9$~GeV will be performed in the near
future at SLAC, in order to find out whether the present
``oversaturation'' of the sum rule will be removed by high-energy
contributions. Such an experiment will be quite important, because
it will test both our understanding of soft Regge physics in the
spin-dependent forward Compton amplitude and the validity of
high-energy extrapolations of DIS data at large $Q^2$ to the real
photon point.
\newline
\indent For the neutron, the convergence of the GDH sum rule is
less clear at the moment because of a lack of data. Theoretical
estimates based on our present knowledge of pion photoproduction
multipoles yield only about 85~\% of the sum rule value. This may
point to systematic deficiencies in these multipoles, which have
mostly been obtained from experiments on a deuteron target, or to
large contributions from higher intermediate states, such as two
pions. It is therefore of extreme interest to see the outcome of
dedicated experiments on the neutron which are planned in the near
future at MAMI and GRAAL.
\newline
\indent Extending the sum rules to forward scattering of spacelike
virtual photons, we have shown how to relate nucleon structure
quantities to the inclusive electroproduction cross sections. The
unpolarized cross section (weighted with $1/\nu^2$) leads to a
generalization of Baldin's sum rule, whereas the polarized cross
sections (weighted with $1/\nu^3$) lead to two nucleon spin
polarizabilities. We estimated these quantities at low and
intermediate $Q^2$ by a phenomenological model (MAID), and at
large $Q^2$ by the corresponding moments of DIS structure
functions. As a result, we find that a transition occurs around
$Q^2 \simeq 1 - 2$~GeV$^2$ from a resonance-dominated description
at lower $Q^2$ to a partonic description at larger $Q^2$.
Furthermore, we also studied the generalized GDH integrals, using
very recent experimental results at intermediate $Q^2$ values
measured at SLAC, HERMES, JLab/CLAS, and JLab/HallA. In
particular, the JLab/CLAS data for the helicity difference cross
section $\sigma_{1/2} - \sigma_{3/2}$ of the proton in the range
$Q^2 \simeq 0.15 - 1.2$~GeV$^2$, clearly demonstrate a sign change
from a large negative value at low $Q^2$, where $\sigma_{3/2}$
dominates due to resonance excitation, to the positive DIS value
at larger $Q^2$, where $\sigma_{1/2}$ survives. We have shown that
this transition can also be nicely understood in a quantitative
way. For the proton-neutron difference, where isospin 3/2
resonances such as the $\Delta$ drop out, the validity of chiral
perturbation theory (ChPT) extends towards somewhat larger $Q^2$,
and there is hope to bridge the gap between ChPT and perturbative
QCD, which eventually leads to the well established Bjorken sum
rule at large $Q^2$.
\newline
\indent In Sec.~\ref{sec:rcs}, we extended the dispersion
formalism for forward scattering to real Compton scattering (RCS)
on the nucleon for all angles. At low photon energies, this
process has a well-known low-energy limit, the Thomson term, which
is determined by the total mass and electric charge of the system.
Moving to larger photon energies, one can identify the higher
order terms in a low energy expansion (LEX) with the response of
the nucleon to an external electromagnetic field, parametrized by
dipole and higher order nucleon polarizabilities. However, such a
LEX is only valid up to about 80 MeV photon energy, and a direct
experiment in this energy range would have to be extremely precise
to disentangle the small effects due to nucleon polarizabilities.
In practice, one also has to include experiments at higher
energies, up to and above the pion threshold, and to rely on
dispersion relations to analyze the experiments. We have reviewed
and compared several such dispersion relation formalisms for RCS.
\newline
\indent
In the literature, most of the recent experiments have been
analyzed using unsubtracted fixed-$t$ dispersion relations for the
six RCS amplitudes. In such an approach, one has to estimate the
asymptotic contributions for two of the six RCS amplitudes for
which the unsubtracted dispersion integrals do not converge. These
asymptotic contributions can be parametrized as energy independent
$t$-channel poles. In such parametrizations, the most important
contributions are the $\pi^0$-pole for one of the spin-dependent
amplitudes and a ``$\sigma$''-pole for a spin-independent
amplitude. This procedure is relatively safe for the $\pi^0$-pole
which is well established both experimentally and theoretically.
However, since the $\sigma$-meson mass and coupling constants
enter as free parameters in such a formalism, the
``$\sigma$''-pole introduces a considerable model dependence.
Instead we replace the $\sigma$-meson by existing physical
information on the $I = J = 0$ part of the two-pion spectrum,
within the formalism of fixed-$t$ dispersion relations. This has
been achieved by subtracting the fixed-$t$ dispersion relations
(at $\nu = 0$) and by evaluating the subtraction functions through
a dispersion relation in the variable $t$. The absorptive parts
entering the $t$-channel dispersion integrals can be saturated by
$\pi \pi$ intermediate states in the reaction $\gamma \gamma \to
\pi \pi \to N \bar N$, constructed by means of the
phenomenological information on the $\gamma \gamma \to \pi \pi$
and $  \pi \pi \to N \bar N$ subprocesses. In this way we found
that a consistent description for Compton scattering data at low
energy can be achieved in both formalisms. Going to higher
energies and backward scattering angles, a large part of the
integration range lies outside the physical region, and the full
amplitude has to be constructed by an analytical continuation of
the partial wave expansion. Since this expansion converges only in
a limited range, the application of fixed-$t$ dispersion relations
is restricted in practice to energies up to the $\Delta$-resonance
and to forward angles. In order to overcome this shortcoming, we
also studied fixed-angle dispersion relations, in which case the
integration range of the $s$-channel contribution falls into the
physical region. The $t$-channel dispersion integrals can be
reconstructed from a partial wave expansion which converges for
angles $\gtrsim 100^o$. Furthermore, such fixed-angle dispersion
relations can quantitatively explain the large value for the
difference of the electric and magnetic dipole polarizabilities,
$\alpha - \beta$, without invoking a $\sigma$-meson contribution.
Evaluated at $\theta = 180^o$, the predictions are $\alpha - \beta
= (10.7 \pm 0.2) \cdot 10^{-4} \, \mathrm{fm}^3$, and $\gamma_\pi
= (-38.8 \pm 1.8) \cdot 10^{-4} \, \mathrm{fm}^4$ for the backward
spin polarizability.
\newline
\indent
In conclusion, fixed-$t$ and fixed angle subtracted dispersion
relations nicely complement each other, the former being
convergent at small scattering angles and the latter at large
scattering angles. We applied this combined formalism to all
existing data. Below pion threshold, we found that all methods
nicely agree. This comparison proves that the polarizabilities can
indeed be extracted with a minimum of model dependence for the
energy range below the $\Delta$ resonance. However, subtracted
dispersion relations also provide a quantitative description of
the data through the $\Delta$ resonance.
\newline
\indent 
We have furthermore shown that the sensitivity to the
backward spin polarizability $\gamma_\pi$ can be substantially
increased by an experiment with polarized photons
hitting a polarized proton target.
Such an experiment, although challenging to perform, could
become feasible in the near future, and can teach us more about
the spin response of the nucleon to a static electromagnetic
field.
\newline
\indent 
In Sec.~\ref{sec:vcs}, we have extended the dispersion
relation formalism to virtual Compton scattering (VCS) off a
proton target, as a tool to extract generalized polarizabilities
(GPs) from VCS observables over a large energy range. The way we
evaluated our dispersive integrals using $\pi N$ intermediate
states, allows us to apply the present formalism for VCS
observables through the $\Delta(1232)$-resonance region. The
presented dispersion relation framework, when applied at a fixed
value of $Q^2$, involves two free parameters, which can be
expressed in terms of the electric and magnetic GPs, and which are
to be extracted from a fit to VCS data.
\newline
\indent We confronted our dispersive calculations with existing
VCS data taken at MAMI and JLab below pion threshold. Our
dispersion relation formalism yields results consistent with the
low-energy expansion analysis for photon energies up to about 100
MeV. When increasing the photon energy, the dispersive
calculations show that the region between pion threshold and the
$\Delta$-resonance peak displays an enhanced sensitivity to the
GPs. We also compared our dispersion relation calculations to JLab
data taken at higher photon energies, through the
$\Delta(1232)$-resonance region, and found a good agreement. The
extraction of GPs from the preliminary JLab data below and above
pion threshold yields consistent results. These data indicate a
$Q^2$ dependence of the electric GP similar to a dipole form
factor, whereas the magnetic GP follows a more complicated $Q^2$
behavior. As was already shown for RCS, the magnetic dipole
transition involves a strong cancellation between a diamagnetic
mechanism due to pion cloud effects and a paramagnetic
contribution due to nucleon resonance excitation. Since the cloud
effects have a considerably longer range in space than the
resonance structures, the $Q^2$ behavior of the magnetic GP is
able to disentangle both physical mechanisms, which is already
displayed in the existing data. Given this initial success, future
experiments to measure VCS observables in the $\Delta$-energy
region hold the promise to extract GPs with an enhanced precision,
within the dispersion relation formalism presented here.
\newline
\indent
Besides the VCS experiments without polarization degrees of freedom,
which give access to a combination of only 3 of 6 GPs,
we investigated the potential of double polarization VCS observables.
In fact, a first double polarization experiment is now underway at
MAMI. Although such investigations will be challenging indeed,
they are prerequisite to access and quantify the full set of scalar
and spin GPs of the nucleon.
\newline
\indent In conclusion we find that dispersion relations are indeed
a powerful tool to analyze real and virtual Compton scattering
processes, linking low-energy structure quantities to the
excitation spectrum of the nucleon. Though the experiments with
virtual photons have only become feasible very recently, they have
opened up a new and systematic way to map out, in quantitative
detail, the transition from hadronic degrees of freedom at low
virtuality to partonic degrees of freedom at large virtuality. We
are looking forward to increasing theoretical and experimental
activities in the fields of both real and virtual Compton
scattering, and hope that the present review will be useful to
stimulate and analyze such further work.


\section*{\center{Acknowledgements}}

This work was supported by the Deutsche Forschungsgemeinschaft
(SFB 443), and the European Centre for
Theoretical Studies in Nuclear Physics and Related Areas (ECT*).
We also like to thank for the hospitality of the ECT* (Trento) and its director
W. Weise for hosting two Collaboration meetings related to the subjects of
this paper, ``Real and Virtual Compton Scattering off the Nucleon'' in
2001 and ``Baryon structure probed with quasistatic electromagnetic
fields'' in 2002.
These meetings provided an excellent and stimulating atmosphere with
lively discussions which shaped much of the material presented here.
\newline
\indent
We would like to express our gratitude to
M. Gorchtein,
B. Holstein,
S. Kamalov,
C.W. Kao,
A. Metz,
T. Spitzenberg,
and L. Tiator,
in collaboration with whom some of the results, that are reviewed in
this work, were obtained.
\newline
\indent
Furthermore we would also like to thank
J. Ahrens,
H.J. Arends,
P.Y. Bertin,
V. Burkert, 
J.P. Chen,
N. d'Hose,
G. Dodge, 
H. Fonvieille,
H. Griesshammer,
P.A.M. Guichon,
D. Harrington,
T. Hemmert,
R. Hildebrandt,
C. Hyde-Wright,
G. Laveissi\`ere,
A. L'vov,
H. Merkel,
Z.-E. Meziani,
S. Scherer,
R. Van de Vyver,
L. Van Hoorebeke,
T. Walcher,
and W. Weise,
for many useful and stimulating discussions.

\begin{appendix}
\section{$t$-channel $\pi \pi$ exchange}
\label{sec:app1}

We express the invariant amplitudes $A_i (\nu, t)$ ($i$ = 1,...,6) 
in terms of the $t$-channel helicity amplitudes $T_{\lambda_N
  \lambda_{\bar N}, \, \lambda'_{\gamma} \lambda_\gamma}^t(\nu, t)$, 
for which we have found the expressions
\begin{eqnarray}
A_1 &=& {{1} \over{t \, \sqrt{t - 4 M^2}}} 
\left\{ \; \left[ T_{{1 \over 2}\,{1 \over 2},1\,1}^t \;+\;
T_{{1 \over 2}\,{1 \over 2},-1\,-1}^t \right]
\;-\; {{2 \, \nu \, \sqrt{t}} \over {\sqrt{su - M^4}}}\; 
T_{{1 \over 2}\,-{1 \over 2},1\,1}^t \right\} \;, \nonumber \\
A_2 &=& {{1} \over{t \, \sqrt{t}}} 
\left\{ \;- \left[ T_{{1 \over 2}\,{1 \over 2},1\,1}^t \;-\;
T_{{1 \over 2}\,{1 \over 2},-1\,-1}^t \right]
\;-\; {{2 \, \nu \, \sqrt{t - 4 M^2}} \over {\sqrt{su - M^4}}}\; 
T_{{1 \over 2}\,-{1 \over 2},1\,1}^t \right\} \;, \nonumber \\
A_3 &=& {{M^2} \over {su - M^4}} \; {{1} \over{\sqrt{t - 4 M^2}}} 
\left\{ \;2 \; T_{{1 \over 2}\,{1 \over 2},1\,-1}^t 
\;+\; {{\sqrt{su - M^4}} \over {\nu \,\sqrt{t}}} 
\;\left[ T_{{1 \over 2}\,-{1 \over 2},1\,-1}^t \;+\;
T_{{1 \over 2}\,-{1 \over 2},-1\,1}^t \right] \right\} \;, \nonumber\\
A_4 &=& {{M^2} \over {su - M^4}} \; {{1} \over{\sqrt{su - M^4}}} 
\left\{ \;M \left[- \; T_{{1 \over 2}\,-{1 \over 2},1\,-1}^t 
\;+\;T_{{1 \over 2}\,-{1 \over 2},-1\,1}^t \right] \right. \nonumber\\
&&\hspace{3.5cm}\left.
\;+\; {{\sqrt{t} \, \sqrt{t - 4 M^2}} \over {4 \, \nu}} 
\;\left[ T_{{1 \over 2}\,-{1 \over 2},1\,-1}^t \;+\;
T_{{1 \over 2}\,-{1 \over 2},-1\,1}^t \right] \right\} \;, \nonumber\\
A_5 &=& {{\sqrt{t - 4 M^2}} \over {4 \, \nu \, \sqrt{t} \, \sqrt{su - M^4}}} 
\left\{ \;-2 \; T_{{1 \over 2}\,-{1 \over 2},1\,1}^t \right\} \;, \nonumber\\
A_6 &=& {{\sqrt{t - 4 M^2}} \over {4 \, \nu \, \sqrt{t} \, \sqrt{su - M^4}}} 
\left\{ \; \left[ T_{{1 \over 2}\,-{1 \over 2},1\,-1}^t \;+\;
T_{{1 \over 2}\,-{1 \over 2},-1\,1}^t \right] \right\} \;.
\label{eq:thelampl}
\end{eqnarray}

We decompose
the $t$-channel helicity amplitudes for $\gamma \gamma
\rightarrow N \bar N$ into a partial wave series,
\begin{equation}
T_{\lambda_N \lambda_{\bar N}, \, \lambda'_{\gamma} \lambda_\gamma}^t 
(\nu, t) \;=\; \sum_J {{2 J + 1} \over {2}} \;
T_{\lambda_N \lambda_{\bar N}, \, \lambda'_{\gamma}
  \lambda_\gamma}^{J}
(t)\; d^J_{\Lambda_N \Lambda_\gamma} (\theta_t) \;,
\label{eq:pwgagannbar}
\end{equation}  
where $d^J_{\Lambda_N \Lambda_\gamma}$ are Wigner $d$-functions and
$\theta_t$ is the scattering angle in the $t$-channel, which is
related to the invariants $\nu$ and $t$ by 
$\cos \theta_t = {4 \, M \, \nu} / {\sqrt{t} \, \sqrt{t - 4 M^2}}$. 
We calculate 
the imaginary parts of the $t$-channel helicity amplitudes 
$T_{\lambda_N \lambda_{\bar N}, \, \lambda'_{\gamma} \lambda_\gamma}^t
(\nu, t)$ through the unitarity equation by inserting $\pi \pi$
intermediate states, which should give the dominant
contribution below $K \bar K$ threshold,
\begin{equation}
2\,\mathrm{Im} T_{\gamma \gamma \, \rightarrow \, N \bar N} \;=\; 
{1 \over { {(4 \pi)}^2 } } 
\;{ | {\vec {p}}_{{\pi}} | \over {\sqrt {t}} } 
\; \int {d \Omega}_{{\pi}}\,
\left [ \, T_{\gamma \gamma \, \rightarrow \, \pi \pi}\, \right ] 
\; \cdot \; 
\left [ \, T_{\pi \pi \, \rightarrow \, N \bar N}\, \right ] ^ { \ast }.
\end{equation}
Combining the partial wave expansion for $\gamma \gamma \rightarrow \pi
\pi$~,
\begin{equation}
T^{\gamma \gamma \, \rightarrow \, \pi \pi}
_{{\Lambda}_{\gamma}} (t, \theta_{\pi \pi}) \;=\;
\sum_{J even} \, {{2J+1} \over 2} \; 
T^{J \;(\gamma \gamma \, \rightarrow \, \pi \pi)}
_{{\Lambda}_{\gamma}}(t) \, \cdot \,
\sqrt {(J-\Lambda_{\gamma})! \over {(J+\Lambda_{\gamma})!}} \, \cdot \,
P_J^{\Lambda_{\gamma}}(\cos \theta_{\pi \pi}),
\label{eq:partialgagapipi}
\end{equation}
and the partial wave expansion for $\pi \pi \rightarrow N \bar N$,
\begin{equation}
T^{\pi \pi \, \rightarrow \, N \bar N}
_{\Lambda_N} (t, \Theta) \;=\;
\sum_{J} \, {{2J+1} \over 2} \; 
T^{J \; (\pi \pi \, \rightarrow \, N \bar N)}
_{\Lambda_N}(t) \, \cdot \,
\sqrt {(J-\Lambda_N)! \over {(J+\Lambda_N)!}} \, \cdot \,
P_J^{\Lambda_N}(\cos \Theta) \;.
\label{eq:partialpipinnbar}
\end{equation}
We can now construct the imaginary parts of the
Compton $t$-channel partial waves, 
\begin{equation}
2\,\mathrm{Im} T^{J \;(\gamma \gamma \, \rightarrow \, N \bar N)}
_{\lambda_N \lambda_{\bar N}, \, \lambda'_{\gamma} \lambda_\gamma}(t)\;=\; 
{1 \over {(8 \pi)} } 
\; {p_{\pi} \over \sqrt{t}}
\left [ \, T^{J \;(\gamma \gamma \, \rightarrow \, \pi \pi)}
_{\Lambda_\gamma}(t)\, \right ] 
\left [ \, T^{J \;(\pi \pi \, \rightarrow \, N \bar N)}
_{\Lambda_N }(t)\, \right ] ^ { \ast }.
\label{eq:partialtunit}
\end{equation}
The partial wave amplitudes $T^{J \;(\pi \pi \, \rightarrow \, N
  \bar N)} _{\Lambda_N}$ of
Eq.~(\ref{eq:partialpipinnbar}) are related to the
amplitudes $ f^J_{\pm}(t)$ of Frazer and Fulco
\cite{Frazer60} by the relations
\begin{eqnarray}
T^{J \;(\pi \pi \, \rightarrow \, N \bar N)}_{\Lambda_N = 0}(t)\;
&&= \frac{16 \pi}{p_N}\,(p_N \; p_\pi)^J \, \cdot \,
f^J_+ (t)\ ,\nonumber\\
T^{J \;(\pi \pi \, \rightarrow \, N \bar N)}_{\Lambda_N = 1}(t)\;
&&= 8 \pi \,\frac{\sqrt {t}}{p_N}\,(p_N \; p_\pi)^J \, \cdot \,
f^J_- (t)\ ,
\label{eq:frazerfulco}
\end{eqnarray}
with $p_N$ and $p_{\pi}$ the c.m. momenta of nucleon and pion
respectively ($p_N = \sqrt{t/4 - M^2}$ and $p_\pi = \sqrt{t/4 - m_\pi^2}$).
For the reaction $\gamma \gamma \, \rightarrow \, \pi \pi$, we will
use the partial wave amplitudes 
$F_{J\,\Lambda_\gamma}(t)$, which are
related to those of Eq.~(\ref{eq:partialgagapipi}) by
\begin{equation}
T^{J \;(\gamma \gamma \, \rightarrow \, \pi \pi)} _{\Lambda_\gamma}(t)
\;=\; \frac{2}{\sqrt{2J+1}} \cdot F_{J\,\Lambda_\gamma}(t) \;.
\label{eq:deffgagapipi}
\end{equation}

Inserting the partial-wave expansion of Eq.~(\ref{eq:pwgagannbar}) into
Eq.~(\ref{eq:thelampl}), we can finally express the $2 \pi$ $t$-channel
contributions Im$_t A_i (\nu , t)^{2 \pi}$ by the partial wave
amplitudes for the reactions $\gamma \gamma \, \rightarrow \, \pi \pi$
and $\pi \pi \, \rightarrow \, N \bar N$,
\begin{eqnarray}
&&\mathrm{Im}_t A_1 (\nu , t)^{2 \pi} = {{1} \over {t\,\sqrt{t}}} \, 
{{p_\pi} \over { p_N^2}}\,
\left[
F_{J=0\,\Lambda_\gamma = 0}(t) \, f^{J=0\,*}_+(t) \,
\right.
\nonumber
\\
 & &
\left.
+\frac{\sqrt{5}}{2}\frac{p_\pi^2}{t}(8M^2\nu^2-su+M^4)
F_{J=2\,\Lambda_\gamma = 0}(t) \, f^{J=2\,*}_+(t) \,
-\sqrt{\frac{15}{2}}M\nu^2 \,p_\pi^2\,
F_{J=2\,\Lambda_\gamma = 0}(t) \, f^{J=2\,*}_-(t)\right] 
\;, \nonumber\\
&&\mathrm{Im}_t A_2 (\nu , t)^{2 \pi} = - \sqrt{\frac{15}{2}}
\frac{p_\pi^3}{t^{2}\,\sqrt{t}}4M\nu^2
F_{J=2\,\Lambda_\gamma = 0}(t) \, f^{J=2\,*}_-(t)
\;, \nonumber\\
&&\mathrm{Im}_t A_3 (\nu , t)^{2 \pi} = \,
\frac{\sqrt{5}}{2}
\frac{M^2}{t\sqrt{t}}\frac{p_\pi^3}{p_N^2}
\left[\sqrt{\frac{3}{2}}
 F_{J=2\,\Lambda_\gamma = 2}(t) \, f^{J=2\,*}_+(t)
-M  F_{J=2\,\Lambda_\gamma = 2}(t) \, f^{J=2\,*}_-(t)\right]
\;, \nonumber\\
&&\mathrm{Im}_t A_4 (\nu, t)^{2 \pi} = 0
\;, \nonumber\\
&&\mathrm{Im}_t A_5 (\nu , t)^{2 \pi} = 
-\,\sqrt{\frac{15}{2}}\frac{M}{t \sqrt{t}}\,p_\pi^3
F_{J=2\,\Lambda_\gamma = 0}(t) \; f^{J=2*}_-(t) \;, \nonumber\\
&&\mathrm{Im}_t A_6 (\nu , t)^{2 \pi} = 
\,-
\frac{\sqrt{5}}{2}
\frac{M}{t\sqrt{t}}\,p_\pi^3
F_{J=2\,\Lambda_\gamma = 2}(t) \; f^{J=2*}_-(t). \;
\end{eqnarray}
We note that the s-wave $(J=0)$ component of the $2\pi$ intermediate
states contributes only to $A_1$ and that only waves with
$J\ge4$ contribute to the amplitude $A_4$.


\section{Tensor basis}
\label{sec:app2}

In writing down a gauge-invariant tensor basis for VCS, we use the
combinations
of the four-momenta given in Eq.~(\ref{DDeq3.2.3}),
\begin{equation}
P = {1 \over 2} \left( p + p' \right) \;, \hspace{1cm} K = {1
\over 2} \left( q + q' \right)\;. \label{eq:pk}
\end{equation}

The 12 independent tensors $\rho^{\mu \nu}_i$ entering the VCS
amplitude of Eq.~(\ref{eq:nonborn}) and introduced in
Ref.~\cite{DKK98}, are given by~:
\begin{eqnarray}
\rho_{1}^{\mu\nu} & = & -\qqs g^{\mu\nu} + q^{\prime \mu} q^{\nu}
\vphantom{\frac{1}{1}} , \nonumber \\ \rho_{2}^{\mu\nu} & = & - (2
M \nu)^2 g^{\mu\nu} - 4 \qqs P^{\mu} P^{\nu} + 4 M \nu \, \Big(
P^{\mu} q^{\nu} + P^{\nu} q^{\prime \mu} \Big)
\vphantom{\frac{1}{1}} , \nonumber \\ \rho_{3}^{\mu\nu} & = & - 2
M \nu Q^2 g^{\mu\nu} - 2 M \nu \, q^{\mu} q^{\nu} + 2 Q^2 P^{\nu}
q^{\prime \mu} + 2 \qqs \, P^{\nu} q^{\mu} \vphantom{\frac{1}{1}}
, \nonumber \\ \rho_{4}^{\mu\nu} & = & 8 P^{\mu} P^{\nu} \kdagger
- 4 M \nu \, \Big( P^{\mu} \gamma^{\nu} + P^{\nu} \gamma^{\mu}
\Big) + i \, 4 M \nu \, \gamma_5 \, \varepsilon^{\mu \nu \alpha
\beta}
 K_{\alpha} \gamma_{\beta}
\vphantom{\frac{1}{1}} , \nonumber \\ \rho_{5}^{\mu\nu} & = &
P^{\nu} q^{\mu} \kdagger - \frac{Q^2}{2} \, \Big( P^{\mu}
\gamma^{\nu}
 - P^{\nu} \gamma^{\mu} \Big )
- M \nu \, q^{\mu} \gamma^{\nu} -\frac{i}{2} \, Q^2 \, \gamma_5 \,
\varepsilon^{\mu\nu\alpha\beta}
 K_{\alpha} \gamma_{\beta} ,
\nonumber \\ \rho_{6}^{\mu\nu} & = & - 8 \qqs P^{\mu} P^{\nu} + 4
M \nu \, \Big( P^{\mu} q^{\nu} + P^{\nu} q^{\prime \mu} \Big) + 4
M \qqs \, \Big( P^{\mu} \gamma^{\nu} + P^{\nu} \gamma^{\mu} \Big)
\vphantom{\frac{1}{1}} \nonumber\\ &&- 4 M^2 \nu \, \Big(
q^{\prime \mu} \gamma^{\nu}
 + q^{\nu} \gamma^{\mu} \Big)
+ i \, 4 M \nu \, \Big( q^{\prime \mu} \sigma^{\nu\alpha}
K_{\alpha}
   - q^{\nu} \sigma^{\mu\alpha} K_{\alpha} + \qqs \sigma^{\mu\nu} \Big)
\vphantom{\frac{1}{1}} \nonumber \\ && + i \, 4 M \qqs \, \gamma_5
\, \varepsilon^{\mu\nu\alpha\beta}
  K_{\alpha} \gamma_{\beta}
\vphantom{\frac{1}{1}} , \nonumber \\ \rho_{7}^{\mu\nu} & = &
\Big( P^{\mu} q^{\nu} - P^{\nu} q^{\prime \mu} \Big) \kdagger -
\qqs \,\Big( P^{\mu} \gamma^{\nu} - P^{\nu} \gamma^{\mu} \Big) + M
\nu \, \Big( q^{\prime \mu} \gamma^{\nu}
  - q^{\nu} \gamma^{\mu} \Big)
\vphantom{\frac{1}{1}} , \nonumber \\ \rho_{8}^{\mu\nu} & = & M
\nu \, q^{\mu} q^{\nu} + \frac{Q^2}{2} \Big( P^{\mu} q^{\nu} -
P^{\nu} q^{\prime \mu} \Big) - \qqs \, P^{\nu} q^{\mu} - M q^{\mu}
q^{\nu} \kdagger + M \qqs \, q^{\mu} \gamma^{\nu} \nonumber \\ & &
+ \frac{M}{2} Q^2 \Big( q^{\prime \mu} \gamma^{\nu}
  - q^{\nu} \gamma^{\mu} \Big)
-\frac{i}{2} \, Q^2 \Big( q^{\prime \mu} \sigma^{\nu\alpha}
K_{\alpha}
   - q^{\nu} \sigma^{\mu\alpha} K_{\alpha} + \qqs \sigma^{\mu\nu} \Big)
, \nonumber \\ \rho_{9}^{\mu\nu} & = & 2 M \nu \, \Big( P^{\mu}
q^{\nu} - P^{\nu} q^{\prime \mu} \Big) - 2 M \qqs \, \Big( P^{\mu}
\gamma^{\nu} - P^{\nu} \gamma^{\mu} \Big) + 2 M^2 \nu \, \Big(
q^{\prime \mu} \gamma ^{\nu}
  - q^{\nu} \gamma^{\mu} \Big)
\vphantom{\frac{1}{1}} \nonumber \\ & & + i \, 2 \qqs \, \Big(
P^{\mu} \sigma^{\nu\alpha} K_{\alpha}
   + P^{\nu} \sigma^{\mu\alpha} K_{\alpha} \Big)
-i \,2 M \nu \, \Big( q^{\prime \mu} \sigma^{\nu\alpha} K_{\alpha}
   + q^{\nu} \sigma^{\mu\alpha} K_{\alpha} \Big)
\vphantom{\frac{1}{1}} , \nonumber \\ \rho_{10}^{\mu\nu} & = & - 4
M \nu \, g^{\mu\nu} + 2 \, \Big( P^{\mu} q^{\nu} + P^{\nu}
q^{\prime \mu} \Big) + 4 M \, g^{\mu\nu} \kdagger - 2 M \, \Big(
q^{\prime \mu} \gamma^{\nu}
  + q^{\nu} \gamma^{\mu} \Big)
\vphantom{\frac{1}{1}} \nonumber \\ & & -2 \, i \, \Big( q^{\prime
\mu} \sigma^{\nu\alpha} K_{\alpha}
  - q^{\nu} \sigma^{\mu\alpha} K_{\alpha} + \qqs \sigma^{\mu\nu} \Big)
\vphantom{\frac{1}{1}} , \nonumber \\ \rho_{11}^{\mu\nu} & = & 4
\, \Big( P^{\mu} q^{\nu} + P^{\nu} q^{\prime \mu} \Big) \kdagger -
4 M \nu \, \Big( q^{\prime \mu} \gamma^{\nu}
  + q^{\nu} \gamma^{\mu} \Big)
+ i \, 4 \qqs \, \gamma_5 \, \varepsilon^{\mu\nu\alpha\beta}
   K_{\alpha} \gamma_{\beta}
\vphantom{\frac{1}{1}} , \nonumber \\ \rho_{12}^{\mu\nu} & = & 2
Q^2 P^{\mu} P^{\nu} + 2 M \nu \, P^{\nu} q^{\mu} - 2 M Q^2 P^{\mu}
\gamma^{\nu} -2 M^2 \nu \, q^{\mu} \gamma^{\nu} \,+ i \, 2 M \nu
\, q^{\mu} \sigma^{\nu\alpha} K_{\alpha} \vphantom{\frac{1}{1}}
\nonumber \\ & & + i \, Q^2 \, \Big( P^{\mu} \sigma^{\nu\alpha}
K_{\alpha}
  + P^{\nu} \sigma^{\mu\alpha} K_{\alpha}
  - M \nu \, \sigma^{\mu\nu} \Big)
- i \, M Q^2 \, \gamma_5 \, \varepsilon^{\mu\nu\alpha\beta}
  K_{\alpha} \gamma_{\beta}
\vphantom{\frac{1}{1}} , \label{eq:vcstensors}
\end{eqnarray}
where we follow the conventions of Bjorken and Drell~\cite{Bjo65},
in particular  $\sigma^{\mu \nu} = i/2 \, [\gamma^{\mu}, \gamma^{\nu}]$ and
$\varepsilon_{0123}$ = +1.

\end{appendix}

\end{document}